\documentclass[acmsmall, nonacm]{acmart} % USE FOR ARXIV
\usepackage{graphicx}
\usepackage{color}
\usepackage{algorithm}
\usepackage{algorithmic}
\usepackage{adjustbox}
\usepackage{listings}
\usepackage{enumitem}
\usepackage{amsthm}
\usepackage{float}
\usepackage{import}
\usepackage[T1]{fontenc}
\usepackage{qcircuit}
\usepackage{subfig}
 \usepackage[draft=true]{minted}
\usepackage{bm}
\usepackage{blkarray}
\usepackage{epsfig}
\usepackage{booktabs}
\usepackage{array}
\usepackage{microtype}
\usepackage{hyperref}
\usepackage{comment}
\usepackage{tikz}
\usepackage{bbm}
\usepackage{array}

\usetikzlibrary{arrows}

\newlength\myindent
\newcommand\bindent{%
  \begingroup
  \setlength{\itemindent}{\myindent}
  \addtolength{\algorithmicindent}{\myindent}
}
\newcommand\eindent{\endgroup}

\newcommand{\ket}[1]{\ensuremath{\left|#1\right\rangle}}
\newcommand{\bra}[1]{\ensuremath{\left\langle#1\right|}}
\newcommand{\braket}[1]{\ensuremath{\left\langle#1\right\rangle}}
\newcommand{\abs}[2][]{#1| #2 #1|}
\newcommand{\norm}[2][]{#1| \! #1| #2 #1| \! #1|}
\newcommand{\avg}[1]{\langle #1\rangle }
\newcommand{\dyad}[2]{\ket{#1}\!\bra{#2}}        %dyad
\newcommand{\ip}[2]{\langle #1|#2\rangle}      %the inner product
\newcommand{\Tr}{{\rm Tr}}
\newcommand{\ot}{\otimes}
\newcommand{\id}{\mathbbm{1}}
\global\long\def\argmin{\operatornamewithlimits{argmin}}
\global\long\def\argmax{\operatornamewithlimits{argmax}}
\newcommand{\la}{\langle}
\newcommand{\ra}{\rangle}
\newcommand{\be}{\begin{eqnarray}}
\newcommand{\ee}{\end{eqnarray}}

\long\def\ca#1\cb{} %Use for commenting out: \ca...\cb

\newcommand{\V}[1]{\ensuremath{\mathbf #1}}

\AtBeginDocument{%
  \providecommand\BibTeX{{%
    \normalfont B\kern-0.5em{\scshape i\kern-0.25em b}\kern-0.8em\TeX}}}
\acmJournal{TQC}

\makeatletter % USE FOR ARXIV
\let\@authorsaddresses\@empty % USE FOR ARXIV
\makeatother % USE FOR ARXIV

\makeatletter % USE FOR ARXIV              
\def\mdseries@tt{m} % USE FOR ARXIV    
\makeatother % USE FOR ARXIV
\newcommand{\revision}{}

\begin{document}

\title{Quantum Algorithm Implementations for Beginners}

\author{Abhijith J.}
\authornotemark[1]
\author{Adetokunbo Adedoyin}
\author{John Ambrosiano}
\author{Petr Anisimov}
\author{William Casper}
\author{Gopinath Chennupati}
\author{Carleton Coffrin}
\author{Hristo Djidjev}
\author{David Gunter}
\author{Satish Karra}
\author{Nathan Lemons}
\author{Shizeng Lin}
\author{Alexander Malyzhenkov}
\author{David Mascarenas}
\author{Susan Mniszewski}
\author{Balu Nadiga}
\author{Daniel O'Malley}
\author{Diane Oyen}
\author{Scott Pakin}
\author{Lakshman Prasad}
\author{Randy Roberts}
\author{Phillip Romero}
\author{Nandakishore Santhi}
\author{Nikolai Sinitsyn}
\author{Pieter J. Swart}
\author{James G. Wendelberger}
\author{Boram Yoon}
\author{Richard Zamora}
\author{Wei Zhu}
\author{Stephan Eidenbenz}
\authornote{abhijithj@lanl.gov; eidenben@lanl.gov; baertschi@lanl.gov; pcoles@lanl.gov; vuffray@lanl.gov; lokhov@lanl.gov. LA-UR-20-22353}
\author{Andreas B\"{a}rtschi}
\authornotemark[1]
\author{Patrick J. Coles}
\authornotemark[1]
\author{Marc Vuffray}
\authornotemark[1]
\author{Andrey Y. Lokhov}
\authornotemark[1]

\renewcommand{\shortauthors}{Abhijith J., et al.}

\affiliation{\\Los Alamos National Laboratory, Los Alamos, New Mexico 87545, USA}

\begin{abstract}
As quantum computers become available to the general public, the need has arisen to train a cohort of quantum programmers, many of whom have been developing classical computer programs for most of their careers. While currently available quantum computers have less than 100 qubits, quantum computing hardware is widely expected to grow in terms of qubit count, quality, and connectivity. This review aims to explain the principles of quantum programming, which are quite different from classical programming, with straightforward algebra that makes understanding of the underlying fascinating quantum mechanical principles optional. We give an introduction to  quantum computing algorithms and their implementation on real quantum hardware. We survey 20 different quantum algorithms, attempting to describe each in a succinct and self-contained fashion. We show how these algorithms can be implemented on IBM's quantum computer, and in each case, we discuss the results of the implementation with respect to differences between the simulator and the actual hardware runs. This article introduces computer scientists, physicists, and engineers to quantum algorithms and provides a blueprint for their implementations. 
\end{abstract}

\maketitle
% \newpage
\tableofcontents

\section{Introduction}

%- Quantum computing in BML context"

Quantum computing exploits quantum-mechanical effects---in particular superposition, entanglement, and quantum tunneling---to more efficiently execute a computation. Compared to traditional, digital computing, quantum computing offers the potential to dramatically reduce both execution time and energy consumption. These potential advantages, steady advances in nano-manufacturing, and the slow-down of traditional hardware scaling laws (such as Moore's Law) have led to a substantial commercial and national-security interest and investment in quantum computing technology in the 2010s. Recently, Google announced that it has reached a major milestone known as quantum supremacy--the demonstration that a quantum computer can perform a calculation that is intractable on a classical supercomputer \cite{arute2019quantum}. The problem tackled here by the quantum computer is not one with any direct real-world application. Nonetheless, this is a watershed moment for quantum computing and is widely seen as an important step on the road towards building quantum computers that will offer practical speedups when solving real-world problems \cite{preskill2012quantum}.   (See~\cite{DBLP:conf/coco/AaronsonC17} for a precise technical definition of quantum supremacy.)

While the mathematical basis of quantum computing, the programming model, and most quantum algorithms have been published decades ago (starting in the 1990s), they have been of interest only to a small dedicated community. We believe the time has come to make quantum algorithms and their implementations accessible to a broad swath of researchers and developers across computer science, software engineering, and other fields. The quantum programming model is fundamentally different from traditional computer programming. It is also dominated by physics and algebraic notations that at times present unnecessary entry barriers for mainstream computer scientists and other more mathematically trained scientists.

In this review, we provide a self-contained, succinct description of quantum computing and of the basic quantum algorithms with a focus on implementation. 
%We provide open-source implementations of these algorithms in a github code repository~\cite{coderepository}. 
Since real quantum computers, such as IBM Q~\cite{ibmqx}, are now available as a cloud service, we present results from simulator and actual hardware experiments for smaller input data sets. Other surveys of quantum algorithms with a different target audience and also without actual implementations include \cite{quantumzoo,childs2010quantum,mosca2012quantum,santha2008quantum,bacon2010recent,montanaro2016quantum}. Other cloud service based quantum computers are also available from Rigetti and IonQ, but in this review we will focus solely on IBM's quantum computing ecosystem. The code and implementations accompanying the paper can be found at \url{https://github.com/lanl/quantum_algorithms}.

%-#TODO: Report on implementations on real quantum computing hardware (IBM QX4)
%- Pointer to repository

\subsection{The quantum computing programming model}

Here we provide a self-contained description of the quantum computing programming model. We will define the common terms and concepts used in quantum algorithms literature. We will not discuss how the constructs explained here are related to the foundations of quantum mechanics. Interested readers are encouraged to take a look at Ref. \cite{NielsenChuang} for a more detailed account along those lines. Readers with a computer science background are referred  to Refs.~\cite{yonofskyBook2008, rieffulBook2011, liptonBook2014}, for a more comprehensive introduction to quantum computing from a computer science perspective.

Quantum computing basically deals with the manipulation of quantum systems. The physical details of this is dependent on the quantum computer's hardware design. Here we will only talk about the higher level abstractions used in quantum computing: a typical programmer will only be exposed to these abstractions. The state of any quantum system is always represented by a vector in a complex vector space (usually called a Hilbert space). Quantum algorithms are always expressible as  transformations acting on this vector space. These basic facts follow from the axioms of quantum mechanics. Now we will explain some of the basic concepts and terminology used in quantum computing.

%\paragraph{{\bf Qubit}}
\subsubsection{The qubit}

The qubit (short for 'quantum bit') is the fundamental information carrying unit used in quantum computers. It can be seen as the quantum mechanical generalization of a bit used in classical computers. More precisely, a qubit is a two dimensional quantum system. The state of a qubit can be expressed as,
\begin{align}
\label{eq:eqn1}
\ket{\phi} = \alpha \ket{0} + \beta \ket{1}\,.
\end{align}
Here $\alpha$ and $\beta$ are complex numbers such that, $|\alpha|^2 +  |\beta|^2 = 1.$
In the \emph{ket-notation} or the \emph{Dirac  notation}, $\ket{0} = \begin{pmatrix} 1 \\ 0 \end{pmatrix} $ and $\ket{1} = \begin{pmatrix} 0 \\ 1 \end{pmatrix}$ are shorthands for the vectors encoding the two basis states of a two dimensional vector space. So according to this notation, Eq. \eqref{eq:eqn1} expresses the fact that the state of the qubit is the two dimensional complex vector $\begin{pmatrix} \alpha \\ \beta \end{pmatrix}$. Unlike a classical bit the state of a qubit cannot be measured without changing it. Measuring a qubit, whose state given by Eq. \eqref{eq:eqn1}, will yield the classical value of either zero ($\ket{0}$) with probability $|\alpha|^2$ or one ($\ket{1}$) with probability $|\beta|^2.$ Qubit implementations and technologies are a very active area of research that is not the focus of our review, an interested reader is referred to \cite{ladd2010quantum} for a survey.

%\paragraph{{\bf System of qubits}}
\subsubsection{System of qubits}

The mathematical structure of a qubit generalizes to higher dimensional quantum systems as well. The state of any quantum system is a normalized vector (a vector of norm one) in a complex vector space. The normalization is necessary to ensure that the total probability of all the outcomes of a measurement sum to one.

A quantum computer contains many number of qubits. So it is necessary to know how to construct the combined state of a system of qubits given the states of the individual qubits.
The joint state of a system of qubits is described using an operation known as the \emph{tensor product}, $\ot$. Mathematically, taking the tensor product of two states is the same as taking the Kronecker product of their corresponding vectors. Say we have two single qubit states \ket{\phi} = $\begin{pmatrix} \alpha \\ \beta \end{pmatrix}$ and \ket{\phi^\prime} =  $\begin{pmatrix} \alpha^\prime \\ \beta^\prime \end{pmatrix}$. Then the full state of a system composed of two independent qubits is given by,
\begin{equation}
    \ket{\phi} \ot \ket{\phi ^\prime} = ~ \begin{pmatrix} \alpha \\ \beta \end{pmatrix} \ot \begin{pmatrix} \alpha ^\prime \\ \beta ^\prime \end{pmatrix} = \begin{pmatrix} \alpha \alpha^\prime \\ \alpha \beta^\prime \\ \beta \alpha^\prime \\ \beta \beta^\prime \end{pmatrix}
\end{equation}
Sometimes the $\ot$ symbol is dropped all together while denoting the tensor product to reduce clutter. Instead the states are written inside a single ket. For example, $\ket{\phi} \ot \ket{\phi ^\prime}$ is shortened to $\ket{\phi \phi ^\prime} $,  and $\ket{0} \ot \ket{0} \ot \ket{0}$ is shortened to $\ket{000}.$ For larger systems the Dirac notation gives a more succinct way to compute the tensor product using the distributive property of the Kronecker product.
For a system of, say, three qubits with each qubit in the state $\ket{\gamma_j} = \alpha_j \ket{0} + \beta_j \ket{1}$, for $j = 1, 2, 3$, the joint state is,
\begin{align}
\label{eq:eqn2}
\ket{\gamma_1 \gamma_2 \gamma_3}
&=\ket{\gamma_1} \ot \ket{\gamma_2} \ot \ket{\gamma_3}\\
& =
\alpha_1 \alpha_2 \alpha_3 \ket{000} +
\alpha_1 \alpha_2 \beta_3 \ket{001} +
\alpha_1 \beta_2 \alpha_3 \ket{010} +
\alpha_1 \beta_2 \beta_3 \ket{011} \notag\\
&\hspace{6pt}+\beta_1 \alpha_2 \alpha_3 \ket{100} +
\beta_1 \alpha_2 \beta_3 \ket{101} +
\beta_1 \beta_2 \alpha_3 \ket{110} +
\beta_1 \beta_2 \beta_3 \ket{111}
\end{align}
A measurement of all three qubits could result in any of the eight ($2^3$) possible bit-strings associated with the eight basis vectors. One can see from these examples that the dimension of the state space grows exponentially in the number of qubits $n$ and that the number of basis vectors is $2^n$. 

%\paragraph{{\bf Superposition and Entanglement}}
\subsubsection{Superposition and entanglement}
\emph{Superposition} refers to the fact that any linear combination of two quantum states, once normalized, will also be a valid quantum state.  The upshot to this is that any quantum state can be expressed as a linear combination of a few basis states. For example, we saw in Eq.~\eqref{eq:eqn1} that any state of a qubit can be expressed as a linear combination of $\ket{0}$ and $\ket{1}$.  Similarly, the state of any $n$ qubit system can be written as a normalized linear combination of the $2^n$ bit-string states (states formed by the tensor products of \ket{0}'s and \ket{1}'s). The orthonormal basis formed by the $2^n$ bit-string states is called the \emph{computational basis.}

Notice that Eq.~\eqref{eq:eqn2} described a system of three qubits whose complete state was the tensor product of three different single qubit states. But it is possible for three qubits to be in a state that cannot be written as the tensor product of three single qubit states. An example of such a state is,
\begin{equation}\label{eq:GHZ}
\ket{\psi} =  \frac{1}{\sqrt{2}} (\ket{000}~ +~ \ket{111} ).    
\end{equation}
States of a system of which cannot be expressed as a tensor product of states of its individual subsystems are called \emph{entangled states}. For a system of $n$ qubits, this means that an entalged state cannot be written a tensor product of $n$ single qubit states.  The existence of entangled states is a physical fact that has important consequences for quantum computing, and quantum information processing in general. In fact, without the existence of such states quantum computers would be no more powerful than their classical counterparts \cite{vidal2003efficient}. Entanglement makes it possible to create a complete $2^n$ dimensional complex vector space to do our computations in,  using just $n$ physical qubits.

%\paragraph{{\bf  Inner products, Outer products and Measurement}}
\subsubsection{Inner and outer products}
We will now discuss some linear algebraic notions necessary for understanding quantum algorithms. First of these is the \emph{inner product} or  \emph{overlap} between two quantum states. As we have seen before, quantum states are vectors in complex vectors spaces. The overlap between  two states is just the inner product between these complex vectors. For example, take two single qubit states, $\ket{\phi} = \alpha \ket{0} + \beta \ket {1}$ and $\ket{\psi} = \gamma \ket{0} + \delta \ket{1} .$ The overlap between these states is denoted in the ket notation as  $\braket{\psi | \phi}$. And this is given by,
\begin{equation}
    \braket{\psi | \phi} =  \gamma^* \alpha +  \delta^* \beta, 
\end{equation}
where $^*$ denotes the complex conjugate. Notice that,$ \braket{\psi | \phi} =  \braket{\phi | \psi}^*$. The overlap of two states is in general a complex number. The overlap of a state with a bit-string state will produce the corresponding coefficient. For instance from Eq. \eqref{eq:eqn1}, $\braket{0|\phi} = \alpha$ and  $\braket{1|\phi} =  \beta$. And  from Eq. \eqref{eq:eqn2}, $\braket{001|\gamma_1 \gamma_2 \gamma_3} = \alpha_1 \alpha_2 \beta_3$. Another way to look at overlaps between quantum states is by defining what is called a \emph{bra} state. The states we have seen so far are ket states, like $\ket{\phi}$, which represented column vectors. A bra state corresponding to this ket state, written as $\bra{\phi}$, represents a row vector with complex conjugated entries. For instance $\ket{\phi}$ = $\begin{pmatrix} \alpha \\ \beta \end{pmatrix}$ implies that $\bra{\phi}$ = $\begin{pmatrix} \alpha^* & \beta^* \end{pmatrix}.$ The overlap of two states is then the matrix product of a row vector with a column vector, yielding a single number. The reader must have already noticed the wordplay here. The overlap, with its closing angled parenthesis, form a `bra-ket'!

The \emph{outer product} of two states is an important operation that outputs a matrix given two states. The outer product of  the two states we defined above will be denoted by, \ket{\psi}\bra{\phi}. Mathematically the outer product of two states is a matrix obtained by multiplying the column vector of the first state with the complex conjugated row vector of the second state (notice how the ket is written before the bra to signify this). For example,
\begin{equation}
    \ket{\psi}\bra{\phi} =  \begin{pmatrix} \alpha \\ \beta \end{pmatrix} \begin{pmatrix} \gamma^* & \delta^* \end{pmatrix} =  \begin{pmatrix} \alpha \gamma^* & \alpha \delta^* \\ \beta \gamma^* &\beta\delta^* \end{pmatrix}
\end{equation}
In this notation any matrix can be written as a linear combination of outer products  between bit-string states. For a $2\times 2$ matrix,
\begin{equation}
A = \begin{pmatrix} A_{00} & A_{01} \\ A_{10} & A_{11} \end{pmatrix} =  A_{00} \ket{0} \bra{0} +~ A_{01} \ket{0} \bra{1} +~ A_{10} \ket{1}\bra{0}  + ~ A_{11} \ket{1}\bra{1}.
\end{equation}
Acting on a state with a matrix then becomes just an exercise in computing overlaps between states. Let us demonstrate this process:
\begin{align}\label{eq:outer_prod}
A \ket{\phi}  &= A_{00} \ket{0} \braket{0|\phi} +~ A_{01} \ket{0} \braket{1|\phi} +~ A_{10} \ket{1}\braket{0|\phi}  + ~ A_{11} \ket{1}\braket{1|\phi} \nonumber, \\ 
              &=  (A_{00}\alpha  +  A_{01} \beta )\ket{0} + (A_{10}\alpha  ~+ ~ A_{11} \beta )\ket{1} ~ =~  \begin{pmatrix} A_{00}\alpha  +  A_{01} \beta  \\ A_{10}\alpha  +  A_{11} \beta  \end{pmatrix}.  
              \end{align}
This notation might look tedious at first glance but it makes algebraic manipulations of quantum states easily understandable. This is especially true when we are dealing with large number of qubits as otherwise we would have to explicitly write down exponentially large matrices.

The outer product notation for matrices also gives an intuitive input-output relation for them. For instance, the matrix $ \ket{0}\bra{1} + \ket{1}\bra{0}$ can be read as "output 0 when given a 1 and output 1 when given a 0". Likewise,the matrix, $\ket{00}\bra{00} + \ket{01}\bra{01} + \ket{10}\bra{11} + \ket{11}\bra{10}$ can be interpreted as the mapping \{"00" --> "00", "01" --> "01", "11" --> "10", "10" -->
    "11" \}.  But notice that this picture becomes a bit tedious when the input is in a superposition. In that case the correct output can be computed like in Eq. \eqref{eq:outer_prod}.

\subsubsection{Measurements}

Measurement corresponds to transforming the quantum information (stored in a quantum system) into classical information. For example, measuring a qubit typically corresponds to reading out a classical bit, i.e., whether the qubit is 0 or 1. A central principle of quantum mechanics is that measurement outcomes are \textit{probabilistic}. 

Using the aforementioned notation for inner products, for the single qubit state in Eq.~\eqref{eq:eqn1}, the probability of obtaining  $\ket{0}$ after measurement is $|\braket{0|\phi}|^2$ and  the probability of obtaining $\ket{1}$ after measurement is $|\braket{1|\phi}|^2$. So measurement probabilities can be represented as the squared absolute values of overlaps. Generalizing this, the probability of getting the bit string \ket{x_1 \ldots x_n} after measuring an $n$ qubit state, $\ket{\phi}$, is then $|\braket{x_1\ldots x_n |  \phi}|^2$.

Now consider a slightly more complex case of measurement. Suppose we have a three qubit state, $\ket{\psi}$ but we only measure the first qubit and leave the other two qubits undisturbed. What is the probability of observing a $\ket{0}$ in the first qubit? This probability will be given by,
\begin{equation} \label{eq:3qubitm1}
    \sum_{(x_2 x_3) \in \{0,1\}^2} | \braket{0x_2x_3 | \phi}|^2. 
\end{equation}
The state of the system after this measurement will be obtained by normalizing the state,
\begin{equation}\label{eq:3qubitm2}
    \sum_{(x_2 x_3) \in \{0,1\}^2}  \braket{0x_2 x_3 | \phi}  \ket{0x_2 x_3}.
\end{equation}
Applying this paradigm to the state in Eq. \eqref{eq:GHZ} we  see that the probability of getting $\ket{0}$ in the first qubit will be $0.5$, and if this result is obtained, the final state of the system would change to $\ket{000}.$ On the other hand, if we were to measure $\ket{1}$ in the first qubit we would end up with a state $\ket{111}.$ Similarly we can compute the effect of subsystem measurements on any $n$ qubit state.

In some cases we will need to do measurements on a basis different from the computational basis. This can be achieved by doing an appropriate transformation on the qubit register before measurement. Details of how to do this is given in a subsequent section discussing observables and expectation values.

The formalism discussed so far is sufficient to understand all measurement scenarios in this paper. We refer the reader to Ref.~\cite{NielsenChuang} for a more detailed and more general treatment of measurement.

\subsubsection{Unitary transformations and gates}

A qubit or a system of qubits changes its state by going through a series of \emph{unitary transformations}. A unitary transformation is described by a matrix $U$ with complex entries. The matrix $U$ is called unitary if
\begin{align}
\label{eq:eqn3}
 UU^\dagger = U^\dagger U  = I ,
\end{align}
where $U^\dagger$ is the transposed, complex conjugate of $U$ (called its \emph{Hermitian conjugate}) and $I$ is the identity matrix. A qubit state $\ket{\phi} = \alpha \ket{0} + \beta \ket{1}$ evolves under the action of the $2\times 2$ matrix $U$ according to
\begin{align}
\label{eq:single_qubit_gate}
\ket{\phi} \rightarrow U\ket{\phi} = \begin{pmatrix}
   U_{00}   &  U_{01} \\
   U_{10}   &  U_{11}
\end{pmatrix}  \begin{pmatrix} \alpha \\ \beta \end{pmatrix}  = \begin{pmatrix} U_{00}\alpha + U_{01}\beta \\ U_{10}\alpha + U_{11}\beta \end{pmatrix}\,.
\end{align}
Operators acting on different qubits can be combined using the Kronecker product. For example, if $U_1$ and $U_2$ are operators acting on two different qubits then the full operator acting on the combined two qubit system will be given by $U_1 \otimes U_2$.

For an $n$ qubit system the set of physically allowed transformations, excluding measurements, consists of all $2^n \times 2^n$ unitary matrices. Notice that the size of a general transformation increases exponentially with the number of qubits.  In practice a transformation on $n$ qubits is effected by using a combination of unitary transformations that act only on one or two qubits at a time. By analogy to classical logic gates like NOT and AND, such basic unitary transformations, which are used to build up more complicated $n$ qubit transformations, are called \emph{gates}. Gates are  unitary transformations themselves and from Eq. \eqref{eq:eqn3} it is clear that unitarity can only be satisfied if the number of input qubits is equal to the number of output qubits. Also, for every gate $U$ it is always possible to have another gate $U^ \dagger$ that undoes the transformation. So unlike classical gates quantum gates have to be reversible.  \emph{Reversible} means that the gate's inputs can always be reconstructed from the gate's outputs.  For instance, a classical NOT gate, which maps 0 to 1 and 1 to 0 is reversible because an output of 1 implies the input was 0 and vice versa. However, a classical AND gate, which returns 1 if and only if both of its inputs are 1, is not reversible.  An output of 1 implies that both inputs were 1, but an output of 0 provides insufficient information to determine if the inputs were 00, 01, or 10.

But this extra restriction of reversibility does not mean that quantum gates are `less powerful' than classical gates. Even classical gates can be made reversible with minimal overhead. Reversibility does not restrict their expressive power \cite{saeedi2013synthesis}. Quantum gates can then be seen as a generalization of classical reversible gates. 

The most common gates are described in Table~\ref{gate_table}. The $X$ gate is the quantum version of the NOT gate.  The CNOT or ``controlled NOT'' negates a target bit if and only if the control bit is~1. We will use the notation $\text{CNOT}_{ij}$ for a CNOT gate controlled by qubit $i$ acting on qubit $j$. The CNOT gate can be expressed using the outer product notation as,
\begin{equation}
    \text{CNOT}  =  \ket{0}\bra{0} \otimes I + \ket{1}\bra{1} \otimes  X  =   \ket{00}\bra{00} + \ket{01}\bra{01} +  \ket{10}\bra{11} + \ket{11}\bra{10}.
\end{equation}

The Toffoli gate or ``controlled-controlled NOT'' or CCNOT, is a three qubit gate that is essentially the quantum (reversible) version of the AND gate.  It negates a target bit if and only if both control bits are~1. In the outer product notation,
\begin{equation}
    \text{CCNOT} =  \ket{11}\bra{11} \otimes X  +  (I - \ket{11}\bra{11} ) \otimes I.
\end{equation}

Another way to look at the CCNOT gate is as a CNOT gate with an additional control qubit,

\begin{equation}
      \mathrm{CCNOT} = \ket{0}\bra{0} \otimes I + \ket{1}\bra{1} \otimes
\mathrm{CNOT}.
\end{equation}

In general, one can define controlled versions of any unitary gate $U$ as,
\begin{equation}
    CU =  \ket{0}\bra{0} \otimes I +  \ket{1}\bra{1} \otimes U.
 \end{equation}
$CU$ applies $U$ to a set of qubits only if the first qubit (called the control qubit)  is $\ket{1}$.

\begin{table}
\footnotesize
\begin{tabular}{|c|c|}%{|*2{>{\centering\arraybackslash}p{.3\textwidth}|}}
\hline
\textbf{One-qubit gates} & \textbf{Multi-qubit gates}\\
\hline  ~&~ \\
\(\text{Hadamard} = H =  \frac{1}{\sqrt{2}}\left( \begin{array}{cc} 1 & 1  \\ 1 & -1 \end{array}\right)  \)
& \(\text{CNOT} = CX = \left( \begin{array}{cccc} 1 & 0 & 0 & 0  \\ 0 & 1 & 0 & 0 \\ 0 & 0 & 0 & 1\\ 0 & 0 & 1 & 0 \end{array}\right)  \)\\~&~ \\
\hline  ~&~ \\
\(I = \left( \begin{array}{cc} 1 & 0 \\ 0 & 1 \end{array}\right)\), \newline  ~~ \(S = \left( \begin{array}{cc} 1 & 0  \\ 0 & i \end{array}\right) \) & \(CZ = \left( \begin{array}{cccc} 1 & 0 & 0 & 0  \\ 0 & 1 & 0 & 0 \\ 0 & 0 & 1 & 0\\ 0 & 0 & 0 & -1 \end{array}\right) \)\\ ~&~ \\
\hline  ~&~ \\
\(T = \left( \begin{array}{cc} 1 & 0  \\ 0 & e^{i \pi / 4}  \end{array}\right) \) & \( \text{Controlled-}U = CU =  \left( \begin{array}{cccc} 1 & 0 & 0 & 0  \\ 0 & 1 & 0 & 0 \\ 0 & 0 & U_{00} & U_{01}\\ 0 & 0 & U_{10} & U_{11} \end{array}\right) \)\\ ~&~ \\
\hline ~&~ \\
 \(\text{NOT} = X = \left( \begin{array}{cc} 0 & 1  \\ 1 & 0 \end{array}\right) \)
& \( \text{SWAP} = \left( \begin{array}{cccc} 1 & 0 & 0 & 0  \\ 0 & 0 & 1 & 0 \\ 0 & 1 & 0 & 0\\ 0 & 0 & 0 & 1 \end{array}\right) \)\\~&~ \\
\hline  ~&~ \\
\(Y = \left( \begin{array}{cc} 0 & -i  \\ i & 0 \end{array}\right) \), \newline ~~~~\(Z = \left( \begin{array}{cc} 1 & 0 \\ 0 & -1 \end{array}\right) \)  & ~ 
\(  \underset{(\text{CCNOT})}{\text{Toffoli}}   = \left( \begin{array}{cccccccc} 1 & 0 & 0 & 0 & 0 & 0 & 0 & 0  \\ 0 & 1 & 0 & 0 & 0 & 0 & 0 & 0 \\ 0 & 0 & 1 & 0 & 0 & 0 & 0 & 0 \\ 0 & 0 & 0 & 1 & 0 & 0 & 0 & 0 \\ 0 & 0 & 0 & 0 & 1 & 0 & 0 & 0 \\ 0 & 0 & 0 & 0 & 0 & 1 & 0 & 0 \\ 0 & 0 & 0 & 0 & 0 & 0 & 0 & 1 \\ 0 & 0 & 0 & 0 & 0 & 0 & 1 & 0 \end{array}\right) \)\\ ~&~ \\
\hline  ~&~ \\
\( R(\theta) =  P(\theta) = \left( \begin{array}{cc} 1 & 0 \\ 0 & e^{i \theta} \end{array}\right) \)
& \(  \)\\ ~&~ \\
\hline 
\end{tabular}

\caption{Commonly used quantum gates.}
\label{gate_table}
\end{table}

A set of gates that together can execute all possible quantum computations is called a \emph{universal gate set}. Taken together, the set of all unary (i.e.,~acting on one qubit) gates and the binary (i.e.,~acting on two qubits) $\text{CNOT}$ gate form a universal gate set. More economically, the set $\{H , T,\text{CNOT}\}$ (Refer Table \ref{gate_table} for definitions of these gates) forms a universal set. Also, the Toffoli gate by itself is universal \cite{NielsenChuang}.

%\paragraph{\bf Observables and expectation values}

\subsubsection{Observables and expectation values}

We have seen that experiments in quantum mechanics are probabilistic. Often in experiments we will need to associate a real number with a measurement outcome. And quantities that we measure in quantum mechanics will always be statistical averages of these numbers. For instance, suppose we do the following experiment on many copies of the single qubit state in Eq.~\eqref{eq:eqn1}: We measure a copy of the state and if we get $\ket{0}$ we record $1$ in our lab notebook , otherwise we record $-1$. While doing this experiment we can never predict the outcome of a specific measurement. But we can ask statistical questions like: ``What will be the average value of the numbers in the notebook?'' From our earlier discussion on measurement we know that the probability of measuring $\ket{0}$ is $|\alpha|^2$ and the probability of measuring $\ket{1}$ is $|\beta|^2$. So the average value of the numbers in the notebook will be,
\begin{equation}
   |\alpha|^2  - |\beta|^2  
\end{equation}
In quantum formalism, there is neat way to express such experiments and their average outcomes, without all the verbiage, using certain operators. For the experiment described above the associated operator would be the  $Z$ gate,
\begin{equation}
    Z = \ket{0}\bra{0} - \ket{1}\bra{1} = \left( \begin{array}{cc} 1 & 0 \\ 0 & -1 \end{array}\right) 
\end{equation}
By associating this operator with the experiment we can write the average outcome of the experiment, on $\ket{\phi}$,  as the overlap between $\ket{\phi}$ and $Z\ket{\phi}$,
\begin{equation}
    \braket{\phi |Z|\phi} = \braket{\phi|0}\braket{0|\phi} - \braket{\phi|1}\braket{1|\phi} =  |\alpha|^2 - |\beta|^2.
\end{equation}
The operator $Z$ is called the \emph{observable} associated with this experiment. And the quantity $ \braket{\phi |Z|\phi}$ is called its \emph{expectation value}. The expectation value is sometimes denoted by $\langle Z \rangle$, when there is no ambiguity about the state on which the experiment is performed.

Here we saw an experiment done in the computational basis. But this need not always be the case. Experiments can be designed by associating real numbers to measurement outcomes in any basis. What would be the observable for such an experiment? For an experiment that associates the real numbers $\{a_i\}$ to a measurement onto a basis set $\{\ket{\Phi_i}\}$, the observable will be,
\begin{equation}
    O \equiv \sum_i a_i \ket{\Phi_i}\bra{\Phi_i}.
\end{equation}
This observable will reproduce the correct expectation value for this experiment done on any state $\ket{\psi}$,
\begin{equation}
    \braket{\psi|O|\psi} = \sum_i a_i \braket{\psi| \Phi_i}\braket{\Phi_i| \psi} = \sum_i a_i |\braket{\Phi_i| \psi}|^2.
\end{equation}
Because the states $\{\ket{\Phi_i}\}$ are orthonormal, we can see that $O$ obeys the following eigenvalue equation,
\begin{equation}
    O \ket{\Phi_j} =  \sum_i a_i \ket{\Phi_i}\braket{\Phi_i|\Phi_j} = a_j \ket{\Phi_j}.
\end{equation}
So $O$ is an operator that has complete set of orthogonal eigenvectors and real eigenvalues. Such operators are called \emph{Hermitian operators.} Equivalently, these operators are equal to their Hermitian conjugates ($O = O^\dagger$). In quantum mechanics, any Hermitian operator is a valid observable. The eigenvectors of the operator give the possible outcomes of the experiment and the corresponding eigenvalues are the real numbers associated with that outcome. 

But can all valid observables be measured in practice? The answer to this depends on the quantum system under consideration. In this tutorial, the system under consideration is an IBM quantum processor. And in these processors only measurements onto the computational basis are  supported natively. Measurements to other basis states can be performed by applying an appropriate unitary transformation before measurement. Suppose that the hardware only lets us do measurements onto the computational basis $\{\ket{i}\}$ but we want to perform a measurement onto the basis set $\{\ket{\Phi_i}\}$. This problem can be solved if we can implement the following unitary transformation,
\begin{equation}
    U =  \sum_i \ket{i}\bra{\Phi_i}. 
\end{equation}
Now measuring $U\ket{\psi}$ in the computational basis is the same as measuring $\ket{\psi}$ in the $\{\ket{\Phi_i}\}$ basis. This can be seen by computing the outcome probabilities on $U\ket{\psi}$,
\begin{equation}
    |\braket{j|U|\psi}|^2 =  |\sum_i \braket{j|i}\braket{\Phi_i|\psi}|^2 =  |\braket{\Phi_j|\psi}|^2.
\end{equation}
So once $U$ is applied, the outcome $\ket{j}$ becomes equivalent to the outcome $\ket{\Phi_j}$ in the original measurement scenario. Now, not all such unitary transformations are easy to implement. So if a quantum algorithm requires us to perform a measurement onto some complicated set of basis states, then the cost of implementing the corresponding $U$ has be taken into account.

%\paragraph{\bf Quantum circuits}
\subsubsection{Quantum circuits}

Quantum algorithms are often diagrammatically represented as circuits in literature. Here we will describe how to construct and read quantum circuits. In the circuit representation, qubits are represented by horizontal lines. Gates are then drawn on the qubits they act on. This is done in sequence from left to right. The initial state of the qubit is denoted at the beginning of each of the qubit lines. Notice that when we write down a mathematical expression for the circuit, the gates are written down from right to left in the order of their application.  

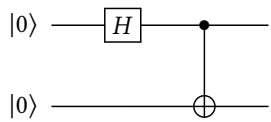
\begin{figure}[H]
\begin{equation*}
\Qcircuit @C=2em @R=2em {
\lstick{\ket{0}}&\gate{H} &\ctrl{1} &\qw \\
\lstick{\ket{0}}&\qw &\targ &\qw
}
\end{equation*}
    \caption{Quantum circuit for preparing a Bell state}
    \label{fig:bell}
\end{figure}

These principles are best illustrated by an example. Given in Fig.~\ref{fig:bell} is a circuit to preparing an entangled two qubit state called a Bell state from $\ket{00}$.

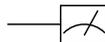
\begin{figure}[H]
\begin{equation*}
\Qcircuit @C=2em @R=2em {
 & \meter
}
\end{equation*}
    \caption{The measurement gate}
    \label{fig:m_gate}
\end{figure}

The circuit encodes the equation,
\begin{equation*}
 \text{CNOT}_{12} ~ (H \otimes I ) \ket{00} = \frac{1}{\sqrt{2}} (\ket{00} + \ket{11}).
\end{equation*}
Let us now carefully go over how the circuit produces the Bell state. We read the circuit from left to right.
The qubits are numerically labelled starting from the top.  First the $H$ gate acts on the top most qubit changing the state of the system to,
\begin{equation*}
H \otimes I \ket{00} =   (H\ket{0} )\otimes (I\ket{0}) = \left( \frac{\ket{0} + \ket{1} }{\sqrt{2}} \right) \otimes \ket{0} =  \frac{1}{\sqrt{2}} (\ket{00} + \ket{10}) .   
\end{equation*}
Then  $\text{CNOT}_{12}$ acts on both of these qubits. The blackened dot on the first qubit implies that this qubit is the control qubit for the CNOT. The $\oplus$ symbol  on the second qubit implies that this qubit is the target of the NOT gate (controlled by the state of the first qubit). The action of the CNOT then gives,
\begin{equation*}
\text{CNOT}_{12} \left(\frac{1}{\sqrt{2}} (\ket{00} + \ket{10})\right) =  \frac{1}{\sqrt{2}} (\text{CNOT}_{12} \ket{00} +  \text{CNOT}_{12} \ket{10}) =  \frac{1}{\sqrt{2}} (\ket{00} + \ket{11}).
\end{equation*}
The measurement of a qubit is also denoted by a special gate with a  meter symbol on it, given in Fig~\ref{fig:m_gate}. The presence of this gate on a qubit means that the qubit must be measured in the computational basis. 

%\paragraph{\bf Quantum algorithms}

\subsubsection{Quantum algorithms}

We have now introduced all the basic elements needed for the discussion of practical quantum algorithms.
A quantum algorithm consists of three basic steps: 

\begin{itemize}
    \item Encoding of the data, which could be classical or quantum, into the state of a set of input qubits.
    \item A sequence of quantum gates applied to this set of input qubits.
    \item Measurements of one or more of the qubits at the end to obtain a classically interpretable result.
\end{itemize}

In this review, we will describe the implementation of these three steps for a variety of quantum algorithms. 

\subsection{Implementations on a real quantum computer}

\subsubsection{The IBM quantum computer}

In this article, we consider IBM's publicly available quantum computers. In most cases, we specifically consider the \verb|ibmqx4|, which is a 5-qubit computer, although in some cases we also consider other quantum processors freely accessible through the IBM Quantum Experience platform. These processors can be accessed by visiting the IBM Quantum Experience website (\url{https://quantum-computing.ibm.com/})  

There are several issues to consider when implementing an algorithm on real quantum computers, for example:
\begin{enumerate}
  \item What is the available gate set with which the user can state their algorithm?
  \item What physical gates are actually implemented?
  \item What is the qubit connectivity (i.e.,~which pairs of qubits can two-qubit gates be applied to)?
  \item What are the sources of noise (i.e.,~errors)?
\end{enumerate}

We first discuss the available gate set. In IBM's graphical interface to the \verb|ibmqx4|, the available gates include:
\begin{align}
\label{eqn5}
\{I, X, Y, Z, H, S, S^{\dag}, T, T^{\dag}, U_1(\lambda), U_2(\lambda, \phi ), U_3(\lambda, \phi, \theta ), \text{CNOT}\}.
\end{align}
The Graphical User Interface (GUI) also provides other controlled gates and operations like measurement and reset. Most of these gates appear in our Table~\ref{gate_table}. The gates $U_1(\lambda)$, $U_2(\lambda, \phi )$, and $U_3(\lambda, \phi, \theta )$ are continuously parameterized gates, defined as follows:
\begin{align}
\label{eqn6}
U_1(\lambda) = \begin{pmatrix}
   1   & 0 \\
   0   &  e^{i\lambda}
\end{pmatrix},
\quad
U_2(\lambda, \phi ) = \frac{1}{\sqrt{2}} \begin{pmatrix}
   1   & - e^{i \lambda } \\
   e^{i\phi }   &  e^{i(\lambda+\phi)}
\end{pmatrix},
\quad
U_3(\lambda, \phi, \theta ) = \begin{pmatrix}
   \cos(\theta/2)   & - e^{i \lambda } \sin(\theta/2) \\
   e^{i\phi } \sin(\theta/2)   &  e^{i(\lambda+\phi)}\cos(\theta/2)
\end{pmatrix}\,.
\end{align}
Note that $U_3(\lambda, \phi, \theta )$ is essentially an arbitrary one-qubit gate.

The gates listed in Eq.~\eqref{eqn5} are provided by IBM for the user's convenience. However these are not the gates that are physically implemented by their quantum computer. IBM has a compiler that translates the gates in \eqref{eqn5} into products of gates from a physical gate set. The physical gate set employed by IBM is essentially composed of three gates \cite{IBMConfig}:
\begin{align}
\label{eqn7}
\{U_1(\lambda), R_X(\pi / 2 ), \text{CNOT}\}\,.
\end{align}
Here, $R_X(\pi / 2 )$ is a rotation by angle $\pi / 2$ of the qubit about it's $X$-axis, corresponding to a matrix similar to the Hadamard:
\begin{align}
\label{eqn8}
R_X(\pi / 2 ) = \frac{1}{\sqrt{2}} \begin{pmatrix}
   1   & -i \\
   -i   &  1
\end{pmatrix}\,.
\end{align}
The reason why it could be important to know the physical gate set is that some user-programmed gates may need to be decomposed into multiple physical gates, and hence could lead to a longer physical algorithm. For example, the $X$ gate gets decomposed into three gates: two $R_X(\pi / 2 )$ gates sandwiching one $U_1(\lambda)$ gate.

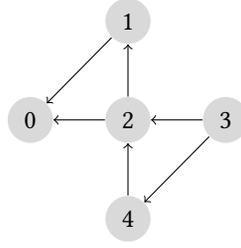
\begin{figure}
    \centering
    \begin{tikzpicture}[scale=1.3]
    \node[circle, fill=gray!30] (b) at (0,0) {2};
    \node[circle, fill=gray!30] (a) at (0,1) {1};
    \node[circle, fill=gray!30] (z) at (-1,0) {0};
    \node[circle, fill=gray!30](c) at (1,0)  {3};
    \node[circle, fill=gray!30](d) at (0,-1) {4};
    \draw [->] (b) -- (a);
    \draw [->] (b) -- (z);
    \draw [->] (a) -- (z);
    \draw [->] (c) -- (b);
    \draw [->] (c) -- (d);
    \draw [->] (d) -- (b);
    \end{tikzpicture}
    \caption{The connectivity diagram of \texttt{ibmqx4}. The circles represent qubits and the arrows represent the ability to apply a physical CNOT gate between the qubits.}
    \label{fig:ibmqx4_layout}
\end{figure}

The connectivity of the computer is another important issue. Textbook algorithms are typically written for a fully-connected hardware, which means that one can apply a two-qubit gate to any two qubits. In practice, real quantum computers may not have full connectivity. In the \verb|ibmqx4|, which has 5 qubits, there are 6 connections, i.e.,~there are only 6 pairs of qubits to which a CNOT gate can be applied (Fig.\ref{fig:ibmqx4_layout}).  In contrast a fully connected 5-qubit system would allow a CNOT to be applied to 20 different qubit pairs. In this sense, there are 14 ``missing connections''.  Fortunately, there are ways to effectively generate connections through clever gate sequences. For example, a CNOT gate with qubit $j$ as the control and qubit $k$ as the target can be reversed (such that $j$ is the target and $k$ is the control) by applying Hadamard gates on each qubit both before and after the CNOT, i.e.,
\begin{align}
\label{eqn9}
\text{CNOT}_{kj} = (H \ot H)\text{CNOT}_{jk}(H \ot H)\,.
\end{align}
Similarly, there exists a gate sequence to make a CNOT between qubits $j$ and $l$ if one has connections between $j$ and $k$, and $k$ and $l$, as follows:
\begin{align}
\label{eqn10}
\text{CNOT}_{jl} = \text{CNOT}_{kl}\text{CNOT}_{jk}\text{CNOT}_{kl}\text{CNOT}_{jk}\,.
\end{align}
Hence, using \eqref{eqn9} and \eqref{eqn10}, one can make up for lack of connectivity at the expense of using extra gates.

Finally, when implementing a quantum algorithm it is important to consider the sources of noise in the computer. The two main sources of noise are typically gate \emph{infidelity} and \emph{decoherence}. Gate infidelity refers to the fact that the user-specified gates do not precisely correspond to the physically implemented gates. Gate infidelity is usually worse for multi-qubit gates than for one-qubit gates, so typically one wants to minimize the number of multi-qubit gates in one's algorithm. Decoherence refers to the fact that gradually over time the quantum computer loses its ``quantumness'' and behaves more like a classical object. After decoherence has fully occurred, the computer can no longer take advantage of quantum effects. This introduces progressively more noise as the quantum algorithm proceeds in time. Ultimately this limits the depth of quantum algorithms that can be implemented on quantum computers. It is worth noting that different qubits decohere at different rates, and one can use this information to better design one's algorithm. The error rates for individual qubits in the IBM processors are listed in the IBM Quantum Experience website. In this tutorial, we will show in many cases how infidelity and decoherence affect the algorithm performance in practice. 

\begin{figure}
    \centering
    \includegraphics[scale=0.5]{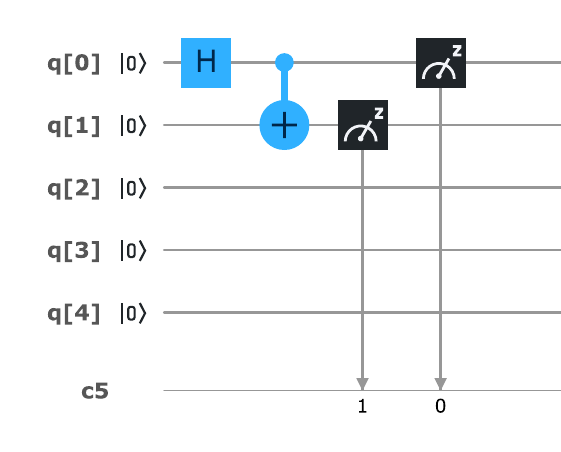}
    \caption{The quantum circuit to prepare a Bell state and measure it in the IBM quantum experience GUI}
    \label{fig:bell_ibmqx4}
\end{figure}

A simple example of programming the IBM quantum computer is given in Fig.~\ref{fig:bell_ibmqx4}, which shows the Bell state preparation circuit Fig.\ref{fig:bell} compiled using the IBM quantum experience GUI. Extra measurement operations at the end serve to verify the fidelity of the implementation.

\subsubsection{Programming the IBM quantum computer: Qiskit library}

Qiskit \cite{Qiskit} is an open-source quantum computing library developed under the aegis of IBM. Qiskit allows users to write and run programs on either IBM's quantum processors or on a local simulator, without the use of the graphical interface. This is an important feature because the graphical interface becomes impractical as the number qubits become large. At the time of writing, users can use Qiskit to access quantum processors with up to 16 qubits. Smaller processors are also accessible. Qiskit is a very powerful software development kit (SDK) which has multiple  elements in it that tackle a variety of problems associated with practical quantum computing. Qiskit is further split into four modules called: Terra, Aer, Aqua, and Ignis. Each of these modules deal with a specific part of quantum software development. In this section we will only give a brief overview of programming simple quantum circuits with Qiskit. For a comprehensive overview of  Qiskit and its various capabilities, the reader is encouraged to visit the official website ( \url{www.qiskit.org} ) \cite{Qiskit}.   

For our purposes, Qiskit can be viewed as a Python library for quantum circuit execution. A basic Qiskit code has two parts, designing the circuit and running it. In the circuit design phase, we create an instance of  \verb| QuantumCircuit| with the required number of qubits and classical bits. Then gates and measurements are added to this blank circuit. Gates and measurements are implemented in Qiskit as methods of the \verb|QuantumCircuit| class. After the circuit has been designed we need to choose a backend to run the circuit. This can be either be a simulator called the \verb|qasm_simulator| or it can be one of IBM's quantum processors. To use a quantum processor, you will need to load your IBM Q account information into Qiskit. Given in Fig. \ref{fig:bell_qiskit} is a simple code to construct the Bell state. This is the Qiskit version of the circuit in Fig. \ref{fig:bell} with measurement added at the end to verify our results.

%\lstinputlisting[language=Python]{bell.py}
\begin{figure}[H]
    \centering
\begin{minted}{python}
### Quantum circuit for preparing the Bell state ####

import numpy as np
from qiskit import QuantumCircuit, execute, Aer

# Create a Quantum Circuit with two qbits and 2 classical bits
circuit = QuantumCircuit(2,2)

# Add a H gate on qubit 0
circuit.h(0)

# Add a CX (CNOT) gate on control qubit 0 and target qubit 1
circuit.cx(0,1)

# Map the quantum measurement to the classical bits
circuit.measure([0,1],[0,1])


# Use Aer's qasm_simulator
simulator = Aer.get_backend('qasm_simulator')

# Execute the circuit on the qasm simulator
job = execute(circuit, simulator, shots=1000)

# Grab results from the job
result = job.result()

# Returns counts
counts = result.get_counts(circuit)
print("\nTotal count for 00 and 11 are:",counts)
\end{minted}
    \caption{Qiskit code to create and measure a Bell state. Source: \url{www.qiskit.org}}
    \label{fig:bell_qiskit}
\end{figure}

In Fig.\ref{fig:bell_qiskit} we are running the circuit on the simulator for $1000$ independent runs. The final output was \verb|{'11': 493, '00': 507}|. This is what we expect from measuring the Bell state ($\frac{\ket{00} + \ket{11}}{\sqrt{2}}$), up to statistical fluctuations. While running the same code on the $14$ qubit \verb|ibmq_16_melbourne| processor for $1024$ runs  gave  $\ket{11}$ with  probability  $0.358$  and $\ket{00}$ with probability $0.54$. The remaining probability was distributed over $01$ and $10$, which should not be a part of the Bell state. As we discussed before, this phenomenon is due to errors inherent to the quantum processor. As the backend technology improves we expect to get better results from these trials. Often, we will also present a circuit using OpenQASM (Open Quantum Assembly Language). OpenQASM provides an intermediate  representation of a program in the form of a quantum circuit, that is neither the actual program written by the programmer nor the machine instructions seen by the processor. OpenQASM `scores' we show in this paper will be simple sequence of gates and measurements, with the corresponding registers that they act on. The syntax of these scores will be self explanatory.

\subsection{Classes of quantum algorithms}

In this review, we broadly classify quantum algorithms according to their area of application. We will discuss quantum algorithms for graph theory, number theory, machine learning and so on. The complete list of algorithms discussed in this paper, classified according to their application areas, can be found in Table ~\ref{alg_table}.  The reader is also encouraged to take a look at the excellent Quantum Algorithm Zoo website \cite{quantumzoo}  for a concise and comprehensive list of quantum algorithms.

In classical computing, algorithms are often designed by making use of one or more algorithmic paradigms like dynamic programming or local search, to name a few. Most known quantum algorithms also use a combination of algorithmic paradigms specific to quantum computing. These paradigms are the Quantum Fourier Transform (QFT), the Grover Operator (GO), the Harrow-Hassidim-Lloyd (HHL) method for linear systems, variational quantum eigenvalue solver (VQE), and direct Hamiltonian simulation (SIM). The number of known  quantum algorithmic paradigms is much smaller compared to the number of known classical paradigms.  The constraint of unitarity on quantum operations and the impossibility of non-intrusive measurement make it difficult to design quantum paradigms from existing classical paradigms. But researchers are constantly in search for new paradigms and we can expect this list to get longer in the future. Table~\ref{alg_table} also contains information about the paradigms used by the algorithms in this article.

\begin{table}
\scriptsize
\begin{tabular}{lllll}
\toprule
Class & Problem/Algorithm &  Paradigms used  & Hardware & Simulation Match \\
\midrule
Inverse Function Computation
        &       Grover's Algorithm                      & GO                    & QX4   & med   \\
        &       Bernstein-Vazirani                      & n.a.                  & QX4, QX5      & high  \\ \midrule
Number-theoretic Applications
        & Shor's Factoring      Algorithm       & QFT                   & QX4   & med   \\
 %       & Subset Sum                                    & QFT                   & none  & n.a.          \\
\midrule
Algebraic Applications
        & Linear Systems                                & HHL                   & QX4   & low   \\
        & Matrix Element Group Representations  & QFT                   & ESSEX   & low   \\
        & Matrix Product Verification   & GO                    & n.a.   & n.a.  \\
        & Subgroup Isomorphism                  & QFT                   & none  & n.a.  \\
  %      & Persistent Homology                   & GO, QFT               & QX4   & med-low       \\
\midrule
Graph Applications
        & Quantum Random Walk & n.a.                    & VIGO & med-low   \\
        & Minimum Spanning Tree                 & GO                   & QX4   & med-low       \\
        & Maximum Flow                                  & GO                   & QX4   & med-low       \\
        & Approximate Quantum Algorithms& SIM                   & QX4   & high  \\
\midrule
Learning Applications
        & Quantum Principal Component Analysis (PCA)    & QFT                   & QX4   & med   \\
        & Quantum Support Vector Machines (SVM) & QFT           & none  & n.a.  \\
        & Partition Function                    & QFT                   & QX4   & med-low       \\
\midrule
Quantum Simulation
        & Schr\"{o}dinger Equation Simulation      & SIM           & QX4   & low   \\
        & Transverse Ising Model Simulation     & VQE           & none   & n.a.  \\
\midrule
Quantum Utilities
        & State Preparation                             & n.a.          & QX4   & med   \\
        & Quantum Tomography                    & n.a.          & QX4   & med   \\
        & Quantum Error Correction              & n.a.          & QX4   & med   \\ \bottomrule
\end{tabular}
\caption{Overview of studied quantum algorithms. Paradigms include Grover Operator (GO), Quantum Fourier Transform (QFT), Harrow-Hassidim-Lloyd (HHL), Variational Quantum Eigenvalue solver (VQE), and direct Hamiltonian simulation (SIM). The simulation match column indicates how well the hardware quantum results matched the simulator results}
\label{alg_table}
\end{table}

The  rest of the paper  presents each of the algorithms shown in Table~\ref{alg_table}, one after the other. In each case, we first discuss the goal of the algorithm (the problem it attempts to solve). Then we describe the gate sequence required to implement this algorithm. Finally, we show the results from implementing this algorithm on IBM's quantum computer\footnote{The code and implementations for most of the algorithms can be found at \url{https://github.com/lanl/quantum_algorithms}.}.

\revision{The list of algorithms in Table~\ref{alg_table} is by no means exhaustive. These algorithms have been chosen due to their relative importance and to provide an overview of the field. Many interesting quantum algorithms like those  for topological data analysis \cite{LloydNature}, spatial search \cite{szegedy2004quantum}, supervised learning \cite{schuld2015introduction}, etc., have not been covered in this review. Nevertheless the tools and ideas elucidated in this paper will help the reader  understand and implement many quantum algorithms that are not included here.}

% Inverse Function Computation
\section{Grover's Algorithm}

\subsection{Problem definition and background}
Grover's algorithm as initially described~\cite{grover} enables one to find (with probability $>1/2$) a specific item within a randomly ordered database of $N$ items using $O(\sqrt{N})$ operations.
By contrast, a classical computer would require $O(N)$ operations to achieve this.
Therefore, Grover's algorithm provides a quadratic speedup over an optimal classical algorithm.
It has also been shown~\cite{bennett1997strengths} that Grover's algorithm is optimal in the sense that no quantum Turing machine can do this in less than $O(\sqrt{N})$ operations.

While Grover's algorithm is commonly thought of as being useful for searching a database, the basic ideas that comprise this algorithm are applicable in a much broader context.
This approach can be used to accelerate search algorithms where one could construct a ``quantum oracle''  that distinguishes the needle from the haystack.
The needle and hay need not be part of a database.
For example, it could be used to search for two integers $1<a<b$ such that $ab=n$ for some number $n$, resulting in a factoring algorithm.
Grover's search in this case would  have worse performance than Shor's algorithm ~\cite{shor1994algorithms,Shor1997} described below, which is a specialised algorithm to solve the factoring problem.
Implementing the quantum oracle can be reduced to constructing a quantum circuit that flips an ancillary qubit, $q$, if a function, $f(\mathbf{x})$, evaluates to 1 for an input $\mathbf{x}$. We use the term \emph{ancilla} or \emph{ancillary qubit} to refer to some extra qubits that are used by the algorithm. 

The function $f(\mathbf{x})$ is defined by
\begin{equation}\label{eq:grov_f}
	f(\mathbf{x})=
	\begin{cases}	
		1 & \text{if}~\mathbf{x}=\mathbf{x}^*\\
		0 & \text{if}~\mathbf{x}\ne\mathbf{x}^*
	\end{cases}
\end{equation}
where $\mathbf{x}=x_1x_2 \ldots x_n$ are binary strings and $\mathbf{x}^*$ is the specific string  that is being sought.
It may seem paradoxical at first that an algorithm for finding $\mathbf{x}^*$ is needed if such a function can be constructed.
The key here is that $f(\mathbf{x})$ need only recognize $\mathbf{x}^*$ -- it is similar to the difference between writing down an equation and solving an equation.
For example, it is easy to check if the product of $a$ and $b$ is equal to $n$, but harder to factor $n$. In essence, Grover's algorithm can invert an arbitrary function  with binary outputs, provided we have a quantum oracle that implements the function. Grover's algorithm has been used, with appropriate oracles, to solve problems like finding triangles in a graph \cite{magniez2005}, finding cycles \cite{circella2006}, and finding maximal cliques \cite{wie2017}. For the analysis of Grover's algorithm, the internals of the oracle is typically considered a black-box. Often, the oracle operator for the problem at hand has to be constructed as a quantum circuit. But, keep in mind that an inefficient oracle construction can nullify any practical advantages gained by using Grover's search.

Here we implement a simple instance of Grover's algorithm.
That is, the quantum oracle we utilize is a very simple one.
Let $\mathbf{x}= x_1x_2$ and we wish to find $\mathbf{x}^*$ such that $x^*_1=1$ and $x^*_2=1$.
While finding such an $x^*$ is trivial, we don a veil of ignorance and proceed as if it were not.
This essentially means that our function $f(\mathbf{x})$ is an AND gate. But AND gate is not reversible and cannot be a quantum gate.
However the Toffoli gate, that was introduced in the previous section, is a reversible version of the classical AND gate.
The Toffoli gate takes three bits as input and outputs three bits.
The first two bits are unmodified.
The third bit is flipped if the first two bits are 1. The unitary matrix corresponding to the Toffoli gate can be found in Table \ref{gate_table}.
In other words, the Toffoli gate implements our desired quantum oracle where the first two inputs are $x_1$ and $x_2$ and the third bit is the ancillary bit, $q$.
The behavior of the oracle in general is $\ket{\mathbf{x}}\ket{q}\rightarrow\ket{\mathbf{x}}\ket{f(\mathbf{x})\bigoplus q}$, where $\bigoplus$ is the XOR operation . Here we will only discuss the case where $\mathbf{x^*}$ is unique. Grover's algorithm can also be used to search for multiple items in a database.

\subsection{Algorithm description}
\begin{figure}
	\includegraphics[width=\textwidth]{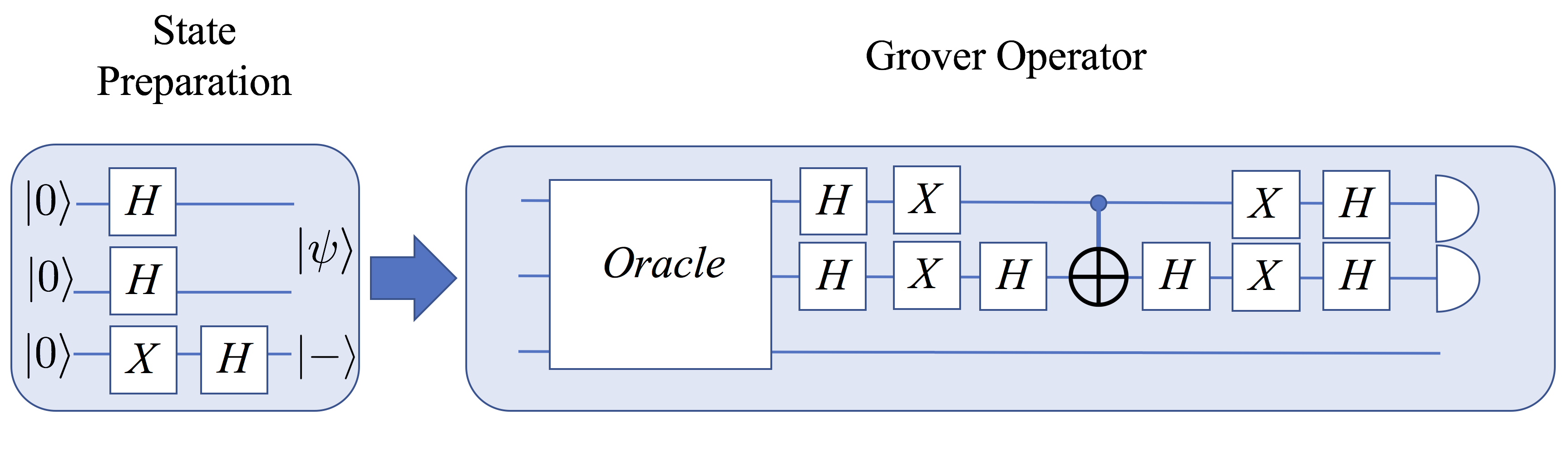}
	\caption{A schematic diagram of Grover's algorithm is shown.
	Note that in this case, one application of the Grover operator is performed.
	This is all that is necessary when there are only two bits in $\mathbf{x}$, but the Grover operator should be applied repeatedly for larger problems.}
	\label{fig:abstractcircuit}
\end{figure}

Here we present a brief introduction to Grover's algorithm.
A more detailed account can be found in Nielsen and Chuang~\cite{NielsenChuang}. Let $N$ be the number of items (represented as bit strings) amongst which we are performing the search. This number will also be equal to the dimension of the vector space we are working with. 
An operator, called the Grover operator or the diffusion operator, is the key piece of machinery in Grover's algorithm.
This operator is defined by
\begin{equation}
	G=(2\ket{\psi}\bra{\psi} - I)O
\end{equation}
where $\ket{\psi} =  \frac{1}{\sqrt{N}} \sum_i \ket{i} $ is the uniform superposition over all the basis states and $O$ is the oracle operator (see Fig.~\ref{fig:abstractcircuit} for a representation of this operator in the case where $\mathbf{x}$ consists of 2 bits).
The action of $(2\ket{\psi}\bra{\psi} - I)$ on an arbitrary state, given by $\sum_{i} a_i \ket{i}$,  when decomposed over the basis states is,
\begin{equation}
	(2\ket{\psi}\bra{\psi} - I) \sum_{i} a_i \ket{i} = \sum_{i} \left( 2\left<a\right> - a_i \right)\ket{i}
	\label{eq:inversion}
\end{equation}
where $\left<a\right>=\frac{\sum_i a_i}{N}$ is the average amplitude in the basis states.
From Eq.~\eqref{eq:inversion} one can see that the amplitude  of each $\ket{i}$-state ($a_i$) is flipped about the mean amplitude.

In order to use the Grover operator to successfully perform a search, the qubit register must be appropriately initialized.
The initialization is carried out by applying a Hadamard transform to each of the the main qubits ($H^{\otimes n}$) and applying a Pauli X transform followed by a Hadamard transform ($HX$) to the ancilla.
This leaves the main register in the uniform superposition of all states, $\ket{\psi}$, and the ancilla in the state $\frac{\ket{0}-\ket{1}}{\sqrt{2}}$.
After performing these operations, the system is in the state $\ket{\psi}\frac{\ket{0}-\ket{1}}{\sqrt{2}}$.
Using Eq.~\eqref{eq:inversion}, we can now understand how the Grover operator works.
The action of the oracle operator on $\ket{\mathbf{x}^*}\frac{\ket{0}-\ket{1}}{\sqrt{2}}$ reverses the amplitude of that state
\begin{equation}\label{eq:phase_kickback}
	O\ket{\mathbf{x}^*}\frac{\ket{0}-\ket{1}}{\sqrt{2}}\rightarrow\ket{\mathbf{x}^*}\frac{\ket{f(\mathbf{x}^*)\bigoplus0}-\ket{f(\mathbf{x}^*)\bigoplus1})}{\sqrt{2}} = \ket{\mathbf{x}^*}\frac{\ket{1}-\ket{0}}{\sqrt{2}} = -\ket{\mathbf{x}^*}\frac{\ket{0}-\ket{1}}{\sqrt{2}}
\end{equation}
A similar argument shows that all other states are unmodified by the oracle operator.
Combining this with Eq.~\eqref{eq:inversion} reveals why the Grover operator is able to successfully perform a search.
Consider what happens on the first iteration:
The oracle operator makes it so that the amplitude of $\ket{\mathbf{x}^*}$ is below $\left< a \right>$ (using the notation of Eq.~\eqref{eq:inversion}) while all the other states have an amplitude that is slightly above $\left<a\right>$.
The effect of applying $2\ket{\psi}\bra{\psi} - I$ is then to make $\ket{\mathbf{x}^*}$ have an amplitude above the mean while all other states have an amplitude below the mean.
The desired behavior of the Grover operator is to increase the amplitude of $\ket{\mathbf{x}^*}$ while decreasing the amplitude of the other states.
If the Grover operator is applied too many times, this will eventually stop happening.
The Grover operator should be applied exactly $\left\lceil\frac{\pi \sqrt{N}}{4}\right\rceil$ times after which a measurement will reveal $\mathbf{x^*}$ with probability close to $1$. 
In the case where $\mathbf{x}$ has two bits, a single application of Grover's operator is sufficient to find $\mathbf{x}^*$ with certainty (in theory). Below is a high level pseudocode for the algorithm.

\begin{algorithm}[H]
\caption{Grover's algorithm}
\begin{algorithmic} 
    \STATE \textbf{Input:}
    \bindent
        \STATE $\bullet$ An Oracle operator effecting the transformation $\ket{x}\ket{q} \rightarrow  \ket{x}\ket{q \oplus f(x)}$.
    \eindent
    \STATE \textbf{Output:}
    \bindent
        \STATE $\bullet$ The unique bit string $\mathbf{x^*}$ satisfying Eq. \eqref{eq:grov_f}
    \eindent
    \STATE \textbf{Procedure:}
    \bindent
        \STATE \textbf{Step 1.} Perform state initialization $\ket{0\ldots0} \rightarrow \ket{\psi}(\frac{\ket{0} - \ket{1}}{\sqrt{2}})$
        
        \STATE \textbf{Step 2.} Apply Grover operator  $\left\lceil\frac{\pi \sqrt{N}}{4}\right\rceil$ times
        \STATE \textbf{Step 3.} Perform measurement on all qubit except the ancillary qubit.
    \eindent
\end{algorithmic}
\end{algorithm}

\subsection{Algorithm implemented on IBM's 5-qubit computer}
\begin{figure}
	\includegraphics[width=\textwidth]{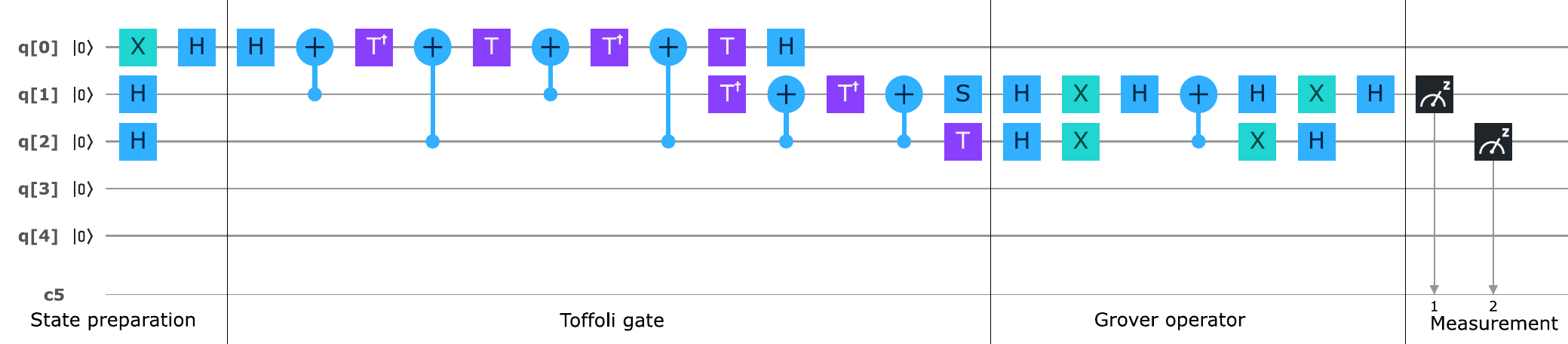}
	\caption{The circuit that was executed on IBM's 5-qubit quantum computer.
	The first two time slots correspond to the state preparation.
	The next 13 time slots implement a Toffoli gate.
	The next 7 time slots implement the $2\ket{\psi}\bra{\psi} - I$ operator, and the final two time slots are used for observing $x_1$ and $x_2$.}
	\label{fig:ibmcircuit}
\end{figure}

Fig.~\ref{fig:ibmcircuit} shows the circuit that was designed to fit the \verb|ibmqx4| quantum computer. The Toffoli gate is not available directly in \verb|ibmqx4| so it has to be constructed from the available set of gates given in Eq. \ref{eqn5}.

The circuit consists of state preparation (first two time slots), a Toffoli gate (the next 13 time slots), followed by the $2\ket{\psi}\bra{\psi} - I$ operator (7 time slots), and measurement (the final 2 time slots). 
We use $q[0]$ (in the register notation from Fig.~\ref{fig:ibmcircuit}) as the ancillary qubit, and $q[1]$ and $q[2]$ as $x_1$ and $x_2$ respectively. 
Note that the quantum computer imposes constraints on the possible source and target of CNOT gates.

Using the simulator, this circuit produces the correct answer $\mathbf{x}=(1,1)$ every time.
We executed 1,024 shots using the \verb|ibmqx4| and $\mathbf{x}=(1,1)$ was obtained 662 times with $(0,0)$, $(0, 1)$, and $(1, 0)$ occurring 119, 101, and 142 times respectively.
This indicates that the probability of obtaining the correct answer is approximately 65\%.
The deviation between the simulator and the quantum computer is due to the inherent errors in \verb|ibmqx4|. This deviation will get worse for circuits of larger size.

We also ran another test using CNOT gates that did not respect the underlying connectivity of the computer. 
 This resulted in a significantly deeper circuit and the results were inferior to the results with the circuit in Fig.~\ref{fig:ibmcircuit}.

This implementation used a Toffoli gate with a depth of 23 (compared to a depth of 13 here) and obtained the correct answer 48\% of the time.

\section{Bernstein-Vazirani Algorithm}
%\section{Bernstein-Vazirani Algorithm in IBM Quantum Experience}

\subsection{Problem definition and background}

Suppose we are given a classical Boolean function, $f : \{0,1\}^n \mapsto \{0,1\}$.
It is guaranteed that this function can always be represented in the form, $f_\V{s}(\V{x}) = \bigoplus_{i} s_i x_i \equiv \la \V{s}, \V{x} \ra $. 
Here, \V{s} is an unknown bit string, which we shall call a \emph{hidden string}. Just like in Grover's algorithm we assume that we have a quantum oracle that can compute this function.

The Bernstein-Vazirani (BV) algorithm then finds the hidden string with just a single application of the oracle. The number of times the oracle is applied during an algorithm algorithm is known as its \emph{query complexity}. The BV algorithm has a query complexity of one. From our earlier discussions we saw that Grover's algorithm has a query complexity of $O(\sqrt{N}).$ 

In the classical case each call to  $f_\V{s}(\V{x})$ produces just $1$ bit of information, and since an arbitrary hidden string \V{s} has $n$-bits of information, the classical query complexity is seen to be $n$. Even with bounded error, there is no way that this classical complexity can be brought down, as can be seen using slightly more rigorous information-theoretic arguments.

The quantum algorithm to solve this problem was developed by Bernstein and Vazirani~\cite{BV93} building upon the earlier work of Deutsch and Jozsa~\cite{DJ92}. Their contribution was a quantum algorithm for the hidden string problem, which has a non-recursive quantum query complexity of just $1$. This constitutes a polynomial $\mathcal{O}(n)$ query-complexity separation between classical and quantum computation. They also discovered a less widely known recursive hidden-string query algorithm, which shows a $\mathcal{O}(n^{\log{n}})$ separation between classical and quantum query-complexities. These developments preceded the more famous results of Shor and Grover, and kindled a lot of early academic interest in the inherent computational power of quantum computers.

One thing to note about the BV algorithm is that the \emph{black-box} function $f_\V{s}(\cdot)$ can be very complex to implement using reversible quantum gates. For an $n$-bit hidden string, the number of simple gates needed to implement $f_\V{s}(\cdot)$ scales typically as $\mathcal{O}(4^n)$\cite{NielsenChuang}. Since the black box is a step in the algorithm, its serial execution time could in the worst-case even scale exponentially. The real breakthrough of this quantum algorithm lies in speeding up the query complexity and not the execution time per~se.

\subsection{Algorithm description}
Let us explore the BV algorithm in more detail. Let $U_s$ be the oracle for the function $f_{\V{s}}(\V{x})$. It acts in the usual way and computes the value of the function onto an ancilla qubit,

\begin{equation}
    U_s \ket{\V{x}}\ket{q} = \ket{\V{x}}\ket{q \oplus \la \V{s}, \V{x} \ra}
\end{equation}

 Denoting $\ket{-} = {(\ket{0}-\ket{1})}/{\sqrt{2}}$, we can easily verify from Eq. \eqref{eq:phase_kickback} that,
\begin{equation}
     U_s \ket{\V{x}}\ket{-} = (-1)^{\la \V{s}, \V{x} \ra} \ket{\V{x}}\ket{-}.
\end{equation}

Also, note that the $n$-qubit Hadamard operator, which is just $n$ single qubit $H$ operators applied in parallel, can be expanded as,
\begin{equation}\label{eq:H_identity}
H^{\otimes n} = \frac{1}{\sqrt{2^n}}\sum_{\V{x},\V{y} \in \{0,1\}^n} \left(-1\right)^{\braket{\V{x},\V{y}}}\ket{\V{y}}\bra{\V{x}}
\end{equation}

The reader may verify this identity by applying $H^{\otimes n}$ to the computational basis states.

$U_s$ and $H^{\otimes n}$ are the only two operators needed for the BV algorithm. The pseudocode for the algorithm is given in Algorithm \ref{alg:BV}. Notice that the initialization part is identical to that of Grover's algorithm. This kind of initialization is a very common strategy in quantum algorithms.

\begin{algorithm}[H]
\caption{Bernstein-Vazirani algorithm}
\begin{algorithmic} \label{alg:BV}
    \STATE \textbf{Input:}
    \bindent
        \STATE $\bullet$ An oracle operator, $U_s$, effecting the transformation $\ket{x}\ket{q} \rightarrow  \ket{x}\ket{q \oplus \braket{\V{s}, \V{x}}}$.
    \eindent
    \STATE \textbf{Output:}
    \bindent
        \STATE $\bullet$ The hidden string $\V{s}.$
    \eindent
    \STATE \textbf{Procedure:}
    \bindent
        \STATE \textbf{Step 1.} Perform state initialization on $n+1$ qubits, $\ket{0\ldots0} \rightarrow \ket{\psi} \ket{-}$
        
        \STATE \textbf{Step 2.} Apply $U_s$ .
        \STATE \textbf{Step 3.} Apply $H^{\otimes n}$ to the first $n$ qubits.
        \STATE \textbf{Step 4.} Measure all qubits except the ancillary qubit.
    \eindent
\end{algorithmic}
\end{algorithm}

\begin{figure}[!t]
    \centering
\includegraphics[width=0.90\textwidth]{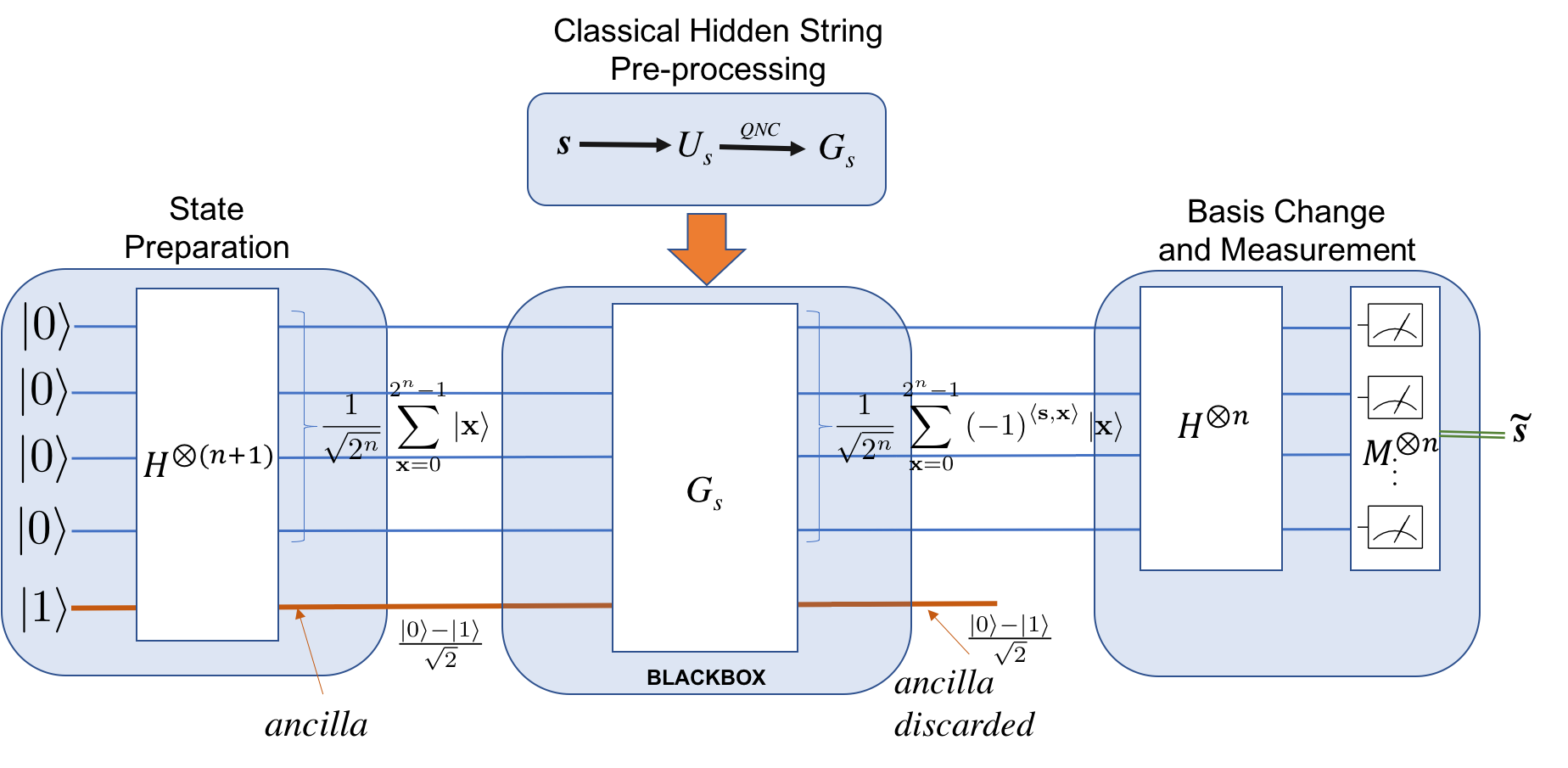}
    \caption{Bernstein-Vazirani hidden string discovery quantum algorithm. The hidden string \V{s} is discovered with just a single query. The measurement result $\widetilde{\V{s}}$ gives the hidden string.\label{BV:Fig:Block}}
\end{figure}

The final measurement will reveal the hidden string, $\V{s}$, with probability $1$. Let us now delve into the algorithm to see how this result is achieved. The entire circuit for the BV algorithm is represented in Figure~\ref{BV:Fig:Block}. This circuit can be analyzed as follows,
\begin{eqnarray}
    \ket{0}^n\ket{1} & \xrightarrow{H^{\otimes(n)} \otimes H} & \frac{1}{\sqrt{2^n}} \sum_{\V{x}=0}^{2^n-1} \ket{\V{x}} \otimes  \ket{-}
                     \quad \xrightarrow{U_s} \quad \frac{1}{\sqrt{2^n}} \sum_{\V{x}=0}^{2^n-1} \left(-1\right)^{\braket{\V{s}, \V{x}}}\ket{\V{x}} \otimes \ket{-} \nonumber\\
                     & \xrightarrow{H^{\otimes n}} & \frac{1}{\sqrt{2^n}} \sum_{\V{x},\V{y}=0}^{2^n-1} \left(-1\right)^{\braket{\V{s}, \V{x}} \oplus \braket{\V{x}, \V{y}}} \ket{\V{y}} \otimes \ket{-}
                     \quad \equiv \quad \ket{\V{s}} \otimes \ket{-}.
\end{eqnarray}

Here we have crucially used the identity  for $H^{\otimes n}$ given in Eq.\eqref{eq:H_identity}.

%\vspace*{-3ex}
\subsection{Algorithm implemented on IBM's 5-qubit and 16-qubit computers}
From the BV algorithm description in the previous section, we see that in any practical implementation of this algorithm, the main ingredient is the construction of the oracle $U_{\V{s}}$ given a binary hidden string \V{s}. Let us see how this is done using an example binary hidden string ``$01$''.  Equation~(\ref{BV:Tab:Perm}) below shows how the $3$-qubit operator maps the $2^3 = 8$ basis vectors onto themselves.  The first line is the input binary vector (in the order $ x_1, x_0,q$), and the second line is the output binary vector.
\begin{equation}
    U_{01} = \begin{pmatrix}
        000 & 010 & 100 & 110 & 001 & 011 & 101 & 111 \\
        000 & 011 & 100 & 111 & 001 & 010 & 101 & 110
    \end{pmatrix}
    \label{BV:Tab:Perm}
\end{equation}
This mapping,  $U_{01} : \ket{\V{x}}\ket{ q} \mapsto \ket{\V{x}}\ket{ \braket{01, \V{x}} \oplus q}$, is unitary. The next task in implementation is to lower the unitary matrix operator $U_{01}$ to primitive gates available in the quantum computer's architecture given in Eq \eqref{eqn5}. The time cost of applying these gates can be accessed from IBM's published calibration models~\cite{IBM:QX:Models} for the primitive hardware gates.

In order to decompose arbitrary unitary matrices to the primitive gates, we need to first perform a unitary diagonalization of the $2^{(n+1)} \times 2^{(n+1)}$ matrix using multi-qubit-controlled single-qubit unitary Given's rotation operations. Such multi-qubit-controlled single-qubit operations can be decomposed further to primitive gates using standard techniques~\cite{NielsenChuang} to the hardware primitive gates. Even after this step we will be left with arbitrary CNOT gates that do not respect the topology of the underlying quantum processor. Since both \verb|ibmqx4|, \verb|ibmqx5| computers have restricted CNOT connectivity between qubits, we will need to decompose the CNOT gates further into available CNOT gates using the method discussed in the introductory section. As we saw in the Grover's algorithm section, such decompositions will further degrade the quality of our results. As the overall primitive gate counts scale as $\mathcal{O}(4^n)$ for arbitrary $n$-qubit unitary operators, these decompositions quickly becomes hard to do by hand. To address this we wrote a piece of software called \emph{Quantum Netlist Compiler (QNC)}~\cite{NS:QNC2017} for performing these transformations automatically. QNC can do much more than convert arbitrary unitary operators to OpenQASM-2.0 circuits---it has specialized routines implemented to generate circuits to do state-preparations, permutation operators, Gray coding to reduce gate counts, mapping to physical machine-topologies, as wells as gate-merging optimizations. Applying QNC tool to the unitary matrix $U_{\V{s}}$ gives us a corresponding quantum gate circuit $G_{\V{s}}$ as shown in Figure~\ref{BV:Fig:Block} for a specific bit-string \V{s}.

\begin{figure}[!hb]
    \centering
\includegraphics[width=\textwidth]{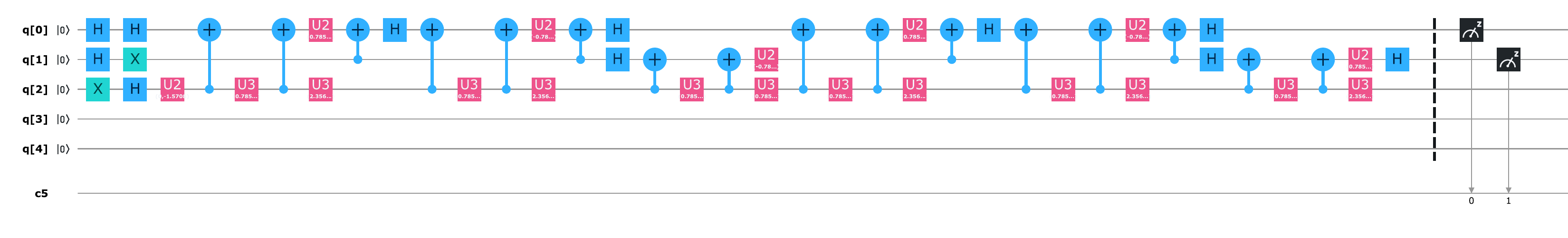}
    \caption{Quantum circuit for BV algorithm with hidden string ``01'' targeting the \texttt{ibmqx4} architecture.\label{BV:Fig:Score}}
\end{figure}

QNC generated black-box circuits with following gate-counts for the non-trivial $2$-bit hidden-strings: ``01'': $36$, ``11'': $38$, ``10'': $37$, with estimated execution time\footnote{\label{note1} These times are estimated using the data available from IBM at the time of writing. These values will change as the hardware improves.}  for critical path ${\sim}17\mu s$ on an ideal machine with all-to-all connection topology. For the $5$-qubit \verb|ibmqx4| machine the corresponding gate-counts where: ``01'': $42$, ``11'': $43$, ``10'': $41$, with estimated execution time for critical path ${\sim}15\mu s$, and for the $16$-qubit \verb|ibmqx5|, they were: ``01'': $66$, ``11'': $67$, ``10'': $67$, with estimated execution time for critical path ${\sim}28\mu s$. In all these cases, QNC used a specialized decomposition of $U_{01}$, considering its permutation matrix nature, and therefore was able to reduce gate-counts by $5\times$ over the case when this special structure was ignored. Considering that the machines' observed coherence times are of the order of ${\sim}60 \mu s$, these QNC optimizations were crucial to the feasibility of the resulting score.
The quantum score (circuit) generated by QNC for $U_{01}$ for \verb|ibmqx4| is shown in Figure~\ref{BV:Fig:Score}. A similarly prepared score for $3$-bit hidden-string ``111'' had a gate-count of $428$ in the \verb|ibmqx4| architecture with an estimated execution time of $153 \mu s$ which was well above the machines' coherence times.

\begin{figure}[H]
    \centering
\includegraphics[width=0.26\textwidth]{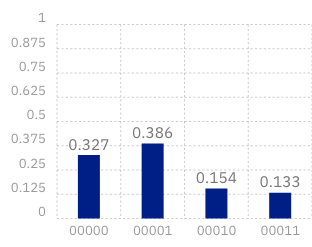}
    \hspace*{2ex}
\includegraphics[width=0.26\textwidth]{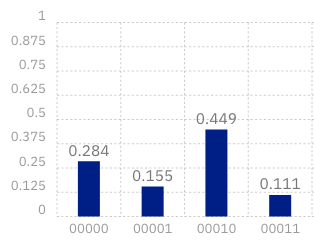}
    \hspace*{2ex}
\includegraphics[width=0.26\textwidth]{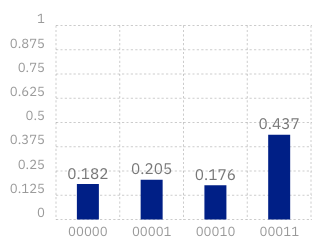}
    \caption{Results from running the BV algorithm for $8192$ shots on $2$-bit hidden-strings ``01'', ``10'' and ``11'' respectively (left to right) on \texttt{ibmqx4}. The y-axis here is the probability of obtaining the hidden string, which theoretically should be $1$.\label{BV:Fig:Results}}
\end{figure}

We tested the QNC generated quantum scores for all non-trivial $1$-qubit, $2$-bit and $3$-bit strings using the IBM-Qiskit based local simulator. In all cases, the simulator produced the exact hidden-string as the measurement result, $100\%$ of the trials. We then tested all $1$-bit and $2$-bit strings on both the $5$-qubit \verb|ibmqx4| and the $16$-qubit ibmqx5 machines. The results are shown in Figure~\ref{BV:Fig:Results}. For $2$-bit strings, the worst case noise was observed for the string ``01'' on \verb|ibmqx4| when the qubits $q_0, q_1, q_2$ where used for $x_0, x_1, y$ respectively. Since the estimated critical path times exceeded the machines' coherence times for $3$-bit strings, we did not run those scores on the physical machines. Even for $2$-bit strings, the scores were quite long, and the results were quite noisy even with $8192$ machine-shots.

\section{Linear Systems}

\subsection{Problem definition and background}

Solving linear systems is central to a majority of science, engineering, finance and economics applications. For example, one comes across such systems while solving differential or partial differential equations or while performing regression.
The problem of solving a system of linear equations is the following: Given a system ${A} \vec{x} = \vec{b}$, find $\vec{x}$ for a given matrix $\vec{A}$ and vector $\vec{b}$.
Here we assume that $A$ is a Hermitian matrix, in that it is self-adjoint.
To represent $\vec{x}$, $\vec{b}$ as quantum states $\ket{x}$, $\ket{b}$, respectively, one has to rescale them as unit vectors, such that $\norm{\vec{x}} = \norm{\vec{b}} = 1$.
Thus, one can pose the problem as finding $\ket{x}$ such that 
\begin{align}
A\ket{x} = \ket{b},
\end{align} 
with the solution $\ket{x}$ being
\begin{align}
\ket{x} = \frac{A^{-1} \ket{b}}{\norm{A^{-1} \ket{b}}}.
\end{align}

\subsection{Algorithm description}
The quantum algorithm for the linear system was first proposed by Harrow, Hassidim, and Lloyd (HHL)~\cite{harrow2009quantum}. The HHL algorithm has been implemented on various quantum computers in~\cite{2013PhRvL.110w0501C,zheng2017solving,barz2014two}.
The problem of solving for $\vec{x}$ in the system $A \vec{x} = \vec{b}$ is posed as obtaining expectation value of some operator $M$ with $\vec{x}$, $\vec{x}^{\dag} M \vec{x}$, instead of directly obtaining the value of $\vec{x}$.
This is particularly useful when solving on a quantum computer, since one usually obtains probabilities with respect to some measurement, typically, these operators are Pauli's operators $X$, $Y$, $Z$.
These probabilities can then be translated to expectation values with respect to these operators. 

The user has to keep in mind certain caveats while using the HHL algorithm. The algorithm requires that the elements of $\vec{b}$ be accessible in superposition. Also, the solution vector is given as a quantum state which collapses after every measurement. This means that the HHL algorithm involves additional overheads for loading and reading data from a quantum computer \cite{aaronson2015read}. Recently, classical algorithms inspired by HHL have been developed that, while having assumptions similar to HHL, considerably reduce the complexity of solving linear systems on classical computers \cite{chia_et_al:LIPIcs:2020:13391}

The main idea of the HHL algorithm is as follows. Let $\{ \ket{u_j} \}$ and $\{ \lambda_j \}$ be the eigenvectors and eigenvalues of $A$, respectively, with the eigenvalues rescaled such that $0 < \lambda_j < 1$. Then the state $\ket{b}$, can be written as a linear combination of the eigenvectors $\{ \ket{u_j} \}$, $\ket{b} = \sum_{j=1}^{N} \beta_j \ket{u_j}$.
The goal of the HHL algorithm is to obtain $\ket{x}$ in the form $\ket{x} = \sum_{j=1}^{N} \beta_j \frac{1}{\lambda_j} \ket{u_j}$.
By decomposing $A = R^{\dag} \Lambda R$, the HHL algorithms in a nutshell involves performing a set of operations that essentially performs the three steps: 
\begin{align}
R^{\dag} \Lambda R \ket{x} = \ket{b} \overset{\mathrm{Step 1}}{\implies} \Lambda R \ket{x} = R \ket{b} \overset{\mathrm{Step 2}}{\implies} R \ket{x} = \Lambda^{-1} R \ket{b} \overset{\mathrm{Step 3}}{\implies} \ket{x} = R^{\dag} \Lambda^{-1} R \ket{b} 
\label{eq:decomp}
\end{align}

This procedure requires us to find the eigenvalues of $A.$ This can be done using a quantum subroutine called  {\it phase estimation.}  We will discuss this subroutine in some detail as it is a common ingredient in many quantum algorithms.

\subsection{Phase estimation}
\label{subsec:phase_estim}
Phase estimation is a quantum subroutine that lets us find the eigenvalues of a unitary matrix $U$ given the ability to apply it to a quantum register as a controlled gate. Let $\ket{u}$ be an eigenvector  of $U$ such that, $U \ket{u} =  e^{2 \pi i \lambda_u} \ket{u}$. Then the phase estimation subroutine effects the following transformation,
\begin{equation}
    \ket{0} \ket{u} \xrightarrow{} \ket{\tilde{\lambda}_u }\ket{u}.
\end{equation}

Here $\tilde{\lambda}_u$ is an estimate for $\lambda_u$. This subroutine makes use of an important transformation called the Quantum Fourier Transform (QFT)

\begin{figure}[htbp]
\includegraphics{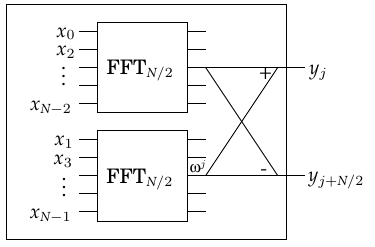}
\caption{Fast Fourier Transform circuit, where $j$ denotes a row from the top half of the circuit and $\omega^j$ denotes that the corresponding value is multiplied by $\omega^j$. The plus and minus symbols indicate that the corresponding values have to be added or subtracted, respectively.}
\label{fig:FFT}
\end{figure}

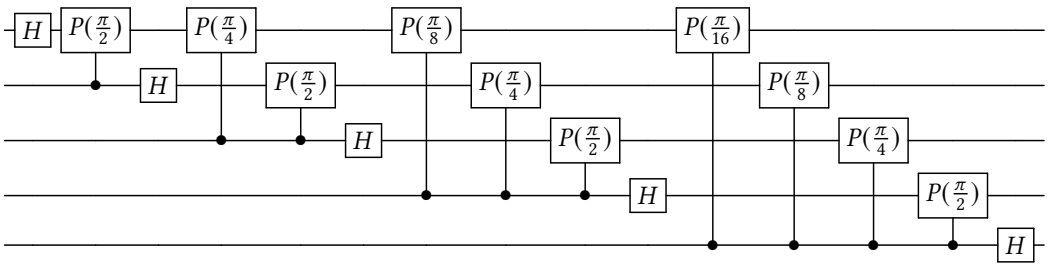
\begin{figure}[htbp]
\centerline{
\Qcircuit @C=0.39em @R=0.39em {
& \gate{H} & \gate{P (\frac{\pi}{2})} & \qw      & \gate{P (\frac{\pi}{4})} & \qw                      & \qw      & \gate{P (\frac{\pi}{8})} & \qw                & \qw                      &\qw      & \gate{P (\frac{\pi}{16})} &\qw & \qw &\qw &\qw &\qw\\
& \qw      & \ctrl{-1}                & \gate{H} & \qw                      & \gate{P (\frac{\pi}{2})} & \qw      & \qw                               & \gate{P (\frac{\pi}{4})} & \qw                      &\qw      & \qw & \gate{P (\frac{\pi}{8})} & \qw &\qw &\qw &\qw     \\
& \qw      & \qw                      & \qw      & \ctrl{-2}                &\ctrl{-1}                 & \gate{H} & \qw                      & \qw                      & \gate{P (\frac{\pi}{2})} &\qw      &\qw  &\qw & \gate{P (\frac{\pi}{4})} &\qw &\qw &\qw  \\
& \qw      & \qw                      & \qw      & \qw                      & \qw                      & \qw      & \ctrl{-3}               & \ctrl{-2}                &\ctrl{-1}                 &\gate{H} &\qw  &\qw &\qw & \gate{P (\frac{\pi}{2})} &\qw &\qw    \\
& \qw      & \qw                      & \qw      & \qw                      & \qw                      & \qw      & \qw                     & \qw                      & \qw                      &\qw   & \ctrl{-4} & \ctrl{-3} &\ctrl{-2} &\ctrl{-1} &\gate{H} &\qw
}}
\caption{A Quantum Fourier Transform circuit for five qubits ($n=5$).}
\label{fig:QFT}
\end{figure}

\paragraph*{Quantum Fourier Transform}
The Discrete Fourier Transform (DFT) takes as an input a vector $X$ of size $N$ and outputs vector $Y=WX$ where the \textit{Fourier matrix} $W$ is defined by
\begin{equation*}
W=\frac{1}{\sqrt{N}}
\begin{bmatrix}
    1       & 1 & 1 & \dots & 1 \\
    1       & \omega & \omega^2 & \dots & \omega^{N-1} \\
    1       & \omega^2 & \omega^4 & \dots & \omega^{2(N-1)} \\
    \vdots & \vdots & \vdots & \ddots & \vdots &  \\
    1       & \omega^{N-1} & \omega^{2(N-1)} & \dots & \omega^{(N-1)(N-1)}
\end{bmatrix},
\end{equation*}
where the $ij$-th element of the matrix is $W_{ij}=\omega^{ij}$ and $\omega$ is a primitive $N$-th root of one($\omega^N=1$). A straightforward implementation of the matrix-vector multiplication takes $O(N^2)$ operations, but, by using the special structure of the matrix, the Fast Fourier Transform (FFT) does the multiplication in only $O(N\log N)$ time. The algorithm is recursive and is illustrated on Figure~\ref{fig:FFT}.
The Quantum Fourier Transform (QFT) is defined as a transformation between two quantum states that are determined using the values of DFT (FFT)\@. If $W$ is a Fourier matrix and $X=\{x_i\}$ and $Y=\{y_i\}$ are vectors such that $Y=WX$, then the QFT is defined as the transformation
\begin{equation}
  {\mathit QFT}\left(\sum_{k=0}^{N-1}x_k\ket{k}\right)= \sum_{k=0}^{N-1}y_k\ket{k}.
\end{equation}
The implementation of the QFT mimics the stages (recursive calls) of the FFT, but implements each stage using only $n+1$ additional gates per stage. A single Hadamard gate on the last (least significant) bit implements the additions/subtractions of the outputs from the recursive call and the multiplications by $\omega^j$ are done using $n$ controlled phase gates. The circuit for $n=5$ is shown on Figure~\ref{fig:QFT}.

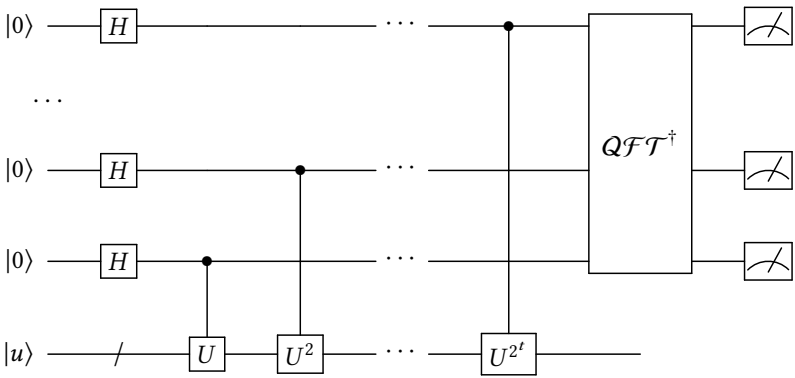
\begin{figure}[H]
\begin{equation*}
\Qcircuit @C=2em @R=2em {
\lstick{\ket{0}}&\gate{H} &\qw                &\qw        &\qw       & \lstick{\cdots}  &\ctrl{4} & \multigate{3}{\mathcal{QFT}^\dagger} &\meter \\
\ldots\\
\lstick{\ket{0}}&\gate{H} &\qw               &\ctrl{2}       &\qw    & \lstick{\cdots}   & \qw &\ghost{\mathcal{QFT}^\dagger}  & \meter \\
\lstick{\ket{0}}&\gate{H} & \ctrl{1}         &\qw                & \qw & \lstick{\cdots}   & \qw &\ghost{\mathcal{QFT}^\dagger} & \meter\\
\lstick{\ket{u}}&  \qw  {/}      & \gate{U} &\gate{U^2} & \qw   &\lstick{\cdots} &\gate{U^{2^t}} &\qw \\
}
\end{equation*}
    \caption{Quantum circuit for phase estimation.}
    \label{fig:phase_estim}
\end{figure}

The phase estimation procedure cleverly uses the QFT operator to  estimate the eigenphases of the operator $U$. 
The circuit for performing phase estimation given in Fig. \ref{fig:phase_estim}. Notice that the QFT is applied in reverse.

The pseudocode for phase estimation is given in Algorithm \ref{alg:PS}. Notice that the algorithm also works if the input state is not an eigenstate. The output in this case can be determined by expanding the input state in terms of the eigenstates and then applying the linearity of quantum operations. In the code, we have numbered the ancillary qubits from the top and $C_i  U$ denotes the unitary controlled by the $i^{\text{th}}$ ancilla qubit acting on the main $n$ qubit register. 

\begin{algorithm}[H]
\caption{Phase estimation subroutine}
\begin{algorithmic} \label{alg:PS}
    \STATE \textbf{Input:}
    \bindent
        \STATE $\bullet$ Controlled unitaries $C_{i} U$ 
        \STATE $\bullet$  An $n$ qubit input state $\ket{\psi} = \sum_u \psi_u \ket{u}$, where $U \ket{u} =  e^{2 \pi i \lambda_u} \ket{u}.$  
    \eindent
    \STATE \textbf{Output:}
    \bindent 
        \STATE $\bullet$ $\sum_u \psi_u \ket{ \tilde{ \lambda}_u} \ket{u} $
    \eindent
    \STATE \textbf{Procedure:}
    \bindent
        \STATE \textbf{Step 1.} Take $t$ ancillary qubits initialized to zero and perform  $H^{\otimes t}$ on them to produce the uniform superposition state over them.
        
         \FOR{ $0 \leq i < t$}
         \bindent
         \STATE \textbf{Step 2.} Apply $C_{t - i -1} U^{2^i}$
         \eindent
         \ENDFOR
        
        \STATE \textbf{Step 3.} Apply $\mathcal{QFT}^\dagger$ .

        \STATE \textbf{Optional} Measure the ancillary qubits to get $\ket{ \tilde{ \lambda}_u} \ket{u}$ with probability $|\psi_u|^2$  
        
    \eindent
\end{algorithmic}
\end{algorithm}

The number of ancillary qubits used in the phase estimation algorithm will determine both its run-time and its accuracy. On the accuracy front, the number of ancillary qubits used is equal to the bit precision of $\tilde{\lambda}_u$ as the answer is stored in this register. The exact complexity of this subroutine is discussed in Ref. \cite{NielsenChuang}.

Now we can discuss the HHL algorithm which makes use of the phase estimation procedure to perform a matrix inversion. The HHL algorithm requires three sets of qubits: a single ancilla qubit, a register of $n$ qubits used to store the eigenvalues of $A$ in binary format with precision up to $n$ bits, and a memory of $O(\mathrm{log}(N))$ that initially stores $\ket{b}$ and eventually stores $\ket{x}$.
Start with a state $\ket{0}_a \ket{0}_r \ket{b}_m$, where the subscripts $a$, $r$, $m$, denote the sets of ancilla, register and memory qubits, respectively. This subscript notation was used in~\cite{zheng2017solving}, and we found it to be most useful in keeping things clear. The HHL algorithm requires us to run the phase estimation procedure on the unitary operator $e^{iA}.$ The phases estimated would be approximations to the eigenvalues of $A$. The problem  of applying the unitary operation $e^{iA}$ given the matrix $A$ is called \emph{quantum simulation}. There are many algorithms in literature that tackle the problem of quantum simulation \cite{berry2007efficient} \cite{georgescu2014quantum} and that will not be our focus in this section.  We will explain the steps of the HHL algorithm below assuming that the quantum simulation part is taken care of. We will also include some mathematical details in the pseudocode given in Algorithm \ref{alg:HHL} . 

\begin{figure}[!htb]
\begin{center}
\includegraphics[width=\columnwidth]{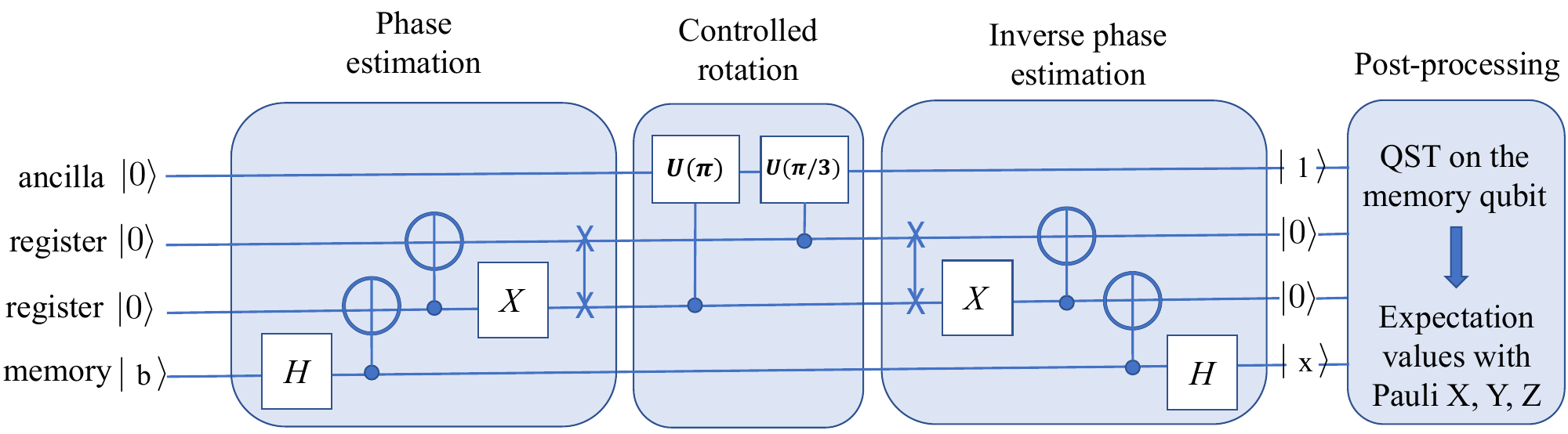}
\caption{Schematic of the circuit for the quantum algorithm for solving a 2 $\times$ 2 linear system. The first step involves phase estimation, which maps the eigenvalues $\lambda_j$ of A into the register in the binary form. The second step involves controlled rotation of the ancilla qubit, so that the inverse of the eigenvalues $\frac{1}{\lambda_j}$ show up in the state. The third step is the inverse phase estimation to disentangle the system, and restores the registers to $\ket{0}$. The memory qubit now stores $\ket{x}$, which is then post-processed to get the expectation values with respect to the Pauli operators $X$, $Y$ and $Z$.}
\label{fig:ls_setup}
\end{center}
\end{figure}

\begin{algorithm}[!htb]
\caption{HHL algorithm}
\begin{algorithmic} \label{alg:HHL}
    \STATE \textbf{Input:}
    \bindent
       \STATE $\bullet$   The state $\ket{b} = \sum_j \beta_j \ket{u_j}$
       \STATE $\bullet$ The ability to perform controlled operations with unitaries of the form $e^{iAt}$
    \eindent
    \STATE \textbf{Output:}
    \bindent 
        \STATE $\bullet$ The quantum state $\ket{x}$  such that $A \vec{x} = \vec{b}.$
    \eindent
    \STATE \textbf{Procedure:}
    \bindent
        \STATE \textbf{Step 1.}  Perform quantum phase estimation using the unitary transformation $e^{iA}$. This maps the eigenvalues $\lambda_j$ into the register in the binary form to transform the system,
        \begin{equation}
        \ket{0}_a \ket{0}_r \ket{b}_m \rightarrow \sum_{j=1}^{N}\beta_j\ket{0}_a \ket{\lambda_j}_r \ket{u_j}_m.
        \end{equation}       
        \STATE \textbf{Step 2.}   Rotate the ancilla qubit $\ket{0}_a$ to $\sqrt{1-\frac{C^2}{\lambda_j^2}} \ket{0}_a + \frac{C}{\lambda_j} \ket{1}_a$ for each $\lambda_j$. This is performed through controlled rotation on the $\ket{0}_a$ ancilla qubit.
    The system will evolve to 
\begin{align}
\sum_{j=1}^{N} \beta_j \left( \sqrt{1-\frac{C^2}{\lambda_j^2}} \ket{0}_a + \frac{C}{\lambda_j} \ket{1}_a\right) \ket{\lambda_j}_r \ket{u_j}_m.
\label{eq:controlled_rotation}
\end{align}
        
        \STATE \textbf{Step 3.}  Perform the reverse of Step 1.
This will lead the system to 
\begin{align}
\sum_{j=1}^{N} \beta_j \left( \sqrt{1-\frac{C^2}{\lambda_j^2}} \ket{0}_a + \frac{C}{\lambda_j} \ket{1}_a\right) \ket{0}_r \ket{u_j}_m.
\end{align}

\STATE \textbf{Step 4.} Measuring the ancilla qubit  will give ,
\begin{equation}
\ket{x} \approx \sum_{j=1}^{N} C\left(\frac{\beta_j}{\lambda_j} \right) \ket{u_j},
\end{equation} if the measurement outcome is $\ket{1}$
        
    \eindent
\end{algorithmic}
\end{algorithm}

These three steps are equivalent to the three steps shown in Eq.~\eqref{eq:decomp}. The algorithm is probabilistic, we get $\ket{x}$ only if the final measurement gives \ket{1}.  But this probability can be boosted using a technique called \emph{amplitude amplification} \cite{brassardEtAl}. This technique is explained in detail in Section VII.

\subsection{Algorithm implemented on IBM's 5 qubit computer}

\begin{figure}
\begin{center}
\includegraphics[width=\textwidth]{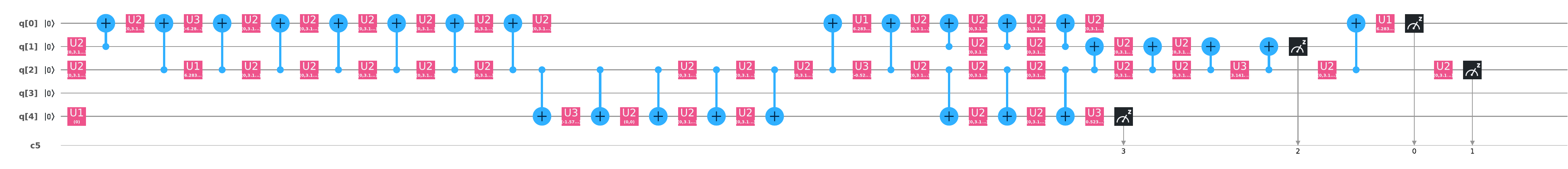}
\caption{Circuit implemented on IBM's 5-qubit \texttt{ibmqx4} quantum computer for the case with $\ket{b}$ set to $\ket{0}$ and with $\langle Z \hspace{1pt} \rangle$ measurement. After implementing the circuit in Fig.~\ref{fig:ls_setup} and setting the coupling map of the \texttt{ibmqx4} architecture, Qiskit-sdk-py re-arranges the qubits to fit the mapping. This circuit represents the outcome of the re-arrangement which was implemented on the \texttt{ibmqx4} quantum computer. }
\label{fig:composercircuit}
\end{center}
\end{figure}

Now we implement the HHL algorithm on a 2 $\times$ 2 system.
For this, we chose $A=\left( \begin{matrix}1.5 & 0.5\\ 0.5 & 1.5\end{matrix}\right)$.
We use four qubits for solving the system -- one ancilla, one memory and two register qubits.
For this case, the eigenvalues of A are $\lambda_1=1$ and $\lambda_2=2$ with the eigenvectors being $\frac{1}{\sqrt{2}}\left(\begin{matrix}1 \\ -1\end{matrix} \right) \equiv \ket{-}$ and $\frac{1}{\sqrt{2}}\left(\begin{matrix}1 \\ 1\end{matrix} \right) \equiv \ket{+}$, respectively.
For this system, the three steps of the HHL algorithm, can be performed by the operations shown in Fig.~\ref{fig:ls_setup}.
For the controlled rotation, we use a controlled $U$ rotation with $\theta = \pi$ for $\lambda_1$ and $\theta=\pi/3$ for $\lambda_2$.
This is done by setting $C=1$ in the Eq.~\eqref{eq:controlled_rotation}.
Both $\lambda$ and $\phi$ are set to zero in these controlled $U$ rotations.
Although the composer on Quantum Experience does not have this gate, in IBM Qiskit-sdk-py, we use \texttt{cu3} function for this purpose.
Three cases are used for b: $\left(\begin{matrix}1 \\ 0\end{matrix} \right)$, $\frac{1}{\sqrt{2}}\left(\begin{matrix}1 \\ -1\end{matrix} \right)$ and $\frac{1}{\sqrt{2}}\left(\begin{matrix}1 \\ 1\end{matrix} \right)$.
We post selected the states with $\ket{1}$ in the ancilla qubit.
The probabilities of these states are normalized such that their sum is one.
Measurements with respect to $\langle X \hspace{1pt} \rangle$, $\langle Y \hspace{1pt} \rangle$, $\langle Z \hspace{1pt} \rangle$ can then be performed  to obtain the expectation values.
QASM code is output from Qiskit-sdk-py and then uploaded on to IBM Quantum Experience.
Figure~\ref{fig:composercircuit} shows the equivalent composer circuit generated from QASM for the measurement in the computational basis (Z measurement).

\begin{figure}[!htb]
\begin{center}
\includegraphics[width=0.4\textwidth]{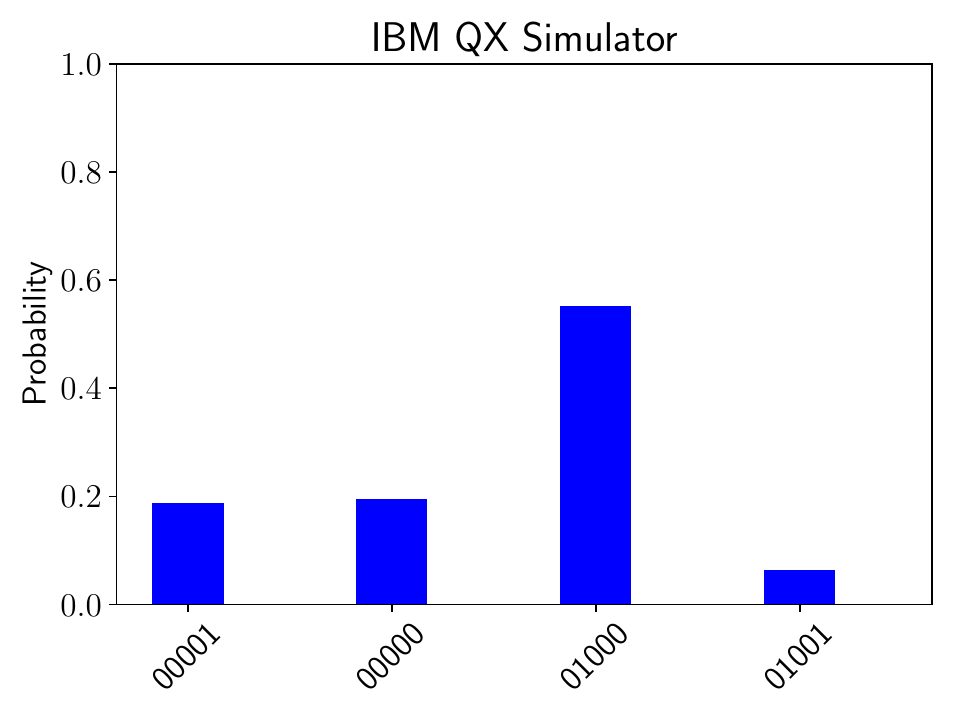}
\includegraphics[width=0.4\textwidth]{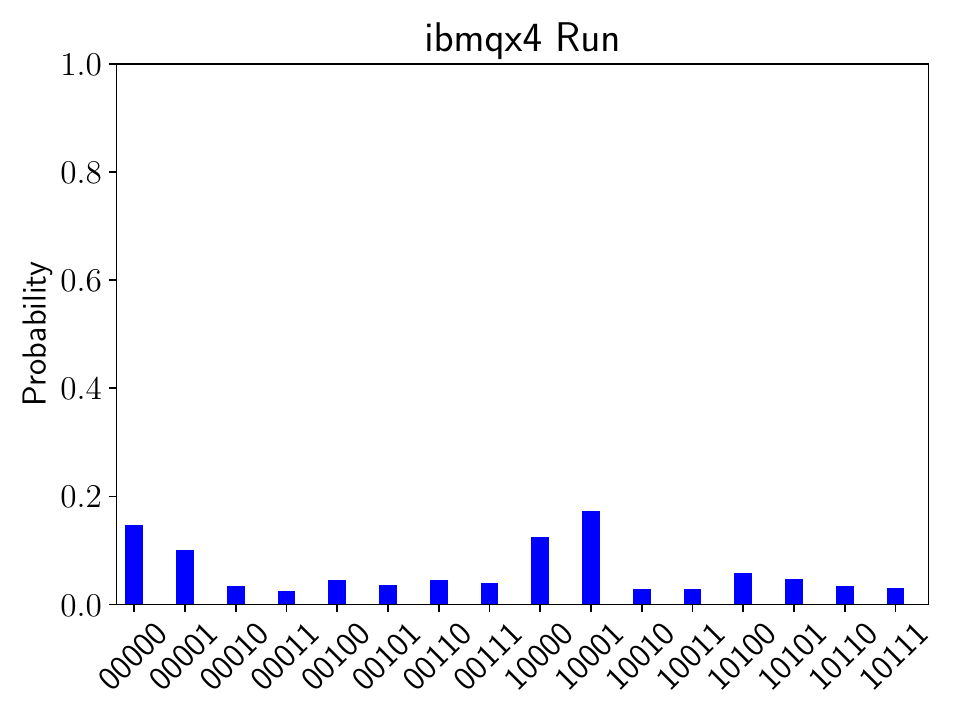}
\caption{Results of the circuit with Z measurement (computational basis  measurement) from the actual run and the simulator on a \texttt{ibmqx4}. 4096 shots were used for both the cases.}
\label{fig:comparison}
\end{center}
\end{figure}

To first test our implementation of the algorithm, we ran nine cases on the local simulator provided by Qiskit-sdk-py --  three b cases and three measurements with respect to the operators $X$, $Y$, $Z$, for each b case.
The comparison between the theoretical expectation values $\langle X \hspace{1pt} \rangle$, $\langle Y \hspace{1pt} \rangle$, $\langle Z \hspace{1pt} \rangle$ and the simulator values are shown in Table~\ref{tab:comparison}.
The simulator expectation values and the theoretical values match well.
This shows that the implementation of the algorithm gives expected results.
Similar expectation values were also seen using the simulator on IBM Quantum Experience instead of the local simulator. 
We then ran the circuit on the quantum computer \verb|ibmqx4|. 
Fig.~\ref{fig:comparison} shows a comparison between the simulator results and the results from the \verb|ibmqx4| with Z measurement on the circuit. 
As can be seen from Fig.~\ref{fig:comparison}, the results from the actual run do not give the expected answer as seen in the simulator results. We remark that recent modifications to the algorithm~\cite{subacsi2019quantum, bravo2019variational} can in some cases allow for larger scale and more accurate implementations on noisy quantum computers.

\begin{table}
\centering
\caption{Comparison between theoretical and simulator values for the expectation values $\langle X \hspace{1pt} \rangle$, $\langle Y \hspace{1pt} \rangle$, $\langle Z \hspace{1pt} \rangle$. T stands for theoretical and S stands for simulator.}
\label{tab:comparison}
\begin{tabular}{| l | l | l | l | l | l | l |}
\hline
 $\ket{b}$ &  T $\langle X \hspace{1pt} \rangle$ &  S $\langle X \hspace{1pt} \rangle$ &  T $\langle Y \hspace{1pt} \rangle$ & S $\langle Y \hspace{1pt} \rangle$ & T $\langle Z \hspace{1pt} \rangle$ & S $\langle Z \hspace{1pt} \rangle$\\
\hline
 $\ket{0}$ & -0.60 & -0.60 & 0.00 & -0.027 & 0.80 & 0.81\\
\hline 
 $\ket{+}$ & 1.00 & 1.00  & 0.00 & -0.06  & 0.00 & 0.02 \\
\hline
 $\ket{-}$ & -1.00 & -1.00  & 0.0060  & 0.000 & -0.02 & 0.00\\
\hline
\end{tabular}
\end{table}

\section{Shor's Algorithm for Integer Factorization}

\subsection{Problem definition and background}
%2/3 page

The integer factorization problem asks, given an integer $N$ as an input, to find integers $1<N_1,N_2<N$ such that $N=N_1N_2$. This problem is hardest when $N_1$ and $N_2$ are primes with roughly the same number of bits. If $n$ denotes the number of bits of $N$, no algorithm with polynomial in $n$ time complexity is known. The straightforward algorithm that tries all factors from 2 to $\sqrt{N}$ takes time polynomial in $N$, but exponential in $n$. The most efficient known  classical algorithm has running time $O\left(\exp \left(\sqrt[3]{\frac{64}{9}n(\log n)^2}\right)\right)$~\cite{Pomerance96atale}. In  practice, integers with 1000 or more bits are impossible to factor using known algorithms and classical hardware. The difficulty of factoring big numbers is the basis for the security of the RSA cryptosystem~\cite{Rivest:1978:MOD:359340.359342}, one of the most widely used public-key cryptosystems.

One of the most celebrated results in quantum computing is the development of a quantum algorithm for factorization that works in time polynomial in $n$. This algorithm, due to Peter Shor and known as Shor's algorithm~\cite{shor1994algorithms}, runs in $O(n^3\log n)$ time and uses $O(n^2\log n\log\log n)$ gates. The first experimental implementation of this algorithm on a quantum computer was reported in 2001, when the number 15 was factored~\cite{2001Natur.414..883V}. The largest integer factored by Shor's algorithm so far is 21~\cite{martin2012experimental}.

In this section we describe Shor's algorithm and its implementation on \verb|ibmqx4|

\subsection{Algorithm description}%1.5 p.
\paragraph{Reducing factorization to period finding}
One way to factor an integer is by using modular exponentiation. Specifically, let an odd integer $N=N_1N_2$ be given, where $1<N_1,N_2<N$. Pick any integer $k<N$ such that $\gcd(k,N)=1$, where $\gcd$ denotes the greatest common divisor. One can show that there exists an exponent $p>0$ such that $k^p \equiv 1 \pmod N$. Recall that, by definition, $x \equiv y \pmod m$ if and only if $m$ divides $x-y$. Assume that $p$ is the smallest such number. If we find such $p$ and $p$ is even, then, by the definition of the modulo operation, $N$ divides
\begin{equation*}
  k^p-1=(k^{p/2}-1)(k^{p/2}+1).
\end{equation*}
But since the difference between $n_1=k^{p/2}+1$ and $n_2=k^{p/2}-1$ is 2, $n_1$ and $n_2$ have no common factor greater than 2. Moreover, both numbers are nonzeros by the minimality of $p$. Since $N=N_1N_2$ was assumed to be odd, then $N_1$ is a factor of either $n_1$ or $n_2$. Assume $N_1$ is a factor of $n_1$. Since $N_1$ is also a factor of $N$, then $N_1$ divides both $n_1$ and $N$ and one can find $N_1$ by computing $\gcd (n_1,N)$.
Hence, if one can compute such a $p$, one can find the factors of $N$ efficiently as $\gcd$ can be computed in polynomial time.

In order to find $p$, consider the modular exponentiation sequence $A=a_0,a_1,\dots$, where $a_i=k^i \pmod N$. Each $a_i$ is a number from the finite set $\{0,\dots,N-1\}$, and hence there exists indices $q$ and $r$ such that $a_q=a_r$. If $q$ and $r$ are the smallest such indices, one can show that $q=0$ and $A$ is periodic with period $r$. For instance, for $N=15$ and $k=7$, the modular exponentiation sequence is $1,7,4,13,1,7,4,13,1,\dots$ with period 4. Since the period 4 is an even number, we can apply the above idea to find
\begin{equation*}
  7^4 \bmod 15 \equiv 1 \Rightarrow 7^4-1 \bmod 15 \equiv 0 \Rightarrow (7^2-1)(7^2+1) \bmod 15 \equiv 0 \Rightarrow \mbox{15 divides } 48\cdot 50,
\end{equation*}
which can be used to compute the factors of 15 as $\gcd(48,15)=3$ and $\gcd(50,15)=5$.

Finding the period of the sequence $A$ is, however, not classically easier than directly searching for factors of $N$, since one may need to check as many as $\sqrt{N}$ different values of $A$ before encountering a repetition. However, with quantum computing, the period can be found in polynomial time using the Quantum Fourier Transform (QFT). The QFT operation was introduced earlier during our discussion of phase estimation.

The property of the QFT that is essential for the factorization algorithm is that it can ``compute'' the period of a periodic input. Specifically, if the input vector $X$ is of length $M$ and period $r$, where $r$ divides $M$, and its elements are of the form
\begin{equation*}
x_i=\begin{cases}
\sqrt{r/M} & \text{if $i \bmod r \equiv s$} \\
0 & \text{otherwise}
\end{cases}
\end{equation*}
for some offset $s<r$, and ${\mathcal QFT}\left(\sum_{i=0}^Mx_i\ket{i}\right)=\sum_{i=0}^My_i\ket{i}$, then
\begin{equation*}
y_i=\begin{cases}
1/\sqrt{r} & \text{if $i \bmod M/r \equiv 0$} \\
0 & \text{otherwise}
\end{cases}
\end{equation*}
i.e.,~the output has nonzero values at multiples of $M/r$ (the values $\sqrt{r/M}$ and $1/\sqrt{r}$ are used for normalization). Then, in order to factor an integer, one can find the period of the corresponding  modular exponentiation sequence using QFT, if one is able to encode its period in the amplitudes of a quantum state (the input to QFT).

\begin{figure}[htbp]
\includegraphics[scale=0.9]{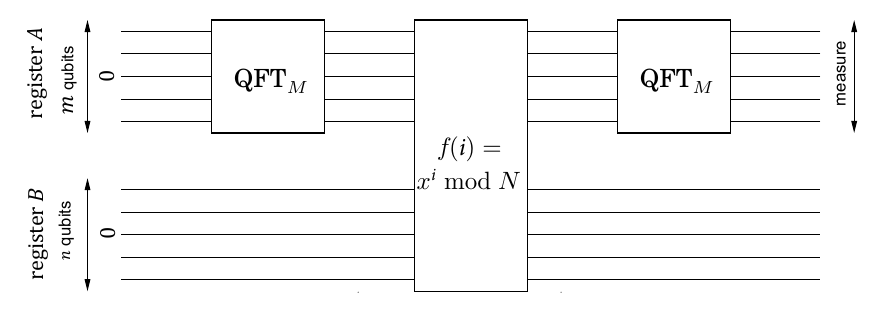}
\caption{Illustration of the period-finding circuit, where $m=2n$ and $M=2^m$.}
\label{fig:period}
\end{figure}

A period-finding circuit for solving the integer factorization problem is shown in Fig~\ref{fig:period}~\cite{Dasgupta:2006:ALG:1177299}.
The first QFT on register~$A$ produces an equal superposition of the qubits from $A$, i.e.,~the resulting state is
\begin{equation*}
  \frac{1}{\sqrt{M}}\sum_{i=0}^M\ket{i,0}.
\end{equation*}
Next is a modular exponentiation circuit that computes the function $f(i)=x^i \pmod N$ on the second register. The resulting state is
\begin{equation*}
  \frac{1}{\sqrt{M}}\sum_{i=0}^M\ket{i,f(i)}.
\end{equation*}
Before we apply the next QFT transform, we do a measurement of register~$B$. (By the principle of deferred measurement~\cite{NielsenChuang} and due to the fact that register~$A$ and~$B$ don't interact from that point on, we don't have to actually implement the measurement, but it will help to understand the final output.) If the value measured is $s$, then the resulting state becomes
\begin{equation*}
  \frac{1}{\sqrt{M/r}}\sum_{\substack{i=0\\f(i)=s}}^M\ket{i,s},
\end{equation*}
where $r$ is the period of $f(i)$.
In particular, register~$A$ is a superposition with equal non-zero amplitudes only of $\ket{i}$ for which $f(i)=s$, i.e.,~it is a periodic superposition with period $r$. Given the property of QFT, the result of the transformation is the state
\begin{equation*}
  \frac{1}{\sqrt{r}}\sum_{i=0}^r\ket{i(M/r),s}.
\end{equation*}
Hence, the measurement of register~$A$ will output a multiple of $M/r$. If the simplifying assumption that $r$ divides $M$ is not made, then the circuit is the same, but the classical postprocessing is a bit more involved \cite{NielsenChuang}.

Period finding can also be viewed as a special case of phase estimation. The reader may refer Nielsen and Chuang \cite{NielsenChuang} for this perspective on period finding.

\subsection{Algorithm implemented on IBM's 5-qubit computer}
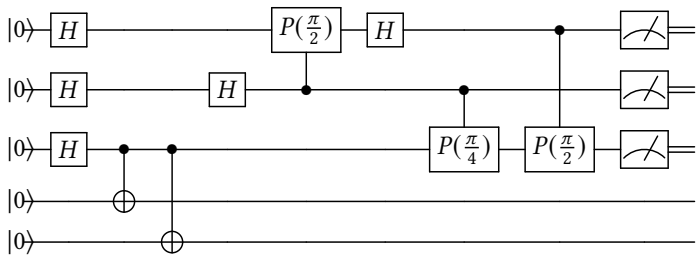
\begin{figure}[htbp]

\centerline{
\Qcircuit @C=1em @R=.7em {
& |0 \rangle ~~~~~& \gate{H} & \qw      &  \qw     & \qw      & \gate{P(\frac{\pi}{2})} &\gate{H} & \qw &\ctrl{2} &\meter & \cw \\ 
& |0 \rangle ~~~~~ & \gate{H} & \qw      &   \qw    & \gate{H} & \ctrl{-1}               &\qw  & \ctrl{1} & \qw &\meter & \cw \\
&|0 \rangle ~~~~~& \gate{H} & \ctrl{1} & \ctrl{2} & \qw      & \qw                     &\qw  & \gate{P(\frac{\pi}{4})} & \gate{P(\frac{\pi}{2})} & \meter &\cw \\
& |0 \rangle ~~~~~& \qw      &  \targ   &  \qw     & \qw      & \qw                     &\qw  & \qw &\qw  & \qw &\qw \\
&|0 \rangle ~~~~~  & \qw      &  \qw     &  \targ   & \qw      & \qw                     &\qw  & \qw &\qw &\qw &\qw
}
}
\caption{Circuit for Shor's algorithm for $N=15$ and $x=11$.}
\label{fig:shor_circuit}
\end{figure}

We implemented the algorithm on \verb|ibmqx4|, a 5-qubit quantum processor from the IBM Quantum Experience, in order to factor number 15 with $x=11$. The circuit as described on Figure~\ref{fig:period} requires 12 qubits and 196 gates, too large to be implemented on \verb|ibmqx4|. Hence, we used an optimized/compiled version from~\cite{2001Natur.414..883V} that uses 5 qubit and 11 gates (Fig~\ref{fig:shor_circuit}).

\begin{figure}[htbp]
\includegraphics[scale=0.6]{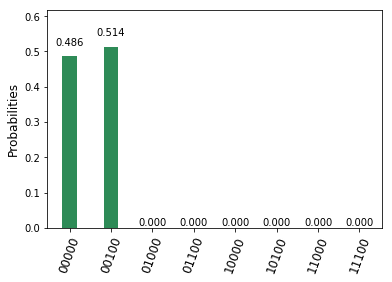}\hfill
\includegraphics[scale=0.6]{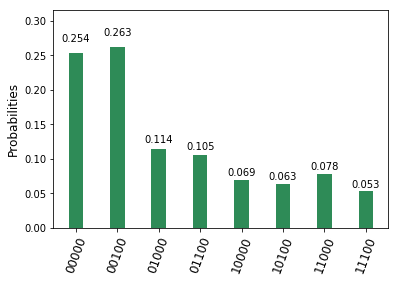}
\caption{Output from the circuit from Figure~\ref{fig:shor_circuit} implemented on the simulator (left) and \texttt{ibmqx4} (right).}
\label{Djidjev:fig:results}
\end{figure}

The results from the measurements are shown on Figure~\ref{Djidjev:fig:results}.

The periods found by the simulator are $p=0$, which is ignored as a trivial period, and $p=4$, which is a good one. Since $M=8$, we can conclude that $r$ divides $M/p=8/4=2$, hence $r=2$. Then $15$ divides
\begin{equation*}
  (x^r-1)=(11^2-1)=(11-1)(11+1)=10 \cdot 12.
\end{equation*}
By computing $\gcd(15,10)=5$ and  $\gcd(15,12)=3$, we find the factors of 15.

The output from \verb|ibmqx4| finds the same periods 0 and 4 with the highest probabilities, but contains much more noise.

\section{Matrix Elements of Group Representations}

\subsection{Problem definition and background}

In this section we will discuss another quantum algorithm that makes use of the QFT operation. In this section we will also introduce a subroutine called the \emph{Hadamard test}, which lets us compute matrix elements of unitary operators. But first, we will require some knowledge of group theory to understand the problem being tackled here. This section follows the work of Jordan in Ref. \cite{Jordan}.

A \emph{Group} ($G$, $\cdot$) or ($G$) is a mathematical object defined by its elements ($g_1$, $g_2$,~\ldots) and an operation between elements ($\cdot$), such that these four properties are satisfied. 
\begin{enumerate}

 \item  Closure: for any two group elements, the defined group operation produces another element, which belongs to the group (for $\forall$ $g_i$,$g_j$ $\in$ $G$,  $g_i\cdot g_j=g_k$ $\in$ $G$). 
 
 \item Associativity:  for $\forall$ $g_i,g_j,g_m$ $\in$ $G$, $g_i\cdot \left(g_j\cdot g_m\right)=\left(g_i\cdot g_j\right)\cdot g_m$. 
\item  Identity element: $e$ $\in$ $G$, such that $e\cdot g_i=g_i\cdot e=g_i$.

\item Inverse element: for $\forall$ $g_i$ $\in$ $G$, there exists $g_p$, such that $g_i\cdot g_p = g_p\cdot g_i=e$.
\end{enumerate}

A group with a finite amount of elements $n$ is called a finite group with order $n$, while a group with an infinite amount of elements is an infinite group. In this section, we will discuss quantum algorithms to solve certain problems related to finite groups. As before, we will also implement them on the IBM machines. Some examples of groups are given below.

\textit{Example 1A}. Abelian group $A_n$ with $n$ elements: $0, 1,~\ldots, n-1$, and the group operation addition  modulo $n$: $g_i \cdot  g_j=(i+j)$mod$(n)$. For instance, for $n=3$: $a_0=0$, $a_1=1$, $a_2=2$. Then, $a_2 \cdot a_2=4$ mod$(3)=1=a_1$, $a_2 \cdot a_1=3$ mod$(3)=0=a_0$, \textit{etc}. The identity element is $a_0=0$ and its inverse is itself. For all other elements the inverse element is, $a_i^{-1}=a_{n-i}$. This group is called Abelian or commutative, because in addition to the four group properties, it has a property of commutativity: $a_i\cdot a_j=a_j\cdot a_i$ for $\forall$ $a_i$, $a_j$ $\in$ $A_n$.

\textit{Example 1S}. Symmetry group $S_n$ with $n!$ group elements, each is a permutation of $n$ objects: $[1,2..,n]$, $[2,1..,n]$,~\dots, $[n,n-1..,2,1]$. Consequent application of two permutations is a group operation. For  instance, for group $S_2$: ($e$,$p$) we have two objects $a$ and $b$. The identity element $e$ is no permutation: $ab\rightarrow ab$, while one permutation $p$ is the second group element: $ab\rightarrow ba$. Then, $p\cdot p=e$, and $p^{-1}=p$. Only $S_1$ and $S_2$ are Abelian  groups. For $n\geq 3$, $S_n$ are not commutative. Let us write elements of group $S_3$ as a permutation of elements $123$ in the next order: $[123]$ $\rightarrow$ $[123]$, $[231]$, $[312]$, $[213]$, $[132]$, $[321]$. Then $s_4\cdot s_2=s_6$, while $s_2\cdot s_4=s_5$.

While group definition is quite simple, it is not straightforward how to operate with group elements in general, especially when defined operations between them is not trivial and/or the group order, $n$, is large. In this case, it is helpful to apply the representation theory to the group. The idea is simple: if we can map a group of unknown objects with nontrivial operations to the group of known objects with some trivial operations, we can gain some information about the unknown group. In general, we introduce a function applied to a group element: $\rho\left(g_i\right)$, which does this mapping between two groups. Such function defines the group representation of $G$ if for $\forall$ $g_i$, $g_j$ $\in$ $G$, $\rho(g_i)*\rho(g_i)=\rho(g_i\cdot g_j)$, where ($*$) can be a different operation from ($\cdot$).

\textit{Example 2A}. Representation of Abelian group $A_n$: $a_j\rightarrow \rho(a_j)=e^{i 2\pi j/N }$, where the original operation ($+$mod$(n)$) is substituted by the new operation of multiplication. Note that the group $S_2$ can be represented in the same way as $A_2$.
  
\textit{Example 2S}. Representation of group $S_3$: $s_j\rightarrow \rho(s_j)=1$, where the original operation is again substituted by the new operation of multiplication. Such representation of the group $S_3$ is trivial, since it does not carry any information about the group, however it satisfies the definition of the group representation. Moreover, [$1$,$1$,~\ldots] is a trivial representation for any group. Another representation of group $S_3$ is, $[1,1,1,-1,-1,-1]\rightarrow[s_1, s_2,~\ldots, s_n]$, where we map odd permutations to $-1$ and even permutations to $1$ . While it carries more information about the initial group than the trivial representation, it does not imply that the group $S_3$ is not Abelian. One cannot construct a one-dimensional representation for group $S_3$ which would retains all its properties. The smallest equivalent representation for $S_3$ is two-dimensional. The multidimensional representations can be easy understood when represented by matrices.  

Most useful representations are often ones which map a group to a set of matrices. When $\rho(g)$ is a $d_{\rho} \times d_{\rho}$ matrix, the representation is referenced as a matrix representation of the order $d_{\rho}$, while ($*$) is the operation of matrix multiplication. All representations of finite group can be expressed as unitary matrices given an appropriate choice of basis. To prove the last fact, we introduce a particular representation called the \emph{regular representation}.

The regular representation of a group of $N$ elements is a matrix representation of order $N$. We will explain the construction of the regular representation using the Dirac notation. First, we associate with each element of the group $g_i$ a ket $\ket{g_i}$. This ket could simply be the basis state $\ket{i}$, since the elements of the group are numbered. This ensures that the kets associated with different group elements are orthonormal by construction, $\braket{g_i | g_j} = \delta_{ij}.$ This also ensures that the identity operator can be expressed as $\sum_{i = 1}^N \ket{g_i} \bra{g_i}.$ The regular representation of $g_k$ is then given by,

\begin{equation}
    R(g_k)=\sum_{j=1}^N \dyad{g_k\cdot g_j}{g_j}.
\end{equation}

The matrix elements of this representation are, $R_{ij}(g_k) \equiv \braket{g_i | R(g_k)| g_j} =  \ip{g_i }{ g_k\cdot g_j }. $ From the defining properties of a group it can be easily seen that multiplying every element in the group by the same element just permutes the elements of the group.   This means that $R(g_k)$ matrices are always permutation matrices and are hence unitary. We can prove that the regular representation is a representation using simple algebra,

\begin{align}
R(g_k)\cdot R(g_m)&=\sum_{i=1}^N \sum_{j=1}^N \ket{g_k\cdot g_i} \ip{g_i}{ g_m\cdot g_j }\bra{g_j}, \notag \\
&=\sum_{i=1}^N \sum_{j=1}^N \ket{g_k\cdot g_m\cdot g_j} \ip{g_i}{ g_m\cdot g_j}\bra{g_j} \notag, \\
&=\sum_{j=1}^N \dyad{g_k\cdot g_m\cdot g_j}{g_j}=R(g_k\cdot g_m).
\end{align}

Here we used orthogonality: $\ip{g_i }{g_m\cdot g_j} =1$ only if $\ket{g_i}= \ket{g_m\cdot g_j }$ and $0$ otherwise, which allowed us to swap these two states. Then, we used the same fact to calculate the sum over $i$. Below we give some explicit examples of regular representations.

\textit{Example 3A}. Regular representation of the Abelian group $A_4$, where each matrix element is calculated using the result derived above $R_{ij}(a_k)=\ip{a_i }{a_k\cdot a_j}$:

\begin{align}
\label{Malyzhenkov:eqn1}
R(a_0) = 
\begin{pmatrix}
1     &  0 & 0 & 0  \\
0     &  1 & 0 & 0  \\
0     &  0 & 1 & 0  \\
0     &  0 & 0 & 1  
\end{pmatrix}\,,\ 
R(a_1) = 
\begin{pmatrix}
0     &  0 & 0 & 1  \\
1     &  0 & 0 & 0  \\
0     &  1 & 0 & 0  \\
0     &  0 & 1 & 0  
\end{pmatrix}\,,\
R(a_2) = 
\begin{pmatrix}
0     &  0 & 1 & 0  \\
0     &  0 & 0 & 1  \\
1     &  0 & 0 & 0  \\
0     &  1 & 0 & 0  
\end{pmatrix}\,,\ 
R(a_3) = 
\begin{pmatrix}
0     &  1 & 0 & 0  \\
0     &  0 & 1 & 0  \\
0     &  0 & 0 & 1  \\
1     &  0 & 0 & 0  
\end{pmatrix}\,.\ 
\end{align}
Commutative property is conserved: $R(a_i)\cdot R(a_j)= R(a_j)\cdot R(a_i)$. 

\textit{Example 3S}. Regular representation of the group $S_3$, where we use the same order of permutations introduced above ($[123]$ $\rightarrow$ $[123]$, $[231]$, $[312]$, $[213]$, $[132]$, $[321]$)

\begin{align}
\label{Malyzhenkov:eqn2}
R(s_1) = 
\begin{pmatrix}
1   &  0 & 0 & 0 & 0 & 0 \\
0     &  1 & 0 & 0 & 0 & 0 \\
0     &  0 & 1 & 0  & 0 & 0\\
0     &  0 & 0 & 1  & 0 & 0\\
0     &  0 & 0 & 0  & 1 & 0\\
0     &  0 & 0 & 0  & 0 & 1
\end{pmatrix}\,,\ 
R(s_2) = 
\begin{pmatrix}
0   &  0 & 1 & 0 & 0 & 0 \\
1     &  0 & 0 & 0 & 0 & 0 \\
0     &  1 & 0 & 0  & 0 & 0\\
0     &  0 & 0 & 0  & 0 & 1\\
0     &  0 & 0 & 1  & 0 & 0\\
0     &  0 & 0 & 0  & 1 & 0
\end{pmatrix}\,,\
R(s_3) = 
\begin{pmatrix}
0   &  1 & 0 & 0 & 0 & 0 \\
0     &  0 & 1 & 0 & 0 & 0 \\
1     &  0 & 0 & 0  & 0 & 0\\
0     &  0 & 0 & 0  & 1 & 0\\
0     &  0 & 0 & 0  & 0 & 1\\
0     &  0 & 0 & 1  & 0 & 0
\end{pmatrix}\,,
\end{align}

\begin{align}
R(s_4) = 
\begin{pmatrix}
0   &  0 & 0 & 1 & 0 & 0 \\
0     &  0 & 0 & 0 & 0 & 1 \\
0     &  0 & 0 & 0  & 1 & 0\\
1     &  0 & 0 & 0  & 0 & 0\\
0     &  0 & 1 & 0  & 0 & 0\\
0     &  1 & 0 & 0  & 0 & 0
\end{pmatrix}\,,\ 
R(s_5) = 
\begin{pmatrix}
	0   &  0 & 0 & 0 & 1 & 0 \\
	0     &  0 & 0 & 1 & 0 & 0 \\
	0     &  0 & 0 & 0  & 0 & 1\\
	0     &  1 & 0 & 0  & 0 & 0\\
	1     &  0 & 0 & 0  & 0 & 0\\
	0     &  0 & 1 & 0  & 0 & 0
\end{pmatrix}\,,\ 
R(s_6) = 
\begin{pmatrix}
0   &  0 & 0 & 0 & 0 & 1 \\
0     &  0 & 0 & 0 & 1 & 0 \\
0     &  0 & 0 & 1  & 0 & 0\\
0     &  0 & 1 & 0  & 0 & 0\\
0     &  1 & 0 & 0  & 0 & 0\\
1     &  0 & 0 & 0  & 0 & 0
\end{pmatrix}\,.\ 
\end{align}

Now we can finally explain the problem of calculating matrix elements of the group representations, which is equivalent to the problem of  calculating an expectation value of an operator $\mathbf{A}$ in respect to the state $\ket{\psi}$ in quantum mechanics: $\avg{\mathbf{A}}=\bra{\psi}\mathbf{A}\ket{\psi}$.

\textit{Example 4A}. Calculating matrix elements of the regular representation of the element $a_2$ from the Abelian group $A_4$ with respect to the state $\psi_{13}$ which is the equal superposition of $\ket{a_1}$ and $\ket{a_3}$. In operator form we find:
\begin{align}
\bra{\psi_{12}}\mathbf{a_2}\ket{\psi_{12}}=\frac{\bra{a_1} + \bra{a_3}}{\sqrt{2}} \left(\sum_{i=0}^{N-1} \dyad{a_2\cdot a_i}{a_i }\right) \frac{ \ket{a_1}+\ket{a_3}}{\sqrt{2}}=\frac{\ip{a_3}{a_2\cdot a_1}\ip{a_1}{a_1}}{2}+\frac{\ip{a_1}{a_2\cdot a_3}\ip{a_3}{a_3}}{2}=1\,.
\end{align}

It is quite obvious that if a quantum computer is capable of finding expectation values of a unitary operator, it will be able to solve the problem of finding the matrix elements of the regular representation of a group element. This will consist of, at least, two stages: the first stage is the state preparation, and the second is applying the unitary operator of the regular representation to that state. The unitary operator of the regular representation of an element of any group $G_n$ can be created using a combination of only two type of operations: qubit  flip ($\ket{0}\rightarrow \ket{1}$) and qubit swap ($\ket{q_jq_i}\rightarrow \ket{q_iq_j}$).    

Up to this point, we have only talked about the regular representation. The regular representation is quite convenient, it is straightforward to find for any group, it carries all the information about the group, and a corresponding unitary operator is easy to construct  using standard quantum circuits. However,  for groups with a large number of elements, it requires matrix multiplication between large matrices. So for many applications, instead of regular representations one is interested in what are known as \emph{irreducible representations}, which  are  matrix representations that cannot be  decomposed into smaller representations. Or in other words, every matrix representation (including the regular representation) can be shown to be equivalent to a direct sum of irreducible representations, up to a change of basis. This lets us reduce the representation theory of finite groups into the study of irreducible representations. The importance of irreducible representations in group theory cannot be overstated. The curious reader may refer these notes by Kaski \cite{kaski2002}.

A result from group theory ensures that the direct sum of all irreducible representations (each has different dimensions $d_{\rho}$ in general) where each irreducible representation appears exactly $d_{\rho}$ times is a block diagonal $N\times N$ matrix (the group has $N$ elements). The Fourier transform pair over this group representation can be introduced by decomposing each irreducible representation over the group elements and \textit{vice versa}.
Moreover, the above defined direct sum of all irreducible representations can be decomposed as a regular representation conjugated  by the direct and inverse Fourier transform operators~\cite{Jordan}. This result lets us find the the matrix elements of the irreducible representations given the ability to implement the regular representation.

\subsection{Algorithm description}
\begin{figure}
	\begin{center}
		\includegraphics[width=\columnwidth]{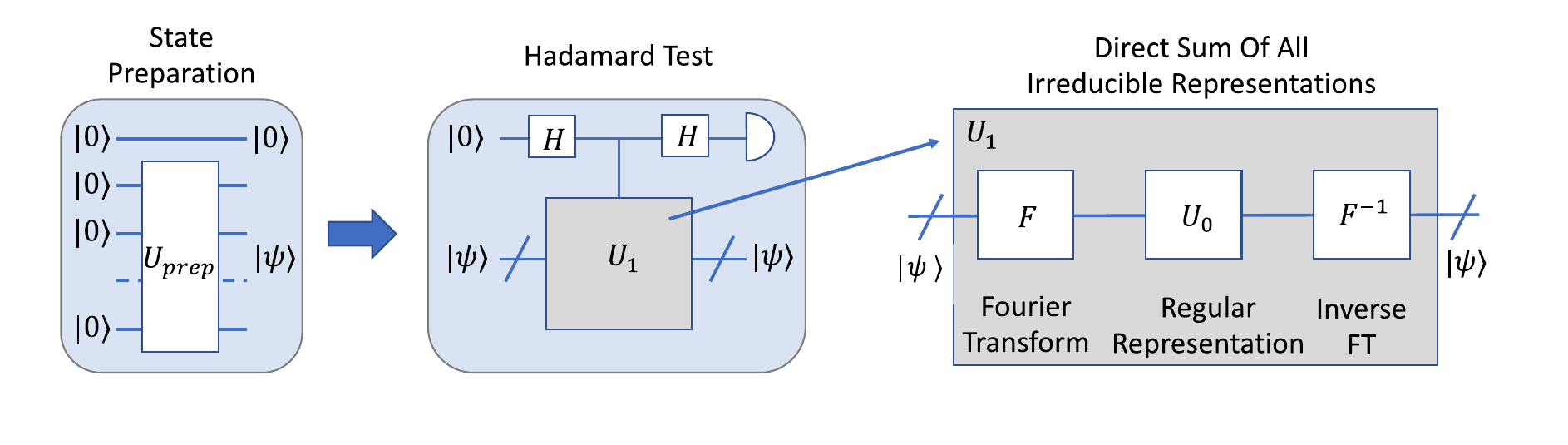}
		\caption{Schematic diagram for the quantum algorithm }
		\label{Malyzhenkov:fgr1}
	\end{center}
\end{figure}

In this section we will describe an algorithm to find the matrix elements of irreducible representations of a group given the ability to apply its regular representations to a quantum register in a  controlled fashion. The quantum algorithm calculating matrix elements $\bra{\psi}\mathbf{U_1}\ket{\psi}$ of a unitary operator $\mathbf{U_1}$ is known as the Hadamard test, which is illustrated on Fig.~\ref{Malyzhenkov:fgr1}.

\begin{algorithm}[!htb]
\caption{Hadamard test}
\begin{algorithmic} 
    \STATE \textbf{Input:}
    \bindent
        \STATE $\bullet$ The controlled unitary $CU.$ 
        \STATE $\bullet$ Input state \ket{0}\ket{\psi}.
    \eindent
    \STATE \textbf{Output:}
    \bindent 
        \STATE $\bullet$ An estimate for the real part of \braket{\psi |U| \psi}
    \eindent
    \STATE \textbf{Procedure:}
    \bindent
        \STATE \textbf{Step 1.} Apply $H$ to the ancilla. This produces the state, $$\frac{\ket{0}+\ket{1}}{\sqrt{2}} \ket{\psi}$$

        \STATE \textbf{Step 2.} Apply $CU$ controlled on the ancilla. This produces the state,
        
        $$\frac{\ket{0}\ket{\psi}+\ket{1}U\ket{\psi}}{\sqrt{2}} $$
        \STATE \textbf{Step 3.} Apply $H$ to the ancilla again. This gives,
        
         $$\frac{\ket{0} (\ket{\psi} + U\ket{\psi} ) +\ket{1}(\ket{\psi} - U\ket{\psi} )}{\sqrt{2}} $$

        \STATE \textbf{Step 4.} Measure the ancillary qubit. Repeat to estimate the probability of obtaining $\ket{0}$ and $\ket{1}.$
        
    \eindent
\end{algorithmic}
\end{algorithm}

 The ancilla qubit should be prepared as  $\frac{\ket{0}-i\ket{1}}{\sqrt{2}}$ to calculate the  imaginary parts of the matrix element. From the pseudocode, we can see that the probability of measuring \ket{0}  is $P_0= ||\frac{\ket{\psi}+ U\ket{\psi}}{\sqrt{2}}||^2=\frac{1+Re\bra{\psi}U\ket{\psi}}{2}$. Hence, we find: $ Re\bra{\psi}U\ket{\psi}=2P_0-1$. The reader is encouraged to work out the same steps for the imaginary part as well.

With the Hadamard test algorithm, the problem of calculating matrix elements of an arbitrary unitary operator is reduced to the problem of effectively implementing it as a controlled gate. For the regular representation of any group $U_0$, where unitary operator is an $N$ x $N$ square matrix with only one non-zero element equal to $1$ in each row, this implementation can be done for any group as a combination of $CNOT$ and $Z$ gates.

At the same time solutions for the direct sum of all irreducible representations $U_1$, which can be decomposed as  $U_1(g)=F_1 U_0 (g^{-1}) F_1^{-1} $,  exists for any group whose Fourier transform over that group can be effectively implemented using quantum circuits. Quantum circuits for the Fourier transform are already known for the symmetric group $S(n)$~\cite{Beals}, the alternating group $A_n$, and some Lie groups: $SU(n)$, $SO(n)$~\cite{Bacon}, while solutions for other groups,  hopefully, will be found in the future.  For Abelian groups this Fourier transform implementation can be efficiently done using the QFT circuit that was discussed in the earlier sections. For non-Abelian groups the implementation is trickier and efficient implementations are not universally known.

\subsection{Algorithm implemented on IBM's 5-qubit computer}

\begin{figure}
	\begin{center}
		\includegraphics[width=\columnwidth]{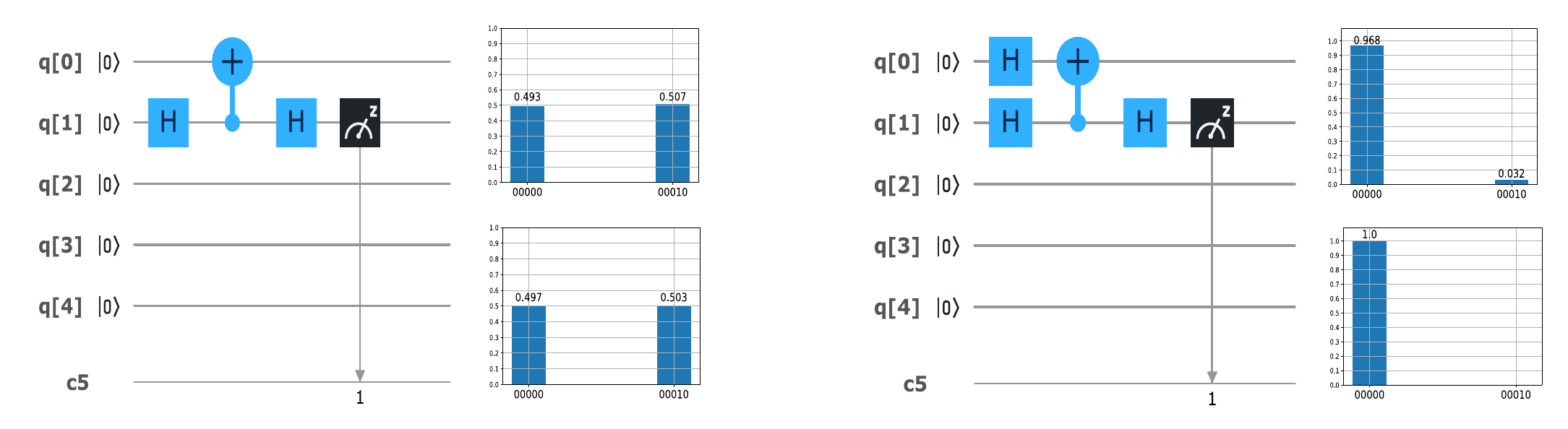}
		\caption{Actual circuit implemented on IBM's 5-qubit computer  for calculating matrix elements of the regular representation for the second element of the group $S_2$ and $A_2$ in respect to the state $\ket{0}$ on the left and $\frac{\ket{0}+\ket{1}}{\sqrt{2}}$ on the right. The expected probabilities to find a final state in the ground state are $(1+0)/2=0.5$ and $(1+1)/2=1$ respectively. The results of the $1024$ runs on the actual chip (on the top) and the simulator (on the bottom) are presented on the right side of each circuit.}
		\label{Malyzhenkov:fgr2}
	\end{center}
\end{figure}

The actual gate sequence that we implemented on IBM's 5-qubit computer (\verb|ibmq_essex|) and IBM's quantum simulator   to find matrix elements of the regular representation of the second element of the group  $S_2$ is shown in Fig.~\ref{Malyzhenkov:fgr2}. The matrix for this representation is simply a $X$ gate. Hence, we have to use one CNOT gate and two Hadamard gates, plus some gates to prepare state $\ket{\psi}$ from the state $\ket{00}$. We  mapped the ancilla qubit to the actual machine $q_1$ qubit instead of $q_0$, because of the machine architecture, where the first qubit can control the zero qubit but not \textit{vice versa}. We could have used the original qubit sequence as in  Fig.~\ref{Malyzhenkov:fgr1}, by realizing  the CNOT gate as a swapped CNOT and four Hadamard gates, but this would add more gates to the circuit and potentially more computational errors rather than just a virtual swap of the qubits.

\begin{figure}
\centering
		\includegraphics[width=\columnwidth]{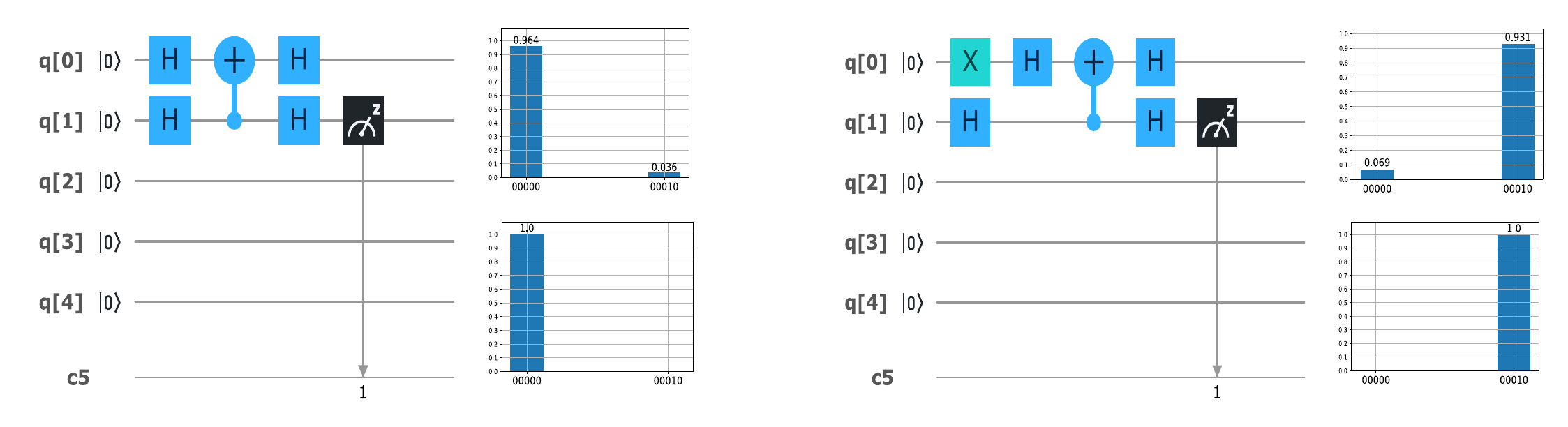}
		\caption{Actual circuit implemented on IBM's 5-qubit computer for calculating matrix elements of the direct sum of the irreducible representations for the second element of the group $S_2$ and $A_2$  with respect to the state $\ket{0}$ on the left and $\ket{1}$ on the right. The expected probabilities to find a final state in the ground state are $(1+1)/2=1$ and $(1-1)/2=0$ respectively. The results of the $1024$ runs on the actual chip (on the top) and the simulator (on the bottom) are presented on the right side of each circuit.}
		\label{Malyzhenkov:fgr3}
\end{figure}

For the irreducible representation of the same element of the group $A_2$ , the  element is represented by the $Z$ gate. Hence the Hadamard test requires implementing a controlled-$Z$ gate, which is not available as an actual gate on the IBM Quantum Experience. However, it can be constructed using two Hadamard and one $CNOT$ gates as shown in Fig.~\ref{Malyzhenkov:fgr3}. Notice that the Hadamard gate is actually the Fourier transform operator over group $S_2$ and $A_2$, while the $X$ gate is a regular representation operator, as we mentioned earlier. Hence, such controlled-$Z$ gate representation is in fact the decomposition of the irreducible representation to the regular representation using Fourier transform over that group.

\section{Quantum Verification of Matrix Products}

\subsection{Problem definition and background}
Matrix multiplication is one of the most important linear algebra subroutines. Most scientific computing algorithms use matrix multiplication in one form or another. Therefore,
the computational complexity of matrix multiplication is a subject of intense study. For two $n \times n$ matrices the computational complexity of the naive matrix multiplication algorithm is $O(n^3)$ 
A faster algorithm for matrix multiplication
implies a considerable performance improvement for a variety of computational tasks.
Strassen~\cite{Strassen1969} first
showed that two $n \times n$ matrices can be multiplied in time $O(n^{2+\alpha})$ ($\alpha<1$).
The best known algorithm to date with
$\alpha\approx 0.376$ was found by Coppersmith and Winnograd~\cite{Winograd1987}.
Despite that, it remains an open problem to determine the
optimal value of $\alpha$.
The so-called problem of matrix verification is defined as,
verifying whether the product of two $n \times n$ matrices
is equal to a third one. So far the best classical algorithm can do this with high probability in time proportional to $n^2$~\cite{Freivalds1979}.

 Ref.~\cite{Ambainis2002} was the first to study matrix verification for quantum computation. The authors use a quantum algorithm based on Grover's
algorithm to verify whether two $n \times n$ matrices equal a third in time $O(n^{7/4})$, thereby improving
the optimal classical bound of Ref.~\cite{Freivalds1979}.
Ref.~\cite{Buhrman2004} presents a quantum algorithm that verifies a product of two $n\times n$ matrices over any integral
domain with bounded error in worst-case time $O(n^{5/3})$ and expected time $O(n^{5/3}/ min(w, \sqrt{n})^{1/3})$,
where $w$ is the number of wrong entries. This further improves the time performance $O(n^{7/4})$ from Ref.~\cite{Ambainis2002}.

\subsection{Algorithm description}

We briefly sketch the quantum algorithm from Ref.~\cite{Ambainis2002}. The presentation here follows from Ref.~\cite{spalek2006quantum}. Before we discuss this algorithm we introduce the concept of \emph{amplitude amplification.}

Many real world algorithms are probabilistic, i.e., independent runs of the algorithm on the same input will not necessarily give the same output. This is because the algorithm uses some source of randomness during its execution. Most quantum algorithms are probabilistic owing to the inherent randomness present in quantum mechanics.

Suppose that the job of our probabilistic classical/quantum algorithm is to return one of a specific set of states. Assume that we also have at our disposal an oracle that can identify the members of this set from other states. An example of this would be polynomial root finding. The set of states in this case would correspond to the roots of the polynomial. Our algorithm should return one of the roots of the polynomial and we can verify if an output is a root by plugging it in to the polynomial. 

Obviously the algorithm is good only if it can return a state that is a member of this set with high probability. But how high of a success probability is good enough? For practical reasons we would like the probability of success to be a constant. That is, it should be a value independent of the problem size and other parameters in the problem.  Any constant value between $0$ and $1$ would work here. The value $\frac23$ is usually used in literature.

But often algorithms won't succeed with constant probability and their success provability will diminish with growing input size. In that case, how can we boost the success probability to the desired level? The classical answer to this question is to repeatedly run the algorithm until we  succeed, i.e., till the algorithm outputs a state from the specific set of states that we want. If the algorithm initially had a success probability of $p$, after $O(\frac{1}{p})$ repetitions we are guaranteed to find the desired state with constant probability.  

For quantum algorithms we can do something better. Let $U$ be a quantum algorithm and suppose that we want this algorithm to return a state from the subspace spanned by the orthogonal states, $\{ \ket{u_i} \} .$ Let $P$ be the projection operator onto this subspace, $P =  \sum_i \ket{u_i} \bra{u_i} .$ The oracle we have is then, $O = I - 2 P$. This oracle will mark the states in the desired subspace. The success probability of our algorithm is $p =  \braket{0\ldots 0| U^\dagger P U |0\ldots 0} $. In this scenario we can use amplitude amplification to boost the success probability to a constant with only $O(\frac{1}{\sqrt{p}})$ repetitions. This is a quadratic speedup over the classical strategy. 

Essentially, amplitude amplification is a generalization of Grover search described in Section II . In Grover search we repeatedly apply the Grover operator, $G = (2 \ket{\psi}\bra{\psi} - I ) O$, where  $\ket{\psi}$ is the uniform superposition state. Amplitude amplification uses a more general operator,

\begin{equation}
    G_U =   U(2 \ket{0}\bra{0} - I)U^\dagger O.
\end{equation}

To get the desired result we apply this to the   $U \ket{0\ldots 0}$ state $O(\frac{1}{\sqrt{p}})$ times. Notice that the original Grover search is a specific case of amplitude amplification with $U =  H \otimes \ldots  \otimes H.$ In that case, the probability of getting the marked state in $\ket{\psi}$ is $\frac{1}{N}$ so we run the algorithm for $O(\sqrt{N})$ steps. The reader is referred to Ref. \cite{brassardEtAl} for more details on amplitude amplification.

The matrix product verification procedure uses amplitude amplification as its outer loop.  The algorithm first splits the full matrix verification problem into smaller matrix verification problems. Then it uses amplitude amplification to search if one of these verifications fail. Each of these smaller verification steps also use a Grover search to look for disagreements. So the complete algorithm uses  one quantum search routine nested inside another quantum search routine. This is a common strategy used while designing quantum algorithms to improve query complexity. The full algorithm is sketched below.

\begin{algorithm}[!htb]
\caption{Matrix product verification \cite{Ambainis2002} \cite{spalek2006quantum}}
\begin{algorithmic} 
    \STATE \textbf{Input:}
    \bindent
        \STATE $\bullet$ $n \times n$ matrices $A, B, C.$ 
        
    \eindent
    \STATE \textbf{Output:}
    \bindent 
        \STATE $\bullet$ Verifies if $AB =  C$ 
    \eindent
    \STATE \textbf{Procedure:}
        \bindent
        \STATE  \textbf{Step 1.} Partition $B$ and $C$ into $\sqrt{n}$ submatrices of size $n \times \sqrt{n}$. Call these $B_i$ and $C_i$ respectively. $AB =C$ if and only if $AB_i =  C_i$ for all $i$.
        
        \STATE \textbf{Step 2.} Use amplitude amplification over $i$ on these steps:

             \STATE \hspace{\algorithmicindent}  \textbf{Step 2a.} Choose a random vector $x$ of dimension $\sqrt{n}.$
        
             \STATE \hspace{\algorithmicindent}  \textbf{Step 2b.} Compute $y = B_i x$ and $z = C_i x$ classically
        
            \STATE\hspace{\algorithmicindent}  \textbf{Step 2c.} Verify equation $Ay =z$ by Grover search. Search for a row $j$ such that $ (Ay -z)_j \neq 0$
        
        \eindent
\end{algorithmic}
\end{algorithm}

The number of qubits  and the circuit depth required for this algorithm is too large for it to be successfully implemented on the IBM machines. But at the heart of this algorithm is the Grover search procedure, which we have already discussed and implemented in Section II

\section{Group Isomorphism}

\subsection{Problem definition and background}

The \textit{group isomorphism} problem, originally identified by Max Dehn in 1911~\cite{Dehn1911}, is a well-known decision problem in abstract algebra.  Simply stated, it asks whether there exists an isomorphism between two finite groups, $G$ and $G'$. Which, according to the standpoint of group theory, means that they are equivalent (and need not be distinguished). At the end of Section 5 we saw an example of two isomorphic groups, $S_2$ and $A_2$. These two are the same group in terms of how the group operation works on the group elements, but are defined in different ways.  More precisely, two groups, $(G_1,\cdot)$ and $(G_2, *)$ are called isomorphic if there is a bijection, $f: G_1 \rightarrow G_2$, between them such that, $f(g_1 \cdot g_2) = f(g_1) * f(g_2).$ 

To solve this problem using a quantum algorithm, we assume that each element can be uniquely identified by an arbitrary bit-string label.  We also assume that a so-called group \textit{oracle} can be used to return the product of multiple elements. That is, given an ordered list of group-element labels, the oracle will return the product label. In practice, this means that we must be able to construct a quantum circuit to implement $U_a : \ket{y} \rightarrow \ket{ay}$, for any $a \in G$.

In this section, we will focus our attention on the abelian group isomorphism problem, because it can be solved using a generalization of Shor's algorithm~\cite{Shor1997}. As we saw before, abelian simply means that the operation ($\cdot$) used to define the group is commutative, such that $a \cdot b = b \cdot a,$ for $a, b \in G$. Although Shor's approach is specifically intended to leverage a quantum period-finding algorithm to reduce the time-complexity of factoring, the procedure effectively solves the group isomorphism problem over cyclic groups. Using this relationship, Cheung and Mosca~\cite{Cheung2001} have developed a theoretical quantum algorithm to solve the abelian group isomorphism problem by computing the decomposition of a given group into a direct product of cyclic subgroups. 

\subsection{Algorithm description}

The procedure presented in Algorithm~\ref{alg:group} assumes the fundamental theorem of finite abelian groups, that they can be decomposed as a direct sum of cyclic subgroups of prime power order. This decomposition can then be used to test if an isomorphism exists between two groups. 

\begin{algorithm}[!htb]
\caption{Decompose($a_1,~\ldots, a_k$, $q$), of Cheung and Mosca~\cite{Cheung2001}}
\begin{algorithmic} \label{alg:group}
    \STATE \textbf{Input:}
    \bindent
        \STATE $\bullet$ A generating set $\{ a_1,~\ldots, a_k \}$ of $G$.
        \STATE $\bullet$ The maximum order, $q$, of the generating set.
    \eindent
    \STATE \textbf{Output:}
    \bindent
        \STATE $\bullet$ The set of elements $g_1,~\ldots, g_l$ from group $G$, with $l \le k$.
    \eindent
    \STATE \textbf{Procedure:}
    \bindent
        \STATE \textbf{Step 1.} Define $g : \mathbb{Z}_q^k \rightarrow G$ by mapping $(x_1,~\ldots, x_k) \rightarrow g(x) = a_1^{x_1} \cdot \cdot \cdot a_k^{x_k}$.
        \STATE Find generators for the hidden subgroup $K$ of $\mathbb{Z}_q^k$ as defined by function $g$.
        \STATE \textbf{Step 2.} Compute a set $y_1,~\ldots, y_l \in \mathbb{Z}_q^k/K$ of generators for $\mathbb{Z}_q^k/K$.
        \STATE \textbf{Step 3.} Output the set $\{ g(y_1),~\ldots, g(y_l) \}$.
    \eindent
\end{algorithmic}
\end{algorithm}

Since the procedure in Algorithm~\ref{alg:group} is mostly classical, we shall treat the task of finding the generators of the hidden subgroup in \textbf{Step 1} as the most critical for us to explore. This task is commonly referred to as the hidden subgroup problem (HSP). This means that, given a function $g$ that maps a finite group $A$ onto a finite set $X$, we are asked to find a generating set for the subgroup $K$. For $K$ to be the so-called \textit{hidden subgroup} of $A$, we require that $g$ is both constant and distinct on the cosets of $K$. On a quantum computer, this problem can be solved using a number of operations that is polynomial in log$\abs{A}$, in addition to one oracle evaluation of the unitary transform $U \ket{a}\ket{h} = \ket{a} \ket{h \oplus g(a)}$. The general procedure needed to complete \textbf{Step 1} of algorithm~\ref{alg:group} is described in algorithm~\ref{alg:hidden}.  

\begin{algorithm}[!htb]
\caption{Solution to the hidden subgroup problem (for finite abelian groups). Based on Ref.~\cite{NielsenChuang}}
\begin{algorithmic} \label{alg:hidden}
    \STATE \textbf{Input:}
    \bindent
        \STATE $\bullet$ Two quantum registers.
        \STATE $\bullet$ Elements of the finite abelian group $A$ (or the generating set).
        \STATE $\bullet$ A function $g$, such that $g: A \rightarrow X$, with $a \in A$ and $h \in X$.
    \eindent
    \STATE \textbf{Output:}
    \bindent
        \STATE $\bullet$ The generating set for the hidden subgroup $K$.
    \eindent
    \STATE \textbf{Procedure:}
    \bindent
        \STATE \textbf{Step 1.} Create initial state.
        
        \STATE \textbf{Step 2.} Create superposition between resisters.
        
        \STATE \textbf{Step 3.} Apply unitary operation ($U$) for function $g(a)$.
        \begin{align} \label{eqn:step3}
        \rightarrow \frac{1}{\sqrt{\abs{A}}} \sum_{a \in A} \ket{a}\ket{g(a)}
        \end{align}
        
        \STATE \textbf{Step 4.} Apply inverse Fourier transform.
        \begin{align} \label{eqn:step4}
        \rightarrow \frac{1}{\sqrt{\abs{A}}} \sum_{l=0}^{\abs{A}-1} e^{2 \pi i l a / \abs{A}} \ket{\hat{g}(l)}
        \end{align}
        
        \STATE \textbf{Step 5.} Measure the phase from first register.
        \begin{align} \label{eqn:step5}
        \rightarrow l / \abs{A}
        \end{align}
        
        \STATE \textbf{Step 6.} Sample $K$ from l / \abs{A}.
        
    \eindent
\end{algorithmic}
\end{algorithm}

Like the period-finding approach used in quantum factorization in Section V, Algorithm~\ref{alg:hidden} is heavily based on the concept of phase estimation. Note that the Fourier transform in Eq.~\ref{eqn:step4} represents $a \in A$ indexed by $l$. The key concept of the procedure is that $\ket{\hat{g}(l)}$ has nearly zero amplitude for all values of $l$, except those which satisfy

\begin{align} \label{eqn:key}
\abs{K} = \sum_{h \in K} e^{-2 \pi i l h / \abs{A}},
\end{align}
and that knowledge of $l$ can be used to determine both the elements and generating set of $K$. As discussed by Nielsen and Chuang~\cite{NielsenChuang}, the final step in algorithm \ref{alg:hidden} can be accomplished by expressing the phase as

\begin{align}
\rightarrow e^{2 \pi i l a / \abs{A}} = \prod_{i=1}^{M} e^{2 \pi i l_i a_i / p_i}.
\end{align}
for $a_i \in \mathbb{Z}_{p_i}$, where $p_i$ are primes, and $\mathbb{Z}_{p_i}$ is the group containing integers $\{ 0, 1,~\ldots, p_i-1 \}$ with the operator being addition modulo $p_i$. 

The quantum circuit needed to solve the HSP is schematically illustrated in Fig.~\ref{Zamora:fgr1}. This simplified circuit includes steps 1-5 of algorithm~\ref{alg:hidden}, and makes it clear that all forms of the HSP (order-finding, period-finding, discrete logarithm, etc.) are extensions of quantum phase estimation.

\subsection{Algorithm implemented using Qiskit}

\begin{figure}[!tb]
\begin{center}
\includegraphics[width=4.5in]{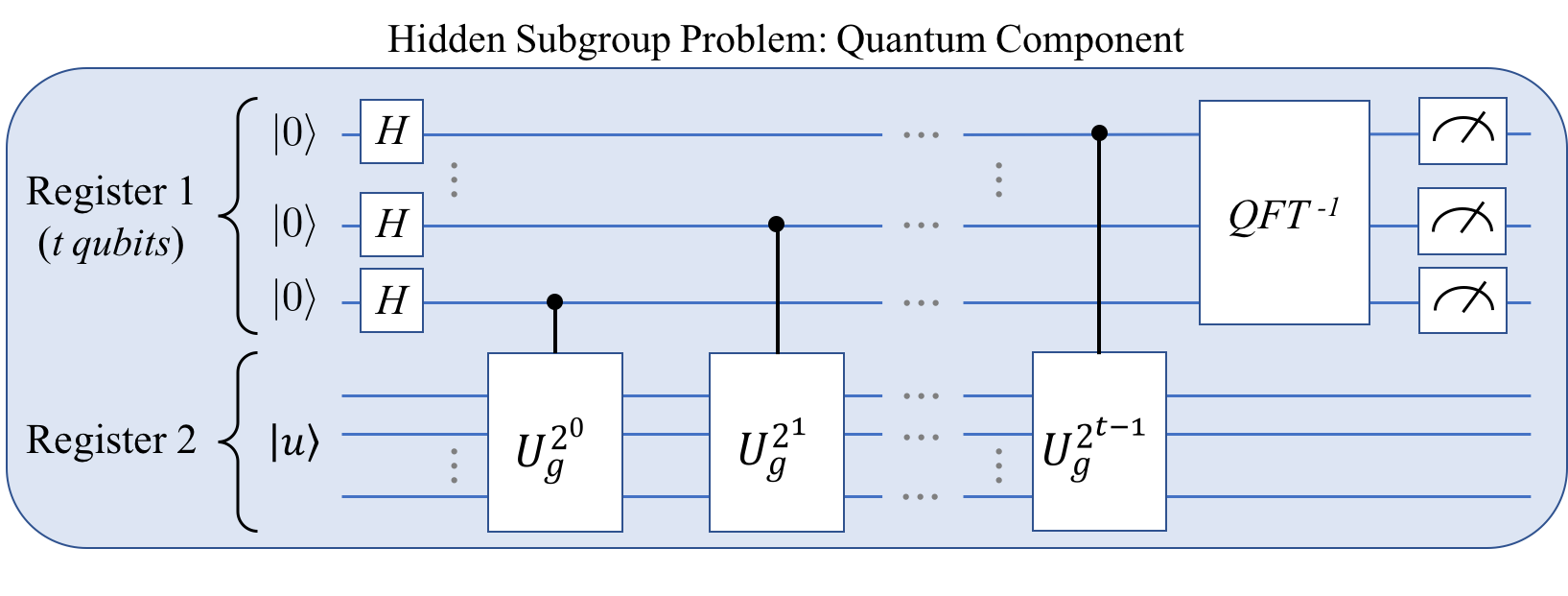}
\caption{Basic phase-estimation quantum circuit needed to solve the general hidden subgroup problem in algorithm~\ref{alg:hidden}. Here, $\ket{u}$ is an eigenstate of the unitary operator $U$.}
\label{Zamora:fgr1}
\end{center}
\end{figure}

Since the generalized group isomorphism problem is somewhat complex, we will focus here on the implementation of the HSP circuit fragment illustrated in Fig.~\ref{Zamora:fgr1}. We also chose a specific instance of the HSP: the problem of finding the period of $a~ \text{mod}~ n$. In Fig.~\ref{Zamora:fgr2}, the basic outline of the code needed for this specific problem is illustrated using the python-based Qiskit interface.

Like most instances of the HSP, one of the most challenging practical tasks of finding the period of $a~ \text{mod}~ n$ on a quantum computer is the implementation of the oracle. The details of the oracle are not explicitly shown in the Qiskit snippet, but for the required $Ca~ \text{mod}~ 15$ operations, one can simply used the circuits developed by Markov and Saeedi~\cite{saeedi2013synthesis}. The code in Fig.~\ref{Zamora:fgr2} also assumes that a function $qft\_inv()$ will return the gates for an inverse quantum Fourier transform, and that a classical \textit{continued fractions} algorithm can be used to convert the end result (a phase) to the desired integer period.

\begin{figure}[!tb]
    \centering
\begin{minted}{python}
#======================================================================#
#----------  Finding period (r) of a % N, with N=15  ------------------#
#======================================================================#
def findperiod(a, N=15, nqubits1, nqubits2):

    # Create QuantumProgram object, and define registers and circuit
    Q_program = QuantumProgram()
    qr1 = Q_program.create_quantum_register("qr1", nqubits1)
    qr2 = Q_program.create_quantum_register("qr2", nqubits2)
    cr1 = Q_program.create_classical_register("cr1", nqubits1)
    cmod15 = Q_program.create_circuit("cmod15", [qr1, qr2], [cr1])

    # Apply a hadamard to each qubit in register 1
    # and prepare state |1> in regsiter 2
    for j in range(nqubits1): cmod15.h(qr1[j])
    cmod15.x(qr2[nqubits2-1])

    # Loop over qubits in register 1
    for p in range(nqubits1):

        # Calculate next 'b' in the Ub to apply
        # ( Note: b = a^(2^p) % N ).
        # Then apply Ub
        b = pow(a,pow(2,p),N)
        CxModM(cmod15, qr1, qr2, p, b, N, nqubits1, nqubits2)

    # Perform inverse QFT on first register
    qft_inv(cmod15, qr1, nqubits1)

    # Measure each qubit, storing the result in the classical register
    for i in range(n_qr1): cmod15.measure(qr1[i], cr1[i])

\end{minted}
    \caption{Simple implementation of the quantum period-finding algorithm in Qiskit}
    \label{Zamora:fgr2}
\end{figure}

Although the specific procedure outlined in Fig.~\ref{Zamora:fgr2} can be directly implemented using the IBM Qiskit interface, the resulting QASM code is not expected to lead to accurate results on the IBMX4 (or IBMX5). This is because the generated circuit is long enough for decoherence error and noise to ultimately dominate the measured state. In other words, the physical hardware requires further optimization to reduce the number of gates used between the initial state preparation and the final measurement.

\ca
\section{Quantum Persistent Homology}

\subsection{Problem definition and background}

Big data analysis often involves large numbers of multidimensional data points. Understanding their structure can lead to insights into the processes that generated them. Data clustering is closely related to spatially connected components. Other features such as holes and voids and their higher dimensional analogs that characterize the distributions of data points are useful for understanding their structure. Persistent homology connects data points across scales to reveal the most enduring features of datasets. Methods from algebraic topology are employed to build simplicial complexes from data points, and the topological features of these simplicial complexes are extracted by linear algebraic techniques.
However, such an investigation on a set of n points leads to storage and computational costs of $O(2^n)$ as there is a combinatorial explosion in the number of simplices generated by n points.
Thus representational and computational efficiency has to be greatly enhanced for viability. Quantum algorithms provide such efficiency by superposing $2^n$ simplex states with only n qubits and implementing quantum parallel computations. The study of such a quantum algorithm, proposed by Lloyd et.\ al~\cite{LloydNature} is the focus of this section.

Data points $P = \{ p_0,\mathellipsis,p_{n-1}\}$ can be envisaged as vertices of a $simplicial$ $decomposition$ of a subset $X$. An oriented $k-$simplex $\sigma_k = [p_{j_0},\mathellipsis,p_{j_k}],0 \leq j_0 < j_1,\mathellipsis, < j_k \leq n-1, $ is the convex hull of $k+1$ points, and the simplicial complex is comprised of all the simplices. Thus, a 0-simplex is a vertex, a 1-simplex is an edge, a 2-simplex is a triangle, a 3-simplex is a tetrahedron, and so on.
\revision{The aim of persistent homology is to find topological features in the data that persist as we view the data at different length scales \cite{Lloyd2014}}. The numbers of various topological features at any scale are obtained from algebraic structures involving the simplices.

\begin{figure}
\begin{center}
\includegraphics[width=\columnwidth]{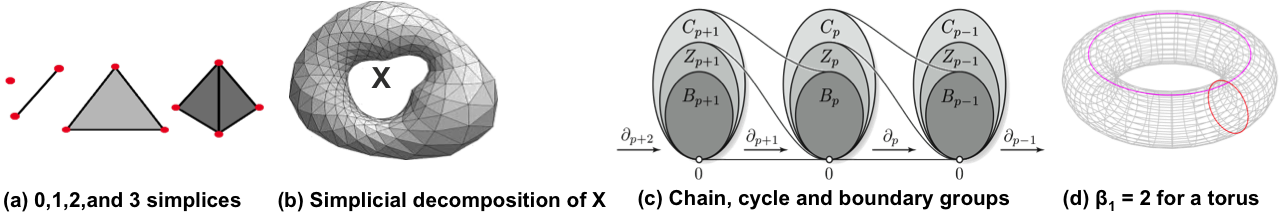}
\caption{Examples of simplices, simplicial decomposition of a topological space $X$, relationships among groups $C_p$ (chains), $Z_p$ (cycles), and $B_p$ (boundaries) under the action of boundary homomorphisms $\partial_p$, and the example of the torus.}
\label{Prasad:fgr1}
\end{center}
\end{figure}

Define the $k$th chain group $C_k$ as the set of all formal integer linear combinations of $k$-simplices: 
$C_k = \{ \sum_{i} a_{i}\sigma^{i}_{k} | a_i \in \mathbb{Z}\}$.
$C_k$ is an abelian group generated by the $k$-simplices. Further, define boundary operators $\partial_{k}:C_k \rightarrow C_{k-1}$ between chain groups as group homomorphisms whose action on a $k$-simplex $\sigma_k = [p_{j_0},\mathellipsis,p_{j_k}]$ is given by $\partial_k\sigma_k = \sum^{k}_{i=0} (-1)^{i}[p_{j_0},\mathellipsis,p_{j_{i-1}},p_{j_{i+1}},\mathellipsis,p_{j_k}]$ (i.e., the ith vertex is omitted from $\sigma_k$, $0\leq i \leq k$, to get the $k+1$ oriented $i$th boundary ($k-1$)-simplex faces). With this, every $k$-chain $c^{i}_{k} \in C_k$ gives rise to a $(k-1)$-chain $c^{i}_{k-1} \in C_{k-1}$. A chain $c \in C_k$ such that $\partial_{k} c = 0$, where $0$ is the null chain, is called a $k$-cycle. Also, $\partial_{k}\partial_{k+1} \equiv 0$. That is to say, the boundary of a boundary is the null chain 0, since the boundary of every $k+1$-chain is a $k$-cycle.  $Z_k = Ker(\partial_k)$ is the subgroup of $C_k$ consisting of all $k$-cycles, and $B_k =  Image(\partial_{k+1})$ is the subgroup of $C_k$ consisting of boundaries of all $(k+1)$-chains in $C_{k+1}$. Clearly, $B_k \subseteqq Z_k$. The relationships between chain groups, cycles and boundaries as established by the boundary homomorphisms is illustrated in Fig.~\ref{Prasad:fgr1}(c). The $k$th $Betti$ $number$ $\beta_k$ of a topological space $X$ is defined as the number of linearly independent $k$-cycles that are not boundaries of $(k+1)$-chains, and characterizes the topological features at dimension $k$. For instance, $\beta_0$ is the number of connected components of $X$, $\beta_1$ is the number of 1-dimensional holes, $\beta_2$ is the number of voids, and so on. The $k$th $Homology$ $Group$ of $X$ is defined as the quotient group $H_{k}(X) = Z_{k}(X)/B_{k}(X)$, whereby $\beta_k$ is the number of generators of $H_k(X)$.

Equivalently, the $k$th Betti number $\beta_k$ is the dimension of the kernel of the combinatorial Laplacian operator, $\Delta_k = \partial^{\dagger}_k\partial_k + \partial_{k+1}\partial^{\dagger}_{k+1}$, $ \beta_k = dim(Ker(\Delta_k))$. This allows the computation of Betti numbers by finding the null space of a linear transformation. The quantum algorithm~\cite{LloydNature} by Lloyd et al. diagonalizes the Laplacian to compute Betti numbers.

\begin{figure}
\begin{center}
\includegraphics[width=3.5in]{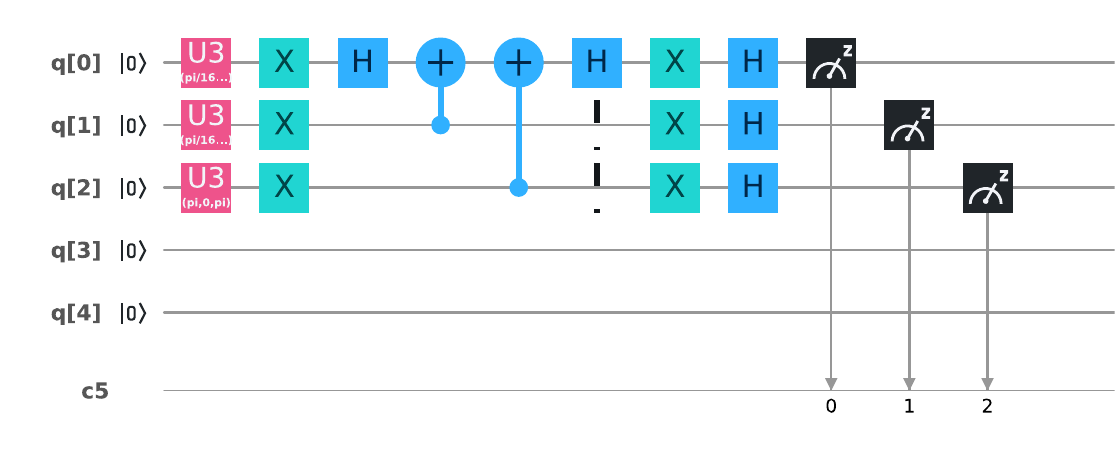}
\caption{Grover's Algorithm circuit implemented on the 5 qubit quantum computer showing 3 qubits being used with the multiple solution version of Grover's Algorithm.  U3 gates are used to input the scaled distances between points. }
\label{Prasad:fgr2}
\end{center}
\end{figure}

\begin{figure}
\begin{center}
\includegraphics[width=1.8in]{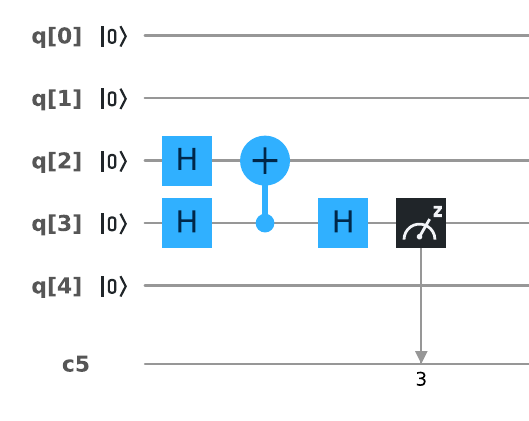}
\caption{Quantum Phase Estimation Algorithm circuit implemented on the 5 qubit quantum computer showing how the quantum density matrix implemented on qubits 0, 1, and 2 would be applied to obtain a classical measurement on qubit 3.}
\label{Prasad:fgr3}
\end{center}
\end{figure}
  
\subsection{Quantum algorithm description}

A quantum algorithm for calculating Betti Numbers is presented in~\cite{LloydNature}. The algorithm uses Grover's search combined with phase estimation to find the dimension $Ker(\Delta_k)$. Grover's algorithm is used to prepare a suitable initial state for the phase estimation. We will demonstrate how phase estimation can be used to estimate the dimension of the kernel of an eigenspace. Suppose that we have an $N \times N$ unitary operator $U = e^{ i 2 \pi H}$ on which we will apply phase estimation. Let $\ket{u_j}$ and $e^{i 2 \pi \lambda_j }$  be the eigenvectors and eigenvalues of $U$. Now given a starting state $\frac{1}{\sqrt{N}}\sum_{j=1}^N\ket{u_j}$ we know that the phase estimation subroutine will effect the following transformation,
\begin{equation}
    \frac{1}{\sqrt{N}}\sum_{j=1}^N\ket{u_j}\ket{0} \rightarrow \frac{1}{\sqrt{N}}\sum_{j=1}^N\ket{u_j}\ket{\tilde{\lambda}_j},
\end{equation}

where $\tilde{\lambda}_j$ are approximations to the original eigenphases. Now consider a measurement on the ancillary register that stores these eigenphases. \revision{If some of the $\lambda_j$ were zero, we can see that the probability of measuring $\ket{0}$ on the second register is equal to $\frac{dim(Ker(H))}{N}$.} Moreover the probability of measuring any $\ket{\tilde{\lambda}_j}$ will similarly be related to the dimension of its eigenspace. So by estimating these probabilities, we can figure out the dimensions of the eigenspaces. Notice that the performance and correctness of the procedure will depend on the precision of $\tilde{\lambda}_j$. This will in turn depend on the number of ancillary qubits used for phase estimation. For this procedure to work it was crucial that we started with the uniform superposition of all the eigenstates. This is a correct but naive way to go about the problem especially  if the dimension of the null space is exponentially small compared to the size of the matrix. But, this technique will work equally well if the input to the procedure was a classical mixture of  some of the eigenstates such that the probability of the null states in the said mixture was related to the dimension of the null space. This would let us recover the dimension of the null space from the measurement probabilities. The algorithm to find Betti numbers uses such a generalization of the naive procedure illustrated above.

\revision{We will give an overview of the full algorithm. The algorithm works with an object called the Vietoris-Rips complex. Given the data points, the  Vietoris-Rips complex of these points at a length scale $\epsilon$ is the set of all $k$ simplices formed from $(k+1)$ data points such that each of these $(k+1)$ points are at a distance of at most $\epsilon$ from each other. This complex is denoted by $S^\epsilon$ and can possibly contain simplices of any size. We denote by $S^\epsilon_k$ the subset of $S^\epsilon$ that contains only $k$-simplices.  We associate with each simplex $\sigma_k$ a computational basis state $\ket{\sigma _k}$ such that the $1$s in $\ket{\sigma_k}$ correspond to the points chosen in $\sigma_k$. Then the algorithm  goes as follows:}
\begin{enumerate}
    \item \revision{ First we construct a uniform superposition  over all the simplex states in $S^\epsilon_k$. The membership of a simplex  to $S^\epsilon_k$ can be checked easily by verifying the number of  in the simplex and querying all the pair wise distances between them. This requires an additional input of scaled distance between points. This membership function can then be used as an oracle in Grover search to compute the desired state given below,}
    \begin{equation}
        \ket{\psi}^{\epsilon}_k =  \frac{1}{\sqrt{|S^{\epsilon}_k|}} \sum_{\sigma_k \in |S^{\epsilon}_k|} \ket{\sigma_k}
    \end{equation}

    \item \revision{Then this state is modified to produce a  mixture over all the simplices in $S^\epsilon$. This mixture is represented as a quantum density matrix $\rho^{\epsilon}.$ Readers unfamiliar with the concept of density matrices should read Section \ref{sec:Tomography} on quantum tomography. This density matrix can be constructed by running Grover search in parallel. First initialize an ancilla register with a uniform superposition over all values of $k$. Then controlled on this ancilla we can run the Grover search to construct  $\ket{\psi}^{\epsilon}_k$. Then measuring the ancilla gives the desired state,}
    
    \begin{align}
        \frac{1}{\sqrt{n 2^n}}\sum_{k,\sigma} \ket{k} \ket{\sigma} ~ &\xrightarrow{\text{Grover search}} \frac{1}{\sqrt{n}}\sum_{k} \ket{k}\ket{\psi}^{\epsilon}_k \\
         &\xrightarrow{\text{Measure ancilla}}~~ \rho^{\epsilon} =  \frac{1}{n} \sum_{k}\ket{k}\bra{k} \otimes  \ket{\psi}^{\epsilon}_k \bra {\psi}^{\epsilon}_k 
    \end{align}
    
    \item \revision{A phase estimation procedure can be performed on $\rho^{\epsilon}$ to estimate the dimensions of the kernels of the Laplacian at each $k$ and from them the Betti numbers. Now by looking at how these numbers change as we vary $\epsilon$, we can study topological features of the data that persist over length scales. The phase estimation procedure is described in subsection \ref{subsec:phase_estim}}.
\end{enumerate}

\revision{The above high level overview hides several caveats of the algorithm.  The reader should refer Lloyd et al. \cite{Lloyd2014} for further details and complexity  analysis of the algorithm.}

Working with only 5 qubits implies that the largest number of points $n$ that could be processed at once is constrained by $n(n-1)/2 \leq 5$. Thus only 3 points at a time could be processed on the 5 qubit quantum computer.

Calculating the quantum density matrix could not be accomplished by the IBM machine due to the lack of Quantum RAM needed for the algorithm.  Options considered to circumvent this problem included implementing a quantum algorithm for computing the outer product to form an 8x8 quantum density matrix.  This was abandoned as the sheer size of quantum algorithms to implement four qubit addition~\cite{Viamontes} was well beyond the 5 qubits available on the quantum computer. The Quantum Experience message boards' suggestion to perform Grover's algorithm 64 times and reassemble the output into a density matrix was also not viable due to decoherence.    
Had a quantum density matrix been produced, phase estimation  would have been applied to find the probabilities of eigenvalues of the boundary operator.  This would then be multiplied times the number of simplices and used as input to find the Betti Number.  Hence, at the time of writing,  the main bottleneck preventing the implementation of this algorithm is seen to be the coherence time of the computer.

\begin{figure}
\begin{center}
\includegraphics[width=3.5in]{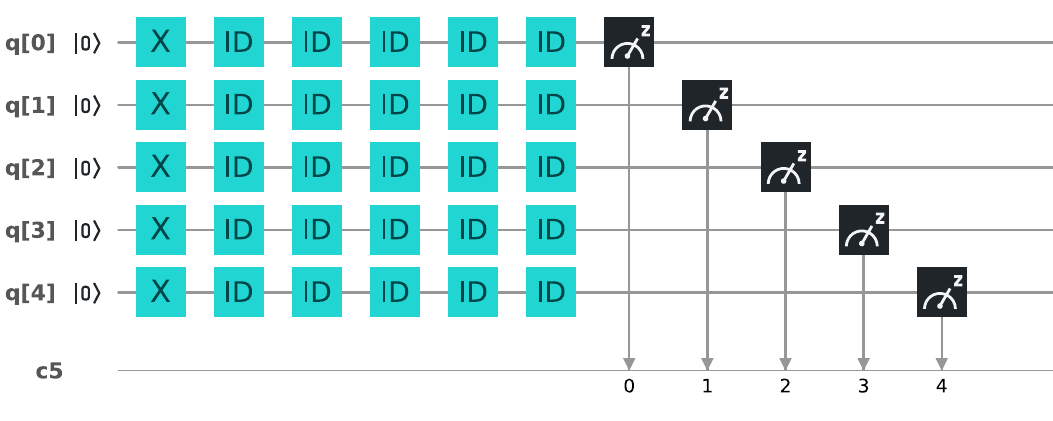}
\caption{The quantum algorithm applied to estimate the Decoherence time of the 5 qubit quantum computer as implemented in the Quantum Composer to measure the decoherence after 5 timesteps.  Note that this produces an optimistic estimate as the "id" quantum gates utilize minimal time whereas other gates, such as CNOT and U3 gates, could use more time.}
\label{Prasad:fgr4}
\end{center}
\end{figure}

\ca In order to check the coherence of the quantum 5 qubit computer a study was designed.  All five qubits were flipped in the first timestep and measurements were then taken place approximately every five timesteps throughout the 74 available timesteps in the quantum composer.  The quantum algorithm applied is shown in Fig.~\ref{Prasad:fgr4} showing the use of the Idle gates for 5 timesteps after the qubits are flipped with the X gate.

\begin{figure}[!htb]
\begin{center}
\includegraphics[width=\columnwidth]{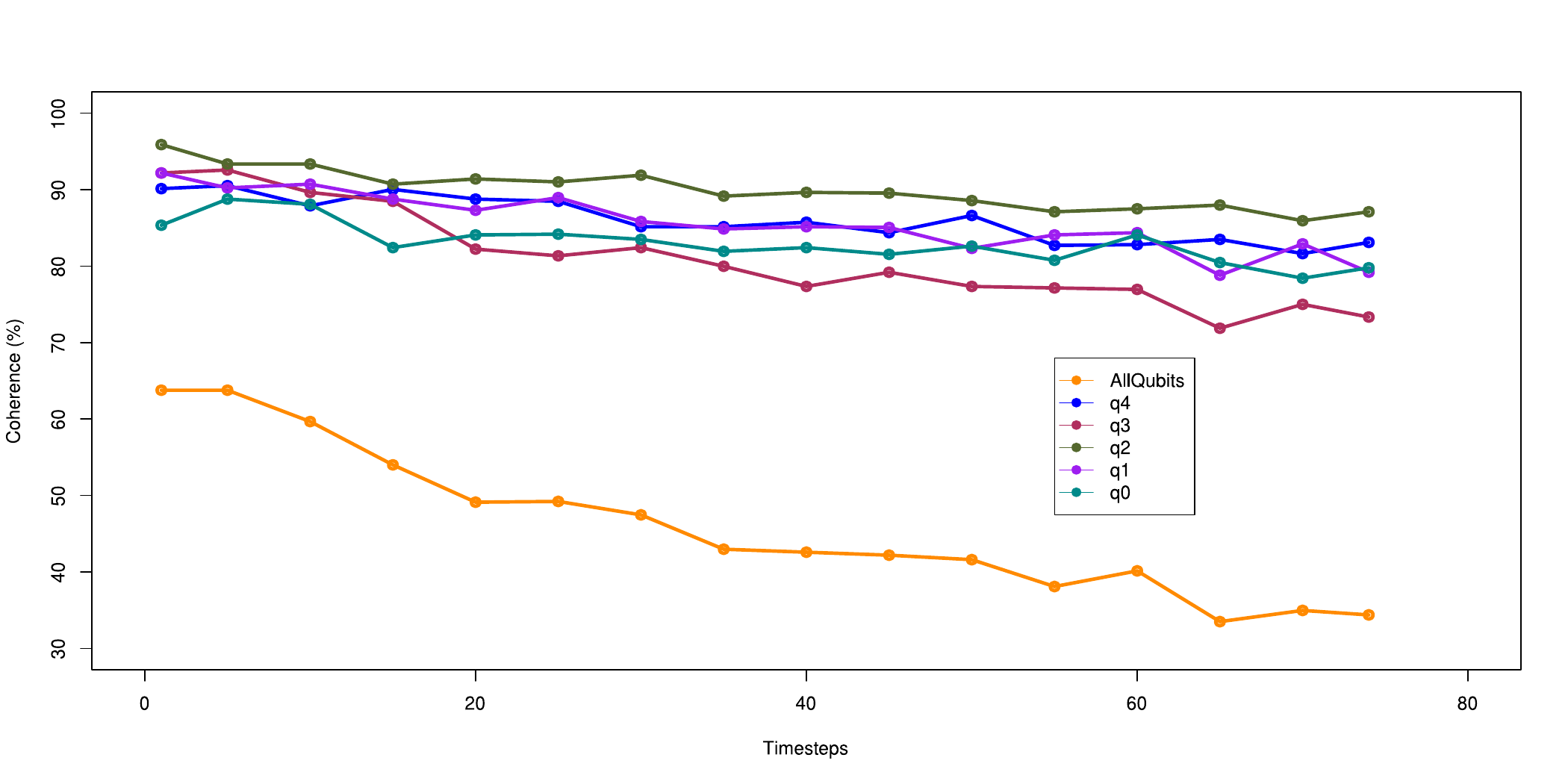}
\caption{Timesteps available with the Quantum Composer are shown on the x axis, ranging up to 74.  Coherence percentage is calculated for all individual qubits and is plotted on the y axis. Coherence percentages of all 5 qubits combined is also shown on the y axis.  Note that the coherence percentage rate falls below 50 percent between the 15th and the 20th timesteps.}
\label{Prasad:fgr5}
\end{center}
\end{figure}

The data was collected for all the timesteps, processed into a form to evaluate the coherence percentages for all individual qubits and for all qubits combined.  The results are depicted in Fig.~\ref{Prasad:fgr5} showing the decoherence rates with quantum composer timesteps. The coherence rates here measure the `quantumness' of the qubits as detailed in Ref. \cite{tsai2010toward}. Note that although the coherence rates are fairly high for individual qubits, the overall qubit coherence percentages are far less.  This is because a single qubit decohering also decoheres the entire 5 qubit quantum computer.  This should be considered to determine how many qubits will actually be usable in an actual machine.  It is also interesting that qubit 3 seems to decohere at a faster rate than the other qubits past timestep 15.

\cb

\section{Quantum Random Walks}

\subsection{Problem definition and background}

Quantum algorithms for graph properties using the adjacency matrix (as part of an oracle)
have been published for minimum spanning tree, shortest path, deciding if a graph is bipartite,
detecting cycles, finding subgraphs (such as a triangle), maximal clique, and many more.
Each typically involves the use of Grover's search~\cite{grover} with an oracle 
constructed from the adjacency matrix.

But for some problems Grover's algorithm is insufficient to achieve optimal query complexity. In such cases, a quantum random walk can sometimes be useful in reducing the query complexity of an algorithm further. An example of this is the quantum algorithm for element distinctness by Ambainis \cite{ambainis2007quantum}.
Additionally, quantum walk algorithms can also be used to search and find graph properties~\cite{lovett2010,douglas2009,kendon2011,codsil2017,magniez2011,kempe2003}.
Quantum random walks can be seen as a quantum mechanical generalization of  classical random walks.
Quantum random walk algorithms come in two forms, discrete time quantum walks and continuous time quantum walks~\cite{kempe2003}.
The discrete form operates in a step-wise fashion, requiring multiple copies of a set of gates per step.
The continuous form uses a transition matrix that is expressed as a Hamiltonian, whose time evolution is then simulated.
Quantum random walks can be used to walk a graph ~\cite{douglas2009,kendon2011}, search
for marked vertices~\cite{magniez2011}, and to solve s-t connectivity~\cite{kendon2011}. An excellent survey of this approach to quantum search can be found in Ref. \cite{santha2008quantum}.

Most quantum algorithms that solve graph problems requires an oracle that knows the properties of the underlying graph. A graph properties oracle can be assembled  as a circuit based
on the adjacency matrix of the graph and linear algebra transformations.
For example, a quantum circuit for finding maximal cliques in a graph with $n$ nodes,
requires an oracle workspace of $n^2$ data qubits and $n^2$ ancilla qubits 
(see~\cite{wie2017}). Each oracle call requires execution of $6n^2$ Toffoli gates 
and $2n$ CNOT gates. An oracle such as this can be run on a simulator, but requires too many qubits to run on actual qubit hardware. Quantum algorithms for finding a triangle, quadrilateral, longer cycles, and arbitrary 
subgraphs~\cite{circella2006} typically use the adjacency matrix  to create the
oracles. Here we will not get into using quantum random walks to solve such problems. Instead we will demonstrate how to implement a simple quantum random walk on a quantum computer.

\subsection{Example of a quantum random walk}
Quantum random walks or simply quantum walks are quantum  analogues of classical random walks and
Markov chains. Unlike the continuous time quantum walk, the discrete time quantum walk algorithm
requires the use of one or more coin qubits representing the 
number of movement choices from each graph vertex. These extra coin degrees of freedom are necessary to ensure unitarity of the quantum walk. An appropriate unitary transformation on these coin qubits then acts like the quantum version of  a random coin toss, to decide the next vertex for the walker.

Intuitively, the quantum walk is very similar to its classical cousin. In a classical walk, the walker observes some random process, say a coin toss, and decides on his next step conditioned on the output of this random process. So for a classical random walk, the walker is given a probability to make a transition. In a quantum walk, on the other hand, the random process is replaced by a quantum process. This quantum process is the application of the coin operator, which is a unitary matrix. So the next step of the walker is controlled by a complex amplitude rather than a probability. This generalization, from positive numbers to complex numbers, makes quantum walks more powerful than classical random walks.

The full Hilbert space for the discrete quantum walk on a cycle with $N = 2^n$ nodes  can then be constructed as follows.  We use  an $n$ qubit register to represent the nodes of the graph as bit strings. For the cycle every node has only two neighbours, so the coin space only needs a dimension of $2$. Hence, only one extra coin qubit is required. The basis vectors of the coin ($\ket{0}$ and $\ket{1}$) will denote the right and left neighbours. So a basis state in the full Hilbert space will have the form $\ket{k, q}$, where $k$ is represents a node on the cycle and $q$ is a single bit representing the coin state.

The quantum walk is then a product of two operators, the shift operator ($S$) and the coin operator ($C$). As we mentioned before the coin operator only acts on the coin qubit. The coin operator can be in principle any unitary that mixes the coin states, but here we will use the Hadamard coin which is just the $H$ gate on the coin qubit,

\begin{equation}\label{eq:coin}
    C \ket{k, q} = I \otimes H \ket{k,q} =  \frac{\ket{k,0} + (-1)^q \ket{k,1}}{\sqrt{2}}.
\end{equation}

The shift operator acts on both the registers. It moves the walker to the left or right depending on the coin state and then flips the coin state,
\begin{equation}\label{eq:shift}
    S \ket{k,q} = \ket{ k +  (-1)^q ,  q \oplus 1}
\end{equation}

The quantum walk  then proceeds by applying these two operators in alternation. A $p$ step quantum walk is just the operator $(SC)^p.$ This type of a walk was first introduced in Ref. \cite{shenvi2003quantum} and is sometimes referred to as a `flip-flop' quantum walk.

The definition of these operators can change for different types of quantum walk. The coin operator can be a Hadamard gate or a sub-circuit that results in mixing the coin states. The shift operator can be simple as described above or can be a more complicated circuit that selects the next vertex in the path based on the state of the coin. A simple pseudo-code for implementing the quantum walk is given in Algorithm \ref{alg:qwalk}.

\begin{algorithm}[H]
\caption{Discrete time quantum walk}
\begin{algorithmic} \label{alg:qwalk}
    \STATE \textbf{Input:}
   \bindent
        \STATE $\bullet$ Two quantum registers. The coin register and the position register.
        \STATE $\bullet$ Number of steps, $T$.
    \eindent
   \STATE \textbf{Output:}
    \bindent
        \STATE $\bullet$ State of the quantum walk after $T$ steps.
    \eindent

    \STATE \textbf{Procedure:}
    \bindent
        \STATE \textbf{Step 1.} Create the initial state. The initial state depends on the application. For instance, in quantum search algorithms, the initial state is the uniform superposition state.

        \STATE \FOR{ $0 \leq k < T $ }
        \bindent
          \STATE \textbf{Step 2a.} Apply the coin operator, $C$, to the coin register.
             \STATE \textbf{Step 2b.} Apply the shift operator, $S$. This shifts the position of the walker controlled on the coin state.
        \eindent
        \ENDFOR
  
         \STATE \textbf{Step 3.} (Optional) Measure the final state.

    \eindent
\end{algorithmic}
\end{algorithm}

\subsection{Algorithm implementation using Qiskit on IBM Q}
\begin{figure}[H]
\begin{center}
\includegraphics[width=0.8in]{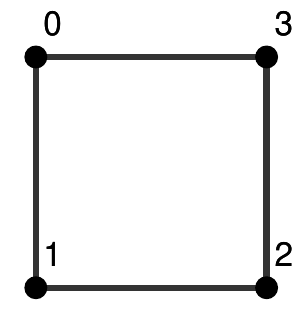}
\caption{A graph of 4 nodes in the form of a square is used for the random walk algorithm. The starting vertex is labeled $0$. The vertex labels are converted to  binary for input into the quantum circuit. The quantum walk algorithm will walk around the graph.}
\label{fig:square}
\end{center}
\end{figure}

In this section we will implement a simple quantum walk on Qiskit and execute it on both the simulator and \verb|ibmq_vigo|, which is a $5$ qubit machine available on IBM Q. We will test the quantum walk on a simple $4$ vertex cycle with the vertices labels as given in Fig. \ref{fig:square}.

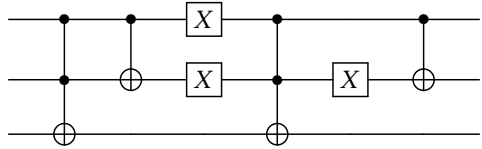
\begin{figure}[H]
\begin{equation*}
\Qcircuit @C=1.7em @R=1em {
&\ctrl{1} &\ctrl{1} & \gate{X} & \ctrl{1} &\qw    &\ctrl{1} &\qw  \\
&\ctrl{1} &\targ    & \gate{X} & \ctrl{1} &\gate{X} &\targ     & \qw\\
&\targ    &\qw      & \qw      &  \targ   &\qw        &\qw   &\qw
}
\end{equation*}
    \caption{Quantum circuit for the shift operation on the $4$ vertex cycle. The top qubit is the coin qubit.}
    \label{fig:shift_circ}
\end{figure}

The coin operator in Eq.~\eqref{eq:coin} is just the $H$ gate acting on the coin qubit. The shift operator defined in Eq.~\eqref{eq:shift} is more non-trivial. We can implement it by the circuit given in Fig.~\ref{fig:shift_circ}.  

Running the walk for multiple steps requires us to apply the shift operator circuit many times. So it would be tedious to implement the quantum walk on the IBM Q graphical interface. Instead we can use Qiskit to design the shift operator as a user defined gate and then run the walk for multiple steps using a simple \texttt{for} loop. The Qiskit code for this is given in Fig.~\ref{fig:qwalk_qiskit}. 

\begin{figure}[!htb]
\centering
\begin{minted}{python}
from qiskit import QuantumCircuit, QuantumRegister, ClassicalRegister
from qiskit import Aer, execute

n_steps =  4          #Number of steps

#Defining the shift gate

shift_q =  QuantumRegister(3)                               #3 qubit register
shift_circ = QuantumCircuit (shift_q, name='shift_circ')    #Circuit for shift operator
shift_circ.ccx (shift_q[0],  shift_q[1], shift_q[2])        #Toffoli gate
shift_circ.cx ( shift_q[0], shift_q[1] )                    #CNOT gate
shift_circ.x ( shift_q[0] )
shift_circ.x ( shift_q[1] )
shift_circ.ccx (shift_q[0],  shift_q[1], shift_q[2])
shift_circ.x ( shift_q[1] )
shift_circ.cx ( shift_q[0], shift_q[1] )

shift_gate = shift_circ.to_instruction()          #Convert the circuit to a gate

q = QuantumRegister (3, name='q')                 #3 qubit register
c = ClassicalRegister (3, name='c')               #3 bit classical register
circ = QuantumCircuit (q,c)                       #Main circuit
for i in range(n_steps):
    circ.h (q[0])                                 #Coin step
    circ.append (shift_gate, [q[0],q[1],q[2]])    #Shift step
    
circ.measure ([q[0],q[1],q[2]], [c[0],c[1],c[2]]) 







\end{minted}
    \caption{Qiskit code to implement the quantum walk on a $4$ vertex cycle.}
    \label{fig:qwalk_qiskit}
\end{figure}

 We ran this Qiskit code for $4$ steps of the quantum walk. We chose $4$ steps since, a simple calculation shows that, starting from $\ket{000}$ and applying $(SC)^4$  will concentrate all the probability to the state $\ket{100}.$ This is confirmed by running the Qiskit code on the simulator. But running the same code on \verb|ibm_vigo| gave $\ket{100}$ with only $21.7\%$  probability. The rest of the probability was distributed among the other basis states, but $\ket{100}$ was still the state with the largest probability. This poor performance is due to the circuit having large depth. We can expect to get better results by running the quantum walk for a single step. After a single step, starting from $\ket{000}$, the state of the system is $\frac{\ket{111} + \ket{010}}{\sqrt{2}}.$ This is again confirmed by the simulator. Running on \verb|ibm_vigo|, we got \ket{111} with $ 33.5 \%$ probability  and \ket{010} with $28.5\%$ probability.

\section{Quantum Minimal Spanning Tree}

\subsection{Problem definition and background}
\begin{figure}[htb]
\centering
\subfloat[][The weighted graph model.]{%
\includegraphics[scale=.33]{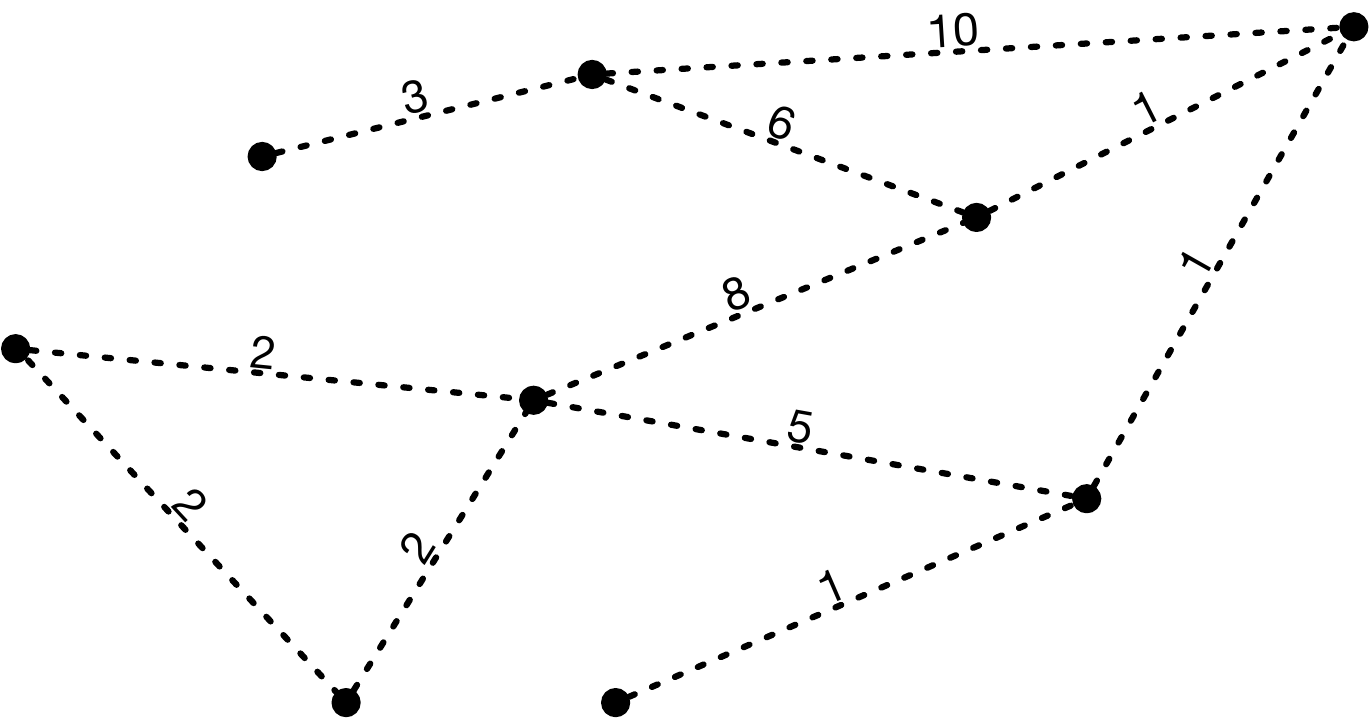}}
\qquad
\subfloat[][A minimal spanning tree.]{%
\includegraphics[scale=.33]{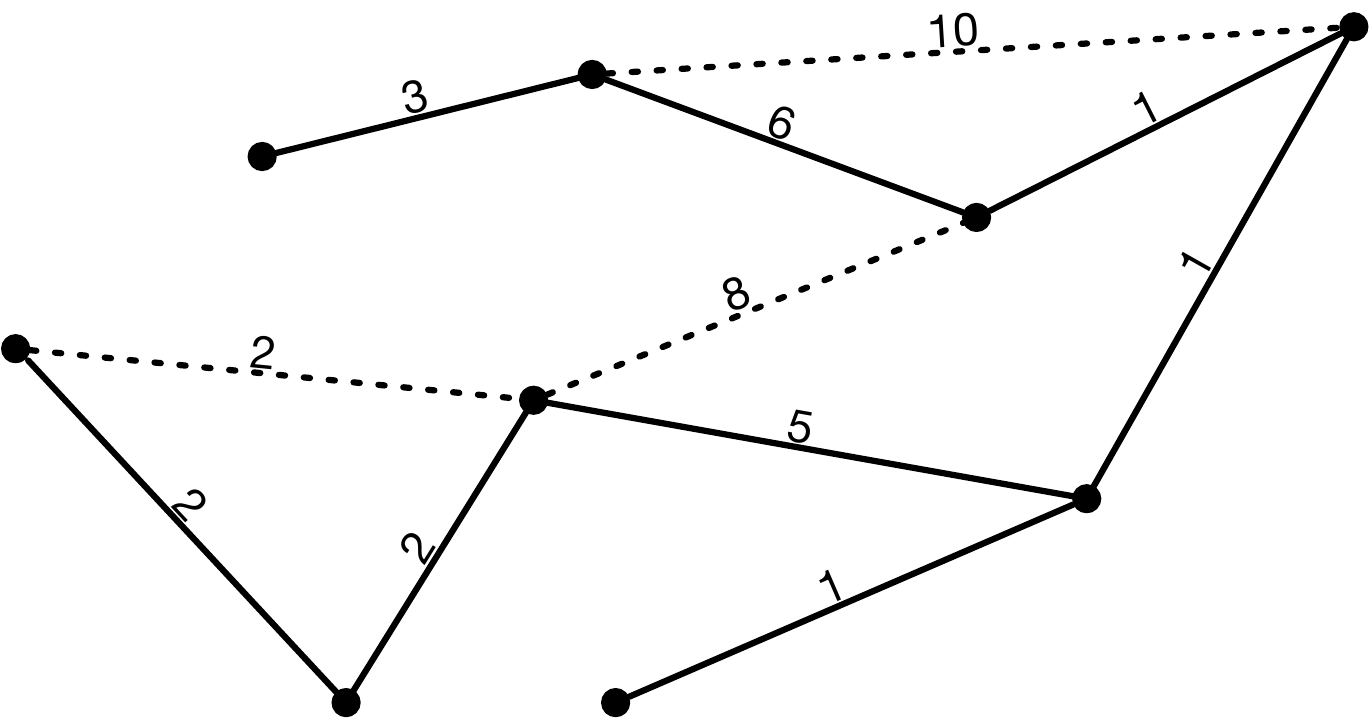}
}
\caption{A graph modeling repair costs of a simple transportation network (a) together with (b) its minimal spanning tree (the solid edges).  The sum of the weights of the edges in the minimal spanning tree is 21.}\label{fig:example}
\end{figure}

A common problem in network design is to find a minimum spanning tree.  Suppose we are responsible for maintaining a simple network of roads.  Unfortunately, each segment needs repair and our budget is limited.  What combination of repairs will guarantee the network remains connected?  Fig~\ref{fig:example} shows a model of a simple road network as a graph, together with a minimal spanning tree.

Formally, a graph $G=(V,E)$ consists of a set $V$ (the nodes) and a set $E$ consisting of pairs of nodes.  A graph is connected if between any two nodes there exists a path.  A spanning tree of a connected graph $G=(V,E)$ is the graph $T=(V,E_T)$ where $E_T\subset E$ and $T$ contains no cycles (i.e.,~there is exactly one path between any two vertices).  It is not hard to see that a graph $T$ is a spanning tree if and only if $T$ is connected and has $n$ nodes and $n-1$ edges.  A weighted graph is a graph $G=(V,E,w)$ where $w$ is a map on the edges $w:E\rightarrow \mathbb{R}$.  A minimal spanning tree of a graph $G$ is then a spanning tree $T=(V,E_T)$ which minimizes 
\begin{equation}
\sum_{e\in E_T} w(e).
\end{equation}
\begin{figure}[bht]
\centering
\includegraphics[scale=.33]{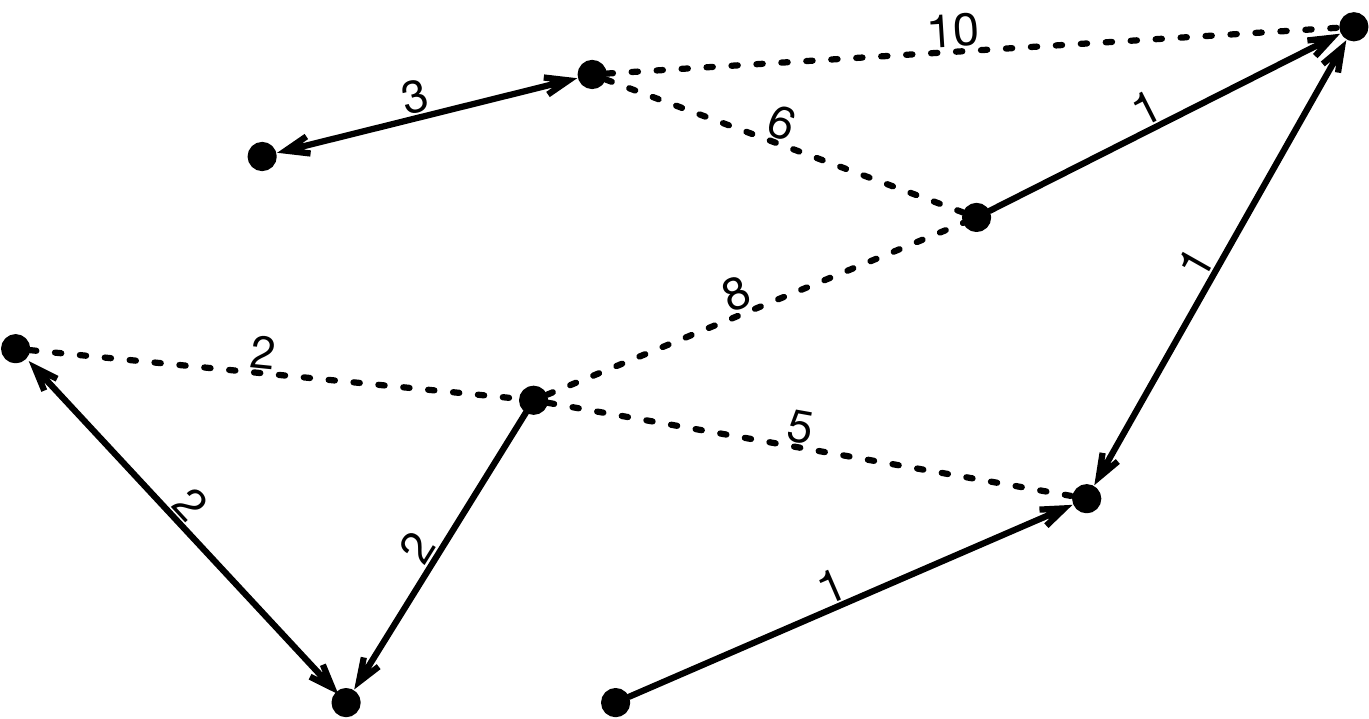}
\caption{The first two steps of Bor\r{u}vka's algorithm.  Starting with each node as a distinct tree, find the minimal weighed edge between each tree and the rest of the trees.  The direction of the solid edges indicates the edge is the minimal weighted edge for the source node.  The components connected by solid edges (disregarding the directions) will form the trees at the start of the second run of step (2) of Bor\r{u}vka's algorithm}\label{fig:boruvka}
\end{figure}

\subsection{Algorithm description}
Algorithmically, a graph is usually presented in one of two ways: either as a list of edges or as an adjacency matrix.  We consider the case where $G$ is presented as a list of edges.  A quantum algorithm for finding a minimal spanning tree of an input graph is given in~\cite{Durr_Quantum_2006}.  This algorithm requires only $O(\sqrt{nm})$ queries where $n$ is the number of nodes and $m$ the number of edges in the graph.  Classically, the best algorithms run in time $O(m\log n)$. In particular, this is the time complexity of Bor\r{u}vka's algorithm~\cite{boruvka}.  The quantum algorithm combines Bor\r{u}vka's algorithm together with the quantum search algorithm of Grover~\cite{grover}.

Bor\r{u}vka's algorithm builds a collection of disjoint trees (i.e.,~a forest) and successively merges by adding minimal weight edges.  The first two steps of the algorithm are shown in Fig~\ref{fig:boruvka}.  Formally, we have 
\begin{enumerate}
\item Let $\mathcal{T}$ be the collection of $n$ disjoint trees, each consisting of exactly one node from the graph $G$.
\item Repeat:
\begin{enumerate}
\item For each tree $T_i$ in $\mathcal{T}$ find the minimal weighted edge, $e_i$, of $G$ between $T_i$ and the other trees of $\mathcal{T}$.
\item Merge the trees $\{T_i\cup\{e_i\}\}$ so that they are disjoint: set this new collection to $\mathcal{T}$.
\end{enumerate} 
\end{enumerate} 
If there are $k$ trees in $\mathcal{T}$ at a given iteration of Step (2), then the algorithm performs $k$ searches for the respective minimal weighted edges. As the trees are disjoint, we can perform the $k$ searches in one sweep by inspecting each of the $m$ edges of $G$ once.  As there will be at most $\log n$ iterations of Step (2), this results in a running time of $O(m\log n)$.  The quantum algorithm takes advantage of the Grover search algorithm, to speed up the searches in Step (2).

In the previous sections we used Grover search to look for a single item in a list of $N$ elements. But the search algorithm will work even if there are $M$ elements in the list that are marked by the oracle. One of these marked elements can then be found using $O(\sqrt{\frac{N}{M}})$ queries to the oracle.

  In the algorithm above, we need to find the minimal element of an appropriate list.  Clearly this can not be implemented directly as an oracle without actually inspecting each of the list elements.  Luckily, there is a simple work around given by Durr et al~\cite{Durr_Quantum_2006} which involves multiple calls to the Grover algorithm as described in Algorithm \ref{alg:min_finding}.

\begin{algorithm}[H]
\caption{Minima finding algorithm}
\begin{algorithmic} \label{alg:min_finding}
    \STATE \textbf{Input:}
    \bindent
    \STATE $\bullet$ A unitary implementation a function $F$ on a list of $N$ elements,
    $$U_F \ket{x}\ket{y}  =  \ket{x}\ket{y \oplus F(x)}.$$
    \eindent
    \STATE \textbf{Output:}
    \bindent
     \STATE $\bullet$ \ket{x^*} such that $F(x^*)$ is the minimum of the function over the list.
    \eindent    

    \STATE \textbf{Procedure:}
    \bindent
        \STATE \textbf{Step 1.} Pick a random $j$ from the list.
        
        \STATE \FOR{ $0 \leq k < T $ }
        \bindent
            \STATE \textbf{Step 2a.} Do Grover search \cite {boyer1998tight} with the oracle for function $f_j$  such that, 
            \begin{equation*}
            f_j(i)= 
	       \begin{cases}	
		    1 & \text{if}~ F(i) \leq F(j)\\
		    0 & \text{if}~ F(i) > F(j) \\
	       \end{cases} 
	       \end{equation*}
        \STATE \textbf{Step 2b.} Update $j$ with the result of Grover search.
        \eindent
        \ENDFOR

    \eindent
\end{algorithmic}
\end{algorithm}

A  probabilistic analysis shows that $T =  22.5 \sqrt {N} +  1.4~ \text{log}^2_2 (N)$ suffices to find the minimal element with high probability \cite{durr1996quantum} . The inner loop of the algorithm uses a Grover search routine with potentially multiple marked items. But the number of marked items is not known beforehand. This poses problem as Grover search being a unitary algorithm needs to be stopped exactly at the right number of iterations to give the correct answer. Running the procedure for longer deteriorates the quality of the answer. If the number of marked items is unknown the stopping criterion of the algorithm is also unknown. This problem can be rectified using some extra steps by a technique given in Boyer et al \cite{boyer1998tight}. We have to use this modified version of Grover search in the inner loop.
%\subsection{Algorithm implementation on IBM's 5-qubit computer}
%\begin{figure}[bht]
%\centering
%\includegraphics[scale=.5]{schematic}
%\caption{A schematic of Grover's %Algorithm}\label{fig:schematic}
%\end{figure} 

\begin{figure}[htb]
\centering
\subfloat[][IBM Q Implementation]{%
\includegraphics[scale=.5]{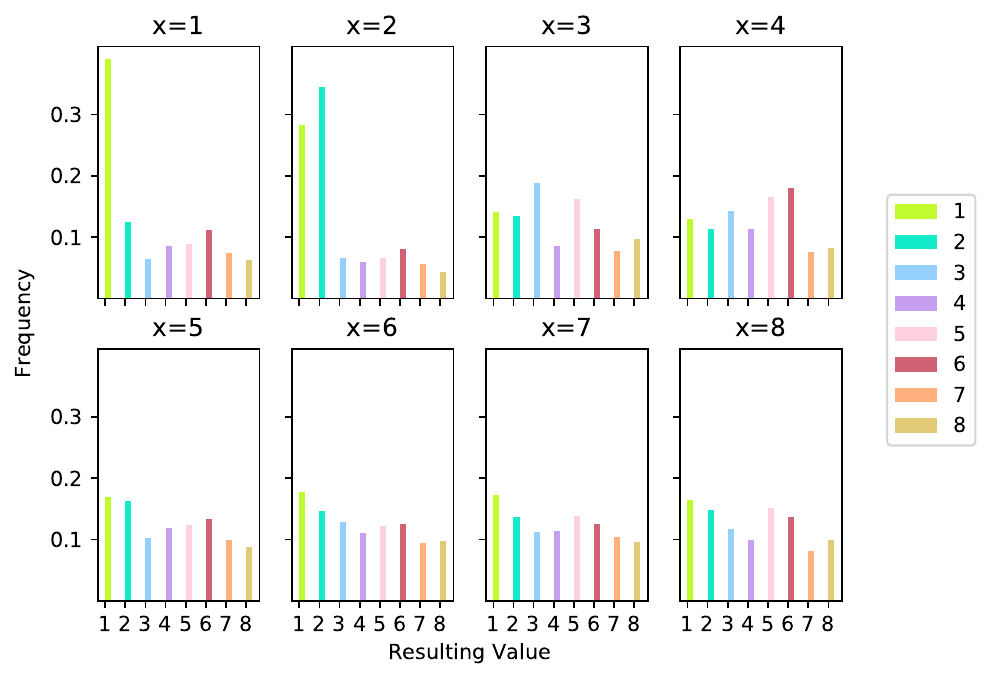}
\label{fig:weighted}}
\qquad
\subfloat[][Simulator Implementation]{%
\includegraphics[scale=.5]{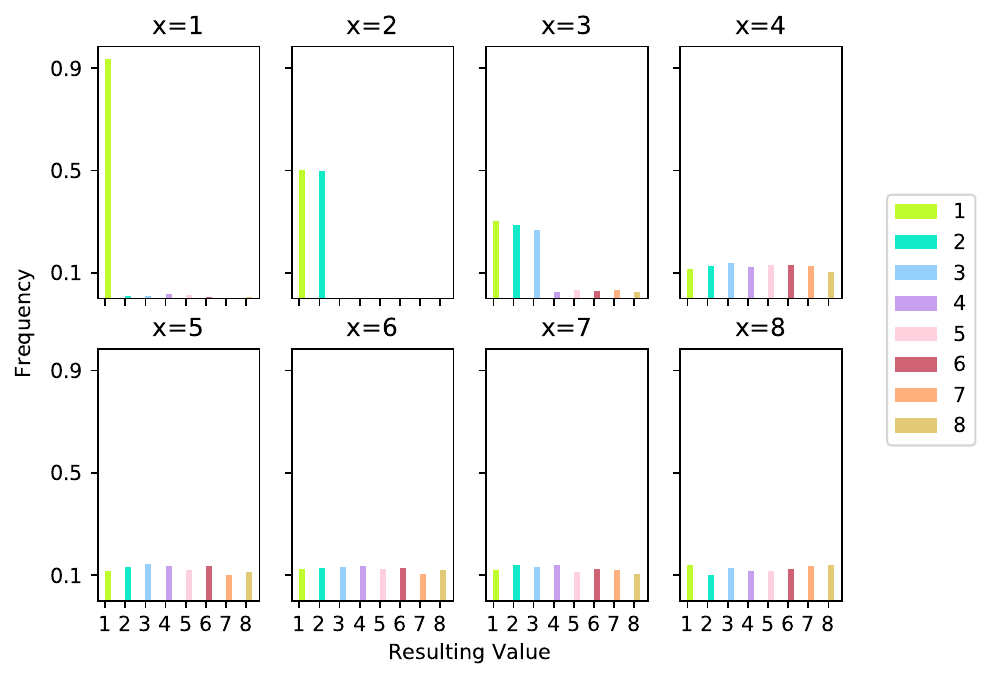}
}
\caption{The results of running 1000 trials of the minima finding algorithm on both (a) the \texttt{ibmqx4} chip and (b) the IBM simulator to find values less than or equal to the input $x$.}\label{Lemons:fig:results}
\end{figure}

We did not implement the full algorithm due to space constraints on the IBM computer.  Even to successfully implement a minima finding algorithm, at least 6 qubits would be necessary to compare two 3-bit numbers.  Therefore we implemented the minima finding algorithm by hard coding the oracle  for each of eight possible functions $f_x$:  $\{f_x(i)=1 \textrm{ if } F(i)\leq F(x)\}$.  The results are shown in Figure~\ref{Lemons:fig:results}.  The QASM code for implementing $f_2(i)=1\textrm{ if } F(i)\leq F(2)$ required just under 100 lines of code (more than 100 individual gates.)  The results, even when using the simulator are not good  when $k\geq N/4$ elements are marked.  A typical way to get around this is to double the length of the list by adding $N$ extra items which will evaluate to 0 under $f_x$, which however requires an extra qubit.

\section{Quantum Maximum Flow Analysis}

\subsection{Problem definition and background}

Network flow problems play a major role in computational graph theory and operations research (OR). Solving the max-flow problem is the key to solving many important graph problems, such as finding a minimum cut set, and finding a maximal graph matching. The Ford-Fulkerson algorithm~\cite{fordFulkerson} is a landmark method that defines key heuristics for solving the max flow problem. The most important of these heuristics include the construction of a residual graph, and the notion of augmenting paths. For integer-capacity flows, Ford-Fulkerson has complexity $O(fm)$ for $m$ edges and max flow $f$. The Edmonds-Karp variant has complexity $O(nm^2)$ for $n$ vertices and $m$ edges. The quantum-accelerated classical algorithm discussed here~\cite{ambainisEtAl} claims complexity  $O(n^{7/6}\sqrt{m})$.

The best classical implementations of the max-flow solver involve several important improvements~\cite{edmondsKarp}, especially that of using breadth-first search to find the shortest augmenting path on each iteration. This is equivalent to constructing layered subgraphs for finding augmenting paths.

\begin{figure}
\begin{center}
\includegraphics[width=\columnwidth]{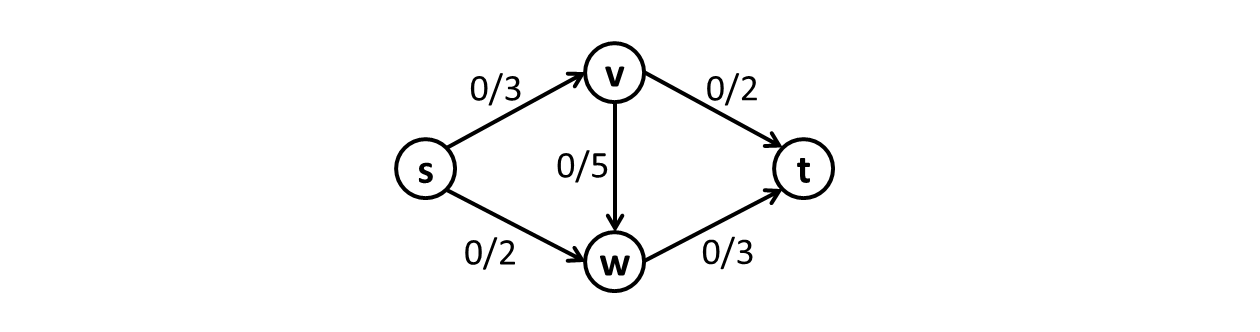}
\caption{A simple directed graph representing flows and capacities. Conventionally, the state of the flow problem is indicated by the current flow relative to the capacity on any directed link using the notation f/c.}
\label{Ambrosiano:fgr1}
\end{center}
\end{figure}

\begin{figure}
\begin{center}
\includegraphics[width=\columnwidth]{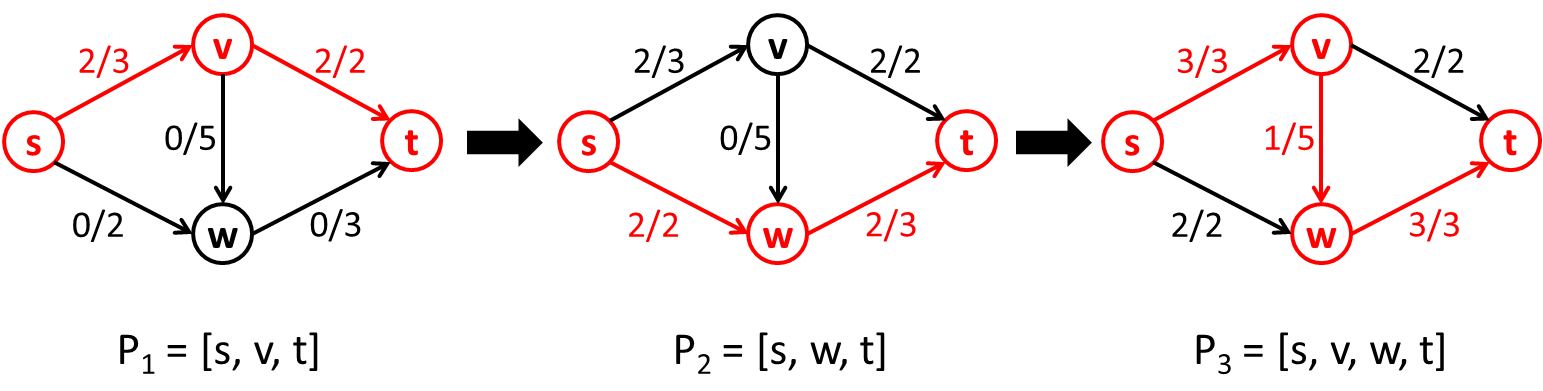}
\caption{The Ford-Fulkerson solution to the max-flow problem in three steps. Each step represents the application of an augmenting path to the previous flow state.}
\label{Ambrosiano:fgr2}
\end{center}
\end{figure}

An illustration of the essential method introduced by Ford and Fulkerson can be described using Figures~\ref{Ambrosiano:fgr1} and~\ref{Ambrosiano:fgr2}. At each link in the network, the current flow $f$ and the capacity $c$ are shown. Typically, the state of flow on the graph is designated by $f/c$, with the residual capacity implicitly given by $c-f$. In Figure~\ref{Ambrosiano:fgr1}, the initial flow has been set to zero.

The basic steps in the solution to the max-flow problem are illustrated by Figure~\ref{Ambrosiano:fgr2}. The algorithm begins on the left by considering the path [s,v,t]. Since 2 is the maximum capacity allowed along that path, all the flows on the path are tacitly set to that value. Implicitly, a reverse flow of -2 is also assigned to each edge so that the tacit flow may be ``undone'' if necessary. Next, the middle of the figure examines the lower path [s,w,t]. This path is constrained by a maximum capacity on the edge [s,w] of again 2.  Finally, the path [s,v,w,t] is the only remaining path. It can only support the residual capacity of 1 on edge [s,v]. We can then read off the maximum flow result at the sink vertex $t$ since the total flow must end there. The maximum flow is seen to be 5.

While this method seems straightforward, without the efficiencies provided by the improvements of Edmonds and Karp, convergence might be slow for integer flows on large graphs, and may not converge at all for real-valued flows. The modification of always choosing the shortest next path in the residual network to augment, is what makes the algorithm practical. To see this, consider what would have happened if the path [s,v,w,t] had been chosen first. Since augmenting that path blocks the remaining paths, flows would have to be reversed before the algorithm could proceed.

Choosing the shortest path requires performing a breadth-first search whenever new flow values have been assigned to the residual graph. This is equivalent to building a layered set of subgraphs to partition the residual graph. This is the step that leads to the $m^2$ complexity of Edmonds-Karp, and it is this step that is speeded up in the ``quantized'' version of the algorithm, leading to a complexity term of $\sqrt{m}$ instead of $m^2$.

\subsection{Algorithm description}

The Quantum algorithm described by Ambainis and Spalek is a ``quantized'' version of the Edmonds-Karp algorithm, that is, the classical algorithm with quantum acceleration. The key quantum component is a generalized version of Grover's search algorithm that finds $k$ items in an unsorted list of length $L$~\cite{boyer1998tight}. The algorithm is used in creating a layered subgraph data structure that is subsequently used to find the shortest augmenting path at a given iteration. Like in  Section XI, we will be oblivious to the number of marked items Grover's algorithm is searching for. So once again we have to use techniques from Ref.\cite{boyer1998tight} while performing the search.

Here we will describe how to build a layered graph partition. In a layered graph partition each vertex in the graph is assigned to thew $i$-th layer such that edges of the graph only connect between $i$-th and $(i +1)$-th layers. 
The key to ``quantization'' lies in using Grover's search to build a layered graph partition by computing layer numbers for all vertices. The layers are represented by an array $\mathcal{L}$ indexing the vertices of the graph, and assigning to each element a subgraph layer number. The sink vertex at vertex zero is set to zero. The the algorithm proceeds according to the following pseudo-code  described in Algorithm \ref{alg:layered_subgraph}.

\begin{algorithm}[H]
\caption{Layered graph partitioning}
\begin{algorithmic} \label{alg:layered_subgraph}
    \STATE \textbf{Input:}
    \bindent
      \STATE $\bullet$ Adjacency information of the  graph (Adjacency matrix, list of edges,etc.) 
      \STATE $\bullet$ Source vertex $s$.
      \eindent
    \STATE \textbf{Output:}
    \bindent
     \STATE $\bullet$ $\mathcal{L}$ such that $\mathcal{L}[i]$ is the layer number of the $i$-th vertex.
    \eindent    

    \STATE \textbf{Procedure:}
    \bindent
        \STATE \textbf{Step 1.} Set $\mathcal{L}[s] = 0$ and $\mathcal{L}[x] = \infty$ for $x \neq 0$
        
        \STATE \textbf{Step 2.}  Create a one-entry queue $W = \lbrace s \rbrace \quad (x=0)$
        
        \WHILE {$W \neq \phi$}
        \bindent
            \STATE \textbf{Step 3a.} Take the first vertex $x$ from $W.$
            
            \STATE \textbf{Step 3b.} Find by Grover search all its neighbors $y$ with $\mathcal{L}[y] = \infty.$
            
            \STATE \textbf{Step 3c.} Set $\mathcal{L}(y) = \mathcal{L}[x] + 1$,  append $y$ into $W$, and remove $x$ from $W$
            
        \eindent
        \ENDWHILE

    \eindent
\end{algorithmic}
\end{algorithm}

Notice that the oracle for Grover search required for this algorithm is one that marks all the neighbours of $x$ whose layer number is currently set to $\infty$. Grover's search speeds up the layers assignment of the vertices by quickly finding all the entries in the layer array $\mathcal{L}$ that contain the value $\infty$. In practical terms, $\infty$ might simply be the largest value reachable in an n-qubit machine. The generalized Grover search would look for all such values without a priori knowing the number of such values.  The size of a circuit required to do a full layered graph partitioning makes it impractical to implement it on the IBM machine. But the heart of the algorithm is Grover search, which we have already implemented earlier.

\section{Quantum Approximate Optimization Algorithm}

\subsection{Problem definition and background}

Combinatorial optimization problems are pervasive and appear in applications such as hardware verification, artificial intelligence and compiler design, just to name a few.  Some examples of combinatorial optimization problems include Knapsack, Traveling Saleman, Vehicle Routing, and Graph Coloring problems. A variety of combinatorial optimization problems including MaxSat, MaxCut, and MaxClique can be characterized by the following generic unconstrained discrete maximization problem,
\begin{equation}
%\tag{MINLP}
\label{eq:max_clause}
\begin{aligned}
    \mbox{maximize:}	&	\sum^m_{\alpha=1} C_\alpha(x) \\
    x_i \in \{0, 1\}	&	\ \forall i \in \{1,~\ldots, n\} 
\end{aligned}
\end{equation}
In this generic formulation, there are $n$ binary variables, $x$, and $m$ binary functions of those variables, $C(x)$, called clauses.  The challenge is to find the assignment of $x$ that maximizes the number of clauses that can be satisfied\revision{, i.e. that can be evaluated to 1}.  In case each clause is an OR of literals (positive or negated variables), this is the so-called MaxSat problem, which is NP-Hard in general~\cite{Krentel1986}, and is an optimization variant of the well-known satisfiability problem, which is  NP-Complete~\cite{Cook1971}.  Hence, solving an instance of Eq.~\eqref{eq:max_clause} in practice can be computationally challenging\revision{, meaning that there is no algorithm which can solve all instances of the problem in time polynomial in their input size $(n,m)$, unless P=NP}.

\subsubsection*{The Maximum Cut Problem}

\begin{figure}[t!]
\begin{center}
\begin{minipage}{0.64\linewidth}
	\includegraphics[width=\linewidth]{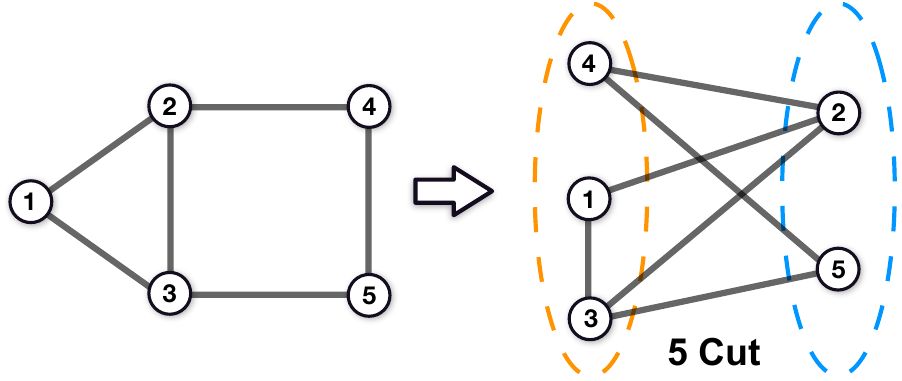}
\end{minipage}
\hfill
\begin{minipage}{0.31\linewidth}
\begin{tabular}{l|c}
	$x_1..\,x_5$	&	Val		\\
	\hline
	00000		&	0		\\
	00001		&	2		\\
	00010		&	2		\\
	00011		&	2		\\
	00100		&	3		\\
	00101		&	3		\\
	00110		&	5		\\
	00111		&	3		
\end{tabular}
\hfill
\begin{tabular}{l|c}
	$x_1..\,x_5$		&	Val		\\
	\hline
	01000		&	3		\\
	\color{orange}{0}\color{cyan!50!blue}{1}\color{orange}{00}\color{cyan!50!blue}{1}	&	\textbf{5}		\\
	01010		&	3		\\
	01011		&	3		\\
	01100		&	4		\\
	01101		&	4		\\
	01110		&	4		\\
	01111		&	3		
\end{tabular}
\end{minipage}
\caption{An illustration of the MaxCut problem: 
(left) input Graph $G$, 
(middle) a solution of maximum value 5, 
(right) values of all possible cuts; note that swapping 0/1 for all variables would result in the same cut sizes.}
\label{fig:mc_example}
\end{center}
\end{figure}

To provide another concrete example of Eq.~\eqref{eq:max_clause}, let us consider the MaxCut problem. As input, the MaxCut problem takes a graph $\mathit{G} = (\mathit{V}, \mathit{E})$, which is characterized by a set of $n$ nodes $\mathit{V}$ and a set of $m$ undirected edges $\mathit{E}$.  The task is to partition the nodes into two sets, such that the number of edges crossing these sets is maximized. Figure~\ref{fig:mc_example} provides an illustrative example, in which a graph with five nodes and six edges is partitioned into two sets that result in a cut of size five.
In general, the MaxCut problem is characterized by the following unconstrained discrete maximization problem,

\begin{equation}
%\tag{MINLP}
\label{eq:max_cut_bool}
\begin{aligned}
	\mbox{maximize:}	&	\sum_{\left\{ u,v \right\} \in {\mathit E}} \mathrm{XOR}(x_u, x_v)	
				&	= \sum_{\left\{ u,v \right\} \in {\mathit E}} x_u + x_v - 2x_u x_v	\\
    	x_i \in \{0, 1\}	&	\quad \forall i \in {\mathit n} 
\end{aligned}
\end{equation}

It is clear that Eq.~\eqref{eq:max_cut_bool} conforms to the structure of Eq.~\eqref{eq:max_clause}: There is one binary variable $x_i \in \left\{ 0,1 \right\}$ for each node in the graph, indicating which set it belongs to. The objective function consists of one term for each edge in the graph. This term is 0 if the the nodes of that edge take the same value and 1 otherwise. Consequently, the optimal solution of \eqref{eq:max_cut_bool} will be a maximal cut of the graph ${\mathit G}$.
In foresight, we also reformulate Eq.~\eqref{eq:max_cut_bool} in terms of spin variables $z_i \in \left\{ -1, +1 \right\}$
as a sum $\sum_\alpha C_\alpha(z)$, using the linear transformation $x_i = (z_i+1)/2$:
\begin{equation}
%\tag{MINLP}
\label{eq:max_cut}
\begin{aligned}
    	\mbox{maximize:}	&	\sum_{\left\{ u,v \right\} \in {\mathit E}} \frac{1}{2}(1-z_u z_v)	
				&	= \quad \frac{m}{2} \quad - \sum_{\left\{ u,v \right\} \in {\mathit E}} \frac{z_u z_v}{2}	\\	
    	z_i \in \{-1, 1\}	&	\quad \forall i \in {\mathit n} 
\end{aligned}
\end{equation}

Interestingly, the form of Eq.~\eqref{eq:max_cut} also highlights that finding a maximal cut of ${\mathit G}$ is equivalent to finding a ground state of the antiferromagnet of ${\mathit G}$ in an Ising model interpretation. 
We will use Eq.~\eqref{eq:max_cut} later in the next subsection to formulate a quantum problem Hamiltonian by replacing spin variables $z_i$ with Pauli $Z$-operators acting on qubit $i$, $Z_i = \underbrace{\mathit{Id} \otimes \ldots \otimes \mathit{Id}}_{i-1} 
\otimes \ ( \begin{smallmatrix} 1 & 0 \\ 0 & -1 \end{smallmatrix} ) \otimes \underbrace{\mathit{Id} \otimes \ldots \otimes \mathit{Id}}_{n-i}$.

\subsubsection*{Heuristics and approximation algorithms}
Given that the decision version of the Maximum Cut problem is NP-hard~\cite{garey-johnson}, the Maximum Cut optimization problem is often approached by \emph{heuristic algorithms}~\cite{maxcut-heuristics}. 
\revision{These are algorithms for which one cannot provide guarantees on their performance, but which are often observed to perform well on typical instances.}
A simple example would be the search for local improvements, where nodes are moved from one set of the cut to the other, if this strictly (or monotonically) increased the cut size. While heuristics may perform very well on most instances, it is generally difficult to tell when they get stuck in a local optimum (or end up in a loop, respectively).
\revision{As heuristics do not always achieve the perfect solution, one can measure their performance on an individual instance $I$ via an approximation ratio, defined as the algorithm's solution value $A(I)$ versus the optimal value $\textsc{OPT}(I)$,
$\tfrac{A(I)}{\textsc{OPT}(I)}$.}

A different approach is the design of a polynomial-time $r$-approximation algorithm $A$, for which one proves an approximation ratio $r$ 
\revision{over \emph{all} possible input instances $I$}: 
$r = \min_I \tfrac{A(I)}{\textsc{OPT}(I)} < 1$ for maximization problems (or $r = \max_I \tfrac{A(I)}{\textsc{OPT}(I)} > 1$ for minimization problems).  
In the case of randomized (classical or quantum) algorithms, the deterministic value $A(I)$ in the definition of the approximation ratio is replaced by the expected value $\mathbb{E}[A(I)]$ of the solution given by the algorithm.
A prominent example for the Maximum Cut problem is the $0.878..$-approximation algorithm by Goemans and Williamson~\cite{goemans-williamson}, which first solves a semi-definite programming relaxation of the problem followed by a randomized hyperplane rounding of the SDP solution. Assuming the unique games conjecture, this is the best possible polynomial-time approximation ratio~\cite{maxcut-UGC} for MaxCut, which on the other hand is only known to be NP-hard to approximate within $0.941..+\varepsilon$~\cite{maxcut-hastad}.
Interestingly, for graphs of maximum degree $3$, it is possible to improve on Goemans-Williamson by combining SDP relaxation \& rounding with the aforementioned local improvement heuristic, taylored to low-degree nodes, achieving a polynomial-time approximation ratio of $0.932..$~\cite{maxcut-cubic}.

When we assume -- as is commonly believed -- that quantum computers cannot solve NP-hard problems in polynomial time (i.e. $\text{NP} \nsubset \text{BQP}$), then there are two routes for the design of quantum algorithms for combinatorial optimization problems: polynomial speedups for existing approximation algorithms or new quantum heuristics. 
The Quantum Approximate Optimization Algorithm (QAOA) studied in this section is such a quantum heuristic. 
Only in rare cases an approximation ratio was shown for QAOA, among them a $0.692..$-approximation for the MaxCut problem on 3-regular graphs (where all nodes have degree 3)~\cite{1411.4028}.

\subsection{Algorithm description}

\begin{figure}[!tb]
\begin{center}
\includegraphics[width=\linewidth]{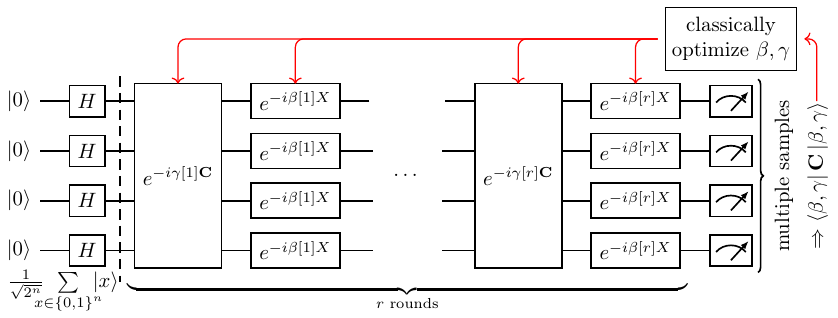}
\caption{A high-level view of the hybrid quantum-classical Quantum Approximate Optimization Algorithm: 
	Starting from a uniform superposition over all computational basis states, the quantum subroutine alternatingly applies the quantum problem Hamiltonian $\mathbf{C}$ and a transverse field $\mathbf{B} = \sum_j X_j$ for times $\gamma[1:r]$ and $\beta[1:r]$, respectively, to prepare the state $\ket{\beta,\gamma}$.
Collecting multiple samples from $\ket{\beta,\gamma}$, one can estimate the expectation $\bra{\beta,\gamma}\mathbf{C}\ket{\beta,\gamma}$ and use a classical optimizer 
to adjust the the parameters $\beta,\gamma$. This eventually results in a state $\ket{\beta,\gamma}$ of high expectation value, from which one can sample \revision{solutions of high objective value}.
}
\label{fig:qaoa_circuit}
\end{center}
\end{figure}

The Quantum Approximate Optimization Algorithm (QAOA) as proposed in~\cite{1411.4028} is a hybrid quantum-classical heuristic algorithm. It leverages gate-based quantum computing for finding \revision{candidate} solutions to combinatorial optimization problems that have the form of Eq.~\eqref{eq:max_clause}, using a variational circuit with parameters tuned in a classical outer loop (Fig.~\ref{fig:qaoa_circuit}).

To apply the quantum subroutine, the user first translates the clause functions $C_\alpha(z)$ into equivalent quantum clause Hamiltonians $\mathbf{C}_\alpha$, 
which give rise to the quantum problem Hamiltonian $\mathbf{C} = \sum_\alpha \mathbf{C}_\alpha$. Additionally, one defines a transverse field mixing Hamiltonian $\mathbf{B} = \sum_i X_j$ with Pauli $X$-operators $(\begin{smallmatrix} 0 & 1 \\ 1 & 0 \end{smallmatrix})$ acting on qubits $j$. Finally, the user defines a number of rounds $r \geq 1$ and two angles per round, $0 \leq \beta[k] \leq \pi$ and $0 \leq \gamma[k] \leq 2 \pi$ for the $k$-th round. 
Starting in a uniform superposition $\ket{+}^{\otimes n}$ of all $n$-bit computational basis states, the quantum subroutine then prepares a state $\ket{\beta,\gamma}$ by alternatingly applying the Hamiltonians $\mathbf{C}$ and $\mathbf{B}$ for times $\gamma[k], \beta[k]$. The former operation corresponds to a unitary $e^{-i\gamma[k] \mathbf{C}}$ which phases each basis state by an angle proportional to its objective value and proportional to $\gamma[k]$. It can be implemented by applying each clause Hamiltonian $\mathbf{C}_\alpha$ on its own. The second operation corresponds to unitaries $e^{-i\beta[k] X_j}$ applied to each qubit $j$ (see pseudocode in Algorithm~\ref{alg:QAOA}). The goal is to prepare a state $\ket{\beta, \gamma}$ such that one can indeed sample \revision{candidate solutions of objective value as close to the optimum value as possible}. This, however, heavily depends on the choice of the parameters $r,\beta,\gamma$.

\begin{algorithm}[!tb]
\caption{Quantum subroutine of the Quantum Approximate Optimization Algorithm}
\begin{algorithmic} \label{alg:QAOA}
    \STATE \textbf{Input:}
    \bindent
      \STATE $\bullet$ Number of rounds of optimization $r$
      \STATE $\bullet$ Two size $r$ array of angles, $\mathbf{\gamma}$  and  $\mathbf{\beta}$.
      \STATE  $\bullet$ Hamiltonians $\mathbf{C_\alpha}$ corresponding to the clauses of the optimization problem.
    \eindent
    \STATE \textbf{Output:}
    \bindent
     \STATE $\bullet$ An approximation to the solution of problem in  Eq.~\eqref{eq:max_clause}.
    \eindent    

    \STATE \textbf{Procedure:}
    \bindent
        \STATE \textbf{Step 1.} Construct the $n$-qubit uniform superposition state by applying $H^{\otimes n}$ to $\ket{0\ldots 0}$

        \FOR {$1 \leq k \leq  r$}
        \bindent

            \STATE \textbf{Step 2a.} Apply $\prod_{\alpha =1}^{m} e^{-i \mathbf{\gamma}[k] \mathbf{C}_{\alpha}}$
            \STATE \textbf{Step 2b.} Apply $\prod_{j =1}^{n} e^{-i \mathbf{\beta}[k] X_j}$            
            
        \eindent
        \ENDFOR
	\STATE \textbf{Step 3.} We call the state so constructed $\ket{\mathbf{\beta}, \mathbf{\gamma}}$.\\ 
	Preparing and measuring $\ket{\beta,\gamma}$ multiple times allows to both estimate the expectation value 
	$\sum_{\alpha =  1}^ m\braket{ \mathbf{\beta},  \mathbf{\gamma} | \mathbf{C}_\alpha | \mathbf{\beta},  \mathbf{\gamma}},$ 
	and to sample an approximate solution to the problem of objective value of at least the expectation minus $1$ (with high probability).
    \eindent
\end{algorithmic}
\end{algorithm}

\subsubsection*{Parameter finding and relation to the adiabatic theorem}
To find parameters $\beta, \gamma$ that enable sampling of \revision{good candidate} solutions to Eq.~\eqref{eq:max_clause} (for a given round $r$), 
one usually resorts to optimize $\beta,\gamma$ for a large expectation value $\bra{\beta,\gamma}\mathbf{C}\ket{\beta,\gamma}$, where
the optimization is done through a classical optimizer outer loop, see Fig.~\ref{fig:qaoa_circuit}.
The reason for this is two-fold: Assuming every solution has an objective value in the range $0,\ldots,m$, 
with high probability $\geq 1 -\tfrac{1}{m}$ the number of samples sufficient to 
(i) sample at least one solution of value at least $(\bra{\beta,\gamma}\mathbf{C}\ket{\beta,\gamma} - 1)$ is $m\log m$~\cite{1411.4028}, and to
(ii) precisely estimate $\bra{\beta,\gamma}\mathbf{C}\ket{\beta,\gamma}$ is $m^3$~\cite{cook2020kVC}, which can be further reduced 
if the distribution of $\mathbf{C}$ is concentrated around the expectation, e.g. to $m^2$ for MaxCut on bounded-degree graphs with a small number $r$ of QAOA rounds~\cite{1411.4028}.

On an ideal noise-free quantum computer, an increase in the number of rounds $r$ should lead to a monotonic increase in the quality of the expectation value $\bra{\beta,\gamma}\mathbf{C}\ket{\beta,\gamma}$, provided that the chosen angles $\beta, \gamma$ are optimal. In fact, any $r$-round QAOA with $\beta[r]=\gamma[r]=0$ corresponds simply to a $(r-1)$-round QAOA with parameters $\beta[1:r-1], \gamma[1:r-1]$, hence the optimal expectation value is non-decreasing in the number of rounds. 
One can also show that with an increasing number of rounds, in the limit $r \rightarrow \infty$ with suitably chosen angles $\beta,\gamma$, the expectation value will converge to the optimum value: $\lim_{r\rightarrow \infty} \max_{\beta,\gamma} \bra{\beta,\gamma}\mathbf{C}\ket{\beta,\gamma} = \max C(z)$.
To this end, we consider a quantum adiabatic algorithm~\cite{farhi-adiabatic} running for time $T$ with time-dependent Hamiltonian $H(t) = (1-t/T)\mathbf{B} + (t/T)\mathbf{C}$. The starting state of QAOA, $\ket{+}^{\otimes n}$, is also the starting state and unique highest energy eigenstate of $H(0) = \mathbf{B}$. Running the quantum adiabatic algorithm sufficiently slow ($T \gg \text{poly}(n)$) thus results in a highest energy eigenstate of $H(T) = \mathbf{C}$, provided the energy difference between the highest and the second highest eigenstate of $H(T)$ is strictly positive for all $t < T$. This is the case by the Perron-Frobenius theorem whenever $\mathbf{C}$ has only non-negative entries since $\mathbf{B}$ is non-negative irreducible matrix, in particular for combinatorial optimization problems.
Quickly alternating between $\mathbf{B}$ and $\mathbf{C}$ with suitable angles $\beta,\gamma$ such that $\sum_r \beta[r]+\gamma[r] = T$ gives a discretization (or so-called Trotterization) of the adiabatic algorithm, with improving precision and improving expectation value $\max_{\beta,\gamma} \bra{\beta,\gamma}\mathbf{C}\ket{\beta,\gamma}$ for increasing $r$.

However, while QOAO may certainly be looked at as inspired by the quantum adiabatic algorithm, there are problems for which the latter fails for subexponential runtimes while QAOA succeeds even in a single round~\cite{1411.4028}. Furthermore, for MaxCut on 3-regular graphs, $1$-round QAOA was shown to give a $0.692..$-approximation. 
No approximation ratios have been shown for more than 1 round and thus in this regime QAOA is purely heuristic. However, there are known limitations for small/constant-round QAOA approaches: for larger node degrees and sublogarithmic number of rounds $r \in o(\log n)$ the approximation ratio is limited by $\approx 0.834$~\cite{qaoa-limitations}. 

So, what strategies of classical optimizers are there to tune the angles $\beta,\gamma$ for a given $r$-round QAOA subroutine? 
In the original QAOA proposal, an exhaustive search over a fine cartesian grid is suggested for small constant $r$~\cite{1411.4028}, where the number of grid points is polynomial in $n$. This works because $\bra{\beta,\gamma}\mathbf{C}\ket{\beta,\gamma}$ does not have narrow peaks that fall between grid points. 
Other approaches are necessary for larger $r$, but we have to be aware that the parameter landscape is non-convex and thus most classical optimization techniques cannot provide a guarantee for confergence to optimum parameters.
Possible optimizers for angle-finding have been extensively studied and are based on techniques such as
gradient descent~\cite{angles-fermionic,angles-gradient}, 
optimal control~\cite{angles-control,angles-control-brady}, 
interpolation of angles for a $r$-round QAOA based on good angles for a $(r-1)$ rounds~\cite{angles-interpolation},
and basin-hopping~\cite{angles-basin,angles-basin-nasa}.
For example, basin-hopping starts with random angles, locally optimizes the solution and then randomly perturbs the found angles more significantly to explore a new basin to try to find a better local optima.
The use of a quantum-variational-eigensolver is also possible~\cite{peruzzo_variational_2014,mcclean_theory_2016}.

\subsubsection*{A closer look at the quantum Hamiltonians}
We discuss the translation of the clauses of a combinatorial problem in the form of Equation~\eqref{eq:max_clause} to quantum clause Hamiltonians by the MaxCut and note generalizations along the way. Recall that we have already discussed how to transform a combinatorial problem $\max \sum_\alpha C_\alpha(x)$ with binary variables $x_i \in \left\{ 0,1 \right\}$ (see Eq.~\eqref{eq:max_cut_bool}) into a combinatorial problem $\max \sum_\alpha C_\alpha(z)$ with spin variables $z_i \in \left\{ -1,1 \right\}$ (see Eq.~\eqref{eq:max_cut}) using the linear transformation $x_i = (z_i+1)/2$.
To formulate a quantum clause Hamiltonian $\mathbf{C}_\alpha$, we replace in the clause $C_\alpha(z)$ 
each constant $1$ with the identity $\mathit{Id}^{\otimes n} = ( \begin{smallmatrix} 1 & 0 \\ 0 & 1 \end{smallmatrix} )^{\otimes n}$ 
and each spin variable $z_i$ with Pauli $Z$-operators $( \begin{smallmatrix} 1 & 0 \\ 0 & -1 \end{smallmatrix} )$ 
acting on qubit $i$, $Z_i = \mathit{Id}^{\otimes i-1} \otimes Z \otimes \mathit{Id}^{\otimes n-i}$.

We note that the MaxCut problem is particularly advantageous for QAOA for the following reasons: (1) all of the clauses in the objective function have the same structure, hence a circuit implementation has only to be found for one unitary $e^{-i \gamma \mathbf{C}_\alpha}$; (2) each clause only involves two decision variables, which keeps the structure of $\mathbf{C}_\alpha$ relativity simple. We note that we have 
$Z_i \cdot Z_j = \mathit{Id}^{\otimes i-1} \otimes Z \otimes \mathit{Id}^{\otimes j-i} \otimes Z \otimes \mathit{Id}^{\otimes n-j}$. 
As an example, for the MaxCut problem on the 2-edge path \textcircled{1}---\textcircled{2}---\textcircled{3},
we get a maximization function $C(z)$ and a quantum problem Hamiltonian $\mathbf{C}$:
\begin{align}
	C(z) & = \frac{1}{2}(1-z_1 z_2) + \frac{1}{2}(1-z_2 z_3) = C_{12}(z) + C_{23}(z)	\notag \\
	\Rightarrow \mathbf{C} &= \frac{1}{2}(\mathit{Id}-Z_1 Z_2) + \frac{1}{2}(\mathit{Id}-Z_2 Z_3)= \mathbf{C}_{12} + \mathbf{C}_{23} \label{eq:mcqh_transform}	\\
	&= \frac{1}{2} \left( \left( \begin{smallmatrix} 
		1 & 0 & 0 & 0 & 0 & 0 & 0 & 0 \\
		0 & 1 & 0 & 0 & 0 & 0 & 0 & 0 \\
		0 & 0 & 1 & 0 & 0 & 0 & 0 & 0 \\
		0 & 0 & 0 & 1 & 0 & 0 & 0 & 0 \\
		0 & 0 & 0 & 0 & 1 & 0 & 0 & 0 \\
		0 & 0 & 0 & 0 & 0 & 1 & 0 & 0 \\
		0 & 0 & 0 & 0 & 0 & 0 & 1 & 0 \\
		0 & 0 & 0 & 0 & 0 & 0 & 0 & 1 
	\end{smallmatrix} \right) - \left( \begin{smallmatrix} 
		1 & 0 & 0 & 0 & 0 & 0 & 0 & 0 \\
		0 & 1 & 0 & 0 & 0 & 0 & 0 & 0 \\
		0 & 0 & -1 & 0 & 0 & 0 & 0 & 0 \\
		0 & 0 & 0 & -1 & 0 & 0 & 0 & 0 \\
		0 & 0 & 0 & 0 & -1 & 0 & 0 & 0 \\
		0 & 0 & 0 & 0 & 0 & -1 & 0 & 0 \\
		0 & 0 & 0 & 0 & 0 & 0 & 1 & 0 \\
		0 & 0 & 0 & 0 & 0 & 0 & 0 & 1 
	\end{smallmatrix} \right) \right )
	+
	\frac{1}{2} \left( \left( \begin{smallmatrix} 
		1 & 0 & 0 & 0 & 0 & 0 & 0 & 0 \\
		0 & 1 & 0 & 0 & 0 & 0 & 0 & 0 \\
		0 & 0 & 1 & 0 & 0 & 0 & 0 & 0 \\
		0 & 0 & 0 & 1 & 0 & 0 & 0 & 0 \\
		0 & 0 & 0 & 0 & 1 & 0 & 0 & 0 \\
		0 & 0 & 0 & 0 & 0 & 1 & 0 & 0 \\
		0 & 0 & 0 & 0 & 0 & 0 & 1 & 0 \\
		0 & 0 & 0 & 0 & 0 & 0 & 0 & 1 
	\end{smallmatrix} \right) - \left( \begin{smallmatrix} 
		1 & 0 & 0 & 0 & 0 & 0 & 0 & 0 \\
		0 & -1 & 0 & 0 & 0 & 0 & 0 & 0 \\
		0 & 0 & -1 & 0 & 0 & 0 & 0 & 0 \\
		0 & 0 & 0 & 1 & 0 & 0 & 0 & 0 \\
		0 & 0 & 0 & 0 & 1 & 0 & 0 & 0 \\
		0 & 0 & 0 & 0 & 0 & -1 & 0 & 0 \\
		0 & 0 & 0 & 0 & 0 & 0 & -1 & 0 \\
		0 & 0 & 0 & 0 & 0 & 0 & 0 & 1 
	\end{smallmatrix} \right) \right ) \notag
\end{align}

\begin{align}
	\mathbf{C} \ = \quad & 
	\begin{blockarray}{ccccccccr}
		|000\rangle & |001\rangle & |010\rangle & |011\rangle & |100\rangle & |101\rangle & |110\rangle & |111\rangle & \\		
		\begin{block}{(cccccccc)r}
			0 & 0 & 0 & 0 & 0 & 0 & 0 & 0 & \ |000\rangle \\
			0 & 1 & 0 & 0 & 0 & 0 & 0 & 0 & \ |001\rangle \\
			0 & 0 & 2 & 0 & 0 & 0 & 0 & 0 & \ |010\rangle \\
			0 & 0 & 0 & 1 & 0 & 0 & 0 & 0 & \ |011\rangle \\
			0 & 0 & 0 & 0 & 1 & 0 & 0 & 0 & \ |100\rangle \\
			0 & 0 & 0 & 0 & 0 & 2 & 0 & 0 & \ |101\rangle \\
			0 & 0 & 0 & 0 & 0 & 0 & 1 & 0 & \ |110\rangle \\
			0 & 0 & 0 & 0 & 0 & 0 & 0 & 0 & \ |111\rangle \\ 
		\end{block}
	\end{blockarray} \label{eq:mcqh_result}
\end{align}

Above, Eq.~\eqref{eq:mcqh_transform} shows the transformation of the maximization function $C(z)$ into the quantum Hamiltonian $\mathbf{C}$, and Eq.~\eqref{eq:mcqh_result} illustrate how the quantum Hamiltonian encodes the inputs and outputs of the different cuts. We note that both clause terms $\mathbf{C}_{12} = \frac{1}{2}(\mathit{Id}-Z_1 Z_2)$ and $\mathbf{C}_{23} = \frac{1}{2}(\mathit{Id}-Z_2 Z_3)$ are diagonal matrices, hence they commute and we have $e^{-i \gamma (\mathbf{C}_{12} + \mathbf{C}_{23})} = e^{-i \gamma \mathbf{C}_{12}}\cdot e^{-i\gamma \mathbf{C}_{23}}$. Similarly, for the mixing Hamiltonian $\mathbf{B} = \sum_j X_j$ the $X_j$ pairwise commute, and we have $e^{-i\beta \mathbf{B}} = \prod_j e^{-i\beta X_j}$ as used in Algorithm~\ref{alg:QAOA}.

Finally, we discuss the influence of the forms of $\mathbf{C}$ and $\mathbf{B}$ on the bounds on the angles $\gamma, \beta$. Since $\mathbf{C}$ is a diagonal matrix with integer eigenvalues, we get that $e^{-i\gamma \mathbf{C}}$ is $2\pi$-periodic in $\gamma$ and we get $0 \leq \gamma \leq 2\pi$. For $\mathbf{B}$ we have $e^{-i(\pi+\beta)X} = e^{-i\pi X}\cdot e^{-i\beta X} = -\mathit{Id} \cdot e^{-i\beta X}$, hence $e^{-i\beta \mathbf{B}}$ is (up to a global phase) $\pi$-periodic in $\beta$ and we get $0 \leq \beta \leq \pi$.

By looking specifically at MaxCut, we can further narrow the angle bounds: we have $e^{-i\pi/2 X} = (\begin{smallmatrix} 0 & -i \\ -i & 0 \end{smallmatrix})$, i.e. increasing $\beta$ in $e^{-i\beta \mathbf{B}}$ by $\pi/2$ only adds a global phase and swaps 0/1 values of all variables and thus the two cut sets. Hence 
$\bra{\beta,\gamma}\mathbf{C}\ket{\beta,\gamma}$ is even only $\tfrac{\pi}{2}$-periodic in $\beta$. Using these periodicities, we observe
\begin{align}
	\bra{\tfrac{\pi}{2}-\beta,2\pi-\gamma}\mathbf{C}\ket{\tfrac{\pi}{2}-\beta,2\pi-\gamma} = \bra{-\beta,-\gamma}\mathbf{C}\ket{-\beta,-\gamma}= \bra{\beta,\gamma}\mathbf{C}\ket{\beta,\gamma},
	\label{eq:qaoa_symmetry}
\end{align}
where the last equality comes from the fact that $\mathbf{C}$ and $\mathbf{B}$ are real valued and satisfy time reversal symmetry. This cuts our angle search space further in halfs, and we can restrict our angle search space to $0 \leq \beta[k] \leq \tfrac{\pi}{2}$ and $0 \leq \gamma[k] \leq \pi$.

\subsection{QAOA MaxCut on \texttt{ibmqx2}}

\begin{figure}[!t]
\begin{center}
\includegraphics[width=\linewidth]{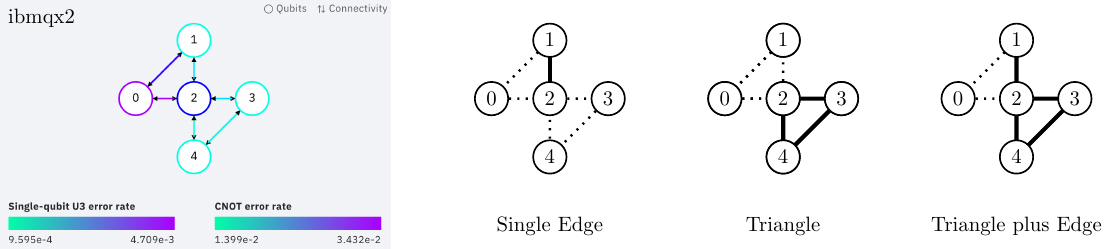}
\caption{The CNOT connectivity and error rates of the \texttt{ibmqx2} Computer (left) followed by the Single Edge (center left), Triangle (center right) and four edge Triangle+Edge (right) graphs considered in the proof-of-concept experiments.}
\label{fig:maxcut-experiments-ibmqx2}
\end{center}
\end{figure}

This section investigates the implementation of the QAOA MaxCut algorithm on the \verb|ibmqx2| quantum computer (Figure~\ref{fig:maxcut-experiments-ibmqx2}). The first challenge is to transform the QAOA algorithm from its mathematical form into a sequence of operations that are available in the IBM Quantum Experience platform.  For the sake of convenience we will mention here the gates we will use in the ensuing discussion,
\begin{align*}
U_1(\lambda)
=
\begin{pmatrix}
  1 & 0 \\
 0 &  e^{ i \lambda}
\end{pmatrix}\,, \quad
U_3(\theta, \phi, \lambda)
=
\begin{pmatrix}
  \cos(\theta/2) & - e^{i \lambda} \sin(\theta/2) \\
   e^{ i \phi} \sin(\theta/2) &  e^{ i (\lambda+\phi )} \cos(\theta/2)
\end{pmatrix} ,\quad 
\text{CNOT} 
= 
\begin{pmatrix}
  1 & 0 & 0 & 0 \\
  0 & 1 & 0 & 0 \\
  0 & 0 & 0 & 1 \\
  0 & 0 & 1 & 0 \\
 \end{pmatrix}\,.
\end{align*}

\begin{figure}[h!]
\begin{center}
\includegraphics[width=1.0in]{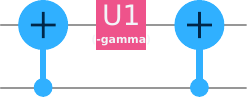}
\caption{An IBM Quantum Experience score illustrating an implementation of the MaxCut edge gate \eqref{eq:mc_edge_gate}.}
\label{fig:edge_gate}
\end{center}
\end{figure}

The inner loop of the algorithm first requires the application of the $\gamma$ angle with the clause Hamiltonians. For the MaxCut Hamiltonian, this can be expanded as follows,
\begin{align}
e^{- i \frac{ \gamma}{2} (\mathit{Id} - Z_1 Z_2)}
= 
e^{- i \gamma
\left( \begin{smallmatrix}
  0 & 0 & 0 & 0 \\
 0 & 1 & 0 & 0 \\
 0 & 0 & 1 & 0 \\
0 & 0 & 0 & 0
 \end{smallmatrix} \right)
}
& =
\begin{pmatrix}
1 & 0 & 0 & 0 \\
0 & e^{-  i \gamma} & 0 & 0 \\
0 & 0 &  e^{-  i \gamma} & 0 \\
0 & 0 & 0 & 1
\end{pmatrix}	
\label{eq:mc_edge_gate}	\\
& =
\underbrace{
\begin{pmatrix}
1 & 0 & 0 & 0 \\
0 & 1 & 0 & 0 \\
0 & 0 & 0 & 1 \\
0 & 0 & 1 & 0
\end{pmatrix}
}_{\text{CNOT}_{12}}
\cdot
\underbrace{
\begin{pmatrix}
1 & 0 & 0 & 0 \\
0 & e^{-  i \gamma} & 0 & 0 \\
0 & 0 & 1 & 0 \\
0 & 0 & 0 & e^{-  i \gamma}
\end{pmatrix}
}_{\mathit{Id} \otimes U_1(-\gamma)}
\cdot
\underbrace{
\begin{pmatrix}
1 & 0 & 0 & 0 \\
0 & 1 & 0 & 0 \\
0 & 0 & 0 & 1 \\
0 & 0 & 1 & 0
\end{pmatrix}
}_{\text{CNOT}_{12}}
\notag
\end{align}

\noindent i.e., we observe that this gate can be implemented as a combination of two CNOT gates and one $U_1(- \gamma)$ gate, as indicated in Figure~\ref{fig:edge_gate}. 
It is also interesting to note the alternate implementation of this gate in~\cite{grove_qaoa}, which leverages a different variety of gate operations~\cite{1608.03355}.
We also remark here that the CNOT gates can be interpreted as computing and uncomputing the parity of qubits 1 and 2 inline, with the phase shift of the $U_1(-\gamma)$ gate applied to odd parities. For higher than quadratic terms of $Z$-operators (such as $Z_iZ_jZ_k$ terms in the MaxE3Lin2 problem~\cite{farhi-e3lin2}), parities can be computed by CNOT gates from all other qubits into a central qubit.

\begin{figure}[!t]
\begin{center}
\vspace{-0.5cm}
\includegraphics[width=\linewidth]{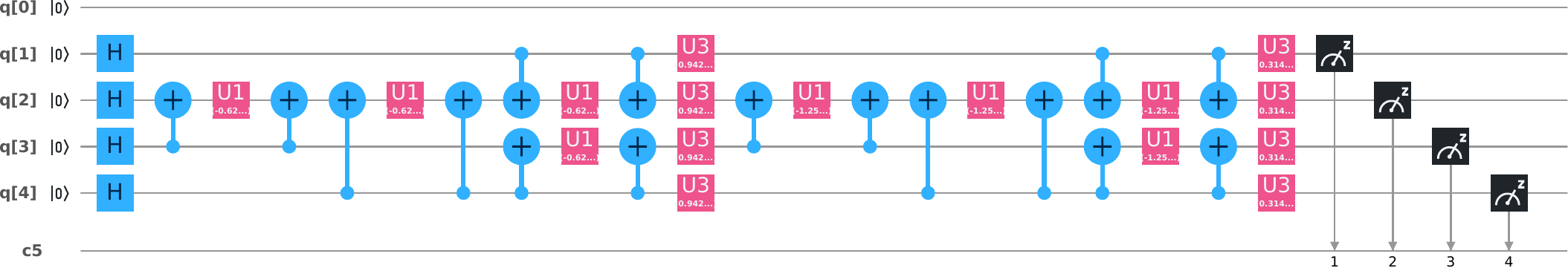}
\vspace{-0.6cm}
\caption{An IBM Quantum Experience circuit for 2-round QAOA of MaxCut on the ``Triangle plus Edge'' graph. We can see the edge gate from Figure~\ref{fig:edge_gate} replicated for the edges of the triangle between qubits $q[2],q[3],q[4]$ as well as the edge between qubits $q[1],q[2]$. Similarly, the vertical layers of $U_3$ gates implement $e^{-i\beta\mathbf{B}}$.}
\label{fig:TplusE_2r}
\end{center}
\end{figure}

The next term in the loop is the application of the $\beta$ angle, which is expanded as follows,
\begin{align}
e^{- i  \beta X}
= 
 e^{-  i  \beta
\begin{pmatrix}
  0 & 1 \\
  1 & 0 
 \end{pmatrix}
}
=
\begin{pmatrix}
  \cos( \beta) & - i \sin( \beta) \\
-i \sin( \beta) & \cos( \beta)
 \end{pmatrix}
 \label{eq:beta_gate}
\end{align} 
Careful inspection of the IBM Quantum Experience gates reveals that this operation is implemented by $U_3(2  \beta_k, - \pi/2, \pi/2)$. So we need to apply this gate to every qubit in the register. 
Putting all of these components together, Figure~\ref{fig:TplusE_2r} presents an IBM Quantum Experience circuit for implementing \revision{the quantum subroutine of} QAOA for MaxCut on the ``Triangle plus Edge'' graph from Figure~\ref{fig:maxcut-experiments-ibmqx2} \revision{with} parameters,
\begin{align*}
 r = 2\colon
 & \gamma_1 = 0.2\cdot\pi = 0.628..,\quad  \beta_1 = 0.15\cdot\pi = 0.471.., \\
 & \gamma_2 = 0.4\cdot\pi = 1.256..,\quad  \beta_2 = 0.05\cdot\pi = 0.157...
\end{align*}

\subsection{A proof-of-concept experiment}

\begin{figure}[b!]
\includegraphics[width=\linewidth]{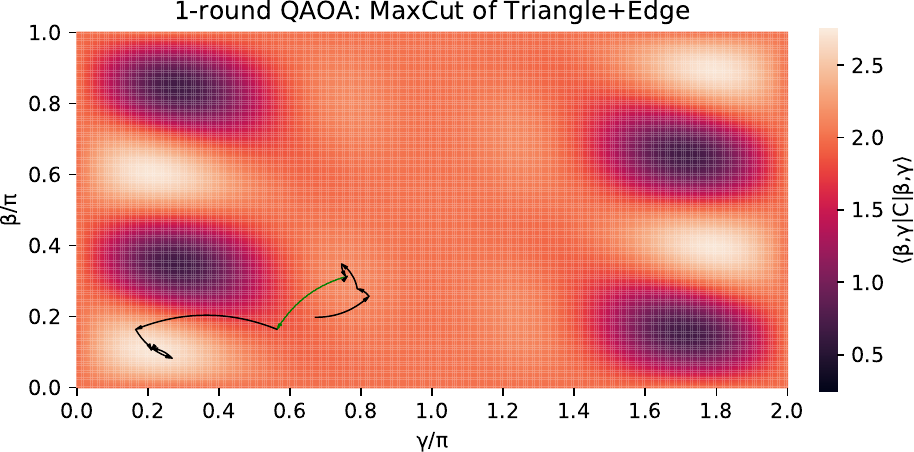}
\caption{(Heatmap) Parameter landscape for a 1-round QAOA for MaxCut on the Triangle+Edge instance. We observe the symmetry
$\bra{\beta,\gamma}\mathbf{C}\ket{\beta,\gamma} = \bra{\tfrac{\pi}{2}-\beta,2\pi-\gamma}\mathbf{C}\ket{\tfrac{\pi}{2}-\beta,2\pi-\gamma}$ 
derived in Eq.~\eqref{eq:qaoa_symmetry}. 
(Overlay) Using a basin-hopping optimizer: A random initialization first explores a basin finding a local maximum; 
a random perturbation next hops to worse-than-random parameters, but the exploration then finds a global optimum.}
\label{fig:qaoa_landscape}
\end{figure}

With a basic implementation of QAOA for MaxCut in qiskit, a preliminary proof-of-concept study is conducted to investigate the effectiveness of QAOA for finding high-quality cuts in the a) Single Edge, b) Triangle and c) Triangle plus Edge graphs presented in Figure~\ref{fig:maxcut-experiments-ibmqx2}.
For both a) and b), we ran a numerical grid search for a for a 1-round QAOA with resolution $\tfrac{1}{8}\pi$ and $\tfrac{1}{10}\pi$, respectively, using qiskit's \verb|statevector_simulator|. The statevector simulator allows for exact evaluation of the expectation value $\bra{\beta,\gamma}\mathbf{C}\ket{\beta,\gamma}$. In both cases, we found parameters $\beta,\gamma$ resulting in a state representing an optimum cut. For c), we ran a grid search on the statevector simulator for both 1- and 2-round QAOA with resolution $\tfrac{1}{1000}\pi$ and $\tfrac{1}{20}\pi$ respectively. The parameter landscape for 1 round is given in Fig.~\ref{fig:qaoa_landscape} and overlayed with a basin-hopping approach.

We then executed the QAOA subroutine for the best-found parameters on \emph{Hardware} by executing the IBM Quantum Experience circuit on the \verb|ibmqx2| device using $4096$ shots, and compared the results to a \emph{Simulation} with the same number of shots using the \verb|qasm_simulator|. For both computations we give the expectation/mean of the returned solutions and the probability to sample the maximum cut and contrast these with the values for a \emph{Random} cut. 
The simulation computation serves to demonstrate the mathematical correctness of the proposed QAOA circuit.  The hardware computation demonstrates the viability of the circuit in a deployment scenario where environmental noise, intrinsic bias, and decoherence can have a significant impact on the results.  The random computation serves to demonstrate that the hardware results are better than what one would expect by chance from pure noise.

a) The first shot experiment considers the Single Edge graph from Figure~\ref{fig:maxcut-experiments-ibmqx2}~(center left) and implements a 1-round QAOA with the parameters
\[ r = 1\colon\ \gamma_1 = 0.5\cdot\pi,\quad  \beta_1 = 0.125\cdot\pi. \]
The results are summarized in Table~\ref{tbl:E_1r}.  The simulation results indicate that the proposed score is mathematically sound and the hardware results indicate similar performance to the simulation, with a few additional errors. The random results indicate that both the simulation and hardware perform significantly better than random chance.
\begin{table}[htbp]
\centering
\caption{MaxCut QAOA with one round on a Single Edge.} 
\begin{tabular}{@{}lrrr@{}} 
\toprule
& Random    & Simulation    & Hardware  \\
\midrule
Expected Size of a sampled cut
& 0.500     & 1.000         & 0.950     \\ 
Probability of sampling a maximum cut
& 0.500     & 1.000         & 0.950     \\ 
\bottomrule
\end{tabular}
\label{tbl:E_1r}
\end{table} 

b) The second shot experiment considers the Triangle graph from Figure~\ref{fig:maxcut-experiments-ibmqx2}~(center right) with parameters
\[ r = 1\colon\ \gamma_1 = 0.8\cdot\pi,\quad  \beta_1 = 0.4\cdot\pi. \]
The results are summarized in Table~\ref{tbl:T_1r}. The simulation results indicate that the proposed circuit is mathematically sound.
Even though the QAOA circuit for a Triangle is longer than the QAOA circuit for a Single Edge, the Hardware performance is better, most likely due to the more favourable distribution of the cuts, also notable in Random. 
\begin{table}[htbp]
\centering
\caption{MaxCut QAOA with one round on a Triangle.} 
\begin{tabular}{@{}lrrr@{}} 
\toprule
& Random    & Simulation    & Hardware  \\
\midrule
Expected Size of a sampled cut
& 1.500     & 1.999         & 1.904     \\ 
Probability of sampling a maximum cut
& 0.750     & 1.000         & 0.952     \\ 
\bottomrule
\end{tabular}
\label{tbl:T_1r}
\end{table} 

c) The third shot experiment considers the Triangle plus Edge graph from Figure~\ref{fig:maxcut-experiments-ibmqx2}~(right).
We run QAOA both in a 1-round and a 2-round scenario, implemented with the following parameters, found through numerical grid searches with a resolution of $\pi/1000$ (1-round) and $\pi/20$, respectively (2-round):
\begin{align*}
 r=1\colon\ 
 & \gamma_1 = 0.208\cdot\pi,\quad  \beta_1 = 0.105\cdot\pi.  
 & r=2\colon\   & \gamma_1 = 0.2\cdot\pi,\quad  \beta_1 = 0.15\cdot\pi,   \\
 &&             & \gamma_2 = 0.4\cdot\pi,\quad  \beta_2 = 0.05\cdot\pi.
\end{align*}
The results are summarized in Table~\ref{tbl:T_1-2r}, the 2-round circuit is shown in Figure~\ref{fig:TplusE_2r}. Simulation and Hardware outperform Random both on 1-round and 2-round QAOA. However, the gains made by Simulation in 2-round over 1-round QAOA almost vanish on the Hardware. 
This degradation in performance is likely due to the double length in the circuit, making the experiment more susceptible to gate errors, environmental noise and qubit decoherence.
\begin{table}[htbp]
\centering
\caption{MaxCut QAOA with several rounds on a Triangle plus Edge graph.} 
\begin{tabular}{@{}lrrrrrr@{}} 
\toprule
&           & \multicolumn{2}{c}{1-round QAOA}  && \multicolumn{2}{c}{2-round QAOA} \\
\cmidrule{3-4}\cmidrule{6-7}
& Random    & Simul.	    & Hardw.	&& Simul.		& Hardw.  \\
\midrule
Expected Size of a sampled cut
& 2.000     & 2.720         & 2.519     && 2.874         & 2.570  \\
Probability of sampling a maximum cut
& 0.375     & 0.744         & 0.652     && 0.895         & 0.727  \\
\bottomrule
\end{tabular}
\label{tbl:T_1-2r}
\end{table} 

\subsubsection*{Towards practical relevance}
When going from the toy problems studied in this subsection towards relevant problems, many problems and issues present themselves. A recent study explored these in detail~\cite{qaoa-google}, here we just note a few of these questions: 

For example, the studied graphs would likely not fit the (planar) hardware connectivity of an actual device; instead they might come from a random Erdős–Rényi, a random $D$-regular graph, or even a dense graph such as in the Sherrington-Kirkpatrick model (a MaxCut problem for complete graphs with edge weights). In the case of sparse graphs, one could resort to heuristic compilers implementing SWAP operations to bring qubits representing adjacent nodes together. For dense graphs (or hypergraphs), one can use (generalized) swap networks instead, which swap qubits such that all pairs (sets) are adjacent at some point during its execution~\cite{swap-networks}. All of these choices affect the quality of sampled solutions in the presence of noise.

This gives rise to the question of up to which problem size and number of rounds the quality of sampled solutions can be increased without gains disappearing due to noise. (A similar question also arises even for idealized quantum devices, if one limits the number of evaluations a classical optimizer can take; in this case a QAOA with fewer rounds might perform better due to the smaller and thus better explored parameter search space~\cite{cook2020kVC}.) 

As of now, quantum advantage for QAOA has not been achieved yet. Current devices are still too noisy and the possible instances too small. 
Whether QAOA can achieve a significant speedup for combinatorial optimization problems, or whether it enables better provable approximation ratios, remains open.

\section{Quantum Principal Component Analysis}\label{sec:QPCA}

\subsection{Problem definition and background}

In data analysis, it is common to have many features, some of which are redundant or correlated. As an example, consider housing prices, which are a function of many features of the house, such as the number of bedrooms, number of bathrooms, square footage, lot size, date of construction, and the location of the house. Often, one is interested in reducing the number of features to the few, most important features. Here, by important, we mean features that capture the largest variance in the data. For example, if one is only considering houses on one particular street, then the location may not be important, while the square footage may capture a large variance.

Determining which features capture the largest variance is the goal of Principal Component Anaylsis (PCA)~\cite{Pearson1901}. Mathematically, PCA involves taking the raw data (e.g., the feature vectors for various houses) and computing the covariance matrix, $\Sigma$. For example, for two features, $X_1$ and $X_2$, the covariance is given by
\begin{align}
\label{Coles:eqn1}
\Sigma = 
\begin{pmatrix}
 \mathbf{E}(X_1 * X_1)     &  \mathbf{E}(X_1 * X_2)  \\
  \mathbf{E}(X_2 * X_1)    &  \mathbf{E}(X_2 * X_2)
\end{pmatrix}\,,
\end{align}
where $\mathbf{E}(A)$ is the expectation value of $A$, and we have assumed that $\mathbf{E}(X_1)=\mathbf{E}(X_2)=0.$ Next, one diagonalizes $\Sigma$ such that the eigenvalues $e_1 \geq e_2 \geq e_3 \geq \cdots$ are listed in decreasing order. Again, for the two-feature case, this becomes
\begin{align}
\label{Coles:eqn2}
\Sigma = 
\begin{pmatrix}
 e_1     &  0  \\
  0    &  e_2
\end{pmatrix}\,.
\end{align}
Once $\Sigma$ is in this form, one can choose to keep the features with $n$-largest eigenvalues and discard the other features. Here, $n$ is a free parameter that depends on how much one wants to reduce the dimensionality. Naturally, if there are only two features, one would consider $n=1$, i.e., the single feature that captures the largest variance.

As an example, consider the number of bedrooms and the square footage of several houses for sale in Los Alamos. Here is the raw data, taken from www.zillow.com, for 15 houses:
\begin{align}
\label{Coles:eqn3}
X_1 &= \text{number of bedrooms} = \{4, 3, 4, 4, 3, 3, 3, 3, 4, 4, 4, 5, 4, 3, 4 \} \notag\\
X_2 &= \text{square footage} = \{3028, 1365, 2726, 2538, 1318, 1693, 1412, 1632, 2875, 3564, 4412, 4444, 4278, 3064, 3857 \}\,.
\end{align}
Henceforth, for scaling purposes, we will divide the square footage by 1000 and subtract off the mean of both features. Classically, we compute the covariance matrix and its eigenvalues to be the following:
\begin{align}
\label{Coles:eqn4}
\Sigma = 
\begin{pmatrix}
 0.380952     &  0.573476  \\
  0.573476    &  1.29693
\end{pmatrix}\,,\quad
e_1 = 1.57286\,,\quad
e_2 = 0.105029 \,.
\end{align}

We now discuss the quantum algorithm for doing the above calculation, i.e., for finding the eigenvalues of $\Sigma$.

\subsection{Algorithm description}

Before we discuss the algorithm we will provide a quick introduction to the concept of a \textit{density matrix.} Density matrices are used to represent  probabilistic mixtures of quantum states. Suppose that there is a quantum system whose state is not known, rather we know that it can be in one of $M$ states, $\ket{\psi_i}$, each occurring with probability $p_i$. The state of this system is then represented by a density matrix $\rho$, defined as
 \begin{equation}\label{eq:density_matrix}
     \rho = \sum_{i=1}^{M} p_i \ket{\psi_i}\bra{\psi_i}. 
 \end{equation}
If the state of a system is known (with probability 1) to be $\ket{\psi}$, then the density matrix would just be $\ket{\psi}\bra{\psi}$ and the system is said to be in a \textit{pure state}. Otherwise, the system is said to be  in  a \textit{mixed state}. So the density matrix can be seen as a generalization of the usual state representation with the extra ability to represent a probabilistic mixture of quantum states. From the definition of the density matrix it can be seen that it is a positive semi-definite matrix with unit trace. In fact, any matrix that satisfies these two properties can be interpreted as a density matrix. More details on the definition and interpretation of density matrices are given in the quantum tomography section (Section~\ref{sec:Tomography}).

 Density matrices are clearly more expressive than state vectors as state vectors can only represent pure states. But, even a system in a mixed state can be seen as a part of a larger system that is in a pure state. This process of converting a mixed state into a pure state of an enlarged system is called \textit{purification.} A mixed state of an $n$ qubit system can be purified by adding $n$ more qubits and working with the $2n$ qubit system. Once purified, the joint system of $2n$ qubits will be in a pure state while the first  $n$ qubits will still be in the original mixed state. We will not discuss the transformations required to purify a state. Interested readers are referred to Ref.~\cite{NielsenChuang} for a complete discussion.

The quantum algorithm for performing PCA presented in Ref.~\cite{Lloyd2014} uses the density matrix representation. The algorithm discussed there has four main steps: (1) encode $\Sigma$ in a quantum density matrix $\rho$ (exploiting the fact that $\Sigma$ is a positive semi-definite matrix), (2) prepare many copies of $\rho$, (3) perform the exponential SWAP operation on each copy and a target system, and (4) perform quantum phase estimation to determine the eigenvalues. For an implementation of this quantum PCA algorithm on a noisy simulator, we refer the reader to Ref.~\cite{larose2018}, which also gives a short-depth compilation of the exponential SWAP operation.

However, given the constraint of only 5 qubits on IBM's computer, preparing many copies of $\rho$ is not possible. Hence, we consider a simpler algorithm as follows. In the special case where there are only two features, $\Sigma$ and $\rho$ are $2\times 2$ matrices (one qubit states), and $\rho$ can be purified to a pure state $\ket{\psi}$ on two qubits. Suppose one prepares two copies of $\ket{\psi}$, which uses a total of 4 qubits, then the fifth qubit (on IBM's computer) can be used as an ancilla to implement an algorithm that determines the purity $P:=\Tr(\rho^2)$ of $\rho$. This algorithm was discussed, e.g., in Ref.~\cite{Johri2017}. It is then straightforward to calculate the eigenvalues of $\Sigma$ from $P$, as follows:
\begin{align}
\label{Coles:eqn5}
e_1 &= \Tr(\Sigma) * (1+\sqrt {1-2(1- P)})/2 \\
\label{Coles:eqn6}
e_2 &= \Tr(\Sigma) * (1-\sqrt {1-2(1- P)})/2\,.
\end{align}
We remark that recently (after completion of this review article), a simpler algorithm for computing purity $P$ was given in Ref.~\cite{cincio2018learning}. While the results presented in what follows use the approach in Ref.~\cite{Johri2017}, the approach in Ref.~\cite{cincio2018learning} could lead to more accurate results.

\begin{figure}[!tb]
\begin{center}
\includegraphics[width=\textwidth]{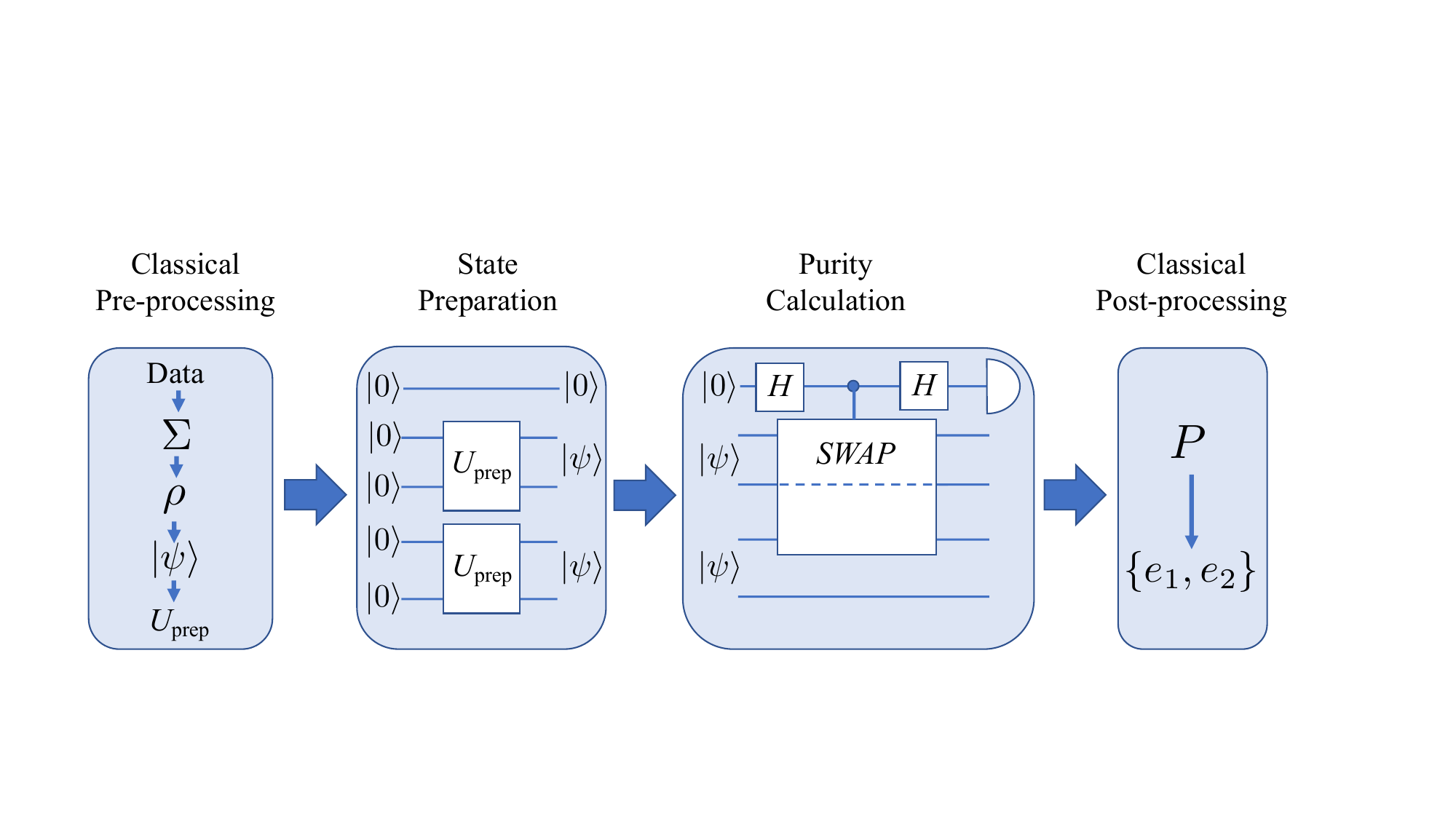}
\caption{Schematic diagram for the quantum algorithm for PCA, in the special case of only two features. The first step is classical pre-processing: transforming the raw data into a covariance matrix $\Sigma$, then normalizing to compute $\rho =\Sigma/\Tr(\Sigma)$, then purifying $\rho$ to a two-qubit pure state $\ket{\psi}$, and finally determining the unitary $U_{\text{prep}}$ needed to prepare $\ket{\psi}$. The second step is to prepare two copies of $\ket{\psi}$ by implementing $U_{\text{prep}}$ on a quantum computer. The third step is purity calculation, which is the bulk of the quantum algorithm. This involves doing a Hadamard on an ancilla, which then is used to implement a controlled-SWAP gate on two qubits (from different copies of $\ket{\psi}$), and then another Hadamard on the ancilla. Finally measuring $\avg{Z}$ on the ancilla gives the purity $P =\Tr(\rho^2)$. The last step is to classically compute the eigenvalues using Eqs.~\eqref{Coles:eqn5}-\eqref{Coles:eqn6}.}
\label{Coles:fgr1}
\end{center}
\end{figure}

As depicted in Fig.~\ref{Coles:fgr1}, this simple algorithm is schematically divided up into four steps: (1) classical pre-processing, (2) state preparation, (3) quantifying the purity, and (4) classical post-processing. 

In the first step, the classical computer converts the raw data vectors into a covariance matrix $\Sigma$, then normalizes this matrix to form $\rho =\Sigma/\Tr(\Sigma)$, then purifies it to make a pure state $\ket{\psi}$, and finally computes the unitary $U_{\text{prep}}$ needed to prepare $\ket{\psi}$ from a pair of qubits each initially in the $\ket{0}$ state. 

In the second step, the quantum computer actually prepares the state $\ket{\psi}$, or in fact, two copies of $\ket{\psi}$, using $U_{\text{prep}}$, which can be decomposed as follows: 
\begin{align}
\label{Coles:eqn7}
U_{\text{prep}} = (U_{A} \ot U_{B}) \text{CNOT}_{AB}(U'_{A} \ot \id_B)\,.
\end{align}
Note that $U_{\text{prep}}$ acts on two qubits, denoted $A$ and $B$, and $\text{CNOT}_{AB}$ is a CNOT gate with $A$ the control qubit and $B$ the target. The single qubit unitaries $U_{A}$, $U_{B}$, and $U'_{A}$ can be written in IBM's standard form:
\begin{align}
\label{Coles:eqn8}
\begin{pmatrix}
  \cos (\theta /2)    &  -e^{i\lambda} \sin (\theta /2)  \\
 e^{i\phi} \sin (\theta /2)  &  e^{i\lambda + \phi}\cos (\theta /2)
\end{pmatrix}\,,
\end{align}
where the parameters $\theta$, $\lambda$, and $\phi$ were calculated in the previous (classical pre-processing) step.

The third step is purity calculation, which makes up the bulk of the quantum algorithm. As shown in Fig.~\ref{Coles:fgr1}, first one does a Hadamard on an ancilla. Let us denote the ancilla as $C$, while the other four qubits are denoted $A$, $B$, $A'$, and $B'$. During the state preparation step, qubits $A$ and $B$ were prepared in state $\ket{\psi}$ with the state of $A$ being $\rho$. Likewise we have the state of $A^\prime$ to be $\rho$. Next, qubit $C$ is used to control a controlled-SWAP gate, where the targets of the controlled-SWAP are qubits $A$ and $A'$. Then, another Hadamard is performed on $C$. Finally, $C$ is measured in the $Z$ basis. One can show that the final expectation value of $Z$ on qubit $C$ is precisely the purity of $\rho$, i.e.,
\begin{align}
\label{Coles:eqn9}
\avg{Z}_C = p_0 - p_1 = \Tr (\rho^2) = P \,,
\end{align}
where $p_0$ ($p_1$) is the probability for the zero (one) outcome on $C$.

The fourth step is classical post-processing, where one converts $P$ into the eigenvalues of $\Sigma$ using Eqs.~\eqref{Coles:eqn5} and \eqref{Coles:eqn6}.

\subsection{Algorithm implemented on IBM's 5-qubit computer}

\begin{figure}[tb]
\begin{center}
\includegraphics[width=\columnwidth]{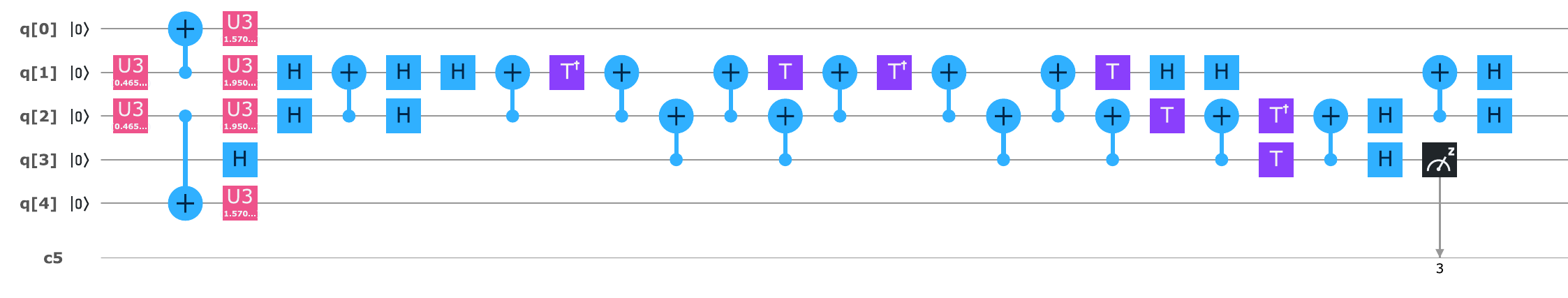}
\caption{Actual circuit for quantum PCA implemented on IBM's 5-qubit simulator and quantum computer. The first three time slots in the score correspond to the state preparation step of the algorithm, and the subsequent time slots correspond to the purity calculation step. Due to connectivity reasons, we chose qubit q[3] as the ancilla and qubits q[1] and q[2] as the targets of the controlled-SWAP operation. We decomposed the controlled-SWAP operation into CNOT gates by first relating it to the Toffoli gate via Eq.~\eqref{Coles:eqn10}, and then decomposing the Toffoli gate into CNOT gates~\cite{Shende2009}.}
\label{Coles:fgr2}
\end{center}
\end{figure}

The actual gate sequence that we implemented on IBM's 5-qubit computer is shown in Fig.~\ref{Coles:fgr2}. This involved a total of 16 CNOT gates. The decomposition of controlled-SWAP into one- and two-qubit gates is done first by relating it to the Toffoli gate:
\begin{align}
\label{Coles:eqn10}
\text{controlled-SWAP}_{CAB} = (\id_C \ot \text{CNOT}_{BA}) \text{Toffoli}_{CAB}(\id_C \ot \text{CNOT}_{BA})
\end{align}
and then decomposing the Toffoli gate, as in Ref.~\cite{Shende2009}.

We note that the limited connectivity of IBM's computer played a signficant role in determining the algorithm. For example, we needed to implement a CNOT from q[1] to q[2], which required a circuit that reverses the direction of the CNOT from q[2] to q[1]. Also, we needed a CNOT from q[3] to q[1], which required a circuit involving a total of four CNOTs (from q[3] to q[2] and from q[2] to q[1]).

Our results are as follows. For the example given in Eq.~\eqref{Coles:eqn3}, IBM's 5-qubit simulator with 40960 trials gave:

\begin{align}
\label{Coles:eqn11}
e_1 = 1.57492\,,\quad e_2 = 0.102965 \quad \text{(IBM's simulator)}\,.
\end{align}
A comparison with Eq.~\eqref{Coles:eqn4} shows that IBM's simulator essentially gave the correct answer. On the other hand, IBM's 5-qubit quantum computer with 40960 trials gave:
\begin{align}
\label{Coles:eqn12}
e_1 = 0.838943 + 0.45396 i\,,\quad e_2 = 0.838943 - 0.45396 i \quad \text{(IBM's Quantum Computer)}\,.
\end{align}
This is a non-sensical result, since the eigenvalues of a covariance matrix must be (non-negative) real numbers. So, unfortunately IBM's quantum computer did not give the correct answer for this problem.

\section{Quantum Support Vector Machine}

Support Vector Machines (SVM) are a class of supervised machine learning algorithms for binary classifications. Consider $M$ data points of $\{(\vec{x}_j, y_j): j=1,2,\ldots,M\}$. Here $\vec{x}_j$ is a $N$-dimensional vector in data feature space, and $y_j$ is the label of the data, which is $+1$ or $-1$. SVM finds the hyperplane $\vec{w}\cdot\vec{x} + b = 0$ that divides the data points into two categories so that $\vec{w}\cdot\vec{x}_j + b \ge 1$ when $y_j = +1$ and $\vec{w}\cdot\vec{x}_j + b \le -1$ when $y_j = -1$, and that is maximally separated from the nearest data points on each category. Least Squares SVM (LS-SVM) is a version of SVM~\cite{Suykens:1999:LSS:326394.326408}. It approximates the hyperplane finding procedure of SVM by solving the following linear equation:
\begin{align}
\begin{bmatrix} 
  0 & \vec{1}^T \\
  \vec{1} & \mathbf{K} + \gamma^{-1}\mathbf{1}
\end{bmatrix} 
\begin{bmatrix} 
  b\\
  \vec{\alpha}
\end{bmatrix} 
= 
\begin{bmatrix}
  0\\
  \vec{y}
\end{bmatrix}\,.
\label{eq:qsvm-lssvm}
\end{align}
Here $\mathbf{K}$ is called the kernel matrix of dimension $M\times M$, $\gamma$ is a tuning parameter, and $\vec{\alpha}$ forms the normal vector $\vec{w}$ where $\vec{w}=\sum_{j=1}^{M}\alpha_j\vec{x}_j$. Various definitions for the kernel matrix are available, but the quantum SVM~\cite{Rebentrost:2013} uses linear kernel: $K_{ij} = \vec{x}_i \cdot \vec{x}_j$. Classically, the complexity of the LS-SVM is $\mathcal{O}\big(M^2(M+N)\big)$.

The quantum version of SVM performs the LS-SVM algorithm using quantum computers~\cite{Rebentrost:2013}. It calculates the kernel matrix using the quantum algorithm for inner product~\cite{2013arXiv1307.0411L} on quantum random access memory~\cite{Giovannetti:2008}, solves the linear equation using a quantum algorithm for solving linear equations~\cite{Giovannetti:2008}, and performs the classification of a query data using the trained qubits with a quantum algorithm~\cite{Rebentrost:2013}. The overall complexity of the quantum SVM is $\mathcal{O}\big(\log NM\big)$.

 The algorithm is summarized below:
\begin{algorithm}[H]
\caption{Quantum SVM \cite{Rebentrost:2013}}
\begin{algorithmic} 
    \STATE \textbf{Input:}
    \bindent
        \STATE $\bullet$ Training data set $\{(\vec{x}_j, y_j): j=1,2,\ldots,M\}$.
        \STATE $\bullet$ A query data $\vec{x}$.
    \eindent
    \STATE \textbf{Output:}
    \bindent
        \STATE $\bullet$ Classification of $\vec{x}$: $+1$ or $-1$.
    \eindent
    \STATE \textbf{Procedure:}
    \bindent
        \STATE \textbf{Step 1.} Calculate kernel matrix $K_{ij} = \vec{x}_i \cdot \vec{x}_j$ using quantum inner product algorithm~\cite{2013arXiv1307.0411L}.
        \STATE \textbf{Step 2.} Solve linear equation Eq.~\eqref{eq:qsvm-lssvm} and find $\vert b, \vec{\alpha}\rangle$ using a quantum algorithm for solving linear equations~\cite{Giovannetti:2008} (training step).
        \STATE \textbf{Step 3.} Perform classification of the query data $\vec{x}$ against the training results $\vert b, \vec{\alpha}\rangle$ using a quantum algorithm~\cite{Rebentrost:2013}.
    \eindent
\end{algorithmic}
\end{algorithm}

The inner product calculation to compute the kernel matrix cannot be done reliably in the currently available quantum processors. The other important part of the algorithm, which is linear system solving, can be quantized and has been dealt with in Section IV.

\section{Quantum Simulation of the Schr\"odinger Equation}

\subsection{Problem definition and background}

The Schr\"odinger's equation describes the evolution of a wave function $\psi(x,t) $ for a given Hamiltonian $ \hat{H} $ of a quantum system:
\begin{equation}
\label{eq:SE}
	i\hbar\frac{\partial}{\partial t}\psi(x,t)
	=\hat{H}\psi(x,t)
	=\left[\frac{\hbar^2\hat{k}^2}{2 m}+V(\hat{x})\right]\psi(x,t),
\end{equation}
where the second equality illustrates the Hamiltonian of a particle of mass $ m $ in a potential $ V(x) $. Simulating this equation starting with a known wave function $\psi(x,0) $ provides knowledge about the wave function at a given time $ t_f $ and allows determination of observation outcomes. For example, $ \left| \psi(x,t_f) \right|^2 $ is the probability of finding a quantum particle at a position $ x $ at time $ t_f $.

Solving the Schr\"odinger's equation numerically is a common approach since analytical solutions are only known for a handful of systems. On a classical computer,
the numerical algorithm starts by defining a wave function on a discrete grid $ \psi(x_i,0) $ with a large number of points $ i \in [1,N] $. The form of the Hamiltonian, Eq.~\eqref{eq:SE}, allows one to split the system's evolution on a single time step $ \Delta t $ in two steps, which are easy to perform:
\begin{equation}
	\psi(x_i,t_{n+1})=e^{-i V(x_i)\Delta t} \emph{QFT} ^\dagger e^{-i k^2\Delta t} \emph{QFT}\ \psi(x_i,t_n),
\end{equation}
where we have assumed that $ \hbar = 1 $ and $m = \frac12$. And $\emph{QFT}$ and $\emph{QFT}^\dagger$ are the  quantum Fourier transform and its inverse. The quantum state evolution thus consists of alternating application of the phase shift operators in the coordinate and  momentum representations. These two representation are linked together by the Fourier Transformation as in the following example of a free space evolution of a quantum particle:
\begin{equation}
	\psi(x_i,t_f)=\emph{QFT}^\dagger\ e^{-i k^2 t_f}\ \emph{QFT}\ \psi(x_i,0),
\end{equation}
where $ V(x)=0 $ for a free particle.

We now discuss the quantum simulation of the Schr\"odinger's equation similar to the one discussed  in~ \cite{benenti2008quantum}, \cite{a:Somma2016} that provides the wave function of the system at a given time $ t_f $. Finding a proper measurement on a quantum simulator that reveals information about the quantum system will however be left out of the discussion. $ \left| \psi(x,t_f) \right|^2 $ will be the only information we will be interested in finding out.

\subsection{Algorithm description}

A quantum algorithm that performs a quantum simulation of one dimensional quantum systems was presented in~\cite{benenti2008quantum}.  The procedure is outlined in  Algorithm \ref{alg:sim1}.

\begin{algorithm}
\caption{Quantum simulation of Schr\"odinger equation  \cite{a:Somma2016}, \cite{benenti2008quantum}}
\begin{algorithmic} \label{alg:sim1}
    \STATE \textbf{Input:}
    \bindent
        \STATE $\bullet$ Initial wave function
        \STATE $\bullet$ Time step size, $\Delta t$, and the number of time steps, $T.$\
        \STATE $\bullet$ The ability to apply phase shifts in the computational basis.
        \STATE $\bullet$ The potential function $V.$
    \eindent
    \STATE \textbf{Output:}
    \bindent
        \STATE $\bullet$ Final wave function at time $t_f =T \delta t $ when evolved using the  Schr\"odinger equation with the potential $V.$
    \eindent
    \STATE \textbf{Procedure:}
    \bindent
        \STATE \textbf{Step 1.}  Encode the wave function on a N-point grid in a quantum state of $ n=\log_2(N) $ qubits. The value of this discretized wavefunction on a grid point is equal to the value of the original wave function at the same point. The constant of proportionality must then be calculated by renormalizing the discretized wavefunction. 
        \FOR{ $1\leq j \leq T$}
        \bindent
        \STATE \textbf{Step 2a.} Apply the Quantum Fourier Transform (QFT) to go to the momentum representation.
        \STATE \textbf{Step 2b.} Apply  a diagonal phase shift of the form $ \ket{x} \rightarrow e^{-i x^2\Delta t} \ket{x} $ in the computational basis.
        \STATE \textbf{Step 2c.} Apply the inverse Quantum Fourier Transform to come back to the position representation.
        \STATE \textbf{Step 2d.} Apply a phase shift of the form $\ket{x} \rightarrow   e^{-i V(x)\Delta t} \ket{x}.$
        \eindent
        \ENDFOR
    \STATE\textbf{Step 3.} Measure the state in the computational basis.    
    \eindent
\end{algorithmic}
\end{algorithm}

\begin{figure}
	\begin{center}
		\includegraphics[width=\columnwidth]{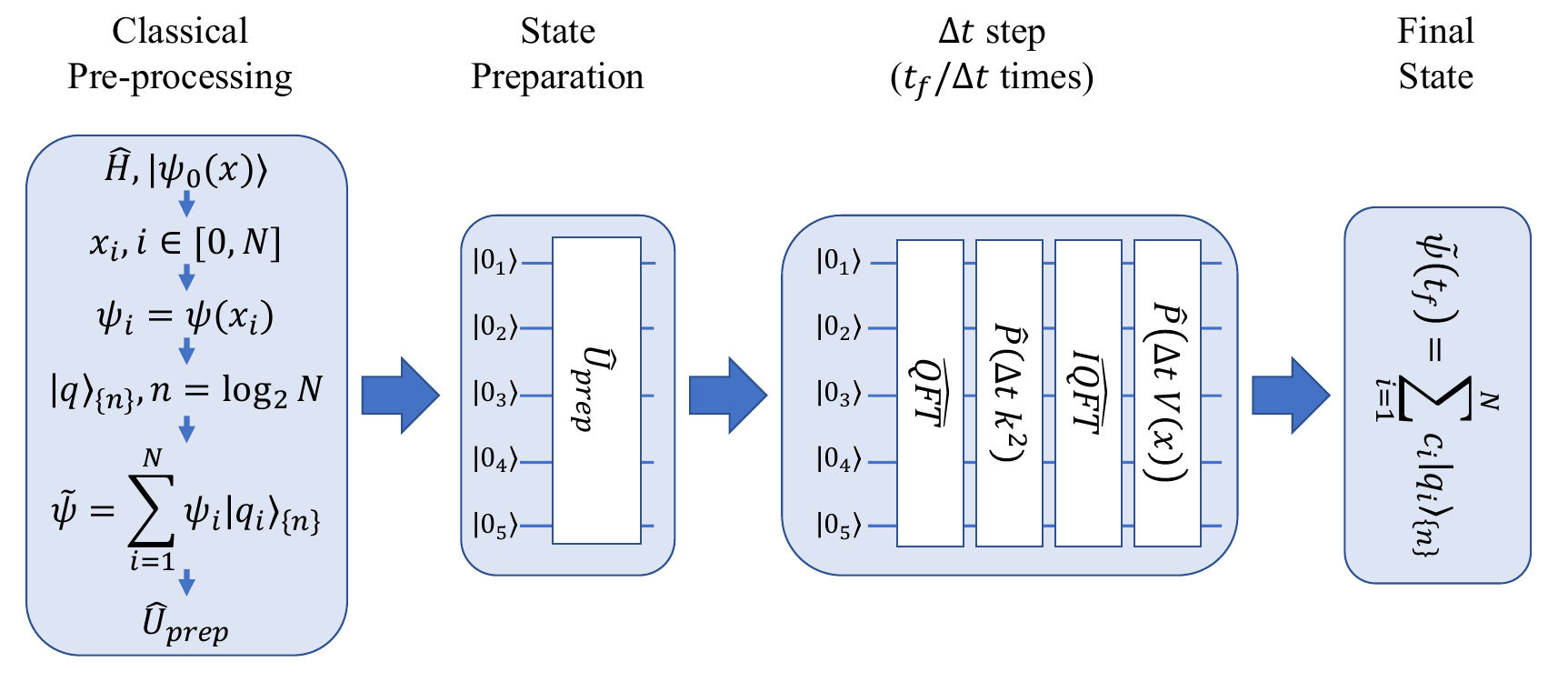}
		\caption{The quantum simulation of the Schr\"odinger's equation. The first stage is a classical pre-processing that encodes the wave function to available qubits and derives a state preparation operator that takes an all-zero state of a quantum computer to a desired state. The second stage prepares an initial state by implementing the state preparation operator $\hat{U}_{\text{prep}}$ on a quantum computer. The third stage is an iterative update looped over $ \Delta t $ steps based on the operator splitting method.}
		\label{fig:1}
	\end{center}
\end{figure}

Figure~\ref{fig:1} shows the following stages of the algorithm. The implementation of QFT was discussed in Section IV. Implementing phase shifts corresponding to arbitrary functions can be done using a series of controlled $Z$  gates or CNOT gates \cite{benenti2008quantum}. Repeating the final measurement step over many independent runs will let us estimate the probabilities $|\psi(x, t_f)|^2$. We will now consider a 2-qubit example of the quantum simulation algorithm in the case of a free particle, $V(x)=0$. 

Our initial wave function is a $\Pi$-function (a rectangular wave), which has $\{0,1,1,0\}$ representation on a $2^n$-point grid for $n=2$ qubits. Its representation by the state of the qubits is proportional to $\left| 0, 1 \right\rangle + \left| 1, 0 \right\rangle $, which can be prepared by constructing the Bell state (see Fig. \ref{fig:bell})  and applying the $X$ gate to the first qubit.

We define the 2-qubit QFT as  $QFT = \text{SWAP}_{12}~H_2 ~ C_2 \left[ \mathcal{P}_1 \left(\frac{\pi }{2}\right) \right]  H_1$, where $C_2\mathcal{P}$ is a phase operator controlled from the second qubit. This transformation applies phase shifts to the probability amplitudes of the qubit states similar to the ones applied by the classical FFT to the function values. Hence, the resulting momentum representation is identical to the classical one in a sense that it is not centered around $k=0$, which can be easily remedied by a single $X_{1}$ gate.

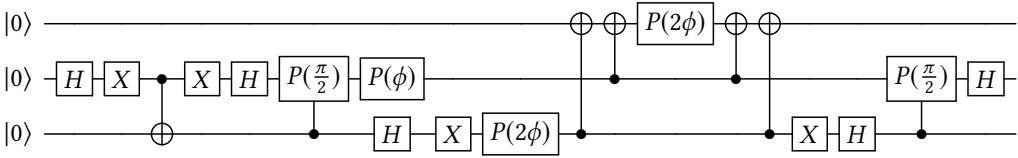
\begin{figure}
	\centerline{  \Qcircuit @C=0.45em @R=0.45em {
& |0 \rangle \quad \quad & \qw      & \qw      &\qw      &\qw      &\qw      &\qw  &\qw &\qw &\qw &\targ &\targ & \gate{P(2 \phi) } & \targ & \targ & \qw & \qw &\qw &\qw &\qw \\
& |0 \rangle \quad \quad & \gate{H} & \gate{X} &\ctrl{1} &\gate{X} &\gate{H} &\gate{P(\frac{\pi}{2})} &\gate{P( \phi)} &\qw &\qw &\qw &\ctrl{-1}  &\qw &\ctrl{-1} &\qw &\qw &\qw &  \gate{P(\frac{\pi}{2})}  &\gate{H} &\qw \\
& |0 \rangle \quad \quad & \qw      & \qw      &\targ    &\qw      &\qw      &\ctrl{-1} & \gate{H} &\gate{X} &\gate{P(2 \phi)} &\ctrl{-2} &\qw &\qw  &\qw &\ctrl{-2} & \gate{X} & \gate{H} &\ctrl{-1} &\qw &\qw
}
    }
		\caption{The quantum circuit implementation of a 2-qubit algorithm that solves the Schr\"odinger's equation on a quantum computer. The initial state preparation is followed by the Quantum Fourier Transform and centering of the momentum representation. The single qubit phase shift transformations are followed by the two-qubit phase shift transformation that uses an ancillary qubit q[0]. The inverse Quantum Fourier Transform preceded by removing the centering operation completes the circuit and returns the wave function to the coordinate representation.}
		\label{fig:2}

\end{figure}

The momentum encoding adopted in this discussion is $k=-\frac{1}{2} \sqrt{\frac{\phi }{\Delta t}} \left(1+\sum_{k=1}^{n} 2^{n-k} Z_{k}\right)$, where $\phi$ is a characteristic phase shift experienced by the state on a time step $\Delta t$. In this representation $-i k^2 \Delta t$ phase shift contains one and two qubit contributions that commute with each other and can be individually implemented. The one qubit phase shift gate has a straightforward implementation but the two qubit phase shift gate requires an ancillary qubit according to Ref.~\cite{NielsenChuang}, which results in a three qubit implementation on a quantum computer. This implementation is captured in Fig~\ref{fig:2} where removing the centering of the momentum representation and the inverse QFT have been added in order to return to the coordinate representation.

\subsection{Algorithm implemented on IBM's 5-qubit computer}

\begin{figure}
	\begin{center}
		\includegraphics[width=\columnwidth]{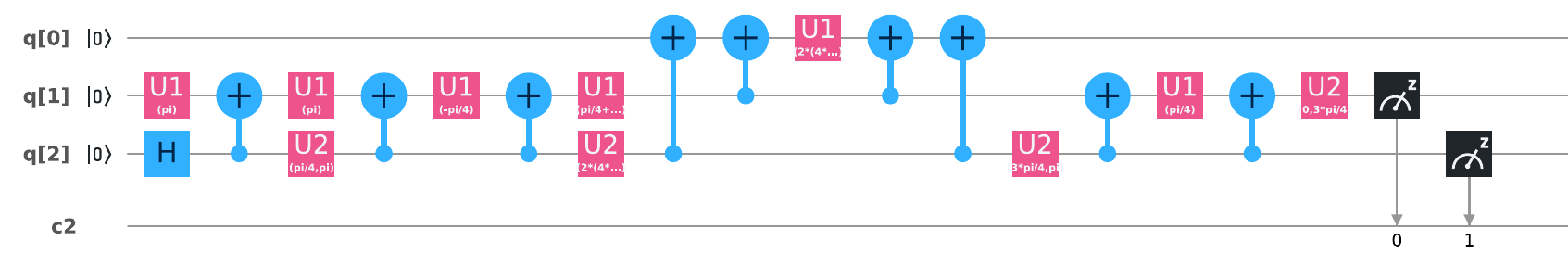}
		\caption{The quantum circuit implementation of a 2-qubit algorithm that solves the Schr\"odinger's equation on the \texttt{ibmqx4} quantum computer.}
		\label{fig:3}
	\end{center}
\end{figure}

The implementation in  Fig.~\ref{fig:3} takes into account the topology of the chip and the availability of the gates such as $U1$ and $U2$. Finally, it performs a consolidation of the single qubit gates in order to reduce the number of physical operations on the qubits. %The resulting QASM score is presented in the Listing~\ref{code}.

The circuit in Fig. \ref{fig:3} was run on the \verb|ibmqx4| quantum chip, where the maximum number of executions in a single run is $ 2^{10} $. The probabilities of observing qubit states in the computational basis was measured for $\phi=0$, $\phi=\pi/2$, $\phi=\pi$, $\phi=3\pi/2$ and $\phi=2\pi$. We expect that as $\phi$ increases from $0$ to $\pi$ the wave function evolves from a $\Pi$-function to a uniform function to a function peaked at the ends of the interval. The consecutive increase returns the wave function back to the $\Pi$-function.

We  started with the $\phi=0$ case that should have reproduced our initial state with ideal probabilities of \{0, 0.5, 0.5, 0\}. However, the observed probabilities were \{0.173, 0.393, 0.351, 0.084\}. Thus it was surprising to see that the $\phi=\pi/2$ case was very close to expected probability of $0.25$ with the observed values of \{0.295, 0.257, 0.232, 0.216\}. This surprise was however short lived as the $\phi=\pi$ case has reverted back large errors for observed probabilities: \{0.479, 0.078, 0.107, 0.335\}.
The final two case had the following observed probabilities \{0.333, 0.248, 0.220, 0.199\} and \{0.163, 0.419, 0.350, 0.068\} respectively.

\section{Ground State of the Transverse Ising Model}
In this section the ground state of the transverse Ising model is calculated using the variational quantum eigenvalue solver, and the result is compared to the exact results. This is a hybrid method that uses alternating rounds of classical and quantum computing.

In the previous section we saw how to simulate the evolution of a single quantum particle. But often, real world phenomena are dependent on the interactions between many different quantum systems. The study of  many-body  Hamiltonians that model physical systems is the central theme of condensed matter physics (CMP). 

Many-body Hamiltonians are inherently hard to study on classical computers as the dimension of the Hilbert space grows exponentially with the number of particles in the system. But using a quantum computer we can study these many-body systems with less overhead as the number of qubits required only grows polynomialy.

\subsection{Variational quantum eigenvalue solver}
A central task in CMP is finding the ground state (lowest energy eigenstate) of a given Hamiltonian, $ \mathcal{H}$,
\begin{align}
    \mathcal{H}|\Psi\rangle=E_g |\Psi\rangle.
\end{align}

Studying the ground state gives us information about the low temperature properties of the system.
Once we know $|\Psi\rangle$, we can deduce the physical properties from the wave function. In this section, we will describe how to use IBM Q to find the ground state energy of the transverse Ising model. We will not be using the \verb|ibmqx4| in this section. This is because the algorithm we use will require many rounds of optimization. Each round requires us to run a circuit on the quantum computer followed by a classical optimization step on a local machine. This process can be automated  easily using Qiskit. But the long queuing times in IBM Q makes repeated runs on the quantum processor impractical. 

To find the eigenvalue of a Hamiltonian, we could use the quantum phase estimation algorithm that was discussed in Section IV. To do this we need the ability to perform  controlled operations with the unitary $U = \exp(-i \mathcal{H} \delta t/\hbar)$, where $\delta t$ is the time step. Then, by preparing different initial states $|\psi_i\rangle$ and repeating  the phase estimation many times one can obtain, in principle, the whole spectrum of the eigenvalues and the corresponding eigenwave functions. For a general Hamiltonian, however, the implementation of a controlled $U$ may be not straightforward. For realistic problems, the quantum phase estimation circuits have large depth. This requires qubits with long coherence times, which are not available at the time of writing. For CMP problems, we are mainly interested in the lowest eigenvalue for most cases.   

\begin{figure}[t]
\psfig{figure=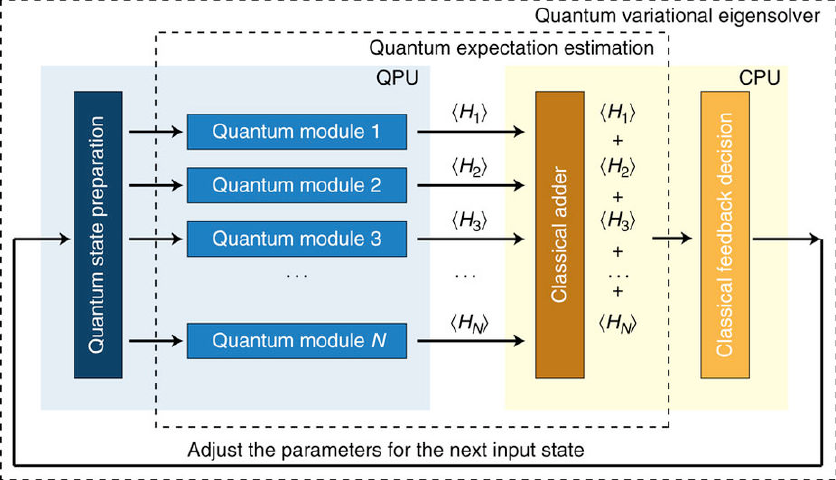,width=0.9\columnwidth}
\caption{Schematic view of the implementation of the variational quantum eigenvalue solver using a hybrid classical and quantum circuit. The figure is adopted from Ref. \cite{peruzzo_variational_2014}.
} \label{f1}
\end{figure}

To overcome these limitations, we use the recently developed variational quantum eigenvalue solver (VQES)~\cite{peruzzo_variational_2014,mcclean_theory_2016}. The basic idea is to take the advantages of both quantum and classical computers, as shown in Fig.~\ref{f1}. It allocates the classically easy tasks to classical computers and the other tasks to quantum computers. The algorithm is summarized as follows:
\begin{enumerate}
\item Prepare a variational state $|\psi(\theta_i)\rangle$ with parameters $\theta_i$. For an efficient algorithm, the number of variational parameters should grow linearly with the system size.
\item Calculate the expectation value of the Hamiltonian using a quantum computer, $E=\langle \psi|\mathcal{H}|\psi\rangle/\langle \psi|\psi\rangle $.
\item Use classical nonlinear optimizer algorithms to find new optimal $\theta_i$. In this report, we will use the relaxation method $\tau_0\partial_t\theta_i=-\partial E/\partial \theta_i$, where $\tau_0$ is a parameter to control the relaxation rate.
\item Iterate this procedure until convergence.
\end{enumerate}

VQES has the following advantage: For most CMP problems, where the interaction is local, we can split the Hamiltonian into a summation over many terms. This means that we can parallelize the algorithm to speed up the computation. The quantum expectation calculations for one term in the Hamiltonian are relatively simple, thus no long coherence times are not required. On the other hand, VQES also has limitations. Because of its variational nature, the trial wave function needs to be prepared carefully. This usually requires physical insights into the problem. The ground state eigenvalue and eigenwave function are biased by the choice of the trial wave functions. In addition, VQES requires communications between classical and quantum computers, which could be a bottleneck for the performance.     

\subsection{Simulation and results}

\begin{figure}[t]
\psfig{figure=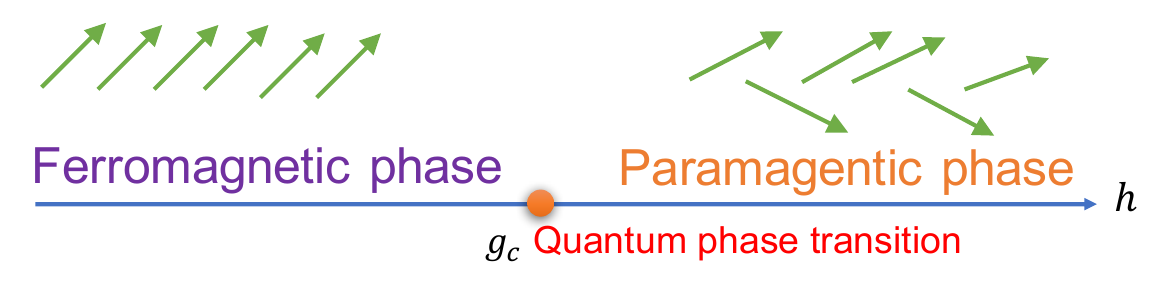,width=.6\columnwidth}
\caption{Schematic view of the quantum phases described by the transverse Ising model. The arrows represent the spin configuration in the ordered and disordered phases. 
} \label{f2}
\end{figure}

We use VQES to find the ground state of the transverse Ising model (TIM) defined by
\begin{align}
    \mathcal{H}=-\sum_i\sigma_i^z\sigma_{i+1}^z-h\sum_i\sigma_i^x,
\end{align}
where $\sigma^z$, $\sigma^x$ are Pauli matrices and $h$ is the external magnetic field. Let us first review briefly the physical properties of this Hamiltonian. This Hamiltonian is invariant under the global rotation of spin along the $x$ axis by $\pi$, $R_x\mathcal{H}R_x^\dagger=\mathcal{H}$, where $R_x(\pi)$ is the rotation operator
\begin{align}
    R_x\sigma^x R_x^\dagger=\sigma^x, \ \ \  R_x\sigma^z R_x^\dagger=-\sigma^z.
\end{align}
The TIM has two phases: When the transverse field $h$ is small, the spins are ordered ferromagnetically and the rotational symmetry associated with $R_x$ is broken. In the ordered phase, the quantum expectation value $\langle\sigma^z\rangle\neq 0$. As $h$ is increased, there is a quantum phase transition from the ordered phase to the disordered phase where $\langle\sigma^z\rangle=0$, as the rotational symmetry is restored. The phase diagram is  shown schematically in Fig.~\ref{f2}.   

Using the phase diagram as a guide, first we propose a product state as a trial wave function. The wave function can be written as
\begin{align}\label{eq4}
  |\psi_i(\theta_i)\rangle=\prod_iU(\theta_i)|0_i\rangle.
\end{align}
Here $U(\theta_i)$ is the unitary operation which describes the spin rotation along the $y$ axis by an angle $\theta_i$, 
\[
U(\theta_i)=\left(
\begin{array}{cc}
 {\cos  (\theta _i/2)}& -\sin  (\theta _i/2) \\
 \sin  (\theta _i/2) & \cos  (\theta _i/2)\\
\end{array}
\right),
\]
where $\theta_i$ are the variational parameters. Here we have used the Bloch sphere representation for a qubit state. For the TIM, we calculate the expectation value of
\begin{align}
E_{J,i}=-\langle\psi|\sigma_i^z\sigma_{i+1}^z|\psi\rangle,\ \ \ E_{Z,i}=-\langle\psi|\sigma_i^x|\psi\rangle.
\end{align}

\begin{figure}[t]
\subfloat[]{%
\includegraphics[scale=.5]{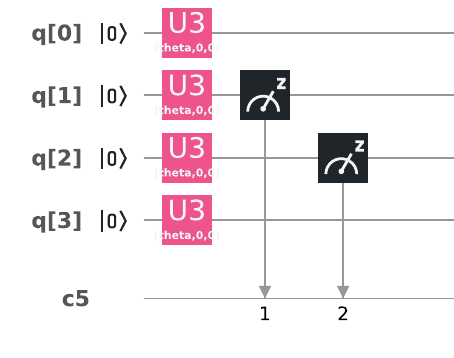}}
\qquad
\subfloat[]{%
\includegraphics[scale=.5]{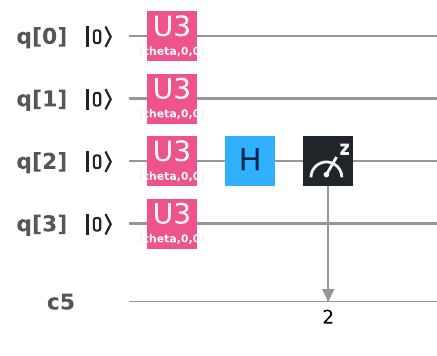}}
\qquad
\subfloat[]{%
\includegraphics[scale=.5]{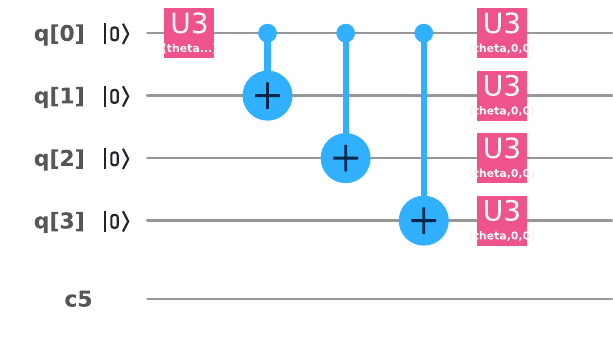}}
\caption{Quantum circuits to prepare  the trial wave-functions. The single qubit unitaries in the text can be implemented using available gates in IBM Q. The first two circuits prepare unentangled trial states. Circuit (a) can be used to measure  $\bra{\psi} \sigma^z_2\sigma^z_3 \ket{\psi}$ . Circuit (b) can be used to measure the  $\bra{\psi} \sigma^x_3 \ket{\psi}$. Circuit (c) prepares the entangled trial state.
} \label{f3}
\end{figure}

\begin{figure}[t]
\psfig{figure=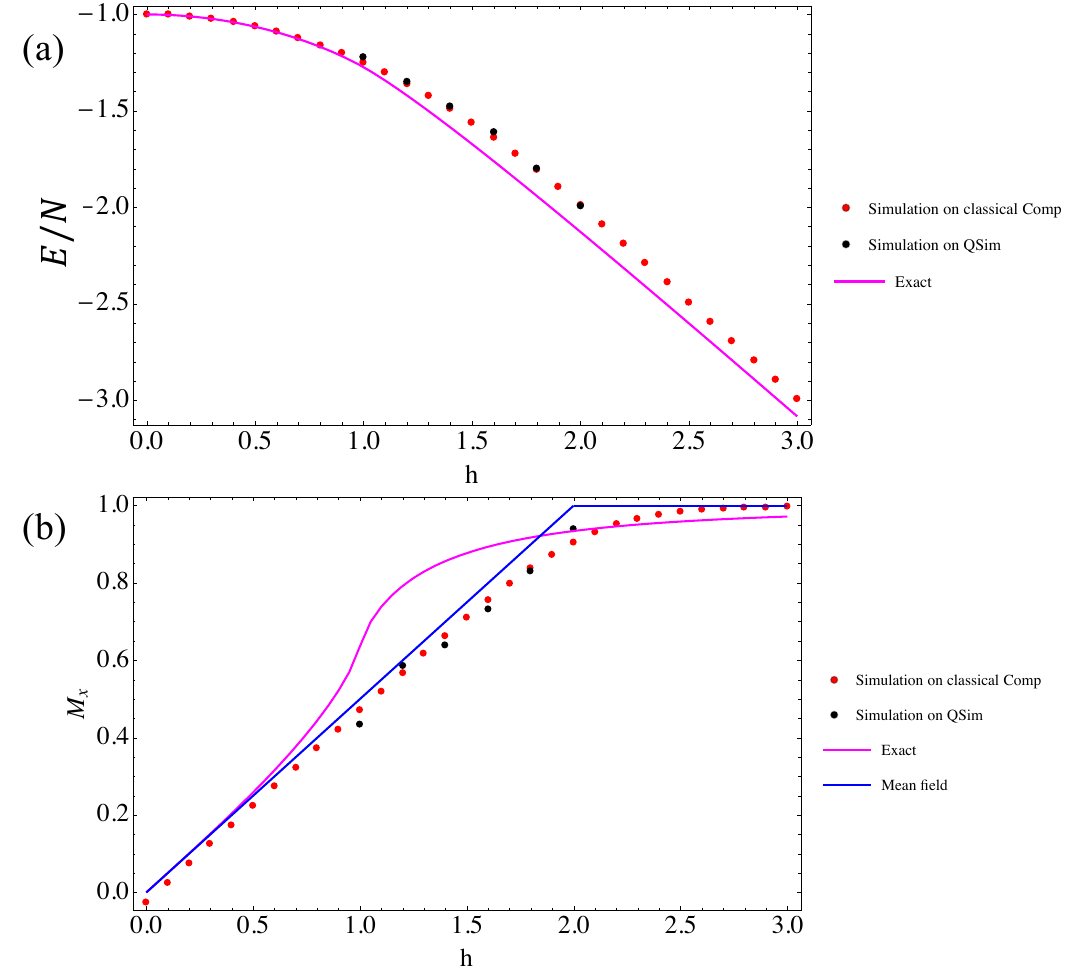,width=.9\columnwidth}
\caption{Comparison of the ground state energy (a) and average magnetization (b) $M_x=\langle\psi|\sum_i\sigma_i^x|\psi\rangle/N$ obtained by using the trial wave functions in Eq. \eqref{eq4} and the exact results. Here we have used the periodic boundary condition. The simulations are run both on the quantum simulator (black symbols) and classical computers (red symbols). The mean-field results (blue line) are also displayed for comparison. 
} \label{f4}
\end{figure}

The quantum circuit to perform the preparation of the state and calculation of the expectations value are shown in Fig.~\ref{f3}(a) andFig.~\ref{f3}(b) . We have 
\begin{align}
E_{J,i}=-[P(q_i=0)-P(q_i=1)][P(q_{i+1}=0)-P(q_{i+1}=1)],
\end{align}
\begin{align}
E_{Z,i}=-[P(q_i=0)-P(q_i=1)],
\end{align}
where $P(q_i=0, 1)$ is the measured probability for the qubit $q_i$ in the $|0\rangle$ or $|1\rangle$ state. As we mentioned before, the communication bottleneck prevented us from implementing this on \verb|ibmqx4|. We ran the code using the quantum simulator in Qiskit. The comparison of the results obtained from quantum simulation and analytical results are shown in Fig.~\ref{f4}. Our trial wave function works very well in the ordered phase, but the simulation results deviate from the exact solution in the quantum phase transition region. This discrepancy is caused by the fact that we have neglected the quantum entanglement in our trial wave function. 

In a second set of experiments, we use a trial wave function that includes quantum entanglement. Because of the symmetry, $|\Psi_i(\theta_i)\rangle$ and $R_x(\pi)|\Psi_i(\theta_i)\rangle$ are two degenerate wave functions with the same energy. The trial wave function can be written as a linear superposition of these two degenerate wave functions
\begin{align}\label{eq8}
    |\psi_i(\theta_i)\rangle=\alpha|\Psi_i(\theta_i)\rangle+\beta R_x(\pi)|\Psi_i(\theta_i)\rangle.
\end{align}
The first step is to prepare $|\psi_i(\theta_i)\rangle$ using quantum circuit. To prepare an arbitrary state in a quantum circuit is not trivial as it requires of the order of $2^n$ CNOT gates, where $n$ is the number of qubits~\cite{PhysRevA.83.032302}. The state in Eq. \eqref{eq8} can be prepared easily using the circuit in Fig.~\ref{f3}(c). Here we consider $4$ spins. The first $U_0(\theta, \phi)$ operation transforms the state into
\[
|0000\rangle \to e^{i\phi}\sin(\theta/2)  |1000\rangle  +\cos(\theta/2) |0000\rangle.
\]
The first CNOT transforms the state into
\[
e^{i\phi}\sin(\theta/2)  |1100\rangle  +\cos(\theta/2) |0000\rangle.
\]
The second CNOT transforms the state into
\[
e^{i\phi}\sin(\theta/2)  |1110\rangle  +\cos(\theta/2) |0000\rangle.
\]
The third CNOT transforms the state into
\[
e^{i\phi}\sin(\theta/2)  |1111\rangle  +\cos(\theta/2) |0000\rangle.
\]
Finally we apply $U(\theta_i)$ rotation and we obtain the desired state in Eq. \eqref{eq8}. Here 
\[
U_0(\theta, \phi)=\left(
\begin{array}{cc}
 {\cos  (\theta _i/2)}& -\sin  (\theta _i/2) \\
 e^{i\phi}\sin  (\theta _i/2) & e^{i\phi}\cos  (\theta _i/2)\\
\end{array}
\right).
\]

\begin{figure}[t]
\psfig{figure=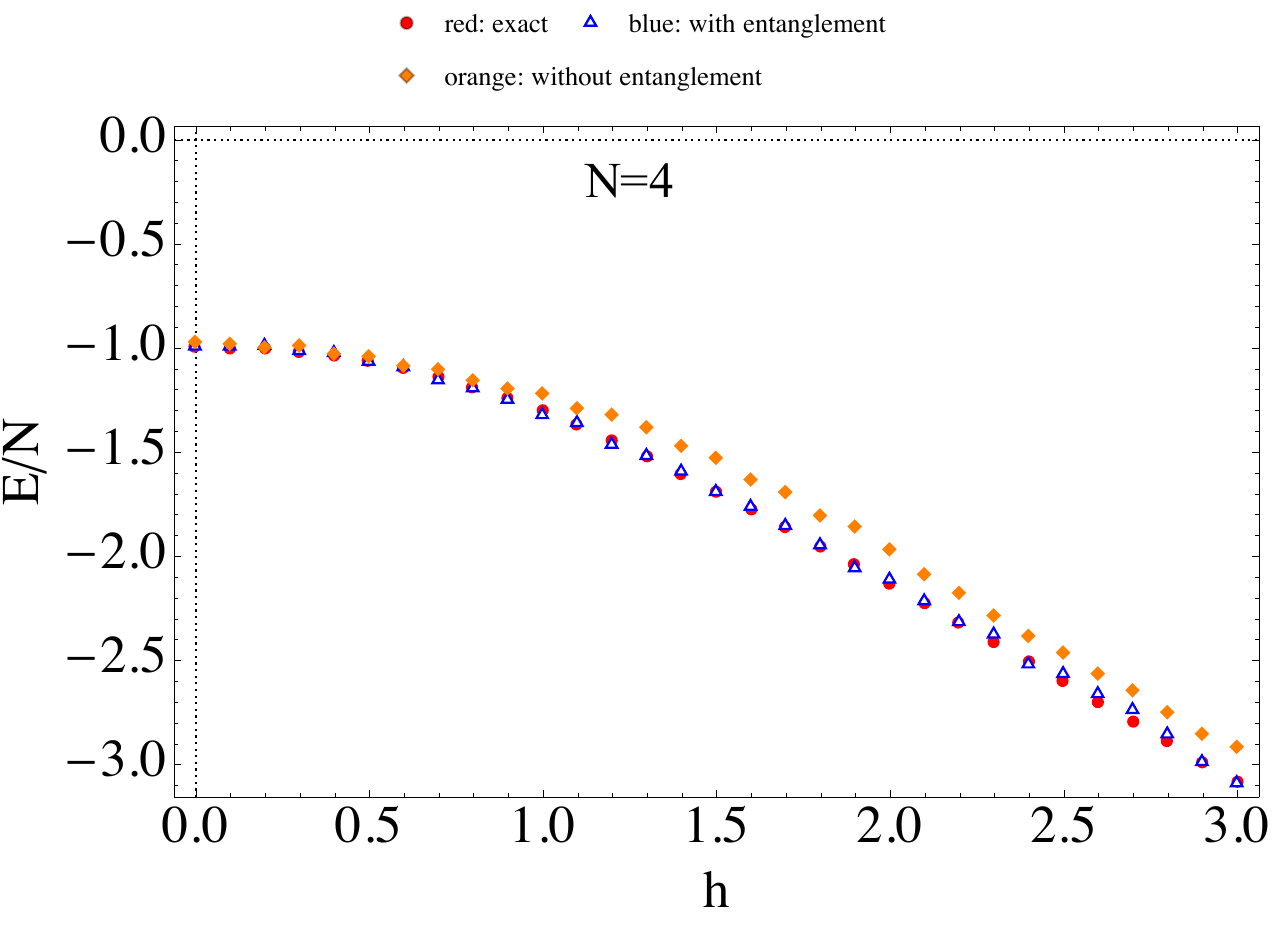,width=.7\columnwidth}
\caption{(color online) Comparison of the ground state energy obtained by using the trial wave functions in Eqs. \eqref{eq4} and \eqref{eq8} and the exact result. Here we have used the periodic boundary condition. The number of spins is 4.} \label{f5}
\end{figure}

We then use VQES to find the ground state energy. As can be seen in Fig.~\ref{f5}, the new trial function nearly reproduces the exact results in the whole magnetic field region and improves upon the product state trial function.

%%%%%%%%%%%%%%%%%%%%%%%%%%%%%%%%%%%%%%%%%%%%%%%%%%%%%%%%%%%%%%%%%%%%%%%%%%%%%%%%%%%%%%%%%%%%%%%%%%%%%%%%%%%%%%%%%%%%%%%%

\section{Quantum Partition Function}

\subsection{Background on the partition function}

Calculation or approximation of the partition function is a sub-step of inference problems in Markov networks~\cite{koller_probabilistic_2009}. Even for small networks, this calculation becomes intractable. Therefore an efficient quantum algorithm for the partition function would make many problems in graphical model inference and learning  tractable and scalable; the same holds for other problems in computational physics~\cite{garnerone2007efficient,geraci2008exact,geraci2008bqp,arad2010quantum}.

The partition function is of particular interest for calculating probabilities from graphical models such as Markov random fields~\cite{koller_probabilistic_2009}. For this article, we consider the graphical model form known as the Potts model. Let $\Gamma = (E, V)$ be a weighted graph with edge set $E$ and vertex set $V$ and $n = |V|$. In the $q$-state Potts model, each vertex can be in any of $q$ discrete states. The Potts model is a generalization of the classical Ising model. In the classical Ising model $q=2$, whereas in the Potts model $q \geq 2$. The edge connecting vertices $i$ and $j$ has a weight $J_{ij}$ which is also known as the interaction strength between corresponding states. The Potts model Hamiltonian for a particular state configuration $\sigma = (\sigma_1, \ldots, \sigma_n)$ is
\begin{equation}
	H(\sigma) = -\sum_{i \sim j} J_{ij} \delta_{\sigma_i, \sigma_j} , 
\end{equation}
where $i \sim j$ indicates that there exists an edge between vertices $i$ and $j$; and where $\delta_{\sigma_i, \sigma_j} = 1$ if $\sigma_i = \sigma_j$ and $0$ otherwise.

The probability of any particular configuration being realized in the Potts model at a given temperature, $T$, is given by the Gibbs distribution:
\begin{equation}
	P(\sigma) = \frac{1}{Z} e^{-\beta H(\sigma)},
\end{equation}
where $\beta = 1 / (k_B T)$ is the inverse temperature in energy units and $k_B$ is the Boltzmann constant. The normalization factor, $Z$, is also known as the \emph{partition function}:
\begin{equation}
	Z = \sum_{\{\sigma\}} e^{-\beta H(\sigma)},
	\label{eq:partition_function}
\end{equation}
where $\{\sigma\}$ means the full set of all possible state configurations. There are $q^n$ possible state configurations, and so this is a sum over a large number of items and is generally intractable as well as difficult to approximate. The calculation of the partition function is \#P-hard (i.e., it is a counting problem which is at least as hard as the NP-hard class of decision problems). There is no known fully polynomial randomized approximation scheme (fpras), and it is unlikely that there exists one~\cite{geraci2008exact}.

\subsection{A simple example}

\begin{figure}
\subfloat[Graph with three vertices and three edges.]{%
	\includegraphics[width=0.12\textwidth]{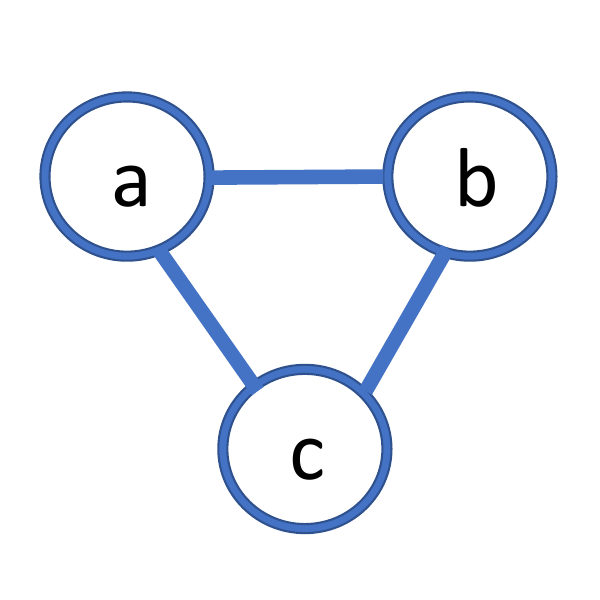}
	\label{fig:three_vertex_graph}
}
\qquad
\subfloat[]{%
	\begin{tabular}{c|l}
		abc & $-H(\sigma)$ \\
		\hline
		000 & $J_{ab} + J_{bc} + J_{ac}$ \\
		001 & $J_{ab}$ \\
		010 & $J_{ac}$ \\
		011 & $J_{bc}$ \\
		100 & $J_{bc}$ \\
		101 & $J_{ac}$ \\
		110 & $J_{ab}$ \\
		111 & $J_{ab} + J_{bc} + J_{ac}$ \\
	\end{tabular}
	\label{table:three_vertex_example}
}
	\caption{(a) Simple example with (b) the enumeration of state configurations and the value of the Hamiltonian for a fully-connected 3-vertex Ising model ($q=2$ Potts model)}
\end{figure}

We give a small example with a graph of $n=3$, $V = \{a, b, c\}$, with edges between all pairs of vertices for three total edges, pictured in Figure~\ref{fig:three_vertex_graph}, and we use $q=2$ for binary states on each vertex. To demonstrate the calculation of the partition function, we first enumerate the configurations as shown in Fig.~\ref{table:three_vertex_example}.

We plug the value of the Hamiltonian for each of the $q^n$ configurations into the partition function given in Eq.~\eqref{eq:partition_function} to get the normalization constant:
\begin{equation}
	Z = 2e^{\beta (J_{ab} + J_{bc} + J_{ac})} + 2e^{\beta J_{ab}} + 2e^{\beta J_{bc}} + 2e^{\beta J_{ac}}. 
\end{equation}
Letting $J_{ij} = 1$ for all $i \sim j$, gives:
\begin{equation}
	Z = 2e^{3 \beta} + 6e^{\beta}.
\end{equation}

\subsection{Calculating the quantum partition function}

\begin{figure}
	\centering
	\includegraphics[width=0.8\textwidth]{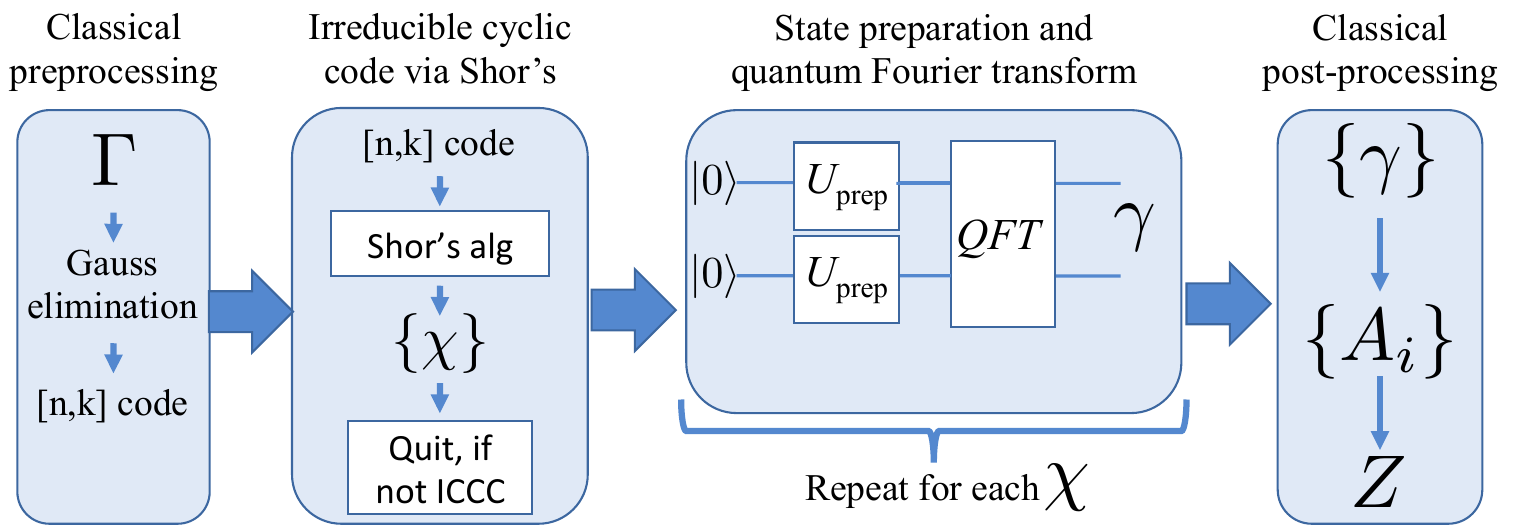}
	\caption{Overview of the quantum partition function algorithm.}
	\label{fig:diagram}
\end{figure}

An efficient quantum algorithm for the partition function is given by~\cite{geraci2008exact} for Potts models whose graph, $\Gamma$, has a topology such that it can be represented with an irreducible cyclic cocycle code (ICCC). This stipulation is non-intuitive and it takes a quantum algorithm to efficiently determine if a given graph meets this requirement. From the graph, $\Gamma$, calculate a cyclic code $C(\Gamma)$ that represents the generating structure of the graph by using Gaussian elimination on the incidence matrix of the graph, and then use Shor's algorithm to determine the irreducible set of code words $\chi$. If the code is not irreducible, then we will not be able to efficiently calculate the partition function for this graph. 

Assuming that the given graph is ICCC, the first step in the partition function algorithm is to calculate the Gauss sum of $G_{\mathbf{F}_{q^k}} = \sqrt{q^k} e^{i\gamma}$, where $\gamma$ is a function of $\chi$. The difficult part is to calculate $\gamma$, which can be done efficiently using the quantum Fourier transform (QFT). Using the set of values, $\{\gamma\}$ for all of the words, $\{\chi\}$ in the code; we calculate the weight spectrum $\{A_i\}$ of the code representing $\Gamma$. From this weights spectrum, the partition function $Z$ can be efficiently calculated using classical computing.

\subsection{Implementation of a quantum algorithm on the IBM Quantum Experience}

\begin{figure}
	\centering
	\includegraphics[width=\textwidth]{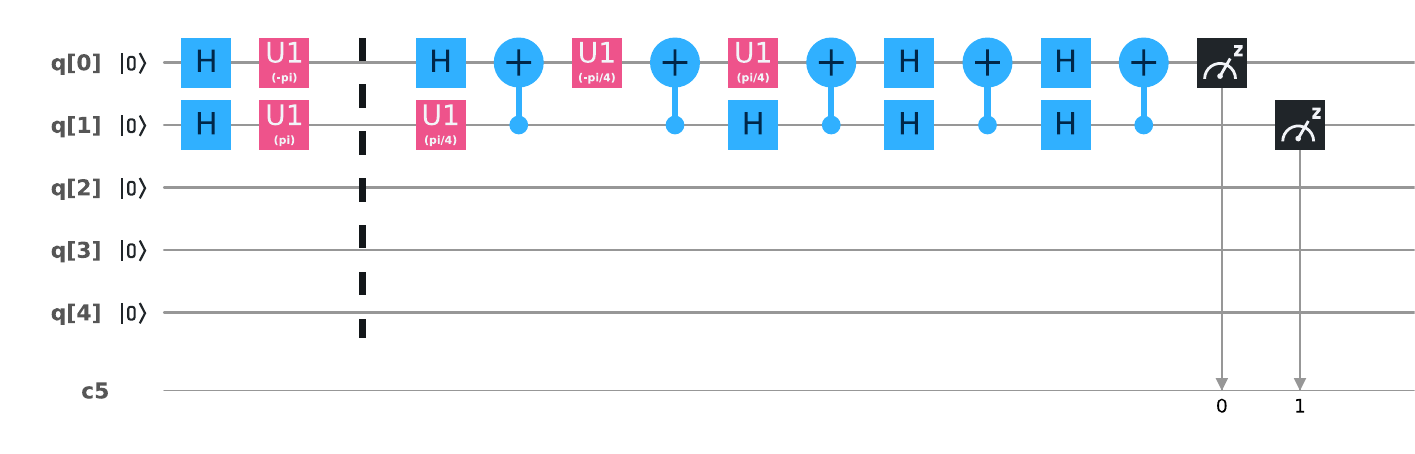}
	\caption{Circuit for preparing the first two qubits and quantum Fourier transform on 2 qubits.}
	\label{fig:score}
\end{figure}

We implemented one step of the full partition function algorithm using the IBM Quantum Experience. The implemented algorithm is the 2-qubit quantum Fourier transform (QFT2), as the first step in actual calculation of the partition function. The input to this step is the irreducible cocyclic code. The irreducible cyclic code for the example problem of a 3-vertex Ising model is $[1, -1]$ with $n = |V| = 3$ and $k= |E| - c(\Gamma) = 2$, where $c(\Gamma)$ is the number of connected components in the graph $\Gamma$. This small example does meet the ICCC requirement (as checked through classical calculation), so we will continue with the calculation of the partition function of the example without implementing the quantum algorithm for checking this requirement. In the fully-connected 3-vertex Ising model example given, the input to QFT2 is $q[0] = \ket{+} = \frac{\ket{0} + \ket{1}}{\sqrt2}$ and $q[1] = \ket{-} = \frac{\ket{0} - \ket{1}}{\sqrt2}$. In the sample score shown in Fig.~\ref{fig:score}, these qubits are prepared before the barrier. The QFT2 algorithm, as given by the Qiskit Tutorial provided by IBM\cite{Qiskit}, is the rest of the code. The output bits should be read in reverse order. Some gates could be added at the end of the QFT2 algorithm to read the gates in the same order as the input. 

%\begin{verbatim}
%include "qelib1.inc";                                                           
%qreg q[5];                                                                      
%creg c[5];                                                                      
%
%h q[0];                                                                                
%h q[1];
%u1(pi) q[0];
%u1(-pi) q[1];                                                                         
%barrier q[0],q[1];                                               
%h q[0];                                                                         
%u1(pi/4) q[1];                                                                  
%cx q[1],q[0];                                                                   
%u1(-pi/4) q[0];                                                                 
%cx q[1],q[0];                                                                   
%u1(pi/4) q[0];                                                                  
%h q[1];                                                                         
%measure q[0] -> c[0];                                                           
%measure q[1] -> c[1];                     
%\end{verbatim}

The result from simulating 1000 shots gives $P(\gamma = 1) = 0.47$ and $P(\gamma = 3) = 0.53$. The results from running on the actual hardware are, $P(\gamma = 0) = 0.077$, $P(\gamma = 1) = 0.462$, $P(\gamma = 2) = 0.075$, and $P(\gamma = 3) = 0.386$. We can threshold the low-probability values of gamma, ensuring that no more than the maximum number (as given in \cite{geraci2008exact}) of distinct values of gamma remain. These gammas are then plugged into the calculation of the weight spectrum and the partition function.

%%%%%%%%%%%%%%%%%%%%%%%%%%%%%%%%%%%%%%%%%%%%%%%%%%%%%%%%

\newcommand{\wt}{\widetilde}
\newcommand{\mxx}[4]{\left(\begin{array}{cc}{#1} & {#2}\\ {#3} & {#4}\end{array}\right)}

\section{Quantum State Preparation}
\noindent The problem of preparing an $n$-qubit state consists first of finding
the unitary transformation that takes the $N$-dimensional vector
(1,0,\ldots 0) to the desired state ($\alpha_1$, \ldots, $\alpha_N$),
where $N=2^n$, and then rendering the unitary transformation into a
sequence of gates.

\subsection{Single qubit state preparation}
\noindent
As discussed before, a single qubit quantum state $\ket\psi$ is represented as a
superposition of $\ket{0}$ and $\ket{1}$ states $\ket\psi = \alpha\ket0 +
\beta\ket1$, where $|\alpha|^2 + |\beta|^2 = 1$.  The sizes
$|\alpha|^2$ and $|\beta|^2$ represent the probability of
$\ket\psi$ being $\ket{0}$ or $\ket{1}$.  Up to a non-observable global phase, we
may assume that $\alpha$ is real, so that $\ket\psi = \cos\theta\,
\ket{0} + e^{i\phi}\sin\theta\,\ket{1}$ for some angles $\theta,\phi$.
In this way, we can represent the state as a point on the unit sphere
with $\theta$ the co-latitude and $\phi$ the longitude.  This is the
well-known \textit{Bloch sphere representation}.  In this way, the problem of
$1$-qubit state preparation consists simply of finding the unitary
transformation that takes the North pole to ($\alpha$, $\beta$).  In
practice, this amounts to finding a sequence of available gates on
actual hardware that will leave the qubit in the desired state, to a
specified desired accuracy.  

To prepare a specified state $\ket\psi$, we must find a $2\times 2$
unitary matrix $U$ taking the vector $\ket0$ to $\ket\psi$. An obvious
simple choice for $U$ is
\begin{align*}
U = \mxx {\cos\theta} {-\sin\theta e^{-i\phi}}
{\sin\theta e^{i\phi}} {\cos\theta}
\end{align*}

This gate is directly available in IBM Q and is implemented in a composite fashion on \texttt{ibmqx4} at the hardware level.  
If our goal is to initialize a base state with the fewest possible
standard gates, this may not be the best choice.  Instead, it makes sense to
consider a more general possible unitary operator whose first column is our
desired base state, and then determine the requisite number of standard gates
to obtain it.

Any
$2\times 2$ unitary matrix may be obtained by means of a product of
three rotation matrices, up to a global phase
$$U = e^{i\alpha} R_z(\beta)R_y(\gamma)R_z(\delta)$$
where here $R_z(\beta) = \text{diag}(e^{i\beta/2},e^{-i\beta/2})$ and
$R_y(\gamma)$ is related to $R_z(\gamma)$ by $R_y(\gamma) = SH
R_z(\gamma) HSZ$.  The rotation matrices $R_y(\gamma)$ and
$R_z(\beta)$ correspond to the associated rotations of the unit sphere
under the Bloch representation.  In this way, the above decomposition
is a reiteration of the standard Euler angle decomposition of elements
of $SO(3)$.  Thus the problem of approximating an arbitrary quantum
state is reduced to the problem of finding good approximations of
$R_z(\gamma)$ for various values of $\gamma$.

There has been a great deal of work done on finding efficient
algorithms for approximating elements $R_z(\gamma)$ using universal
gates to a specified accuracy.  However, these
algorithms tend to focus on the asymptotic efficiency: specifying
approximations with the desired accuracy which are the generically
optimal in the limit of small errors.  From a practical point of view,
this is an issue on current hardware, since representations tend to
involve hundreds of standard gates, far outside the realm of what may
be considered practical. For this reason, it makes sense to ask the
question of how accurately one may initialize an arbitrary qubit with
a specified number of gates.

\begin{figure}[!htb] \minipage{0.32\textwidth}
  \includegraphics[width=\linewidth]{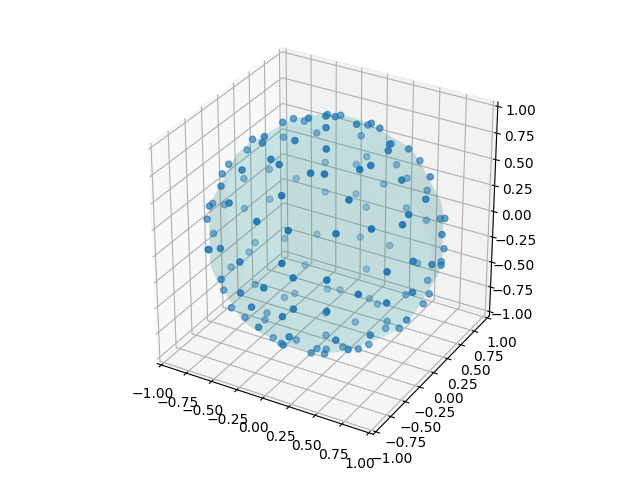}
\endminipage\hfill \minipage{0.32\textwidth}
  \includegraphics[width=\linewidth]{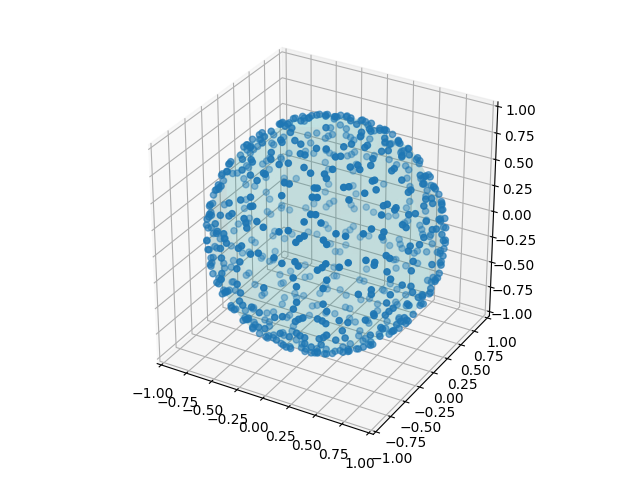}
\endminipage\hfill \minipage{0.32\textwidth}%
  \includegraphics[width=\linewidth]{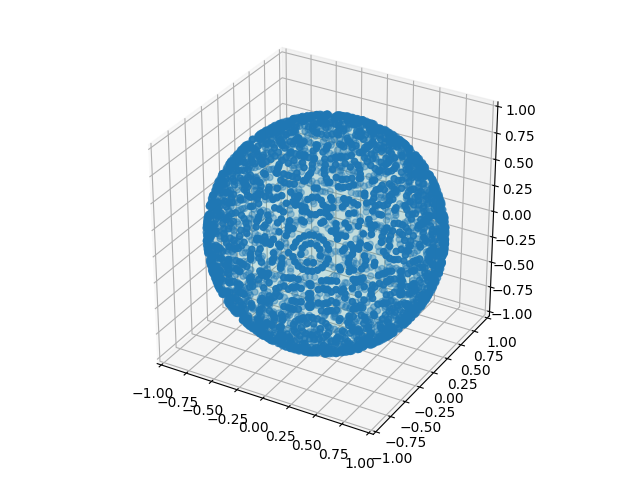}
\endminipage
\caption{Possible exact state initializations using $10,15,$ and $20$
gates.  With $20$ gates, every point on the sphere is within a distance of
approximately $0.072$ of an exactly obtainable state.  With $30$
gates, every point is within $0.024$}
\end{figure} 

We empirically observe that the maximum possible chordal
distance from a point on the Bloch sphere to the set of exact states
decreases exponentially with the number of gates.
With $30$ gates, every point is within a distance of  $0.024$ of a desired gate.
Thus, to within an accuracy of about 2.5\%, we can represent any base state as
a product of about $30$ states.  We do so by preserving the states
generated by $30$ gates, and then for any point finding the closest
exact point.

\subsection{Schmidt decomposition}
The initialization of qubit states using more than one qubit is aided
by the so-called Schmidt decomposition, which we now introduce.
Specifically, the Schmidt decomposition allows one to initialize a
$2n$-qubit state by initializing a single $n$-qubit state, along with
two specific $n$-qubit gates, combined together with $n$ CNOT gates.

Mathematically, an arbitrary $2n$-qubit state $\ket\psi$ may be represented as a superposition
$$\ket\psi = \sum_{i_1,\dots,i_n\in \{0,1\}}\sum_{j_1,\dots,j_n\in \{0,1\}} a_{i_1,\dots,i_n,j_1,\dots,j_n}\ket{i_1i_2\dots i_nj_1j_2\dots j_n}.$$
In a Schmidt decomposition, we obtain such a state by strategically choosing two orthonormal bases $\ket{\xi_j},\ket{\varphi_j}$ for $j=1,\dots,2^n$ of the Hilbert space of $n$-qubit states and then writing $\ket\psi$ as the product
$$\ket\psi = \sum_{i=1}^{2^n} \lambda_i \ket{\xi_i}\ket{\varphi_i},$$
for some well-chosen $\lambda_i$'s.

The bases $\ket{\xi_j}$ and $\ket{\varphi_j}$ may be represented in
terms of two unitary matrices $U,V\in U(2^n)$, while the $\lambda_i$'s
may be represented in terms of a single $n$-qubit state.  We represent
this latter state as $B\ket{00\dots0}$ for some $B\in U(2^n)$.  Then
from a quantum computing perspective, the product in the Schmidt
decomposition may be accomplished by a quantum circuit combining
$U,V,$ and $B$ with $n$ CNOT gates as shown below for $n=6$.

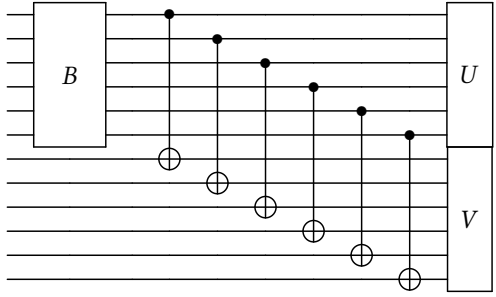
\begin{figure}
  
$$
\Qcircuit @C=1em @R=0em {
& \multigate{5}{\hspace{0.2cm}B\hspace{0.2cm}} & \qw & \ctrl{6} & \qw & \qw & \qw & \qw & \qw & \multigate{5}{U} \\
& \ghost{\hspace{0.2cm}B\hspace{0.2cm}} & \qw & \qw & \ctrl{6} & \qw & \qw & \qw & \qw  & \ghost{U} \\
& \ghost{\hspace{0.2cm}B\hspace{0.2cm}} & \qw & \qw & \qw & \ctrl{6} & \qw & \qw & \qw  & \ghost{U} \\
& \ghost{\hspace{0.2cm}B\hspace{0.2cm}} & \qw & \qw & \qw & \qw & \ctrl{6} & \qw & \qw  & \ghost{U} \\
& \ghost{\hspace{0.2cm}B\hspace{0.2cm}} & \qw & \qw & \qw & \qw & \qw & \ctrl{6} & \qw  & \ghost{U} \\
& \ghost{\hspace{0.2cm}B\hspace{0.2cm}} & \qw & \qw & \qw & \qw & \qw & \qw & \ctrl{6} & \ghost{U} \\
& \qw & \qw & \targ & \qw & \qw & \qw & \qw & \qw & \multigate{5}{V}\\
& \qw & \qw & \qw & \targ & \qw & \qw & \qw & \qw & \ghost{V}\\
& \qw & \qw & \qw & \qw & \targ & \qw & \qw & \qw & \ghost{V}\\
& \qw & \qw & \qw & \qw & \qw & \targ & \qw & \qw & \ghost{V}\\
& \qw & \qw & \qw & \qw & \qw & \qw & \targ & \qw & \ghost{V}\\
& \qw & \qw & \qw & \qw & \qw & \qw & \qw & \targ & \ghost{V}\\
}
$$
  \caption{Schmidt decomposition.}
  
\end{figure}

Let $C_i^j$ denote the CNOT operator with control $j$ and target $i$.  Algebraically, the above circuit may be written as a unitary operator $T\in U(2^{2n})$ of the form
$$T = (U\otimes V)(C_{n+1}^1\otimes C_{n+2}^2\otimes\dots \otimes C_{2n}^n)(B\otimes I).$$
We will use $\ket{e_1},\dots,\ket{e_{2^n}}$ to denote the standard computational basis for the space of $n$-qubit states, in the usual order.
We view each of the elements $e_j$ as a vector in $\{0,1\}^n$.
In this notation, the formation of CNOT gates above acts on simple tensors by sending
$$C_{n+1}^1\otimes C_{n+2}^2\otimes\dots \otimes C_{2n}^n: \ket {e_i}\ket{e_j}\mapsto \ket{e_i}\ket{e_i+e_j},\ \ e_i,e_j\in\{0,1\}^n,$$
where addition in the above is performed modulo 2.
Therefore the action of the operator $T$ associated to the above circuit on the basis vector $\ket{00\dots 0}$ is
\begin{align*}
T\ket{00\dots 0}
  & = (U\otimes V)(C_{n+1}^1\otimes C_{n+2}^2\otimes\dots \otimes C_{2n}^n)(B\otimes I)\ket{00\dots 0}\\
  & = (U\otimes V)(C_{n+1}^1\otimes C_{n+2}^2\otimes\dots \otimes C_{2n}^n)\sum_{i=1}^{2^n} b_{i1}\ket{e_i}\ket{e_1}\\
  & = (U\otimes V)\sum_{i=1}^{2^n} b_{i1}\ket{e_i}\ket{e_i}\\
  & = \sum_{i=1}^{2^n} b_{i1}(U\ket{e_i})(V\ket{e_i}) = \ket\psi.
\end{align*}
Thus we see that the above circuit performs precisely the sum desired from the Schmidt decomposition.

To get the precise values of $U,V,$ and $B$, we write $\ket\psi = \sum_{i,j=1}^{2^n} a_{ij}\ket{e_i}\ket{e_j}$ for some constants $a_{ij}\in\mathbb{C}$ and define $A$ to be the $2^n\times 2^n$ matrix whose entries are the $a_{ij}$'s.
Then comparing this to our previous expression for $\ket\psi$, we see
$$\sum_{i,j=1}^{2^n}a_{ij}\ket{e_i}\ket{e_j} = \sum_{k=1}^{2^n} b_{k1}(U\ket{e_k})(V\ket{e_k}).$$
Multiplying on the left by $\bra{e_i}\bra{e_j}$ this tells us
$$a_{ij} = \sum_{k=1}^{2^n} b_{k1}u_{ik} v_{jk},$$
where here $u_{ik} = \bra{e_i}U\ket{e_k}$ and $v_{jk} = \bra{e_j}V\ket{e_k}$ are the $i,k$'th and $j,k$'th entries of $U$ and $V$, respectively.
Encoding this in matrix form, this tells us
$$V\text{diag}(b_{i1},\dots,b_{in})U^T = A.$$
Then to calculate the value of $U,V$ and the $b_{i1}$'s, we use the fact that $V$ is unitary to calculate:
$$A^\dag A = U^{T\dag}\text{diag}(|b_{i1}|^2,\dots,|b_{in}|^2) U^T.$$
Thus if we let $|\lambda_1|^2,\dots,|\lambda_n|^2$ be the eigenvalues of $A^\dag A$, and let  $U$ to be a unitary matrix satisfying
$$U^{T}A^\dag A U^{T\dag} = \text{diag}(|\lambda_1|^2,\dots,|\lambda_N|^2),$$
let $b_{i1} = \lambda_i$ for $i=1,\dots,n$ and let
$$V = AU^{T\dag}\text{diag}(\lambda_1,\dots,\lambda_n)^{-1}.$$
The matrix $U$ is unitary, and one easily checks that $V$ is therefore
also unitary.  Moreover
$\sum_i |b_{i1}|^2 = \text{Tr}(A^\dag A) = \sum_i |a_{ij}|^2 = 1$, and
so the $b_{i1}$'s are representative of an $n$-qubit state and can be
taken as the first column of $B$. Readers familiar with singular value decompositions (SVD) will recognize that the Schmidt decomposition of a bipartite state is essentially the SVD of the coefficient matrix $A$ associated with the state. The $\lambda_i$ coefficients being the singular values of $A$. 
\subsection{Two-qubit state preparation}
\begin{figure}[H]
$$
\Qcircuit @C=1em @R=.7em { & \gate{B} & \ctrl{1} & \gate{U} & \qw \\ &
\qw & \targ & \gate{V} & \qw }
$$
  \caption{Circuit for two qubit-state preparation. The choice of $U,V,$ and $B$ are covered comprehensively in the Schmidt
decomposition description.}
\label{fig:2stateprep}
\end{figure}
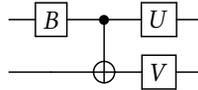

An arbitrary two-qubit state $\ket\psi$ is a linear combination of the
four base states $\ket{00},\ket{01},\ket{10},\ket{11}$ such that the
square sum of the magnitudes of the coefficients is $1$.  In terms of a
quantum circuit, this is the simplest case of the circuit defined
above in the Schmidt decomposition, and may be accomplished with three
$1$-qubit gates and exactly $1$ CNOT gate, as featured in Fig. \ref{fig:2stateprep}.

\subsection{Two-qubit gate preparation}
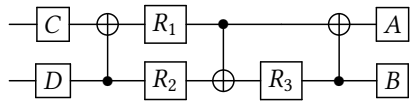
\begin{figure}[H]
$$
\Qcircuit @C=1em @R=.7em {
& \gate{C} & \targ     & \gate{R_1}  & \ctrl{1} & \qw        & \targ     & \gate{A} \\
& \gate{D} & \ctrl{-1} & \gate{R_2}  & \targ    & \gate{R_3} & \ctrl{-1} & \gate{B}
}
 $$
  \caption{Circuit implementation of an arbitrary two qubit gate.}
   \label{fig:two_qubit_op}
\end{figure}

In order to initialize a four-qubit state, we require the
initialization of arbitrary two-qubit gates.  A two-qubit gate may be
represented as an element $U$ of $SU(4)$.  As it happens, any element
of $U(4)$ may be obtained by means of precisely $3$ CNOT gates,
combined with $7$ $1$-qubit gates arranged in a circuit of the form given in Fig.~\ref{fig:two_qubit_op}.
The proof of this is nontrivial and relies on a characterization of
the image of $SU(2)^{\otimes 2}$ in $SU(4)$ using the Makhlin
invariants \cite{shende2004minimal}.  We do not aim to reproduce the proof here.  Instead, we
merely aim to provide a recipe by which one may successfully obtain
any element of $SU(4)$ via the above circuit and an appropriate choice
of the one-qubit gates.

Let $U\in SU(4)$ be the element we wish to obtain.  To choose
$A,B,C,D$ and the $R_i$'s, let $C^i_j$ denote the CNOT gate with
control on qubit $i$ and target qubit $j$ and define
$\alpha,\beta,\delta$ by
$$\alpha = \frac{x+y}{2},\ \beta = \frac{x+z}{2},\ \delta = \frac{y+z}{2}$$
for $e^{ix},e^{iy},e^{iz}$ the eigenvalues of the operator $U(Y\otimes Y)U^T(Y\otimes Y)$.
Then set
$$R_1 = R_z(\delta), R_2 = R_y(\beta), R_3 = R_y(\alpha), E = C^2_1(S_z\otimes S_x)$$
and also
$$V = e^{i\pi/4}(Z\otimes I)C^2_1(I\otimes R_3)C^1_2(R_1\otimes R_2)C^2_1(I\otimes S_z^\dag),$$ where $S_z$ is the single qubit $\pi/2$ rotation around the $z$ axis. 
Define $\wt U,\wt V$ by $\wt U = E^\dag UE$ and $\wt V = E^\dag VE$.
Let $\wt A,\wt B$ be the real, unitary matrices diagonalizing the eigenvectors of $\wt U\wt U^T$ and $V\wt V^T$, respectively.
Set $X = \wt A^T\wt B$ and $Y = V^\dag\wt B^T\wt AU$.
Then $EXE^\dag$ and $EYE^\dag$ are in $SU(2)^{\otimes 2}$ and we choose $A,B,C,D$ such that
$$(AS_z^\dag)\otimes (Be^{i\pi/4}) = EXE^\dag\ \ \text{and}\ \ C\otimes (S_z D) = EYE^\dag.$$
By virtue of this construction, the above circuit is algebraically identical to $U$.

\subsection{Four qubit state preparation} 

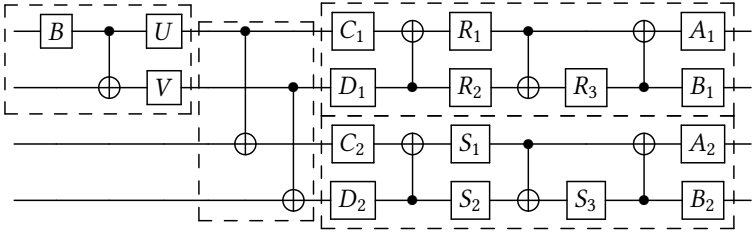
\begin{figure}
$$
    \Qcircuit @C=1em @R=.7em { & \gate{B} & \ctrl{+1} & \gate{U} & \qw & \ctrl{+2}
& \qw & \gate{C_1} & \targ & \gate{R_1} & \ctrl{1} & \qw & \targ &
\gate{A_1} & \qw\\ & \qw & \targ & \gate{V} & \qw & \qw & \ctrl{+2} &
\gate{D_1} & \ctrl{-1} & \gate{R_2} & \targ & \gate{R_3} & \ctrl{-1} &
\gate{B_1} & \qw\\ & \qw & \qw & \qw & \qw & \targ & \qw &
\gate{C_2} & \targ & \gate{S_1} & \ctrl{1} & \qw & \targ & \gate{A_2}
& \qw\\ & \qw & \qw & \qw & \qw & \qw & \targ & \gate{D_2} &
\ctrl{-1} & \gate{S_2} & \targ & \gate{S_3} & \ctrl{-1} & \gate{B_2} &
\qw \gategroup{1}{1}{2}{4}{.7em}{--} \gategroup{1}{5}{4}{7}{.7em}{--}
\gategroup{1}{8}{2}{14}{.7em}{--} \gategroup{3}{8}{4}{14}{.7em}{--} }$$
  
  \caption{Circuit for four qubit-state preparation. The four phases
    of the circuit are indicated in dashed boxes.}
  \label{fig:four_qubit}
\end{figure}

For efficient four qubit state preparation we use the recipe in Ref.\cite{plesch2011quantum}. Results in the previous sections show  that any two-qubit state requires $1$ CNOT gate,
any two-qubit operator requires three CNOT gates, and the Schmidt
decomposition of a four qubit state requires two CNOT gates . From this we see that we should be able to write a circuit
initializing any four-qubit state with only $9$ CNOT gates in
total, along with $17$ one-qubit gates.  This represents the second
most simple case of the Schmidt decomposition, which we write in
combination with our generic expression for $2$-qubit gates as shown in Fig.~\ref{fig:four_qubit}.
The above circuit naturally breaks down into four distinct stages, as
shown by the separate groups surrounded by dashed lines.  During the
first stage, we initialize the first two qubits to a specific state
relating to a Schmidt decomposition of the full $4$ qubit state.
Stage two consists of two CNOT gates relating the first and last
qubits.  Stages three and four are generic circuits representing the
unitary operators associated to the orthonormal bases in the Schmidt
decomposition.

\begin{figure}[!htb]
  \includegraphics[width=0.5\textwidth]{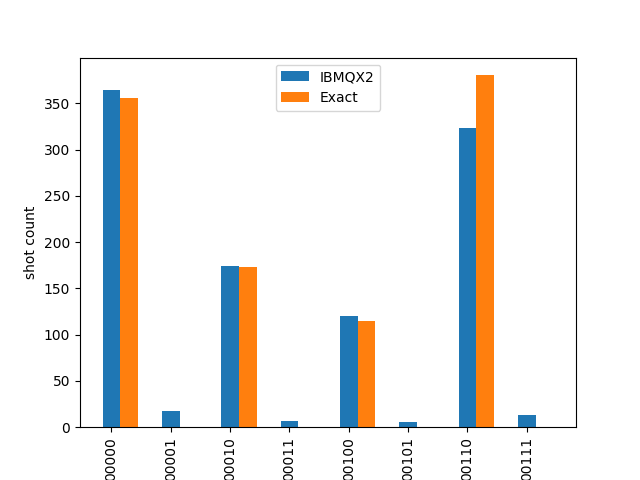}
  \caption{Verification of $4$ qubit state preparation on \texttt{ibmqx2} which is a $5$ qubit machine. The last qubit is not used in the circuit. The above histogram shows that, the state prepared in \texttt{ibmqx2} has nonzero overlaps with basis states that are orthogonal to the target state to be prepared.}
  \label{fig:4q_ibm}
\end{figure}

\begin{figure}[!htb]
\centering
  \includegraphics[width=0.5\textwidth]{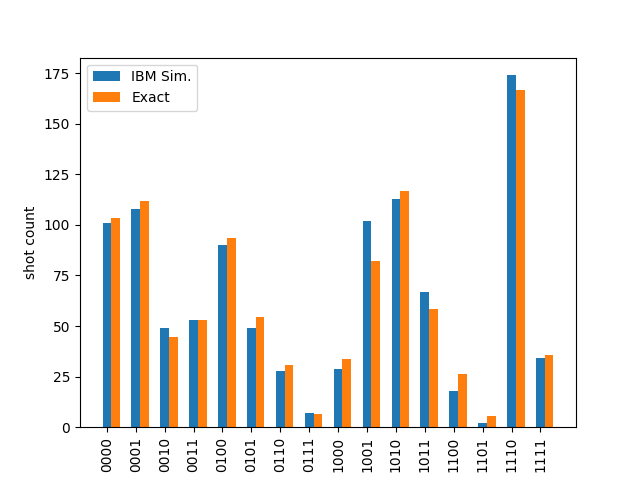}
  \caption{Verification of the quantum circuit for four qubit-state preparation. The differences in the exact and the simulator results are due to statistical fluctuations arising from the probabilistic nature of quantum measurement. They will become closer to each other when the number of samples are increased.}
  \label{fig:4q_sim}
\end{figure}

The results of this circuit implemented on a quantum processor are given in Fig.~\ref{fig:4q_ibm}. While the results when implemented on a simulator are given in Fig.~\ref{fig:4q_sim}.

\section{Quantum Tomography}\label{sec:Tomography}

\subsection{Problem definition and background}
\label{subsec:quantum_tomography_definition}

Quantum state estimation, or tomography, deals with the reconstruction of the state of a quantum system from measurements of several preparations of this state. In the context of quantum computing,
%, a two-qubit example is presented in Figure~\ref{Lokhov:fgr1}. %Imagine
imagine that we start with the state $\ket{00}$, and apply some quantum algorithm (represented by a unitary matrix $U$) to the initial state, thus obtaining a state $\ket{\psi}$. We can measure this state in the computational $z$ basis, or apply some rotation (represented by $V$) in order to perform measurements in a different basis. Quantum state tomography aims to answer the following question: is it possible to reconstruct the state $\ket{\psi}$ from a certain number of such measurements? Hence, quantum tomography is not a quantum algorithm \emph{per se}, but it is an important procedure for certifying the performance of quantum algorithms and assessing the quality of the results that can be corrupted by decoherence, environmental noise, and biases, inevitably present in analogue machines. Moreover similar procedures can be used for certifying the initial state, as well as for measuring the fidelity of gates.

Given a single copy of the state, it is impossible to reconstruct $\ket{\psi}$: for example, there is no quantum measurement that can even distinguish non-orthogonal quantum states, such as $\ket{0}$ and $(\ket{0}+\ket{1})/2$\revision{, with certainty}. However, it is possible to perform quantum tomography when multiple copies of the state is available. It means that one needs to run the quantum algorithm (i.e., apply $U$ to initial state) many times to produce many copies of the quantum state to be able to characterize $\ket{\psi}$.
% A unique identification of state requires a sufficient number of \emph{tomographically complete} measurements, meaning that the algorithm should be run several times.
Unfortunately, because of the noise, in practice it is impossible to obtain the exact same state $\ket{\psi}$ every time; instead, one should see a mixture of different states: $\vert \psi_{1} \rangle$, $\vert \psi_{2} \rangle$, $\ldots$, $\vert \psi_{k} \rangle$. In general, there does not exist a single $\ket{\psi}$ describing this mixture. Therefore, we need a to use the \emph{density matrix} representation of quantum states. We briefly discussed this representation in the context of quantum principal component analysis in Section XIV.

Let us denote $p_i$ the probability of occurrence of the state $\vert \psi_{i} \rangle$. The density matrix of this ensemble is given  by,
\begin{equation}
\rho = \sum_{i} p_i \vert \psi_i \rangle \langle \psi_i \vert.
\end{equation}
Using this more general definition of the state, the expected value of an observable $A$ is given by $\langle A \rangle = \sum\limits_i p_i \langle \psi_i \vert A \vert \psi_i \rangle = \Tr(A\rho)$. The density matrix has the following properties:
\begin{itemize}
\item $\Tr \, \rho = 1$, i.e.,~probabilities sum to one;
\item $\rho = \rho^\dag$, and $\rho \succcurlyeq 0$, i.e.,~all eigenvalues are either positive or zero.
\end{itemize}

The goal of quantum state tomography is to reconstruct the quantum state $\rho$ from many repeated runs of the quantum algorithm, using a set of measurements on $\rho$. Quantum measurements are described by a collection of \emph{measurement operators} $M_i$ that satisfy the completeness relation $\sum_i M^\dag_i M_i = I$. This condition ensures that the probabilities of measurement outcomes $p_i = \Tr (\rho M^\dag_i M_i)$ sum to one for any state $\rho$. In a popular setting for quantum tomography that is closely related to what is currently achievable with modern quantum computers, the measurement operators are chosen in a special class of \emph{projective measurements} -- projectors $P_i$ that satisfy $P_i P_{j} = \delta_{i,j}P_i$ and $\sum_i P_i = I$ given that in this case $P^\dag_i P_i = P_i$, and hence the measurement probability is given by the relation $p_i = \Tr (\rho P_i)$. This choice represents a single instance of a more general measurement formalism that deals with situations when we are only interested in the probabilities of the measurement outcomes, called \emph{Postive Operator-Valued Measures} (POVM). In this introduction we will only deal with projective measurements described above which will be sufficient for our purpose, and refer the reader to~\cite{NielsenChuang} for details on POVMs and general measurement formalism.

% In a popular setting for quantum tomography~\cite{NielsenChuang}, the set of measurement operators $P_i$ are taken as projectors that form several \emph{Postive Operator-Valued Measures} (POVM), i.e.,~they satisfy $\sum_i P_i = I$.
For a single qubit, examples of measurement projectors in the computational basis are given by $P_0 = \vert 0 \rangle \langle 0 \vert$ and $P_1 = \vert 1 \rangle \langle 1 \vert$, and in the $x$-basis by $P_{\pm} = \frac{1}{\sqrt{2}}(\vert 0 \rangle \pm \vert 1 \rangle) \otimes  \frac{1}{\sqrt{2}}(\langle 0 \vert \pm \langle 1 \vert)$. At this point, once could ask: what is the set of projectors that represent a \emph{quorum}, i.e. provides sufficient information to identify the state of the system in the limit of a large number of observations? The answer to this question is important for the quantum state tomography task, as it allows to determine an \emph{informationally complete} set of measurements that is necessary to reconstruct the state. For a single-qubit example, the density matrix $\rho_{\text{qubit}}$ can be decomposed as
\begin{equation}
    \rho_{\text{qubit}} = \frac{\Tr(\rho_{\text{qubit}})I +  \Tr(\rho_{\text{qubit}} X)X + \Tr(\rho_{\text{qubit}} Y)Y + \Tr(\rho_{\text{qubit}} Z)Z}{2},
\end{equation}
where $X$, $Y$, and $Z$ are Pauli matrices, that can be interpreted as projectors in the $x$-, $y$-, and $z$-basis. From this expression, it is easy to see that $I/\sqrt{2}$, $X/\sqrt{2}$, $Y/\sqrt{2}$, $Z/\sqrt{2}$ forms the set of measurement operators that provide sufficient information for reconstructing the state $\rho_{\text{qubit}}$ from measurements $p_i = \Tr(\rho_{\text{qubit}} P_i)$. Using this example, a set of measurements for a general system $\rho$ is said to be informationally complete if the relations $p_i = \Tr(\rho P_i)$ can be inverted to unambiguously reconstruct the state $\rho$. The state $\rho$ can be uniquely expressed using the obtained measurements whenever the matrix $\Tr(P_i P_j)$ is invertible. For a multi-qubit state on $n$ qubits, a simple example of an informationally complete set of measurements is given by a set of tensor products of all possible combinations of Pauli matrices. Notice, however, that a smaller number of measurement operators may be sufficient; the necessary number of measurement operators is related to the number of independent parameters in the density matrix $\rho$.

% Assume that the set of projectors that we take represents a \emph{quorum}, i.e.,~it provides sufficient information to identify the state of the system in the limit of a large number of observations, and that for each subset forming a POVM, $m$ measurements are collected.
In real experiment, a finite number of measurements is collected for each measurement operator. Given the measurement occurrences $m_i$ for each projector $P_{i}$, we define the associated empirical frequency as $\omega_i = m_i / m$. Then the quantum tomography problem can be stated as follows: reconstruct $\rho$ from the informationally complete set of couples of projectors and measurement frequencies $\{P_i,\omega_i\}$. In other words, we would like to ``match'' $\Tr(P_i \rho)$ and $\omega_i$. The next section presents a short overview of most popular general methods for the quantum state estimation.

\subsection{Short survey of existing methods}

Most popular methods for quantum tomography in the general case include:
\begin{enumerate}
\item \textbf{Linear inversion.} In this method, we simply aim at inverting the system of equations $\Tr(P_i \rho) = \omega_i$. Although being fast, for a finite number of measurements thus obtained estimation $\widehat{\rho}$ does not necessarily satisfy $\widehat{\rho} \succcurlyeq 0$ (i.e.,~might contain negative eigenvalues): this happens due to experimental inaccuracies and statistical fluctuations of coincidence counts, which leads to differences between the empirical measurement frequencies $\omega_i$ and the calculated values $\Tr(P_i \rho)$ ~\cite{PhysRevA.64.052312}.
\item \textbf{Linear regression.} This method corrects for the disadvantages of the linear inversion by solving a constrained quadratic optimization problem~\cite{qi2013quantum}:
\begin{equation*}
\widehat{\rho} = \argmin_{\rho} \sum_i [\Tr(P_i \rho) - \omega_i]^2 \quad \text{s.t.} \; \Tr \, \rho = 1 \; \text{and} \; \rho \succcurlyeq 0. 
\end{equation*}
% The advantage of this method is that data does not need to be stored, but only the current estimation can be updated in the streaming fashion.
However, this objective function implicitly assumes that the residuals are Gaussian-distributed, which may hold in a limit of a large number of independent measurements due to the central limit theorem, but does not necessarily apply in practice for a finite number of measurements where deviations from normal distribution can be important.
\item \textbf{Maximum likelihood.} In this by far most popular algorithm for quantum state estimation, one aims at maximizing the log-probability of observations~\cite{hradil1997quantum,PhysRevA.64.052312}:
\begin{equation*}
\widehat{\rho} = \argmax_{\rho} \sum_i \omega_i \ln \Tr(P_i \rho) \quad \text{s.t.} \; \Tr \, \rho = 1 \; \text{and} \; \rho \succcurlyeq 0. 
\end{equation*}
This is a convex problem that outputs a positive semidefinite (PSD) solution $\widehat{\rho} \succcurlyeq 0$. However, it is often stated that the maximum likelihood (ML) method is slow, and several recent papers attempted to develop faster methods of gradient descent with projection to the space of PSD matrices, see e.g.~\cite{shang2017superfast}. Among other common criticisms of this method one can name the fact that ML might yield rank-deficient solutions, which results in an infinite conditional entropy that is often used as a metric of success of the reconstruction.
\item \textbf{Bayesian methods.} This is a slightly more general approach compared to the ML method which includes some prior~\cite{blume2010optimal}, or corrections to the basic ML objective, see e.g.,~the so-called Hedged ML~\cite{blume2010hedged}. However, it is not always clear how to choose these priors in practice. Markov Chain Monte Carlo Methods that are used for general priors are known to be slow.
\end{enumerate}
Let us mention that there exist other state reconstruction methods that attempt to explore a particular known structure of the density matrix, such as compressed-sensing methods~\cite{gross2010quantum} in the case of low-rank solutions, and matrix product states~\cite{cramer2010efficient} or neural networks based approaches~\cite{torlai2017many} for pure states with limited entanglement, \emph{etc}. One of the points we can conclude from this section is that the ultimately best general method for the quantum state tomography is not yet known and certainly depends on the applications. However, it seems that maximum likelihood is still the most widely discussed method in the literature; in what follows, we implement and test ML approach to quantum tomography on the IBM quantum computer.  

% \begin{figure}
% \begin{center}
% \includegraphics[width=0.39\columnwidth]{Tomography_fig/intro3}
% \caption{Quantum tomography in the context of quantum computing: using a certain number of measurements in different basis implemented with $V$, is it possible to reconstruct the system state $\ket{\psi}$ after application of the quantum algorithm $U$, as well as certify the initial state and assess the fidelity of the gates in $U$?}
% \label{Lokhov:fgr1}
% \end{center}
% \vspace{-0.5cm}
% \end{figure}

\vspace{-0.3cm}

\subsection{Implementation of the Maximum Likelihood method on 5-qubit IBM QX}

We present an efficient implementation of the ML method using a fast gradient descent with an optimal 2-norm projection~\cite{smolin2012efficient} to the space of PSD matrices\footnote{The julia implementation of the algorithm is available at \url{http://gitlab.lanl.gov/QuantumProgramming2017/QuantumTomography}}. In what follows, we apply quantum tomography to study the performance of the IBM Q. 

\subsubsection{Warm-up: Hadamard gate}

\begin{figure}
\begin{center}
\includegraphics[width=0.17\columnwidth, angle=0]{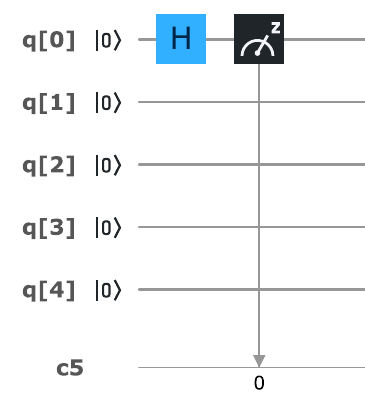} \hfill \includegraphics[width=0.24\columnwidth, angle=0]{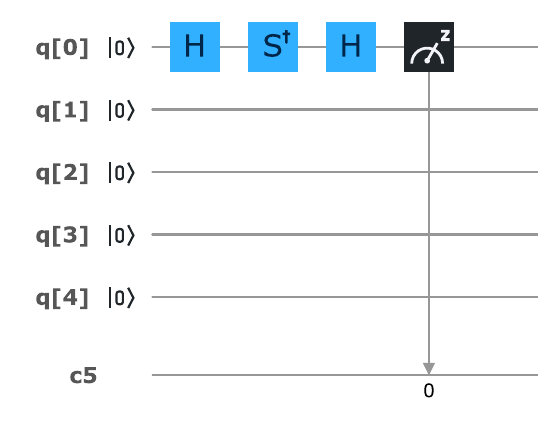} \hfill 
\includegraphics[width=0.2\columnwidth, angle=0]{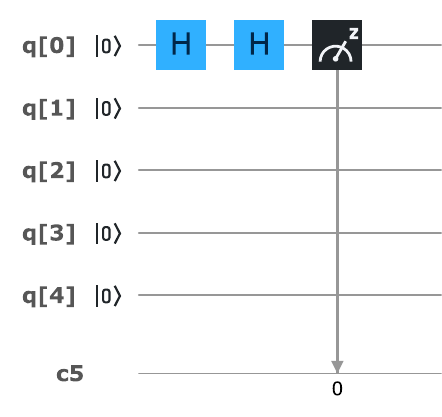} \hfill
\vrule \hfill
\includegraphics[width=0.31\columnwidth, angle=0]{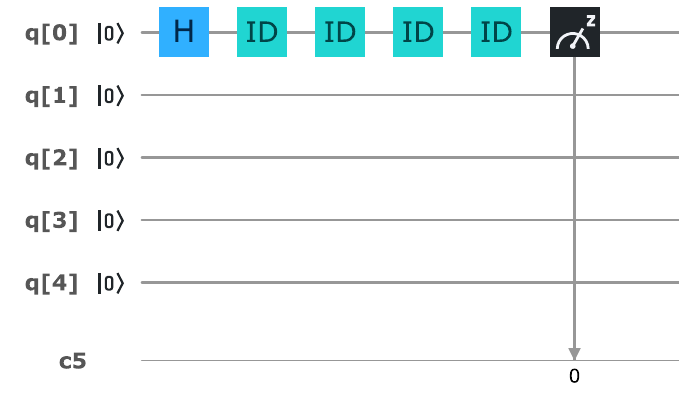}
\caption{Left: measurements of the single qubit state after the application of the Hadamard gate, in $z$, $y$ and $x$ basis. Right: experimental setup for testing the effects of decoherence.}
\label{Lokhov:fgr2}
\end{center}
\vspace{-0.5cm}
\end{figure}

Let us start with a simple one-qubit case of the Hadamard gate, see Fig.~\ref{Lokhov:fgr2}, Left. This gate transforms the initial qubit state $\vert 0 \rangle$ as follows: $H: \, \vert 0 \rangle \rightarrow \vert + \rangle_{x} = \frac{1}{\sqrt{2}}(\vert 0 \rangle + \vert 1 \rangle)$, so that the density matrix should be close to $\rho = \vert + \rangle_{x} \langle + \vert_{x}$. As discussed in section~\ref{subsec:quantum_tomography_definition}, for performing quantum tomography in the single-qubit case, it is sufficient to collect measurements in the $x$, $y$, and $z$ basis. In the limit of a large number of measurements, we expect to see the following frequencies in the $z$, $y$, and $x$ basis (all vector expressions are given in the computational basis):  
\begin{equation*}
\begin{bmatrix}
1\\
0
\end{bmatrix} \rightarrow \frac{1}{2}\,, \quad \begin{bmatrix}
0\\
1
\end{bmatrix} \rightarrow \frac{1}{2}\,, \quad \quad \quad \quad \frac{1}{\sqrt{2}}\begin{bmatrix}
1\\
i
\end{bmatrix} \rightarrow \frac{1}{2}\,, \quad \frac{1}{\sqrt{2}}\begin{bmatrix}
1\\
-i
\end{bmatrix} \rightarrow \frac{1}{2}\,, \quad \quad \quad \frac{1}{\sqrt{2}}\begin{bmatrix}
1\\
1
\end{bmatrix} \rightarrow 1\,, \quad \frac{1}{\sqrt{2}}\begin{bmatrix}
1\\
-1
\end{bmatrix} \rightarrow 0\,.
\end{equation*}

We learn the estimated density matrix $\widehat{\rho}$ from measurements in each basis using the maximum likelihood method, and look at the decomposition:
\begin{equation*}
\widehat{\rho} = \lambda_1 \vert \psi_1 \rangle \langle \psi_1 \vert + \lambda_2 \vert \psi_2 \rangle \langle \psi_2 \vert,
\end{equation*}
which would allow us to see what eigenstates contribute to the density matrix, and what is their weight. Indeed, in the case of ideal observations we should get $\lambda_1 = 1$, with $\vert \psi_1 \rangle = \begin{bmatrix}
\frac{1}{\sqrt{2}} & \frac{1}{\sqrt{2}}
\end{bmatrix}^{T}$, and $\lambda_2 = 0$ with $\vert \psi_2 \rangle = \begin{bmatrix}
\frac{1}{\sqrt{2}} &
-\frac{1}{\sqrt{2}}
\end{bmatrix}^{T}$, corresponding to the original pure state associated with $\vert + \rangle_{x}$.

Instead, we obtain the following results for the eigenvalues and associated eigenvectors after $8152$ measurements (the maximum number in one run on IBM QX) in each basis $(z,y,x)$:
\begin{equation*}
\lambda_1 = 0.968 \rightarrow \begin{bmatrix}
0.715 - 0.012 i\\
0.699
\end{bmatrix}, \quad \quad
\lambda_2 = 0.032 \rightarrow \begin{bmatrix}
0.699 - 0.012 i\\
-0.715
\end{bmatrix},
\end{equation*}
i.e.,~in 96\% of cases we observe the state close to $\vert + \rangle_{x}$, and the rest corresponds to the state which is close to $\vert - \rangle_{x}$. Note that the quantum simulator indicates that this amount of measurements is sufficient to estimate matrix elements of the density matrix with an error below $10^{-3}$ in the ideal noiseless case. In order to check the effect of decoherence, we apply  a number of identity matrices (Fig.~\ref{Lokhov:fgr2}, Right) which forces an additional waiting on the system, and hence promotes decoherence of the state. When applying 18 identity matrices, we obtain the following decomposition for $\widehat{\rho}$
\begin{equation*}
\lambda_1 = 0.940 \rightarrow \begin{bmatrix}
0.727 - 0.032 i\\
0.686
\end{bmatrix}, \quad \quad
\lambda_2 = 0.060 \rightarrow \begin{bmatrix}
0.685 - 0.030 i\\
-0.728
\end{bmatrix},
\end{equation*}
while application of 36 identity matrices results in
\begin{equation*}
\lambda_1 = 0.927 \rightarrow \begin{bmatrix}
0.745 - 0.051 i\\
0.664
\end{bmatrix}, \quad \quad
\lambda_2 = 0.073 \rightarrow \begin{bmatrix}
0.663 - 0.045 i\\
-0.747
\end{bmatrix}.
\end{equation*}
The effect of decoherence is visible in the degradation of the eigenstates, as well as in a more frequent occurrence of the eigenstate close to $\vert - \rangle_{x}$.

\subsubsection{Maximally entangled state for two qubits}

\begin{figure}
\begin{center}
\includegraphics[width=0.35\columnwidth, angle=0]{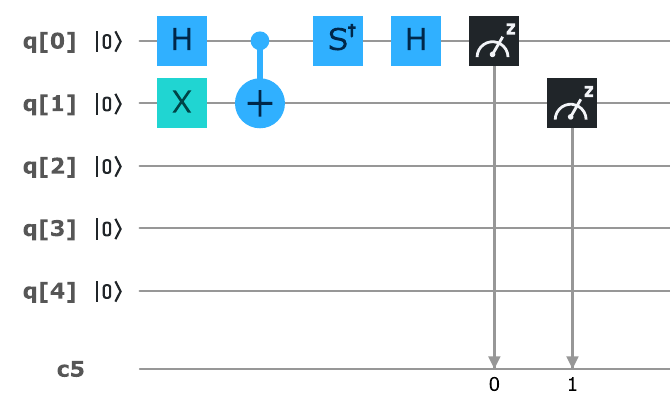} \hfill
\vrule \hfill
\includegraphics[width=0.52\columnwidth, angle=0]{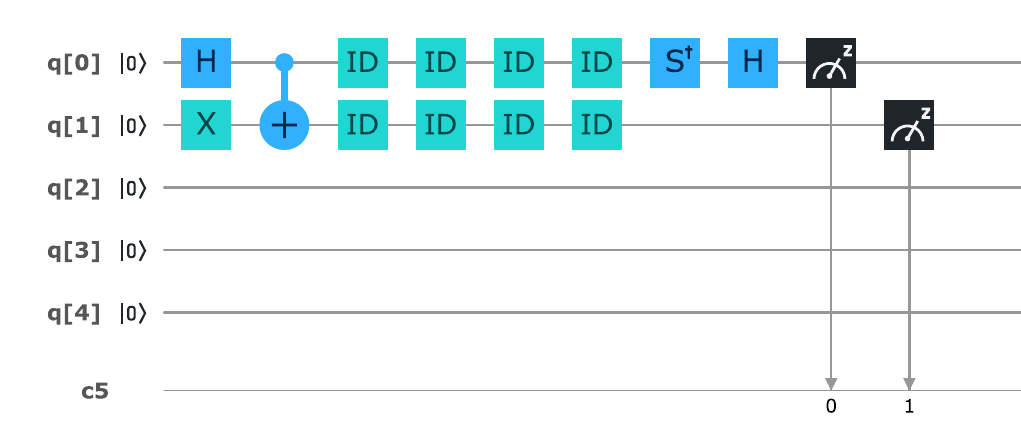}
\caption{Left: example of a measurement of the two-qubit maximally entangled state created with the combination of $H$, $X$ and $CNOT$ gates in the $yz$ basis. Right: experimental setup for testing the effects of decoherence.}
\label{Lokhov:fgr3}
\end{center}
\vspace{-0.3cm}
\end{figure}

Let us now study the two-qubits maximally entangled state, which is an important part of all quantum algorithms achieving quantum speed-up over their classical counterparts. The state $\frac{1}{\sqrt{2}}(\vert 1 0 \rangle + \vert 0 1 \rangle)$ we are interested in is produced by the combination of $H$, $X$ and $CNOT$ gates as shown in Fig.~\ref{Lokhov:fgr3}, Left. We follow the same procedure as in the case of the Hadamard gate, described above, and first estimate the density matrix $\widehat{\rho}$ using $8152$ measurements for each of the $zz$, $yy$, $xx$, $zx$ and $yz$ basis, and then decompose it as   
$\widehat{\rho} = \sum_{i=1}^4 \lambda_i \vert \psi_i \rangle \langle \psi_i \vert$. Once again, ideally we should get $\lambda_1 = 1$ associated with $\vert \psi_1 \rangle = \begin{bmatrix}
0 &
\frac{1}{\sqrt{2}} &
\frac{1}{\sqrt{2}} &
0
\end{bmatrix}^{T}$. Instead, the analysis of the leading eigenvalues indicates that the eigenstate which is close (although significantly distorted) to the theoretical ``ground truth'' $\vert \psi_1 \rangle$ above occurs in the mixture only with probability $0.87$:
\begin{equation*}
\lambda_1 = 0.871 \rightarrow \begin{bmatrix}
-0.025 - 0.024 i\\
0.677\\
0.735\\
-0.029 - 0.017 i
\end{bmatrix}, \quad \quad
\lambda_2 = 0.059 \rightarrow \begin{bmatrix}
0.598\\
0.123 + 0.468 i\\
-0.075  - 0.445 i\\
0.454 - 0.022 i
\end{bmatrix}.
\end{equation*}\\
Our test of decoherence implemented using 18 identity matrices (see Figure~\ref{Lokhov:fgr3}, Right) shows that the probability of the ``original'' entangled state decreases to $0.79$:
\begin{equation*}
\lambda_1 = 0.793 \rightarrow \begin{bmatrix}
-0.025 - 0.012 i\\
0.664\\
0.747\\
-0.017 - 0.008 i
\end{bmatrix}, \quad \quad
\lambda_2 = 0.111 \rightarrow \begin{bmatrix}
0.997\\
-0.002 - 0.058 i\\
0.035 + 0.036 i\\
0.006 + 0.007 i
\end{bmatrix}.
\end{equation*}
Interestingly enough, the second most probable eigenstate changes to the one that is close to $\ket{00}$. This might serve as an indication of the presence of biases in the machine.

The application of the quantum tomography state reconstruction to simple states in the IBM QX revealed an important level of noise and decoherence present in the machine. It would be interesting to check if the states can be protected by using the error correction schemes, which is the subject of the next section.
%%%%%%%%%%%%%%%%%%%%%%%%%%%%%%%%%%%%%%%%%%%%%%%%%%%%%%%

\section{Tests of Quantum Error Correction in IBM Q}
In this section, we  study whether quantum error correction (QEC) can improve computation accuracy in \texttt{ibmqx4}. 
The practical answer to this question seems to be ``No''. Although some error correction effects are observed in \texttt{ibmqx4}, improvements are not exponential and get completely spoiled by errors induced by extra gates and qubits needed for the error correction protocols.

\subsection{Problem definition and background}

As we have seen throughout this review, the quality of computation on actual quantum processors is degraded by errors in the system. This is because currently available chips are not \emph{fault tolerant}. It is widely believed that once the inherent error rates of a quantum processor is sufficiently lowered, fault tolerant quantum computation will be possible using quantum error correction (QEC). The current error rates of the IBM Q machines are not small enough to allow fault tolerant computation. We refer the reader to a survey and introduction on QEC \cite{devitt2013quantum}, while at the same time offering an alternative point of view that we support with a few experiments on the IBM chip. Detailed studies of QEC using more sophisticated error correcting schemes  have been performed on IBM  hardware \cite{qec_takita2016demonstration, qec_wootton2017demonstrating, qec_wootton2018repetition}.  Qiskit also provides some tools that can be used for QEC. The recent work in Ref. \cite{qec_wootton2020benchmarking} introduces some of these capabilities of Qiskit with example code.

The central idea of QEC is to use entanglement to encode quantum superposition in a  manner which is robust to errors. The exact encoding depends upon the kind of errors we want to protect against. In this section we will look at a simple encoding that will protect against bit flip errors. Here  we encode a single qubit state,
\be
|\psi \ra = C_0|0\ra+ C_1 |1\ra , 
\label{sup1}
\ee
using an entangled state, such as 
\be
|\psi \ra =C_0 |0\ra^{\otimes n_q} + C_1 |1\ra^{\otimes n_q},
\label{GHZ}
\ee
where $n_q$ is the number of qubits representing a single qubit in calculations. 

The assumption is that small probability errors will likely lead to unwanted flips of only one qubit (in case when $n_q>3$ this number can be bigger but we will not consider more complex situations here). Such errors produce states that are essentially different from those described by Equation (\ref{GHZ}). Measurements can then be used to fix a single qubit error using, for instance, a majority voting strategy.  More complex errors are assumed to be exponentially suppressed, which can be justified if qubits experience independent decoherence sources.

We question whether QEC can work to protect quantum computations that require many quantum gate operations for the following reason.
The main source of errors then is not spontaneous qubit decoherence but rather the finite fidelity of quantum gates.  When quantum gates are applied to strongly entangled states, such as (\ref{GHZ}), they lead to \emph{highly correlated} dynamics of \emph{all} entangled qubits. We point out that errors introduced by such gates have essentially different nature from random uncorrelated qubit flips. Hence, gate-induced errors may not be treatable by standard error correction strategies when transitions are made between arbitrary unknown quantum state.

To explore this point, imagine that we apply a gate that  rotates a qubit by an angle $\pi/2$. It switches superposition states $|\psi_{\pm}\ra =(|0\ra\pm |1\ra)/\sqrt{2}$ into, respectively, $|0\ra$ or $|1\ra$ in the measurement basis. 
Let the initial state be $|\psi_+ \ra$ but we do not know this before the final measurement. Initially, we know only that initial state can be either $|\psi_+ \ra$ or $|\psi_- \ra$.  To find what it is, we rotate qubit to the measurement basis.
The gate is not perfect, so the final state after the gate application is
\be
|u\ra = \cos(\delta \phi) |0\ra + \sin(\delta \phi) |1\ra,
\label{qub}
\ee 
with some error angle $\delta \phi \ll 1$.  Measurement of this state would produce the wrong answer 1 with probability 
\be
P\approx (\delta \phi)^2.
\label{err2}
\ee
The value $1-P$  is called the fidelity of the gate. In IBM chip it is declared to be 0.99 at the time of writing, which is not much. It means that after about 30 gates we should loose control.  Error correction strategies can increase the number of allowed gates by an order of magnitude even at such a fidelity if we encode one qubit in three.

In order to reduce this error, we can attempt to work with the 3-qubit version of the states in Eq.~\eqref{GHZ}. For example, let us consider the desired gate that transfers states
\be
|\pm\ra = (|000\ra \pm |111\ra)/\sqrt{2}, 
\label{GHZ1}
\ee
into states $|000\ra$ and $|111\ra$ in the measurement basis, respectively. This gate is protected in the sense that a single  unwanted random qubit flip leads to final states that are easily corrected by majority voting. 

However, this is not enough because now we have to apply the gate that makes a rotation by $\pi/2$ in the basis \eqref{GHZ1}. The error in this rotation angle leads to the final state
\be
|u\ra = \cos(\delta \phi) |000\ra + \sin(\delta \phi) |111\ra,
\label{qub2}
\ee 
i.e., this particular error cannot be treated with majority voting using our scheme because it flips all three qubits. On the other hand, this is precisely the type of errors that is most important when we have to apply many quantum gates because basic gate errors are mismatches between desired and received qubit rotation angles irrespectively of how the qubits are encoded. With nine qubits, we could protect the sign in Eq.~\ref{qub2} but this was beyond our hardware capabilities. 

Based on these thoughts, traditional QEC may not succeed in achieving exponential suppression of errors related to non-perfect quantum gate fidelity. The latter is the main source of decoherence in quantum computing that involves many quantum gates. As error correction is often called the only and first application that matters before quantum computing becomes viable at large scale, this problem  must be studied seriously and expeditiously. In the following subsection we report on our experimental studies of this problem with IBM's 5-qubit chip.

\begin{figure}
\begin{center}
\includegraphics[width=4.2in]{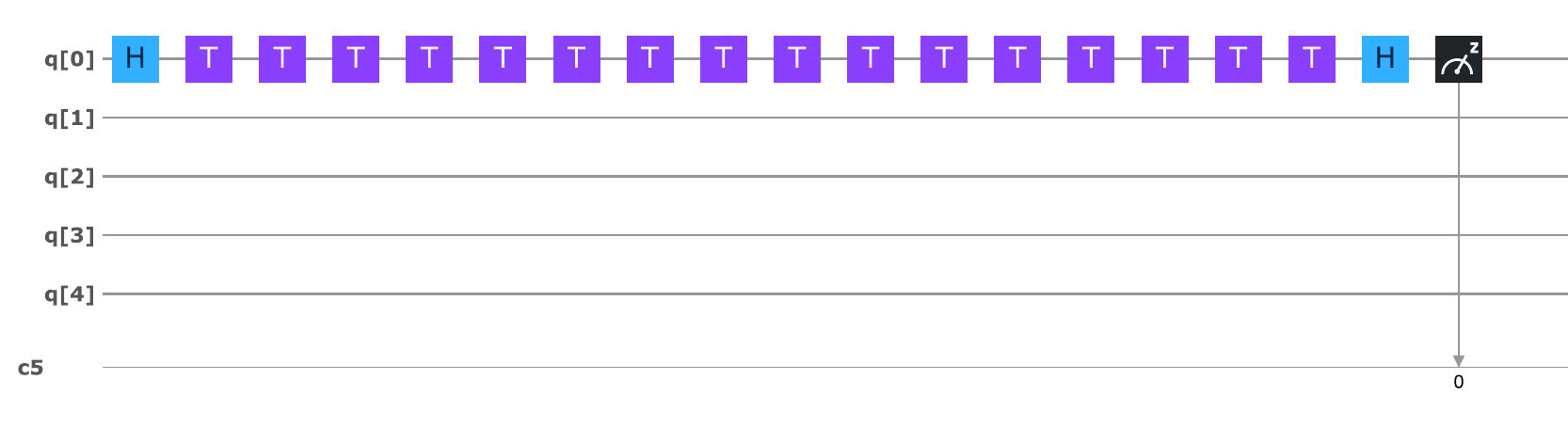}
\caption{Quantum circuit that creates the state  $|+\ra= (|0\ra+ |1\ra)/\sqrt{2}$ then applies 16 T-gates that are equivalent to the identity operation, and then applies the gate that transforms the entangled state into the trivial state $|0\ra$. }
\label{q1-fig}
\end{center}
\end{figure}

\subsection{Test 1: errors in single qubit control}
First, let us perform trivial operation shown in Fig.~\ref{q1-fig}: we create a superposition of two qubit states 
\be
|+\ra= (|0\ra+ |1\ra)/\sqrt{2},
\label{sup2}
\ee
then apply many gates that altogether do nothing, i.e., they just bring the qubit back to the superposition state (\ref{sup2}). We need those gates just  to accumulate some error while the qubit's state is not trivial in the measurement basis. Finally, we apply the gate that transforms its state back to $|0\ra$.

Repeated experiments with measurements then produced wrong answer 13 times from 1000 samples. Thus, we estimate the error of the whole protocol, which did  not use QEC, as 
$$
P_1 = 0.013,
$$
or 1.6\%. This is consistent and even better than declared 1\% single gate fidelity because we applied totally 18 gates.

\subsection{Test 2: errors in entangled 3 qubits control}

\begin{figure}
\begin{center}
\includegraphics[width=4.2in]{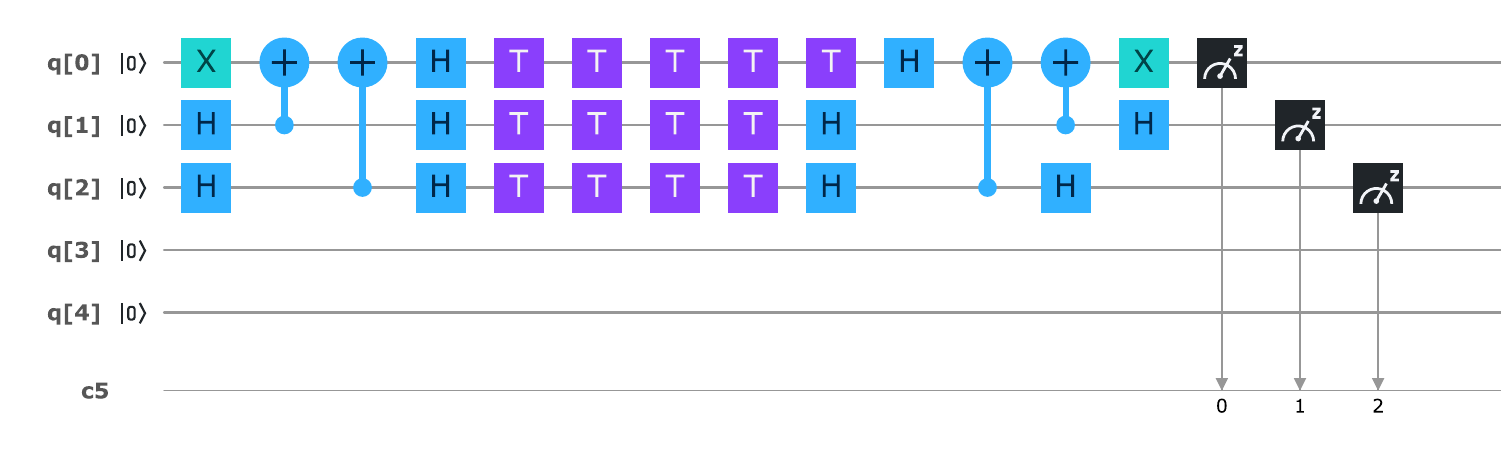}
\caption{Quantum circuit that creates state  $|-\ra= (|000\ra- |111\ra)/\sqrt{2}$ then applies 16 T-gates that are equivalent to the identity operation, and then applies the gate that transforms the entangled GHZ state back into the trivial state $|000\ra$. 
Measurements that return 1 for only one of the three qubits are interpreted as the $|000\ra$ state at the end, while outcomes with two or three units are interpreted as the final state $|111\ra$.
}
\label{q3-fig}
\end{center}
\end{figure}

Next, we consider the circuit  in Fig.~\ref{q3-fig} that initially creates the GHZ state $|-\ra= (|000\ra- |111\ra)/\sqrt{2}$, then applies the same number, i.e. 16, of $T$-gates that lead to the same GHZ state. Then we apply the sub-circuit that brings the whole state back to $|000\ra$. 

Our goal is to quantify the precision of identifying the final result with the state $|000\ra$. 
If a single error bit flip happens, we can interpret results $|100\ra$, $|010\ra$ and $|001\ra$ as $|000\ra$  using majority voting. If needed, we can then apply a proper pulse to correct for this. So,
 in such cases we can consider the error treatable. 
  If the total sum of probabilities of the final state $|000\ra$ and final states with a single bit flipped  is larger than $P_1$ from the previous single-qubit test, then we say that 
the quantum error correction works, otherwise, it doesn't. 
Our experiments showed that probabilities of  events that lead to wrong final interpretation are as follows:
$$
P_{110}=0.006, \quad P_{101}=0.02, \quad P_{011}=0.016, \quad P_{111}=0.005.
$$
Thus, the probability to get the wrong interpretation of the result as the final state $|111 \ra$ of the encoded qubit is 
$$
P_3=P_{110}+P_{101}+P_{011}+P_{111}=0.047,
$$
while the probability to get any error  $1-P_{000}=0.16$. 

\subsection{Discussion}
Comparing results of the tests without and with QEC, we find that the implementation of a simple version of QEC does not improve the probability to interpret the final outcome correctly. The error probability of calculations without QEC gives a smaller probability of wrong interpretation, $P_1=1.3\%$, while the circuit with QEC gives an error probability $P_3=4.7\%$, even though we used majority voting that was supposed to suppress errors by about an order of magnitude.

More importantly, errors that lead to more than one qubit flip are not exponentially suppressed. For example, the probability $P_{101}=0.02$ is close to the probability  of a single bit flip event $P_{010}=0.029$.
We interpret this to mean that errors are not the results of purely random bit flip decoherence effects but rather follow from correlated errors induced by the finite precision of quantum gates. 
The higher error  rate in 3-qubit case could be attributed to the much worse fidelity of the controlled-NOT gate. The circuit itself produces the absolutely correct result $|000\ra$ in 84\% of simulations. If the remaining errors were produced by uncorrelated bit flips, we
would see outcomes with more than one wrong bit flip with total probability less than $1\%$ but we found that such events have a much larger total probability $P_3=4.7\%$.

In defense of QEC, we note that probabilities of single bit flip errors were still several times larger than probabilities of multiple (two or three) wrong qubit flip errors. This means that at least partly QEC works, i.e., it corrects the state to $|000\ra$ with 4.7\% precision, versus the initially 16\% in the wrong state. 
So, at least some part of the errors can be treated. However, an efficient error correction must show \emph{exponential} suppression of errors, which was not observed in this test.  

Summarizing,  this brief test shows no improvements that would be required for efficient quantum error correction. The need to use more quantum gates and qubits to correct errors only leads to a larger probability of wrong interpretation of the final state. 
This problem will likely become increasingly much more important because without quantum error correction the whole idea of conventional quantum computing is not practically useful.  Fortunately, IBM's quantum chips can be used for experimental studies of this problem. 
We also would like to note that quantum computers can provide computational advantages beyond standard quantum algorithms  and using only classical error correction~\cite{sinitsyn-q}. So, they must be developed even if problems with quantum error correction prove detrimental for conventional quantum computing schemes at achievable hardware quality. 

\begin{acks}
We would like to acknowledge the help from numerous readers who pointed out errors and misprints in the earlier version of the manuscript. The code and implementations accompanying the paper can be found at \url{https://github.com/lanl/quantum_algorithms}.
\end{acks}

\bibliographystyle{plain}

\bibliography{rrqp}
\end{document}